\documentclass[a4paper,12pt,twoside,openright,BCOR5mm,bibliography=totoc]{scrreprt}

\usepackage[ngerman,english]{babel}
\usepackage{fancyhdr}
\usepackage{graphicx}
\usepackage{epstopdf}
\usepackage{import}
\usepackage{amsmath}
\usepackage{amstext}
\usepackage{amsfonts}
\usepackage{amssymb}
\usepackage[bookmarks=false]{hyperref}
\usepackage{slashed}
\usepackage{cite}
\usepackage{color}
\usepackage{tikz}
\usetikzlibrary{calc}
\usepackage{overpic}
\usepackage{subfigure}
\usepackage{multirow}
\usepackage{enumerate}

\hypersetup {
pdftitle = {Higgs Effective Field Theories - Systematics and Applications},
pdfauthor = {Claudius Krause}
}

\renewcommand{\chaptermark}[1]%
         {\markboth{\thechapter.\ #1}{}}
\renewcommand{\sectionmark}[1]%
         {\markright{\thesection\ #1}}
\newcommand{\LMUTitle}[9]{
  \thispagestyle{empty}
  \vspace*{\stretch{1}}
  {\parindent0cm
   \rule{\linewidth}{.7ex}}
  \begin{flushright}

    \vspace*{\stretch{1}}
    \sffamily\bfseries\Huge\centering 
    #1\\
    \vspace*{\stretch{1}}
    \sffamily\bfseries\large
    #2
    \vspace*{\stretch{1}}
  \end{flushright}
  \rule{\linewidth}{.7ex}
  \vspace*{\stretch{5}}
  \begin{center}
    \includegraphics[width=3in]{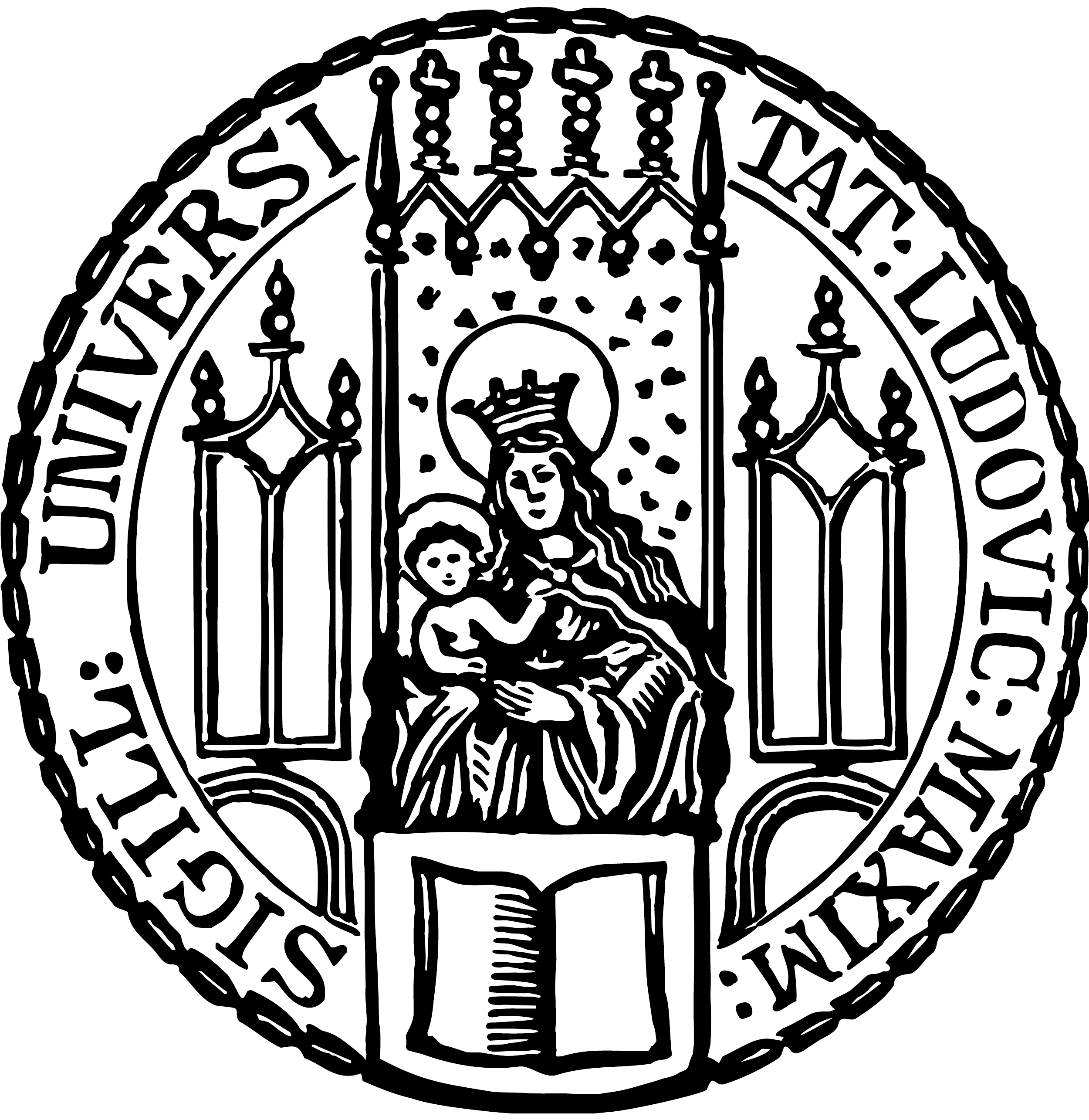} 
  \end{center}
  \vspace*{\stretch{1}}
  \begin{center}\sffamily\LARGE{#5}\end{center}
  \newpage
  \thispagestyle{empty}

  \cleardoublepage
  \thispagestyle{empty}

  \vspace*{\stretch{1}}
  {\parindent0cm
  \rule{\linewidth}{.7ex}}
  \begin{flushright}
    \vspace*{\stretch{1}}
    \sffamily\bfseries\Huge\centering 
    #1\\
    \vspace*{\stretch{1}}
    \sffamily\bfseries\large
    #2
    \vspace*{\stretch{1}}
  \end{flushright}
  \rule{\linewidth}{.7ex}

  \vspace*{\stretch{3}}
  \begin{center}
    \Large Dissertation\\
    \Large an der #4\\
    \Large der Ludwig--Maximilians--Universit\"at\\
    \Large M\"unchen\\
    \vspace*{\stretch{1}}
    \Large vorgelegt von\\
    \Large #2\\
    \Large aus #3\\
    \vspace*{\stretch{2}}
    \Large M\"unchen, den #6
  \end{center}

  \newpage
  \thispagestyle{empty}

  \vspace*{\stretch{1}}

  \begin{flushleft}
    \large Erstgutachter:  #7 \\[1mm]
    \large Zweitgutachter: #8 \\[1mm]
    \large Tag der m\"undlichen Pr\"ufung: #9\\
  \end{flushleft}

  \cleardoublepage
}

\makeatletter
\renewcommand{\@chapapp}{}
\newenvironment{chapquote}[2][2em]
  {\setlength{\@tempdima}{#1}
   \def\chapquote@author{#2}
   \parshape 1 \@tempdima \dimexpr\textwidth-2\@tempdima\relax
   \itshape}
  {\par\normalfont\hfill--\ \chapquote@author\hspace*{\@tempdima}\par\bigskip}

\makeatother

\newcommand{\prob}[2]{\text{prob}( #1 | #2 )}
\newcommand{\cov}[2]{\text{cov}( #1 , #2 )}

\makeatletter
\renewcommand*\env@matrix[1][*\c@MaxMatrixCols c]{%
  \hskip -\arraycolsep
  \let\@ifnextchar\new@ifnextchar
  \array{#1}}
\makeatother

\begin{document}

\pagenumbering{roman}

  \LMUTitle
  {Higgs Effective Field Theories \\
    --- Systematics and Applications ---}           
  {Claudius G. Krause}                                           
  {Zittau}                                                               
  {Fakult\"at f\"ur Physik}                                     
  {M\"unchen, 2016}                                              
  {28. Juli 2016}                                                    
  {Prof. Dr. G. Buchalla}                                         
  {Prof. Dr. A. Ibarra}                                             
  {15. September 2016}                                         

  \tableofcontents{\protect\thispagestyle{fancyplain}}

  \listoffigures{\protect\thispagestyle{fancyplain}}

  \listoftables{\protect\thispagestyle{fancyplain}}
  
  \cleardoublepage

  \addcontentsline{toc}{chapter}{\protect Zusammenfassung}
\chapter*{Zusammenfassung}
\thispagestyle{fancyplain}
 \markboth{Zusammenfassung}{Zusammenfassung}
\selectlanguage{ngerman}
Am 4. Juli 2012 wurde am gro{\ss}en Hadronenbeschleuniger LHC am europ{\"a}ischen Kernforschungszentrum CERN bei Genf die Entdeckung eines neuen Teilchens bekannt gegeben. Die Eigenschaften des Teilchens stimmen, im Rahmen der noch relativ gro{\ss}en experimentellen Unsicherheiten, mit denen des lang gesuchten Higgsbosons {\"u}berein. Teilchenphysiker in aller Welt stellen sich nun die Frage: {\it \glqq Ist es das Standardmodell Higgs-Teilchen, das wir beobachten; oder ist es ein anderes Teilchen mit {\"a}hnlichen Eigenschaften?\grqq }

Effektive Feldtheorien (EFTs) erm{\"o}glichen eine allgemeine, modellunabh{\"a}ngige Be\-schrei\-bung des Teilchens. Dabei benutzen wir wenige minimale Annahmen --- nur Standardmodell Teilchen als Freiheitsgrade und eine Skalenseparation zur neuen Physik --- welche durch aktuelle experimentelle Ergebnisse gest{\"u}tzt werden. Per Konstruktion beschreiben effektive Theorien daher ein physikalisches System nur bei einer bestimmten Energieskala, in unserem Fall der elektroschwachen Skala $v$. Effekte von neuer Physik bei h{\"o}heren Energien, $\Lambda$, werden durch modifizierte Wechselwirkungen der leichten Teilchen parametrisiert. \\

In dieser Dissertation, \glqq Effektive Feldtheorien f{\"u}r das Higgs --- Systematik und Anwendung\grqq, diskutieren wir effektive Feldtheorien f{\"u}r das Higgs Teilchen, welches nicht notwendigerweise das Higgs-Teilchen des Standardmodells ist. Besonderes Augenmerk richten wir auf eine systematische und konsistente Entwicklung der EFT. Diese Systematik ist abh{\"a}ngig von der Dynamik der neuen Physik. Wir unterscheiden zwei verschiedene konsistente Entwicklungen. Zum einen effektive Theorien von Modellen neuer Physik, die bei niedrigen Energien entkoppeln und zum anderen effektive Beschreibungen von nicht entkoppelnden Modellen. Wir diskutieren den ersten Fall, die Standardmodell EFT, kurz, da der Fokus dieser Arbeit auf nicht entkoppelnden effektiven Theorien liegt. Wir erl{\"a}utern, dass die konsistente Entwicklung im zweiten Fall in Quantenschleifen erfolgen muss und f{\"u}hren das dazu {\"a}quivalente Konzept der chiralen Dimensionen ein. Mithilfe der chiralen Dimensionen entwickeln wir die elektroschwache chirale Lagrangedichte bis einschlie{\ss}lich n{\"a}chstf{\"u}hrender Ordnung, $\mathcal{O}(f^{2}/\Lambda^{2})=\mathcal{O}(1/16\pi^{2})$. Wir diskutieren auch den Einfluss verschiedener Annahmen {\"u}ber die sch{\"u}tzende (custodial) Symmetrie im Higgssektor auf die Liste der Operatoren. Wir beenden die Diskussion {\"u}ber die Systematik mit einem Vergleich der entkoppelnden und nicht ent\-kop\-peln\-den EFT. Wir betrachten dabei auch den Fall, dass die neue Physik einen nicht entkoppelnden Sektor bei einer Energieskala $f$ besitzt, welcher deutlich {\"u}ber der elektroschwachen Skala $v$ liegt. Wir diskutieren die Relevanz der daraus resultierenden Doppelentwicklung in $\xi=v^{2}/f^{2}$ und $f^{2}/\Lambda^{2}$ f{\"u}r die Datenanalyse am LHC.\\

Im zweiten Teil dieser Dissertation diskutieren wir Anwendungen der effektiven Theorien, insbesondere der elektroschwachen chiralen Lagrangedichte. Als Erstes verbinden wir die EFT mit expliziten Modellen f{\"u}r neue Physik. Dies illustriert, wie die Vorhersagen des Entwicklungsschemas in einem konkreten Fall realisiert werden. Wir zeigen auch an einem Beispiel, wie verschiedene Parameterbereiche derselben Theorie sowohl eine entkoppelnde als auch eine nicht entkoppelnde EFT generieren. 

Als Zweites nutzen wir die effektive Entwicklung in f{\"u}hrender Ordnung um die aktuellen Higgsdaten des LHCs zu beschreiben. Wir zeigen, dass die aktuelle Parametrisierung der Higgsdaten, welche von den Experimentatoren am CERN verwendet wird (der $\kappa$-Formalismus), sich durch diese Entwicklung quantenfeld-theoretisch begr{\"u}nden l{\"a}sst. Das Ergebnis eines Fits zeigt daher nicht nur, ob das beobachtete Teilchen das Standardmodell Higgs-Teilchen ist, sondern auch, sofern sich Abweichungen manifestieren, welche Art von neuer Physik bevorzugt wird. In unserem konkreten Fall nutzen wir die Daten von 2010--2013. Die effektive Lagrangedichte, die diese Daten beschreibt, l{\"a}sst sich auf sechs freie Parameter reduzieren. Das Ergebnis ist konsistent mit dem Standardmodell, weist aber noch statistische Unsicherheiten von etwa $10\%$ auf. 

\selectlanguage{english}

\cleardoublepage

\addcontentsline{toc}{chapter}{\protect Abstract}
\chapter*{Abstract}
\thispagestyle{fancyplain}
 \markboth{Abstract}{Abstract}
Researchers of the Large Hadron Collider (LHC) at the European Organization for Nuclear Research (CERN) announced on July 4th, 2012, the observation of a new particle. The properties of the particle agree, within the relatively large experimental uncertainties, with the properties of the long-sought Higgs boson. Particle physicists around the globe are now wondering, {\it ``Is it the Standard Model Higgs that we observe; or is it another particle with similar properties?''}

We employ effective field theories (EFTs) for a general, model-independent description of the particle. We use a few, minimal assumptions --- Standard Model (SM) particle content and a separation of scales to the new physics --- which are supported by current experimental results. By construction, effective field theories describe a physical system only at a certain energy scale, in our case at the electroweak-scale $v$. Effects of new physics from a higher energy-scale, $\Lambda$, are described by modified interactions of the light particles.\\

In this thesis, ``Higgs Effective Field Theories --- Systematics and Applications'', we discuss effective field theories for the Higgs particle, which is not necessarily the Higgs of the Standard Model. In particular, we focus on a systematic and consistent expansion of the EFT. The systematics depends on the dynamics of the new physics. We distinguish two different consistent expansions. EFTs that describe decoupling new-physics effects and EFTs that describe non-decoupling new-physics effects. We briefly discuss the first case, the SM-EFT. The focus of this thesis, however, is on the non-decoupling EFTs. We argue that the loop expansion is the consistent expansion in the second case. We introduce the concept of chiral dimensions, equivalent to the loop expansion. Using the chiral dimensions, we expand the electroweak chiral Lagrangian up to next-to-leading order, $\mathcal{O}(f^{2}/\Lambda^{2})=\mathcal{O}(1/16\pi^{2})$. Further, we discuss how different assumptions on the custodial symmetry in the Higgs sector influences the list of operators in the basis. Finally, we compare the decoupling and the non-decoupling EFT. We also consider scenarios in which the new-physics sector is non-decoupling at a scale $f$, far above the electroweak-scale $v$. We discuss the relevance of the resulting double expansion in $\xi=v^{2}/f^{2}$ and $f^{2}/\Lambda^{2}$ for the data analysis at the LHC.\\

In the second part of this thesis, we discuss the applications of the EFTs, especially of the electroweak chiral Lagrangian. First, we connect the EFT with explicit models of new physics. This illustrates how the power counting works in a specific example. We show how different regions of the parameter space of the same model generate a decoupling and a non-decoupling EFT. 

Second, we use the expansion at leading order to describe the current LHC Higgs data. We show how the current parametrization of the Higgs data, which is used by the experimentalists at CERN (the $\kappa$-framework), can be justified quantum field theoretically by the EFT. The result of a fit does therefore not only indicate whether we observe the SM-Higgs, but also, in case there are deviations, what kind of new physics is preferred. In this thesis, we fit the data of Run-1 (2010--2013). The effective Lagrangian describing this data can be reduced to six free parameters. The result of this fit is consistent with the SM. It has, however, statistical uncertainties of about ten percent.

  \cleardoublepage

  \setcounter{page}{1}
  \pagenumbering{arabic}
  \thispagestyle{fancyplain}
  \chapter{Introduction}
\thispagestyle{fancyplain}
\begin{chapquote}{Albert A. Michelson (1852--1931), 1894\cite{Flam:1992jx,university1896annual, wikiquote}}
``\dots most of the grand underlying principles have been firmly established [\dots] the future truths of physical science are to be looked for in the sixth place of decimals.''
 \end{chapquote}
This statement of Albert A. Michelson from 1894 brings a smile to the faces of present-day physicists. As it was noted some years ago in an article in ``Science'' \cite{Flam:1992jx}, {\it ``the next three decades proved among the richest in the history of physics''}. Planck's law for black body radiation (1900),  Einstein's description of the photoelectric effect (1905), special relativity (1905), general relativity (1915), and the development of quantum mechanics in the 1920s are some examples of the scientific advances in these decades. But the progress did not stop after these developments. In 1932, Anderson discovered the first antiparticle, the positron. Soon, more and more particles and interactions were discovered in accelerator machines. The fundamental constituents\footnote{Fundamental as of 2016} and the interactions between them are described in the Standard Model (SM). Together with General Relativity, the Standard Model can explain almost all experimental data collected so far. Some of the measurements, however, hint at effects beyond the Standard Model --- often called new physics. Over the years, scientists have built more and more powerful particle accelerators, to look for these effects and also to precisely determine the Standard Model parameters. \\

\begin{figure}[t]
    \centering
    \includegraphics[width=\textwidth]{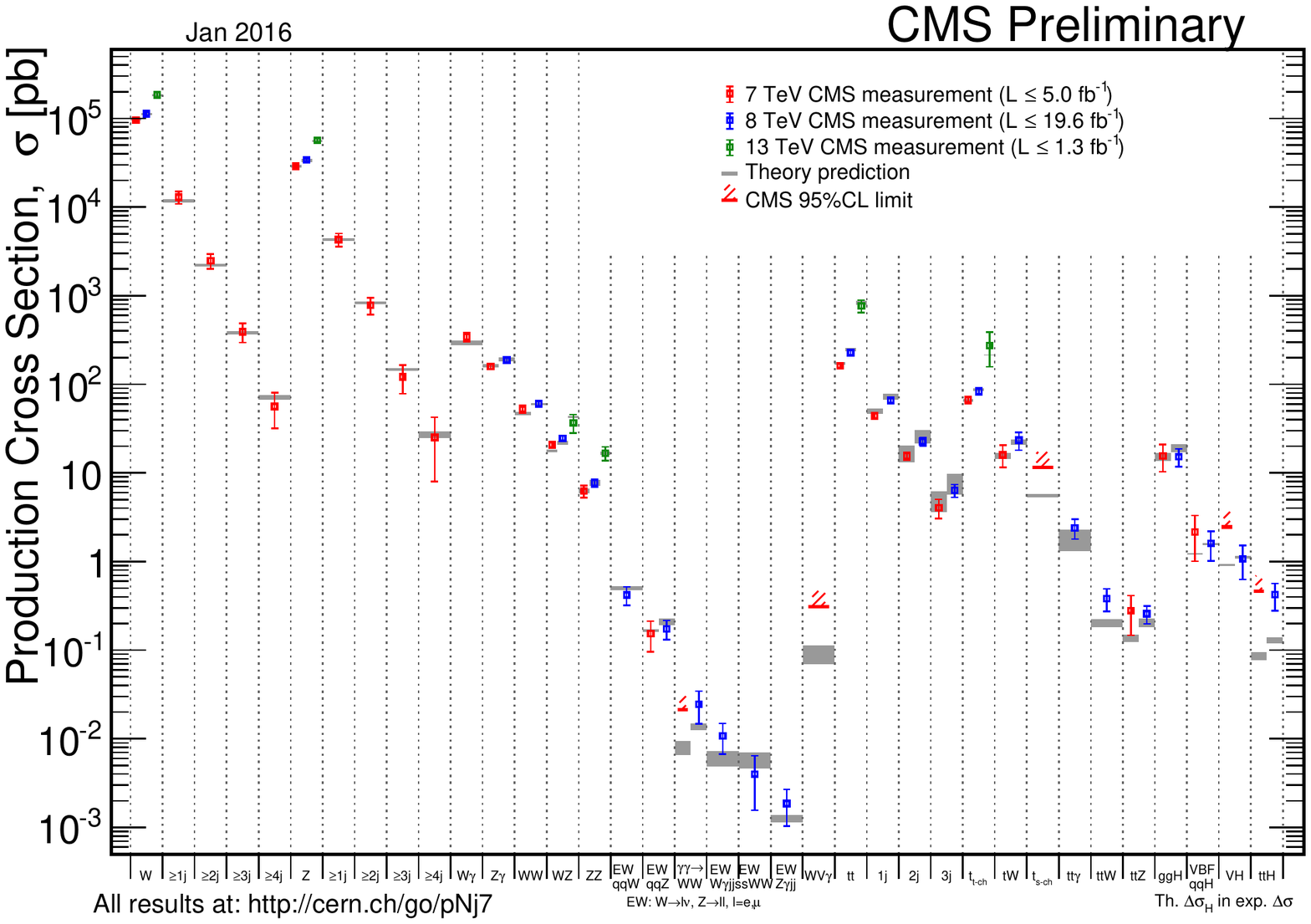}
    \caption[Summary of cross section measurements by the CMS experiment.]{Summary of cross section measurements by the CMS experiment \cite{CMScrosssection}. The agreement between the Standard Model theory prediction and the experimental measurement is remarkable.}
    \label{fig:CMScrosssection}
  \end{figure}

The currently highest center-of-mass energy reached by a laboratory experiment is $13~\text{TeV} (= 2.1\cdot 10^{-6}~\text{J})$. It is reached in proton-proton collisions at the Large Hadron Collider (LHC) at CERN in Switzerland. In Run-1, from 2010 to 2013, the experiments ATLAS and CMS each recorded about $5 ~\text{fb}^{-1}$ of data at $7~\text{TeV}$ and about $ 23~\text{fb}^{-1}$ of data at $8~\text{TeV}$ center-of-mass energy\cite{LHCLumiwebsiteCMS,LHCLumiwebsiteATLAS}. They measured processes and confirmed the predictions of the Standard Model over multiple orders of magnitude. The cross section of proton-proton scattering may serve as an example. Figure \ref{fig:CMScrosssection}\cite{CMScrosssection} shows the experimental results, together with the theory prediction for different final states. The agreement between the experimental measurements and the theory prediction of the Standard Model is remarkable. On top of this confirmation of the Standard Model, the experimental collaborations announced on July 4, 2012 the observation of a scalar particle with couplings compatible with the Standard Model Higgs boson \cite{Chatrchyan:2012xdj,Aad:2012tfa}. Soon, its couplings will be known with better precision and we will know if it is the Standard Model Higgs or only a Higgs-like particle.

Summarizing the experimental results of the LHC, no signs of new physics have been observed so far. This leads us to a conclusion similar to Michelson's: What if there is no new physics to be observed at the LHC? Will we only measure the Standard Model more accurately?  Of course, there are still open questions that the Standard Model cannot answer, but what if these questions are not answered at the LHC?

Given the historical developments that took place after Michelson's comment in 1894, there is no need to be pessimistic today. There may be very exciting decades just ahead of us. The LHC started Run-2 in May 2015 with an increased center-of-mass energy of $13~\text{TeV}$ and more data are being collected as well as analyzed. 

New-physics effects can manifest themselves in two ways in the data. Either new particles are {\it directly} produced and observed at the LHC, or the effects of new physics show up {\it indirectly}, as new interactions or virtual effects of new particles. So far, no direct observation with more than $5\sigma$ significance was made. However, there are some deviations of less than $3 \sigma$ observed \cite{CMS:2015dxe,ATLAS:2015diphoton,Khachatryan:2016hje,Aaboud:2016tru}. 

Motivated by this absence of {\it direct} observations, {\it indirect} searches became increasingly popular. The low-energy impact of high-energy new-physics effects are systematically studied within effective field theories (EFTs). In particular in the bottom-up approach, effective field theories provide a model-independent tool for data analysis.

The Higgs particle received special attention in the context of effective field theories. It is a scalar excitation that is predicted by the Brout-Englert-Higgs mechanism \cite{Englert:1964et,Higgs:1964ia,Higgs:1964pj,Guralnik:1964eu,Higgs:1966ev,Kibble:1967sv}, describing spontaneous breaking of gauge symmetries. The mechanism was proposed in the 1960s as a way to give masses to gauge bosons in a gauge-invariant way by introducing a complex scalar doublet that acquires a vacuum expectation value. This spontaneous breaking of symmetry generates three Goldstone bosons that become the longitudinal degrees of freedom of the $W^{\pm}/Z$ gauge bosons. The remaining fourth degree of freedom is the Higgs particle and it is needed for the unitarity of the theory. The scalar particle that was observed at the LHC \cite{Chatrchyan:2012xdj,Aad:2012tfa} is a good candidate for being the Higgs particle. However, its couplings are only measured up to a precision of the order of ten percent or less, because its discovery was only four years ago. Potentially large new-physics effects can hide in these couplings. Many different ways exist to analyze them. The LHC Higgs Cross Section Working Group (LHCHXSWG) is currently working on a recommendation for analyzing the Higgs couplings, also using effective field theories. 

The LHCHXSWG was founded in 2010 \cite{LHCHXSWGwebsite} in order to produce agreements on Standard Model Higgs observables, like cross sections and branching ratios. It is a joint project of theorists and experimentalists. Later, the group was restructured to discuss also measurements, properties, and beyond-the-Standard-Model scenarios related to the Higgs. Now, in early 2016, the working group is writing the CERN Higgs Yellow Report 4 \cite{YellowReport4,LHCHXSWGwebsite}. Parts of this document are devoted to the use of effective field theories in Higgs analyses. This shows the important role that effective field theories play in present-day high-energy physics.\\

This thesis is based on \cite{mastersthesis,Buchalla:2013rka,Buchalla:2013eza,Buchalla:2014eca,Buchalla:2015wfa,Buchalla:2015qju,Buchalla:2016sop,Buchalla:2016bse} and divided into three parts. In Part I, we review the basic concepts of the Standard Model, especially spontaneous symmetry breaking. We then introduce effective field theories (EFTs) in both the top-down and the bottom-up approach. When applying the concept of bottom-up EFTs to the Standard Model, two different consistent expansions can be formulated, based on different assumptions. 

We explore these two different effective field theories in Part II. We start with the so-called linear or decoupling EFT, sometimes also referred to as Standard Model effective field theory (SM-EFT). The focus of Part II, however, is on the non-linear EFT that is given by the electroweak chiral Lagrangian. In particular, we emphasize its systematics as non-decoupling EFT and its relation to the linear EFT. 

In Part III, we discuss applications of effective field theories. First, we match the effective descriptions to explicit models of high-energy physics. Thereby we explicitly illustrate the relation between the two different expansions. Then, we use the effective Lagrangian to fit data from the LHC. The obtained pattern of coefficients helps us to infer the theory that is underlying the data --- either the Standard Model or some theory beyond the Standard Model. In this context, the statement of Michelson indeed becomes true: Using more and more precise measurements, we will find indirect signs of new physics, and by employing effective field theories, we will be able to infer {\it``the future truths of physical science''}. 
\part{Foundations}
  \chapter[The Standard Model (SM)]{The Standard Model}
\thispagestyle{fancyplain}
\section{The Standard Model Particle Content and Its Symmetries}
\label{ch:SMsym}
The Standard Model (SM) is a consistent model of fundamental particle interactions that was developed in the second half of the $20^{\text{th}}$ century. It builds on the work of many authors, most notably \cite{Glashow1961579,PhysRevLett.10.531,Englert:1964et,Higgs:1964ia,Higgs:1964pj,Guralnik:1964eu,Higgs:1966ev,Kibble:1967sv,PhysRevLett.19.1264,Salam,Veltman1968637,Hooft1971167,PhysRevLett.30.1343,PhysRevLett.30.1346,Fritzsch:1973pi,Fritzsch:1975sr,Marciano1978137,Kobayashi01021973,PhysRevD.8.3633}. The SM is a Quantum Field Theory (QFT) that is able to describe almost all experimental data of particle physics based on only 19 input parameters \cite{Donoghue:1992dd}. We discuss the observations that are not described by the SM in Section~\ref{ch:SM.openQ}. The SM can be summarized in a condensed form in terms of the Lagrangian:
\begin{align}
  \begin{aligned}
    \label{eq:SMunbroken}
  \mathcal{L}_{\text{SM}} &=-\frac{1}{4} B_{\mu\nu}B^{\mu\nu} - \frac{1}{2}\langle W_{\mu\nu}W^{\mu\nu}\rangle - \frac{1}{2}\langle G_{\mu\nu}G^{\mu\nu}\rangle - \theta \frac{g_{s}^{2}}{32\pi^{2}}\langle \widetilde{G}_{\mu\nu}G^{\mu\nu}\rangle\\
  &+i\bar{q}_{L}^{i}\slashed{D}q_{L}^{i} +i\bar{\ell}_{L}^{i}\slashed{D}\ell_{L}^{i} +i\bar{u}_{R}^{i}\slashed{D}u_{R}^{i} +i\bar{d}_{R}^{i}\slashed{D}d_{R}^{i} +i\bar{e}_{R}^{i}\slashed{D}e_{R}^{i}\\
  &+(D_{\mu}\phi^{\dagger})(D^{\mu}\phi) + \frac{1}{2}\mu^{2}\phi^{\dagger}\phi - \frac{\lambda}{4} (\phi^{\dagger}\phi)^{2}\\
  &-\bar{\ell}_{L}^{i}Y_{e}^{ij}\phi e_{R}^{j} - \bar{q}_{L}^{i}Y_{d}^{ij}\phi d_{R}^{j} - \bar{q}_{L}^{i}Y_{u}^{ij} (i\sigma_{2}\phi^{*}) u_{R}^{j} +\text{h.c.}
\end{aligned}
\end{align}
Here, $\langle \dots\rangle$ is the trace. In the rest of this section, we will describe the particles and the symmetry relations they obey in more detail. 

Symmetries are very important in QFTs. For each symmetry, there exists a conserved current $\mathcal{J}^{\mu} $\cite{Noether}, giving rise to a conserved charge $ Q = \int \mathcal{J}^{0} d^{3}x$. Further, invariance under certain symmetries constrains the structure of the particle interactions. 

The fields in $\mathcal{L}_{\text{SM}}$, Eq.~\eqref{eq:SMunbroken}, can be classified in two groups. Particles of integer spin ($s$), bosons, mediate the interactions, with symmetry dictating the interaction's structure. The local $SU(3)_{C}\times SU(2)_{L}\times U(1)_{Y}$ gauge symmetry of the Standard Model induces gauge interactions, mediated by the $s=1$ gauge fields. In Eq.~\eqref{eq:SMunbroken} they are denoted $G$ for the strong $SU(3)_{C}$, $W$ for the left-handed $SU(2)_{L}$, and $B$ for the hypercharge $U(1)_{Y}$ interactions. The scalar ($s=0$) Higgs field $\phi$ participates in the Yukawa interactions together with the fermion fields. These fermions of spin $s=1/2$ form the second group of particles, the matter content of the SM. This group can further be divided into particles that participate in the strong and electroweak interactions (called quarks), and the particles that only interact through the electroweak interactions (called leptons). Each fermion is specified by its representation of the gauge group. Table \ref{tab:SMparticles} summarizes the particle content with its corresponding representation of the gauge group, {\it i.e.} gauge charge. For each representation, three copies of the fermions exist that only differ by their mass. They are called generations. The Standard Model, being a chiral theory, distinguishes particles of different chirality (indicated by the subscripts $L/R$) by different representations of the gauge group in Table \ref{tab:SMparticles}. 
\renewcommand{\arraystretch}{2}
\begin{table}[!ht]
    \centering
    \begin{tabular}[t]{|c|c|c|}
\hline
 \multicolumn{2}{|c|}{Particle} & Representation\\
\hline
\multirow{3}{*}{Quarks} & $q_{L}^{i}\in \left\{\dbinom{u}{d}_{L},\dbinom{c}{s}_{L},\dbinom{t}{b}_{L}\right\}$ & $(3,2,1/6)$\\
\cline{2-3}
& $u_{R}^{i}\in \left\{ u_{R},c_{R},t_{R}\right\}$ & $(3,1,2/3)$\\
\cline{2-3}
& $d_{R}^{i} \in \left\{ d_{R},s_{R},b_{R}\right\}$ & $(3,1,-1/3)$\\
\hline
\multirow{2}{*}{Leptons} & $\ell_{L}^{i}\in \left\{ \dbinom{\nu_{e}}{e}_{L},\dbinom{\nu_{\mu}}{\mu}_{L},\dbinom{\nu_{\tau}}{\tau}_{L} \right\}$ & $(1,2,-1/2)$\\
\cline{2-3}
& $e_{R}^{i}\in \left\{ e_{R},\mu_{R},\tau_{R}\right\}$ & $(1,1,-1)$\\
\hline
\hline
 \multicolumn{2}{|c|}{Higgs $\phi$} & $(1,2,1/2)$\\
\hline
    \end{tabular}
    \caption[Representations of the SM field content.]{Representations of each fermion as well as the Higgs field of the Standard Model, given as $(SU(3)_{C},SU(2)_{L},U(1)_{Y})$. The subscripts $L$ and $R$ indicate the chirality.}
    \label{tab:SMparticles}
  \end{table}
\renewcommand{\arraystretch}{1}

The covariant derivative is constructed in the usual way. For a generic field $\Psi$ we have
\begin{equation}
  \label{eq:2.1.1}
  D_{\mu}\Psi = \partial_{\mu}\Psi+ i g W_{\mu}\Psi + i g' \mathsf{Y}_{\Psi}B_{\mu}\Psi + i g_{s} G_{\mu} \Psi,
\end{equation}
where the gauge fields are contracted with the group generators in the representation of the field $\Psi$. 

The strong interactions of Quantum Chromodynamics (QCD) \cite{Marciano1978137} are governed by the local symmetry called color $SU(3)_{C}$. The quarks and gluons of~\eqref{eq:SMunbroken} are useful degrees of freedom only for energies above $\Lambda_{\text{QCD}}\sim2~\text{GeV}$. Far above $\Lambda_{\text{QCD}}$, they are asymptotically free \cite{PhysRevLett.30.1343,PhysRevD.8.3633,PhysRevLett.30.1346}. At the scale $\Lambda_{\text{QCD}}$, the relevant degrees of freedom change and light mesons, such as pions and kaons, become the propagating degrees of freedom. The fundamental quarks and gluons are confined in the hadrons and only color-neutral states are observed. The electroweak subgroup, $SU(2)_{L}\times U(1)_{Y}$, is at energies at or below the electroweak scale $v$ spontaneously broken to $U(1)_{QED}$ of Quantum Electrodynamics (QED). We discuss spontaneous symmetry breaking and its implications for the SM further in Section \ref{ch:2.2}.\\

In addition to the local symmetries, $\mathcal{L}_{\text{SM}}$ in Eq.~\eqref{eq:SMunbroken} also exhibits global symmetries, some of them being only approximate. For vanishing fermion masses, the three generations would be indistinguishable. This introduces a global $U(3)^{5}$ symmetry, called flavor symmetry \cite{Willenbrock:2004hu}. The Yukawa interactions violate this symmetry, leaving only one $U(1)$ symmetry for the quark sector and one $U(1)$ symmetry for the lepton sector. These symmetries are called Baryon number $B$ and Lepton number $L$, counting the numbers of Baryons and Leptons. If the masses of the neutrinos are vanishing, the lepton number of each family will be conserved separately. A closer inspection reveals that both $B$ and $L$ symmetries are anomalous \cite{Peskin:1995ev}, meaning the symmetry is only conserved at tree-level, but broken by quantum effects. However, the difference $B-L$ is a true global, anomaly-free symmetry of the Standard Model. This is, however, not enforced on the Lagrangian. Rather, it is an accidental symmetry. All terms allowed by gauge and Lorentz symmetry also respect this global symmetry. 

The QCD sector of the Lagrangian, given by Eq.~\eqref{eq:SMunbroken} in the limit $g,g'\rightarrow 0$ and $Y_{\Psi}\rightarrow 0$, is invariant under a global chiral $U(6)_{L}\times U(6)_{R}$ symmetry. This is because for QCD in the limit of vanishing quark masses, there is no distinction between the six quark flavors. The chiral symmetry group is equivalent to $U(1)_{V}\times U(1)_{A}\times SU(6)_{L} \times SU(6)_{R}$, where $V(A)$ refers to the vectorial (axial) combination $L\begin{smallmatrix}\vspace{-2pt}+ \\ (-)\end{smallmatrix}R$. The $U(1)_{V}$ is again the Baryon number $B$. The axial $U(1)_{A}$ is anomalous and thus not a symmetry of the quantum theory. The axial $SU(6)_{A}$ is spontaneously broken by the quark condensate \cite{'tHooft:1979bh}, leaving the $SU(6)_{V}$ as global symmetry of massless QCD. Realistic values of the quark masses exclude this large symmetry group. However, the three lightest quarks, $u$, $d$, and $s$, can be considered massless to a good approximation \cite{Donoghue:1992dd}, giving QCD an approximate $SU(3)_{L}\times SU(3)_{R}\rightarrow SU(3)_{V}$ global invariance. Similar constructions can be made for the two lightest quarks only, giving an approximate $SU(2)_{L}\times SU(2)_{R}\rightarrow SU(2)_{V}$ invariance.

Also the Higgs sector of $\mathcal{L}_{\text{SM}}$ has an approximate global symmetry \cite{Willenbrock:2004hu}. In order to see this, consider the Higgs sector of Eq.~\eqref{eq:SMunbroken}
\begin{equation}
  \label{eq:2.1.2}
  \mathcal{L}_{\text{Higgs}} = (D_{\mu}\phi^{\dagger})(D^{\mu}\phi) + \frac{1}{2}\mu^{2}\phi^{\dagger}\phi - \frac{\lambda}{4} (\phi^{\dagger}\phi)^{2}
\end{equation}
Introducing the Higgs bi-doublet $\Phi$, composed of the doublet $\phi$ and the conjugated doublet $\widetilde{\phi}_{j} = \varepsilon_{ij}\phi_{i}^{*}$, as 
\begin{equation}
\label{eq:2.1.6}
\Phi = \tfrac{1}{\sqrt{2}}(\widetilde{\phi},\phi), 
\end{equation}
we can write the Lagrangian as
\begin{equation}
  \label{eq:2.1.3}
  \mathcal{L}_{\text{Higgs}} = \langle(D_{\mu}\Phi^{\dagger})(D^{\mu}\Phi)\rangle + \frac{1}{2}\mu^{2}\langle\Phi^{\dagger}\Phi\rangle - \frac{\lambda}{4} \langle\Phi^{\dagger}\Phi\rangle^{2}. 
\end{equation}
The covariant derivative of $\Phi$ is given by
\begin{equation}
  \label{eq:2.1.4}
  D_{\mu}\Phi = \partial_{\mu}\Phi + i g T^{a}W_{\mu}^{a}\Phi - i g' B_{\mu}\Phi T^{3},
\end{equation}
where $T^{a}=\sigma^{a}/2$ are the generators of $SU(2)$. Written in this form, the $SU(2)_{L}\times U(1)_{Y}$ transformations act on $\Phi$ as
\begin{equation}
  \label{eq:2.1.5}
  \Phi \rightarrow g_{L}\Phi g_{Y}^{\dagger}, \qquad \text{where } g_{L} \in SU(2)_{L} \text{ and } g_{Y} \in U(1)_{Y}.
\end{equation}
In the limit of vanishing hypercharge interactions, $g'\rightarrow 0$, the Lagrangian $\mathcal{L}_{\text{Higgs}}$ is invariant under the larger symmetry group $SU(2)_{L}\times SU(2)_{R}$. This accidental global symmetry is broken explicitly by hypercharge interactions and also by the different Yukawa couplings for up- and down-type quarks. We will come back to this symmetry after we discussed spontaneous symmetry breaking in the Standard Model. 

The product of the discrete symmetries $\mathcal{C}$ (charge conjugation), $\mathcal{P}$ (parity), and $\mathcal{T}$ (time reversal) is a symmetry of local, hermitean, and Lorentz-invariant Quantum Field Theories \cite{PauliCPT,Luders:1954zz} and therefore also of the Standard Model. In QED, $\mathcal{C}$, $\mathcal{P}$, and $\mathcal{T}$ are separately conserved. QCD also conserves the three symmetries separately if $\theta =0$ in Eq.~\eqref{eq:SMunbroken}. Otherwise, there will be $\mathcal{CP}$ violation in the QCD sector. The electroweak sector violates $\mathcal{C}$ and $\mathcal{P}$ maximally since particles of left- and right-chirality are in different representations of the gauge group $SU(2)_{L}\times U(1)_{Y}$. In addition, there is a $\mathcal{CP}$-violating phase in the CKM matrix that we will introduce below. 
\section{The Standard Model in the Mass-Eigenstate Basis}
\label{ch:2.2}
The Lagrangian in Eq.~\eqref{eq:SMunbroken}, which we discussed in the previous section, was written in terms of the gauge interaction eigenstates. They do not always coincide with the mass eigenstates, which are the propagating degrees of freedom that we observe in the detectors. To connect our theory predictions to experimental observables, it is therefore necessary to rotate Eq.~\eqref{eq:SMunbroken} to the mass-eigenstate basis. We will do so in this section. However, for the discussion it is crucial to have a look at the different fates of symmetries first.
  \subsection{Different Fates of Symmetries}
The symmetries of Eq.~\eqref{eq:SMunbroken} that we discussed so far have different fates \cite{Donoghue:1992dd}. Some of them, such as $B-L$, are indeed symmetries of the particle interactions. Other symmetries, like the combination $B+L$, are anomalous. Even though the Lagrangian is invariant, the measure of the path integral is not. The symmetry will then be broken by quantum effects, {\it i.e.} by loops. It is also possible that the system has only an approximate symmetry, meaning it is only a symmetry in a certain limit. In the full Lagrangian, the symmetry is explicitly broken by a small perturbation. The small masses of up- and down-quarks break the chiral $SU(2)_{L}\times SU(2)_{R}$ symmetry of QCD explicitly. The fourth possible fate of a symmetry of a Lagrangian is that it is not respected by the ground state of the system. The symmetry is then called spontaneously broken. Since spontaneous symmetry breaking is responsible for many phenomena in particle physics, we will discuss it in more detail using the following example, called the linear sigma model. 
\subsubsection{Spontaneous Symmetry Breaking and the Linear Sigma Model}
Consider the linear sigma model. We start from a set of $N$ real scalar fields $\phi_{i}$ with the Lagrangian
\begin{equation}
  \label{eq:2.2.1}
  \mathcal{L} = \frac{1}{2} (\partial_{\mu} \phi_{i}) (\partial^{\mu} \phi_{i}) + \frac{\mu^{2}}{2} |\phi|^{2} - \frac{\lambda}{4}|\phi|^{4}.
\end{equation}
It is invariant under a global $O(N)$ symmetry of the fields $\phi_{i}$. The ground state of this theory is given by the field configuration that minimizes the potential:
\begin{equation}
  \label{eq:2.2.2}
  |\phi_{\text{vac}}|^{2} = \frac{\mu^{2}}{\lambda} \equiv v^{2}. 
\end{equation}
Physical particles are excitations from this vacuum. Rewriting Eq.~\eqref{eq:2.2.1} in terms of the physical fields, $\vec\phi = (\pi_{1},\dots,\pi_{N-1},v+\sigma)^{T}$, yields
\begin{equation}
  \label{eq:2.2.3}
  \mathcal{L} = \frac{1}{2} (\partial_{\mu} \pi_{i}) (\partial^{\mu} \pi_{i}) + \frac{1}{2} (\partial_{\mu} \sigma) (\partial^{\mu} \sigma) - \lambda v^{2} \sigma^{2} -\frac{\lambda}{4}(\pi^{2}+\sigma^{2})^{2}-\lambda v \sigma (\pi^{2}+\sigma^{2}).
\end{equation}
We observe $N-1$ massless fields, the $\pi_{i}$, and one massive $\sigma$ with mass $m_{\sigma} = \sqrt{2\lambda}\mu$ in the spectrum. The global $O(N)$ symmetry of Eq.~\eqref{eq:2.2.1} is hidden in the structure of the interactions in Eq.~\eqref{eq:2.2.3}, only the $O(N-1)$ symmetry of the $\pi_{i}$ is explicit. All these effects are not a coincidence, they appear whenever a global symmetry is spontaneously broken. This was observed by Goldstone \cite{Goldstone:1961eq} in connection with observations of Nambu \cite{PhysRev.117.648,PhysRevLett.4.380,PhysRev.122.345}. He formulated Goldstone's Theorem: For each generator of a global symmetry that is spontaneously broken we observe a massless boson \cite{PhysRev.127.965,Peskin:1995ev} with the quantum numbers of the broken generator. These massless fields are called (Nambu)-Goldstone bosons. From the $(N-1)N/2$ generators of $O(N)$ in our example, only $(N-1)(N-2)/2$ are unbroken in the vacuum, yielding $N-1$ Goldstone bosons. If the spontaneously broken symmetry is also broken explicitly, the Goldstones become massive. They will still be light compared to other particles of the spectrum if the explicit breaking of the symmetry is small. They are called pseudo-(Nambu)-Goldstone bosons in this case. 
\subsubsection{The Non-Linear Sigma Model}
\label{ch:nlsm}
We now consider the non-linear sigma model. We construct it from the linear sigma model in the limit when the mass of $\sigma$ tends to infinity, while the vacuum expectation value $v$  remains constant \cite{Appelquist:1980ae}. As it is impossible to excite the $\sigma$ state, the dynamics of the $\pi_{i}$ are constrained to be on the vacuum manifold with the $O(N-1)$ symmetry. This results in the non-trivial constraint $|\phi| = v$, or equivalently 
\begin{equation}
  \label{eq:2.2.4}
  \pi^{2}+\sigma^{2} = -2 v \sigma.
\end{equation}
The Lagrangian then becomes \cite{PhysRev.166.1568} 
\begin{equation}
  \label{eq:2.2.5}
  \mathcal{L} = \frac{1}{2} (\partial_{\mu} \pi_{i}) (\partial^{\mu} \pi_{i}) + \frac{1}{2}\frac{(\pi_{i}\partial_{\mu}\pi_{i})(\pi_{j}\partial^{\mu}\pi_{j})}{v^{2}-\pi^{2}}.
\end{equation}
As the vacuum manifold is non-linear because of the constraint~\eqref{eq:2.2.4}, the model is called non-linear sigma model. Since the construction of the non-linear sigma model is only based on the structure of the vacuum manifold, it is only the pattern of symmetry breaking that enters here. More information on how the symmetry is broken is not needed for describing the low-energy dynamics of the Goldstone bosons. This makes it very useful in bottom-up effective field theories that we will discuss in the next chapter. 

The constraint that restricts the Goldstone bosons to be on the vacuum manifold can be realized in the Lagrangian in many different ways. This basically corresponds to a choice of coordinate system on the vacuum manifold. The square root representation of Eq.~\eqref{eq:2.2.5} is therefore not the only possible choice. Another convenient representation is the exponential representation, where the Goldstone bosons are written as
\begin{equation}
  \label{eq:2.2.6}
  U  = \exp{\{ i \frac{T^{i}\varphi^{i}}{v}\}},
\end{equation}
with $T^{i}$ being the generators of the coset $O(N)/O(N-1)$ and $\varphi^{i}$ are functions of the $\pi_{i}$. The Lagrangian of Eq.~\eqref{eq:2.2.5} then becomes
\begin{equation}
  \label{eq:2.2.7}
  \mathcal{L} = \frac{v^{2}}{4} \langle \partial_{\mu}U^{\dagger}\partial^{\mu}U\rangle.
\end{equation}
All differently looking non-linear representations give the same results for observables, as they are all related by field redefinitions \cite{Donoghue:1992dd,Haag:1958vt,Coleman:1969sm,Callan:1969sn,Feruglio:1992wf}.\\
\subsubsection{The Higgs Mechanism}
The situation changes when instead of a global symmetry a local symmetry is spontaneously broken. Consider a complex scalar field $\Phi$, gauged under a $U(1)$ symmetry:
\begin{equation}
  \label{eq:2.2.8}
  \mathcal{L} = - \frac{1}{4}F_{\mu\nu}F^{\mu\nu} + D_{\mu}\Phi^{\dagger}D^{\mu}\Phi + \mu^{2} \Phi^{\dagger}\Phi - \lambda (\Phi^{\dagger}\Phi)^{2},
 \end{equation}
where $D_{\mu}\Phi = \partial_{\mu}\Phi + i g A_{\mu} \Phi$ is the covariant derivative of $\Phi$. The vacuum of this theory is given by the condition $|\Phi|^{2} =\mu^{2}/(2\lambda)\equiv v^{2}/2$. The expansion around the ground state is parametrized as $\Phi = (v+h + i \eta)/\sqrt{2}$. The potential from Eq.~\eqref{eq:2.2.8} becomes now, upon neglecting an unphysical constant,
\begin{equation}
  \label{eq:2.2.9}
\mathcal{V} = \lambda v^{2}h^{2}   + \frac{\lambda}{4}(\eta^{2}+h^{2})^{2}+\lambda v h (\eta^{2}+h^{2}).
\end{equation}
This is similar to the case of a spontaneously broken global symmetry in Eq.~\eqref{eq:2.2.3}. The difference arises from the covariant derivative in the kinetic term,
\begin{align}
  \begin{aligned}
    \label{eq:2.2.10}
     D_{\mu}\Phi^{\dagger}D^{\mu}\Phi & = \frac{1}{2} (\partial_{\mu} h) (\partial^{\mu}h)+\frac{1}{2} (\partial_{\mu} \eta) (\partial^{\mu}\eta)- g \eta A_{\mu}(\partial^{\mu}h) + g A_{\mu}(\partial^{\mu}\eta)(v+h)\\
     & + \frac{g^{2}}{2}A_{\mu}A^{\mu}(v^{2} + 2 vh + h^{2} + \eta^{2}).
  \end{aligned}
\end{align}
Again, we have a massless Goldstone boson from spontaneous symmetry breaking, the $\eta$, as well as a massive field, the $h$. Eq.~\eqref{eq:2.2.10}, however, seems to indicate a kinetic mixing between the Goldstone and the gauge field. The situation will clear up once we use the gauge freedom the Lagrangian~\eqref{eq:2.2.8} has. After making the transformations $\Phi\rightarrow \exp{\{-i \eta/v\}}\Phi'$ and $A_{\mu}\rightarrow A'_{\mu} + \partial_{\mu}\eta / (gv) $ we find
\begin{equation}
  \label{eq:2.2.11}
  D_{\mu}\Phi'^{\dagger}D^{\mu}\Phi' = \frac{1}{2} (\partial_{\mu} h) (\partial^{\mu}h) + \frac{g^{2}}{2}A'_{\mu}A'^{\mu}(v^{2} + 2 vh + h^{2}).
\end{equation}
This particular choice of gauge, called unitary gauge, removes the Goldstone bosons completely from the spectrum. We are left with a theory of a massive scalar, $h$, and a gauge field, $A'_{\mu}$, that acquired a mass $m_{A'}= gv$ in Eq.~\eqref{eq:2.2.11}. The mechanism, in which gauge fields get a mass from a spontaneously broken local symmetry, is called Brout-Englert-Higgs mechanism\cite{Englert:1964et,Higgs:1964ia,Higgs:1964pj,Guralnik:1964eu,Higgs:1966ev,Kibble:1967sv}, or Higgs mechanism for short. The Goldstone boson from the spontaneous breaking of the global subgroup is ``eaten'' by the gauge field and becomes its longitudinal degree of freedom. A complex scalar and a massless gauge field have $2+2=4$ degrees of freedom before symmetry breaking. After spontaneous symmetry breaking, we have a massive real scalar and a massive gauge field, giving $1+3=4$ degrees of freedom. The total number of degrees of freedom is therefore unchanged. After seeing the origin of the mass term from spontaneous symmetry breaking explicitly, we also understand why the Lagrangian containing~\eqref{eq:2.2.11} is still gauge invariant. A gauge transformation of $A_{\mu}$, {\it i.e.} $A_{\mu}\rightarrow A_{\mu} +\partial_{\mu} \alpha$, is compensated by a transformation of the longitudinal degree of freedom, {\it i.e.} $\eta\rightarrow\eta -\alpha gv$. The massive gauge field
\begin{equation}
  \label{eq:2.2.12}
  A'_{\mu} = A_{\mu} + \frac{\partial_{\mu} \eta}{gv}
\end{equation}
is then trivially invariant. Equation~\eqref{eq:2.2.12} is sometimes called St{\"u}ckelberg decomposition \cite{Kunimasa:1967zza}. 
  \subsection{The Standard Model After Spontaneous Symmetry Breaking}
\label{ch:SMprod.dec}
We will now see how spontaneous breaking of $SU(2)_{L}\times U(1)_{Y}$ dictates the phenomenology of the Standard Model. This was first described by Glashow \cite{Glashow1961579}, Weinberg \cite{PhysRevLett.19.1264}, and Salam \cite{Salam}, giving this model the name GWS theory. 
\subsubsection{The Gauge and the Higgs Sector}
The Standard Model Higgs potential in Eq.~\eqref{eq:2.1.2} has a non-trivial minimum, giving a vacuum expectation value to the Higgs field
\begin{equation}
  \label{eq:2.2.13}
  (\phi^{\dagger}\phi)_{\text{vac}} = \frac{\mu^{2}}{\lambda}\equiv \frac{v^{2}}{2}.
\end{equation}
The $SU(2)_{L}$ invariance allows us to choose $\phi_{\text{vac}}= (0,v/\sqrt{2})^{T}$. Fluctuations around the vacuum are parametrized by
\begin{equation}
  \label{eq:2.2.14}
  \phi = \begin{pmatrix}\eta_{1} + i \eta_{2} \\ \frac{v+h}{\sqrt{2}} + i\eta_{3}\end{pmatrix},
\end{equation}
where $h$ is the physical Higgs boson and $\eta_{i}$ are the three Goldstone bosons of $SU(2)_{L}\times U(1)_{Y}\rightarrow U(1)_{QED}$. 
Inserting this back into Eq.~\eqref{eq:2.1.2}, we find in unitary gauge
\begin{align}
\begin{aligned}
  \label{eq:2.2.15}
  \mathcal{L}_{\text{Higgs}}  = &\frac{1}{2}(\partial_{\mu}h)(\partial^{\mu}h) + \frac{1}{8}\Big[g^{2}W^{1}_{\mu}W^{1\mu}+g^{2} W^{2}_{\mu}W^{2\mu}\Big](v+h)^{2}\\
  &+ \frac{1}{8}\Big[(gW_{\mu}^{3}-g'B_{\mu})(gW^{3\mu}-g'B^{\mu})\Big](v+h)^{2} - \frac{\lambda}{4}v^{2} h^{2}- \frac{\lambda}{4}v h^{3}- \frac{\lambda}{16}h^{4}.
\end{aligned}
\end{align}
In this expression, we observe a mixing between the $W_{\mu}^{3}$ and the $B_{\mu}$. We define the physical basis, {\it i.e.} the mass-eigenstate basis, {\it via}
\begin{equation}
  \begin{array}{lr}
  \label{eq:2.2.16}
 W^{\pm}_{\mu}=\frac{1}{\sqrt{2}}(W^{1}_{\mu}\mp i W^{2}_{\mu}), & Z_{\mu}= \frac{1}{\sqrt{g^{2}+g'^{2}}}(gW_{\mu}^{3}-g'B_{\mu}) \\
 \text{and } A_{\mu} = \frac{1}{\sqrt{g^{2}+g'^{2}}}(g'W_{\mu}^{3}+gB_{\mu}).
\end{array}
\end{equation}
This gives
\begin{align}
\begin{aligned}
  \label{eq:2.2.17}
\mathcal{L}_{\text{Higgs}} = &\frac{1}{2}(\partial_{\mu}h)(\partial^{\mu}h) + \frac{1}{2}\Big[\frac{g^{2}}{2}W^{+}_{\mu}W^{-\mu}+\frac{(g^{2}+g'^{2})}{4} Z_{\mu}Z^{\mu}\Big](v+h)^{2}\\
&- \frac{\lambda}{4}v^{2} h^{2}- \frac{\lambda}{4}v h^{3}- \frac{\lambda}{16}h^{4},
\end{aligned}
\end{align}
where we can now read off the masses of the physical $W^{\pm}$ and $Z$ boson,
\begin{equation}
  \label{eq:2.2.18}
  m_{W}= \frac{vg}{2}\quad \text{and} \quad m_{Z}=\frac{v}{2}\sqrt{g^{2}+g'^{2}}.
\end{equation}
The fourth gauge field, $A_{\mu}$, remains massless. It is the messenger of the unbroken $U(1)$ --- the photon of QED. The $U(1)_{QED}$ is generated by the combination $T^{3}_{L}+\mathsf{Y}$ of the generators of $SU(2)_{L}\times U(1)_{Y}$, thereby connecting the field's electric charge with its hypercharge and its eigenvalue of the third generator of $SU(2)_{L}$, yielding $Q_{\Psi} = T^{3,\Psi}_{L}+\mathsf{Y}_{\Psi}$. From this, we find the electric charge
\begin{equation}
  \label{eq:2.2.19}
  e = \frac{g g'}{\sqrt{g^{2}+g'^{2}}}.
\end{equation}
Since the transformation in Eq.~\eqref{eq:2.2.16} can be understood as a rotation in field space, it is useful to define the rotation angle, or weak-mixing angle as
\begin{equation}
  \label{eq:2.2.20}
  \cos{\theta_{w}} = \frac{g}{\sqrt{g^{2}+g'^{2}}}.
\end{equation}
We can now return to the matrix notation of the Higgs sector, Eq.~\eqref{eq:2.1.3}. The vacuum expectation value of $\phi$ in Eq.~\eqref{eq:2.2.13} gives the vacuum for $\Phi$ to be
\begin{equation}
  \label{eq:2.2.21}
  \Phi_{\text{vac}} = \frac{1}{2}\begin{pmatrix} v & 0\\0 & v \end{pmatrix}.
\end{equation}
Since $\Phi$ transforms as
\begin{equation}
  \label{eq:2.2.22}
  \Phi \rightarrow g_{L} \Phi g_{R}^{\dagger}, \qquad \text{where } g_{L,R} \in SU(2)_{L,R},
\end{equation}
both symmetries will be broken in the ground state. Only if $g_{L}=g_{R}$, the vacuum will be invariant. This gives the pattern of symmetry breaking $SU(2)_{L}\times SU(2)_{R}\rightarrow SU(2)_{V}$ in the Higgs sector. The $SU(2)_{V}$ symmetry of the vacuum is often called custodial symmetry \cite{Sikivie:1980hm}, as it protects the mass ratio of $W^{\pm}$ and $Z$ from receiving large perturbative corrections. Hypercharge and the difference between up- and down-type Yukawa couplings violate custodial symmetry. Sometimes, violation of custodial symmetry is defined excluding these sources of explicit breaking \cite{Contino:2013kra}. 

\subsubsection{The Yukawa Sector}
The expansion of $\phi$ around its vacuum expectation value, $v/\sqrt{2}$, introduces mass terms for the fermions,
\begin{align}
\begin{aligned}
  \label{eq:2.2.23}
  \mathcal{L}_{\text{Yukawa}} &=  -\bar{\ell}_{L}^{i}Y_{e}^{ij}\phi e_{R}^{j} - \bar{q}_{L}^{i}Y_{d}^{ij}\phi d_{R}^{j} - \bar{q}_{L}^{i}Y_{u}^{ij} (i\sigma_{2}\phi^{*}) u_{R}^{j} +\text{h.c.}\\
  &= -\tfrac{v}{\sqrt{2}}\,\bar{e}_{L}^{i}Y_{e}^{ij} e_{R}^{j} \left(1+\tfrac{h}{v}\right) - \tfrac{v}{\sqrt{2}}\,\bar{d}_{L}^{i}Y_{d}^{ij} d_{R}^{j}\left(1+\tfrac{h}{v}\right) - \tfrac{v}{\sqrt{2}}\,\bar{u}_{L}^{i}Y_{u}^{ij} u_{R}^{j}\left(1+\tfrac{h}{v}\right) +\text{h.c.}
\end{aligned}
\end{align}
Mass terms of Dirac type, $m\bar{\Psi}_{L}\Psi_{R}$, are not allowed in the Standard Model, as left- and right-handed fermions are in different representations of the gauge group. 

Since there is no restriction on the shape of $Y_{\Psi}^{ij}$, the fermions do not need to be in the mass-eigenstate basis. A bi-unitary diagonalization can be done using \cite{Donoghue:1992dd}
\begin{equation}
  \label{eq:2.2.24}
  \begin{pmatrix}m_{\Psi_{1}}&& \\ & m_{\Psi_{2}}&\\ &&m_{\Psi_{3}}\end{pmatrix} = U^{\dagger}_{\Psi} Y_{\Psi} V_{\Psi},
\end{equation}
where there is a unitary transformation $U(V)$ for each left-(right-)handed fermion:
\begin{equation}
  \label{eq:2.2.25}
  \Psi^{\text{gauge}}_{L} = U_{\Psi}\Psi^{\text{mass}}_{L}, \qquad \text{and} \qquad \Psi^{\text{gauge}}_{R} = V_{\Psi}\Psi^{\text{mass}}_{R}.
\end{equation}
Applying this transformation to $\mathcal{L}_{\text{SM}} $ gives us the Lagrangian in the mass-eigenstate basis. Kinetic terms, neutral weak-, electromagnetic-, and strong-gauge currents are diagonal in flavor space. Thus, the unitary matrices cancel to unity and the mass and gauge basis coincide. The charged gauge currents of the weak interactions are different, as they connect an upper and a lower component of an $SU(2)_{L}$ doublet. The unitary matrices will then not cancel, leaving terms like $-g \bar{\Psi}_{L}^{u}\slashed{W}U_{u}^{\dagger}U_{d}\Psi_{L}^{d}$. 

In the lepton sector, this rotation can be absorbed by the neutrinos. As they have no mass term in Eq.~\eqref{eq:2.2.23}, this is not in contradiction to the diagonalization discussed above. The situation is different in the quark sector. There, we do not have the freedom to redefine the fields once we go to the mass basis. The charged currents of the weak interaction will therefore mix the states of different generations. The mixing matrix, usually applied to down-type quarks, is called CKM matrix \cite{PhysRevLett.10.531,Kobayashi01021973},
\begin{equation}
  \label{eq:2.2.26}
  V_{\text{CKM}} = U^{\dagger}_{u}U_{d}.
\end{equation}
It is in general a $3\times 3$, complex-valued matrix. Since it is unitary, it seems to have nine free parameters. However, five of them can be absorbed in the relative phases of the quarks as part of the $U(3)^{5}$ transformation discussed above. An overall common phase does not change the CKM matrix at all \cite{Donoghue:1992dd}. The four remaining free parameters are three mixing angles and one complex phase. The latter is responsible for the $\mathcal{CP}$ violation in the quark sector of the Standard Model. \\

\subsubsection{The Standard Model Lagrangian in the Mass-Eigenstate Basis}
After all the considerations presented above, we can now write down the Lagrangian of the Standard Model in the mass-eigenstate basis in unitary gauge
\begin{align}
  \begin{aligned}
    \label{eq:SMbroken}
  \mathcal{L}_{\text{SM}} &=-\frac{1}{4} F_{\mu\nu}F^{\mu\nu}-\frac{1}{4} Z_{\mu\nu}Z^{\mu\nu} - \frac{1}{2} W_{\mu\nu}^{+}W^{-,\mu\nu} +\mathcal{L}^{\text{triple}}_{\text{gauge}} + \mathcal{L}^{\text{quartic}}_{\text{gauge}}\\
&- \frac{1}{2}\langle G_{\mu\nu}G^{\mu\nu}\rangle - \theta \frac{g_{s}^{2}}{32\pi^{2}}\langle \widetilde{G}_{\mu\nu}G^{\mu\nu}\rangle +\frac{1}{2}(\partial_{\mu}h)(\partial^{\mu}h)\\
  &+i\bar{q}_{L}^{i}\slashed{D}q_{L}^{i} +i\bar{\ell}_{L}^{i}\slashed{D}\ell_{L}^{i} +i\bar{u}_{R}^{i}\slashed{D}u_{R}^{i} +i\bar{d}_{R}^{i}\slashed{D}d_{R}^{i} +i\bar{e}_{R}^{i}\slashed{D}e_{R}^{i}\\
  & + \left[m_{W}^{2}W^{+}_{\mu}W^{-\mu}+\tfrac{m_{Z}^{2}}{2} Z_{\mu}Z^{\mu}\right]\left(1+\tfrac{h}{v}\right)^{2} - \frac{\lambda}{4}v^{2} h^{2}- \frac{\lambda}{4}v h^{3}- \frac{\lambda}{16}h^{4}\\
  &-m_{e}^{i}\,\bar{e}_{L}^{i} e_{R}^{i} \left(1+\tfrac{h}{v}\right) - m_{d}^{i}\,\bar{d}_{L}^{i} d_{R}^{i}\left(1+\tfrac{h}{v}\right) - m_{u}^{i}\,\bar{u}_{L}^{i} u_{R}^{i}\left(1+\tfrac{h}{v}\right) +\text{h.c.}
\end{aligned}
\end{align}
Here, $\mathcal{L}^{\text{triple}}_{\text{gauge}} $ and $ \mathcal{L}^{\text{quartic}}_{\text{gauge}}$ are the non-abelian parts of $\langle W_{\mu\nu}W^{\mu\nu}\rangle$ after the rotation defined in Eq.~\eqref{eq:2.2.16}. The field-strength tensors $X_{\mu\nu}$ for $X\in \{F,Z,W^{\pm}\}$ are then only linear in the gauge fields. The CKM matrix is understood to be implicitly contained in the covariant derivative of the left-handed quarks.\\

We conclude this chapter with some general observations concerning the Standard Model Higgs. Since all of its couplings have the general structure $(v+h)^{n}$, coming from spontaneous symmetry breaking, the SM-Higgs' couplings to the massive gauge fields and fermions are proportional to the masses of the corresponding particles. The Higgs self-coupling is proportional to the mass of the Higgs. 

\begin{figure}[t]
\begin{center}
\subfigure[Vector boson fusion (VBF)\label{fig:SMHiggs.VBF}]{
\begin{overpic}[width=0.45\textwidth]{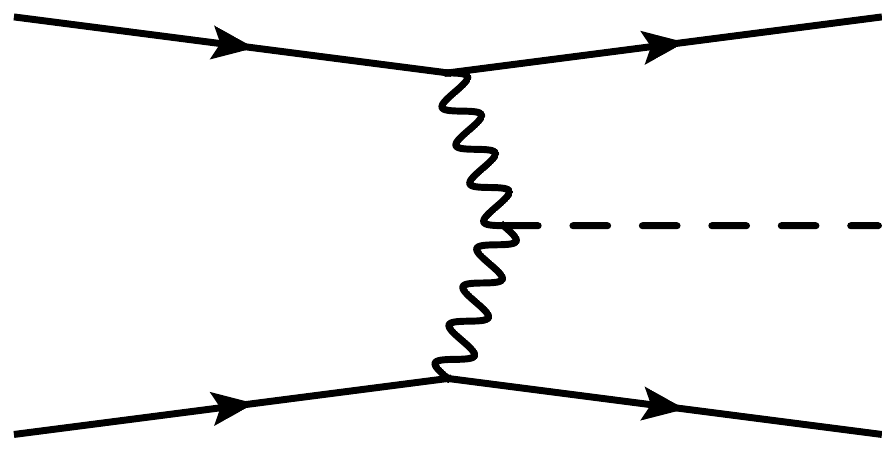}
\put (0,42.5){$q$}
\put (0,5){$q$}
\put (55,35){$W^{\pm}/Z$}
\put (55,10){$W^{\mp}/Z$}
\put (95,27.5){$h$}
\end{overpic}}\hfill
\subfigure[Associated production with vector bosons (WH/ZH)\label{fig:SMHiggs.VH}]{
\begin{overpic}[width=0.45\textwidth]{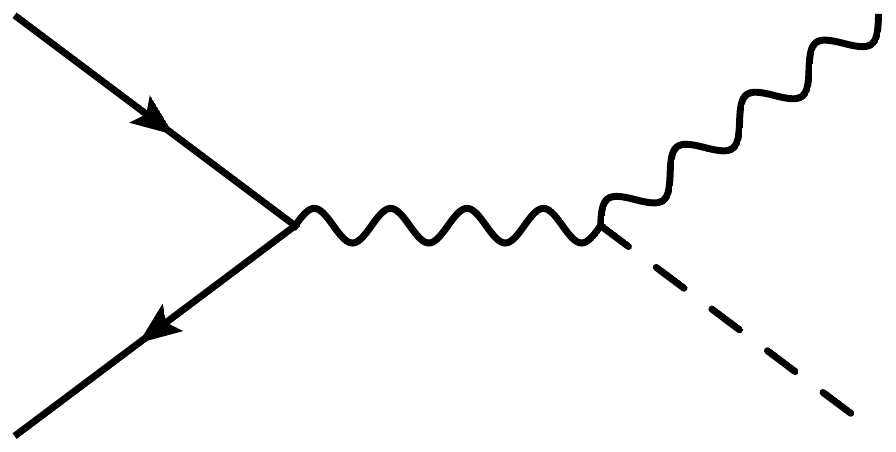}
\put (0,40){$q$}
\put (0,7.5){$\bar{q}$}
\put (50,30){$W^{\pm}/Z$}
\put (95,5){$h$}
\end{overpic}}\\
\subfigure[Associated production with $\bar{t}t$-pairs (ttH)\label{fig:SMHiggs.ttH}]{
\begin{overpic}[width=0.45\textwidth]{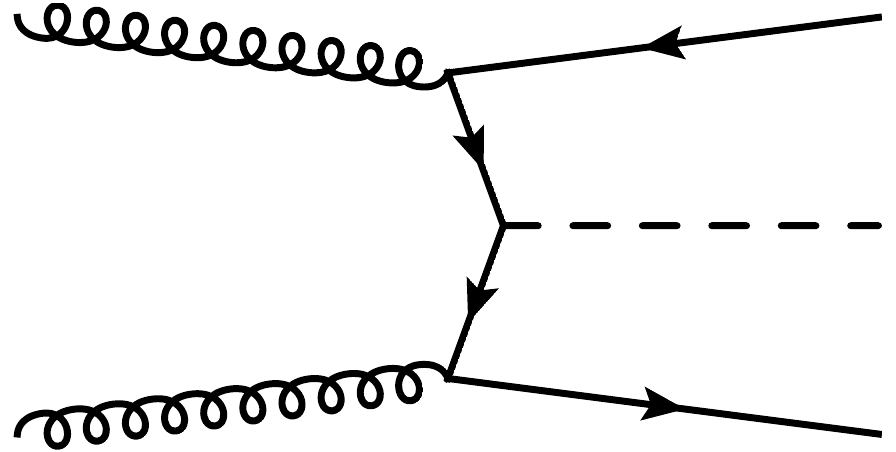}
\put (0,42.5){$g$}
\put (0,5){$g$}
\put (55,35){$t$}
\put (55,10){$\bar{t}$}
\put (95,27.5){$h$}
\put (95,5){$t$}
\put (95,40){$\bar{t}$}
\end{overpic}}\hfill
\subfigure[Gluon fusion (ggF)\label{fig:SMHiggs.ggF}]{
\begin{overpic}[width=0.45\textwidth]{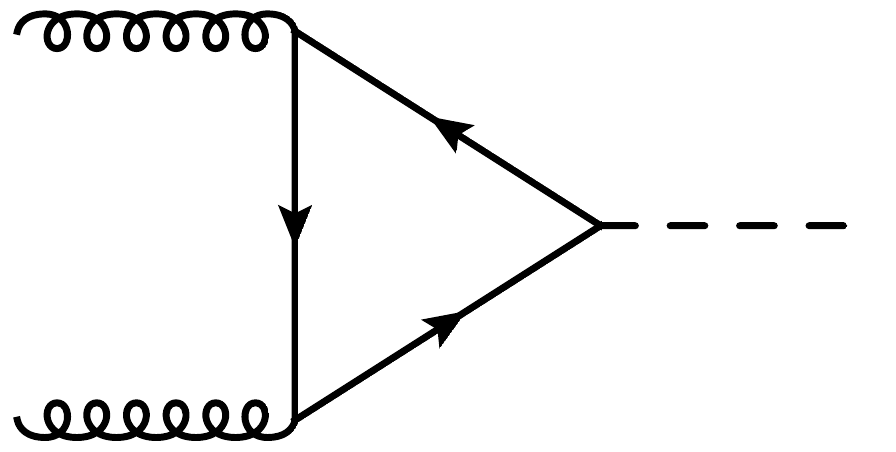}
\put (0,42.5){$g$}
\put (0,5){$g$}
\put (95,27.5){$h$}
\put (35,20){\fontsize{40pt}{48pt} \selectfont $\circlearrowleft$}
\put (41.25,22.5){$t,b$}
\end{overpic}}
\end{center}
\caption[Production modes of the Standard Model Higgs boson.]{Production modes of the Standard Model Higgs boson. At the LHC for a center-of-mass energy of $\sqrt{s}=13~\text{TeV}$ and a Higgs mass $m_{h}=125~\text{GeV}$, the dominant single-Higgs production mode is gluon fusion (ggF). It constitutes $86.1\%$ of the total production \cite{HiggsXS} cross section. Vector boson fusion (VBF) contributes $7.3\%$, whereas associated production with vector bosons (WH/ZH) and top-quark pairs (ttH) only contribute $4.4\%$ and $1\%$, respectively \cite{HiggsXS}.}
\label{fig:SMHiggs.prod}
\end{figure}
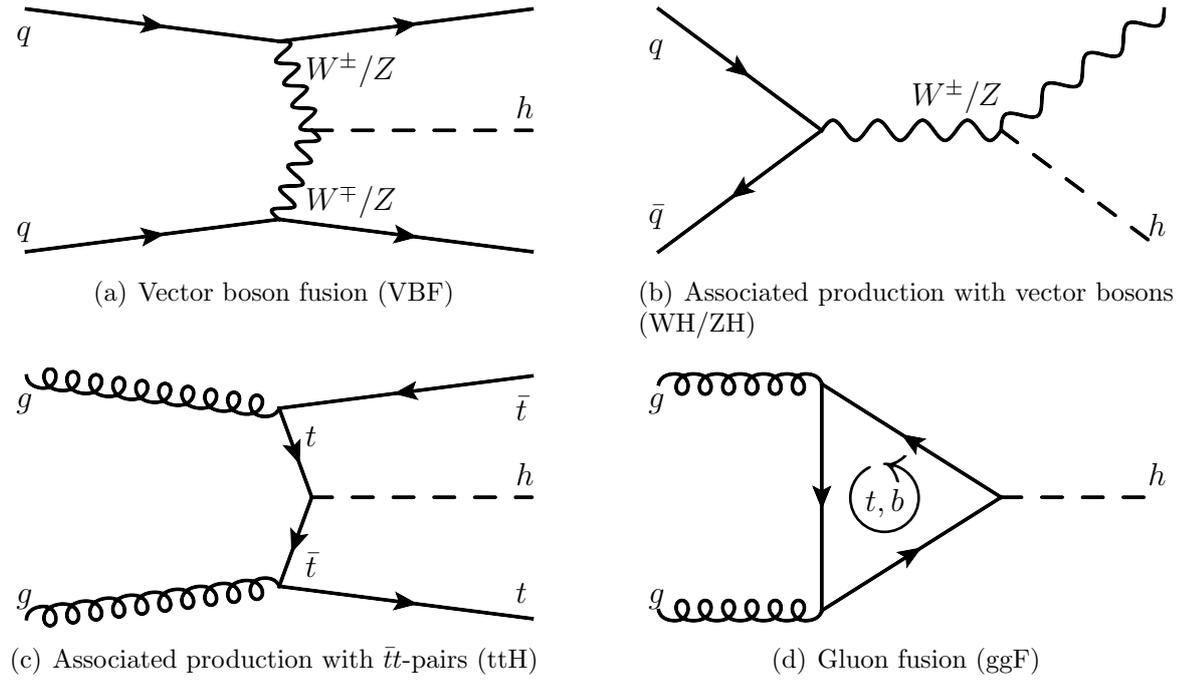 
The Higgs is therefore dominantly produced by either massive vectors in vector boson fusion (VBF, see Fig. \ref{fig:SMHiggs.VBF}) and associated production (WH/ZH, see Fig. \ref{fig:SMHiggs.VH}) or {\it via} heavy quarks. As the parton density function of the latter in the proton is small at the current experimental energies, these quarks are usually not in the initial state. Instead, the Higgs is produced in $\bar{t}t$-associated production (ttH, see Fig. \ref{fig:SMHiggs.ttH}) or in gluon fusion (ggF, see Fig. \ref{fig:SMHiggs.ggF}). At the LHC, gluon fusion is the dominant production mode \cite{HiggsXS}. 

The Higgs decays at tree level to pairs of fermions and $W^{\pm}/Z$ vector  bosons. Since the coupling is proportional to each particle's mass, the branching ratio (BR) is larger for heavier final states. An upper cutoff to the particle's mass of the final state is given by the available energy, {\it i.e.} one half of the Higgs mass. The dominant decay channels are therefore to bottom-quark pairs with $\text{BR}(h\rightarrow\bar{b}b) = 57.7\%$, pairs of $W^{\pm}$ with $\text{BR}(h\rightarrow WW^{*}) = 21.5\%$, pairs of tau-leptons with $\text{BR}(h\rightarrow\tau^{+}\tau^{-}) = 6.3\%$, and pairs of $Z$ with $\text{BR}(h\rightarrow ZZ^{*}) = 2.6\%$ \cite{Heinemeyer:2013tqa}. Similar to the production in gluon fusion, the Higgs can also decay {\it via} a loop of heavy fermions. The loop induced final states include $gg, \gamma\gamma$ and $Z\gamma$. The decay channel to two photons is of high importance for the experimental detection of the Higgs. The very clean signature in the detector compensates for the small branching ratio of $\text{BR}(h\rightarrow \gamma\gamma) = 0.2\%$ \cite{Heinemeyer:2013tqa}.
\section{Open Questions}
\label{ch:SM.openQ}
Even though the Standard Model is tremendously successful in describing experimental data, there are still motivations for beyond-the-Standard-Model (BSM) physics \cite{Gripaios:2015gxa}. These motivations are based on experimental observations and theoretical considerations. In the following section, we present some open questions of the SM.
  \subsection[Experimentally Motivated Hints for Physics Beyond the SM]{Experimentally Motivated Hints for Physics Beyond the Standard Model}
The neutrinos of the Standard Model do not get a mass from the Higgs mechanism and are therefore massless in the SM. Experiments, however, observed an oscillation of propagating neutrinos from one flavor to another \cite{Fukuda:1998mi,Ahmad:2001an,Ahmad:2002jz}, indicating that the flavor basis does not coincide with the basis of propagation, {\it i.e.}~the mass-eigenstate basis. This phenomenon is therefore only possible for massive neutrinos. Currently, no precise value exists for these masses, only upper and lower bounds are reported \cite{Agashe:2014kda}. Furthermore, also the mass hierarchy of the three generations is not determined by experiment and a subject of current research \cite{Patterson:2015xja}. 

Adding a Yukawa interaction to the Lagrangian of the Standard Model, such that the neutrinos acquire a mass {\it via} the Higgs mechanism, requires to introduce right-handed neutrinos. This goes beyond the Standard Model. These neutrinos will be sterile, {\it i.e.} they are uncharged under the $SU(3)_{C}\times SU(2)_{L}\times U(1)_{Y}$ gauge group. They can therefore only be detected {\it via} mixing or gravitational effects. So far, there is no conclusive evidence for right-handed neutrinos. 

Another explanation for the neutrino masses is a Majorana nature of the neutrinos. A Majorana fermion is its own antiparticle, in contrast to the Dirac fermions of the SM that are distinct from their antiparticles. Experimental detection of neutrino-less double-beta decay would confirm the Majorana nature of the neutrinos. So far, it has not been observed \cite{::2016dpe}. \\ 

Rotational curves of galaxies \cite{1933AcHPh...6..110Z,1970ApJ...159..379R} and gravitational lensing observations \cite{1538-4357-648-2-L109} suggest the existence of a type of matter that generates a gravitational potential, but is invisible to electromagnetic radiation. It is called dark matter (DM) and it accounts for approximately $84\%$ \cite{Adam:2015rua,Ade:2015xua} of all the matter in the universe. Assuming a particle-physics explanation for dark matter requires to go beyond the SM, as there is no appropriate candidate for a dark-matter particle. Left-handed neutrinos, the only SM particles that have the right quantum numbers, cannot be used to explain dark matter. Their small mass would yield relativistic (warm/hot) dark matter, in contrast to observations \cite{Gripaios:2015gxa,Adam:2015rua}. Thus, additional particles beyond the Standard Model need to be introduced. \\

A third observation concerns the matter-antimatter asymmetry of the universe \cite{Steigman:1976ev,Kolb:1990vq}. For this to be generated within the Big Bang Theory, the three Sakharov conditions \cite{Sakharov:1967dj} need to be fulfilled. We need Baryon number $B$ violating processes, $\mathcal{C}$- and $\mathcal{CP}$-violating effects, and non-equilibrium conditions. Within the Standard Model, however, the effects are too small \cite{Kolb:1990vq} to account for the observed excess of matter over antimatter. 
  \subsection[Theoretically Motivated Hints for Physics Beyond the SM]{Theoretically Motivated Hints for Physics Beyond the Standard Model}
Many of the theoretically motivated open questions are connected to the notion of naturalness. Naturalness means that at any energy scale $E$ a physical parameter $p(E)$ is only allowed to be small, if replacing $p(E)=0$ increases the symmetry of the system \cite{'tHooft:1979bh}. Recently, this was called ``technical naturalness'' \cite{Hill:2002ap}. The Standard Model is naively expected to be valid up to the Planck-scale, $\Lambda_{Pl} \sim 10^{19}~\text{GeV}$. At these energies, the quantum corrections to gravitational effects of general relativity become dominant and a new, so far unknown theory of quantum gravity is needed. With such a high cutoff, particle masses at the electroweak scale of the SM, $v = 246~\text{GeV}$, seem very unnatural. Setting the fermion masses to zero, however, introduces the chiral symmetry discussed in Section \ref{ch:SMsym}, making these masses natural in the sense of \cite{'tHooft:1979bh}. The scalar Higgs, on the other hand, has no symmetry that protects its mass from corrections of the order of $\Lambda_{Pl}$. These corrections might still cancel to give a value of the order of $v$, but it will be unnatural. This problem is called hierarchy problem. The masses of the gauge bosons do not receive corrections of $\mathcal{O}(\Lambda_{Pl})$, but only $\mathcal{O}(\log{\Lambda_{Pl}})$, due to gauge symmetry \cite{Peskin:1995ev}. The hierarchy problem in the gauge sector is therefore not as severe as in the Higgs sector.

The Higgs sector of the SM gives further motivations for alternative models of electroweak symmetry breaking \cite{Grojean:2009fd}. The renormalization-group (RG) running of the Higgs self-coupling $\lambda$ might, depending on the numerical values of the other SM parameters, yield a Landau pole before the new sector of quantum gravity modifies the dynamics. This so-called triviality problem \cite{Callaway:1988ya,Cabibbo:1979ay,Schrempp:1996fb} indicates that new physics must be present before $\Lambda_{Pl}$. In addition, other configurations for the input parameters would lead to a change of the sign of the self-coupling $\lambda$. Such a value would make the electroweak vacuum unstable, naming this the stability problem \cite{Cabibbo:1979ay,Schrempp:1996fb}. The precise value of the energy scale where this occurs depends strongly on the values of $m_{h}$ and $m_{t}$. Latest experimental results indicate a meta-stable configuration \cite{Agashe:2014kda}, with the lifetime larger than the age of the universe. 

Solutions to the open questions concerning electroweak symmetry breaking extend the Standard Model at the electroweak scale $v$ or above, possibly making these scales natural with respect to $\Lambda_{Pl}$. Yet, there is another scale hierarchy that is highly unnatural. Astrophysical observations \cite{Adam:2015rua} measured the cosmological constant to be $\Lambda_{cc} = (10^{-3}~\text{eV})^{4}$ \cite{Gripaios:2015gxa}, which is 120 orders of magnitude below the Planck-scale. 

Apart from the hierarchy problems discussed above, there are other features only parametrized, but not explained in the Standard Model. The numerical values of couplings, masses, and mixing angles are input values to the SM. So far, there is no theory that predicts the observed pattern: The gauge couplings are all of order one, whereas the Yukawa couplings and therefore the particle's masses span several orders of magnitude. The mixing in the quark sector is small \cite{Agashe:2014kda}, but for the neutrinos it is rather large. 

Many solutions for the problems discussed above have been suggested and looked for. They include proposing new particles, new symmetries, new interactions, unifications of interactions, unifications of field representations, as well as combinations of these proposals. Experimentally, none of them has been observed so far. In the next chapter, we will introduce the concept of effective field theories, which allows us to look for new-physics effects in a model-independent way. 
  \chapter[Effective Field Theories (EFTs)]{Effective Field Theories}
\label{ch:EFT}
\thispagestyle{fancyplain}
In particle physics, we have interesting phenomena coming at many different energy scales, ranging from the sub-eV region of neutrino masses to the TeV region of current experiments, and likely also beyond. Fortunately, we do not need to know the underlying ``theory of everything'' to describe effects at a given (low) energy. Quantum field theories that are only valid in a certain range of energies are called effective field theories (EFTs). Scales much lighter than the given energy are treated as zero, heavier scales are set to infinity to a first approximation \cite{Georgi:1994qn}. Deviations from this simplified picture are treated as perturbations, in which the theory can be systematically expanded.

The influence of heavy particles (UV-physics) on low-energy (IR) observables was analyzed by Appelquist and Carazzone and led to the ``Decoupling Theorem'' \cite{Appelquist:1974tg}. It states that for low-energy observables (at scale $v$) all graphs with internal heavy (of mass $\Lambda$) fields are suppressed with powers of $(v/\Lambda)$ compared to graphs of only light fields. If the low-energy Lagrangian is renormalizable, the influence of the heavy particles to low-energy observables decouples in the limit $\Lambda\rightarrow\infty$, apart from the contribution to renormalization effects. Examples for this decoupling EFT \cite{Ecker:1993ft}  are the Euler-Heisenberg Lagrangian \cite{Heisenberg:1935qt,Pich:1995bw} and Fermi's theory of the weak interaction \cite{Fermi:1934hr}. However, the resulting theory in the limit $\Lambda\rightarrow\infty$ can also be non-renormalizable at leading order, giving a non-decoupling EFT \cite{Ecker:1993ft}. Such non-decoupling effects can arise in the context of spontaneous symmetry breaking, where the heavy and the light degrees of freedom are connected by symmetry. Also mixing effects can introduce non-decoupling effects, as we see in Section \ref{ch:SM+S}. An example for non-decoupling EFTs is chiral perturbation theory. We discuss it in Section \ref{ch:chiPT}. In general, finding a low-energy effective theory for a given UV model is called top-down EFT \cite{Georgi:1994qn}. We discuss it further in Section \ref{ch:EFT.top-down}.

\begin{figure}[ht]
\begin{center}
\tikzstyle{theory} = [draw, text width=10em, text centered, minimum height=1em, rounded corners]
\begin{tikzpicture}
\draw[->,thick] (0,0) -- (0,10em) node [above] {$E$};
\draw[yshift=1em] (-1pt,0) -- (1pt,0) node[left] {$v$};
\draw[yshift=8em] (-1pt,0) -- (1pt,0) node[left] {$\Lambda$};
\node at (9em,8em)(UV1) [theory] {UV Model};
\node at (24em,8em) (UV2) [theory] {???};
\node at (9em,1em) (EFT1) [theory] {Effective Description};
\node at (24em,1em) (EFT2) [theory] {Operator Basis};
\path[->, line width = 1pt] (UV1.south) edge [bend right] node [midway,right] {top-down} (EFT1.north); 
\path[->, line width = 1pt] (EFT2.north) edge [bend right] node [midway,left] {bottom-up} (UV2.south);
\end{tikzpicture}
\caption{Top-down {\it vs.} bottom-up picture of effective field theories.}
\label{fig:EFT.topdownvsbottomup}
\end{center}
\end{figure}
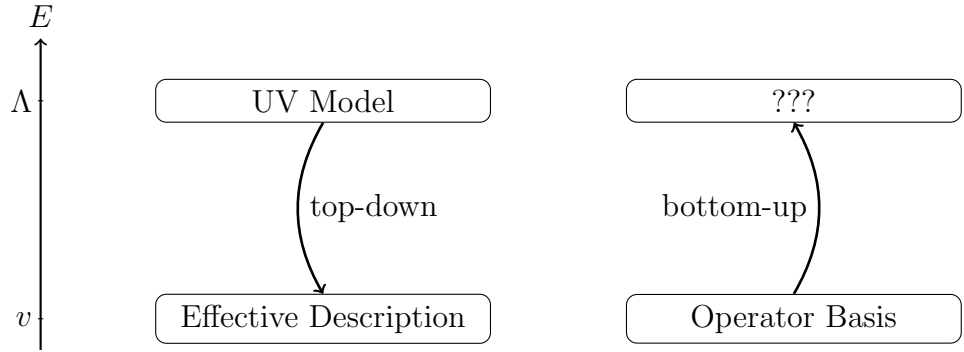

The concept of EFTs cannot only be applied in the top-down approach. Looking from the other side, it can also be used for situations in which the UV theory is unknown, providing a model-independent tool for data analysis. This is the so-called bottom-up approach \cite{Georgi:1994qn}, which we discuss in Section \ref{ch:EFT.bottom-up}. We write down a consistent basis of operators. This enables us to describe effects at a given scale without needing to know what happens at higher scales. Any model of UV-physics can be mapped to the Wilson coefficients of the operators in the bottom-up basis, see {\it e.g.} \cite{Henning:2014wua}. An example for a decoupling bottom-up EFT will be the SM-EFT that we discuss in Chapter \ref{ch:SMEFT}. A bottom-up non-decoupling EFT is the electroweak chiral Lagrangian, which we discuss in Chapter \ref{ch:ewXL}. The bottom-up EFT picture allows us further to interpret non-renormalizable Lagrangians physically \cite{Burgess:2007pt}: For a given accuracy (a given order in $v/\Lambda$), we will need only a finite number of parameters, making the theory predictive \cite{Georgi:1994qn}. Figure \ref{fig:EFT.topdownvsbottomup} illustrates the top-down and bottom-up approach. 
\section{Top-Down and Bottom-Up Approach to Effective Field Theories}
 \subsection{The Top-Down Approach}
\label{ch:EFT.top-down}
In the top-down \cite{Georgi:1994qn} approach to effective field theories, we know the high-energy (UV) theory and we are interested in low-energy (IR) effects only. The resulting EFT simplifies the computations a lot --- making the computations sometimes even feasible in the first place. In this section, we follow closely the arguments of \cite{Burgess:2007pt,Peskin:1995ev}. Experiments at energies below an energy scale $\Lambda$ never produce particles of mass $\Lambda$ as external states. The only contribution of these particles comes through virtual effects. To study these effects systematically, we need the generating functional $\Gamma[\varphi]$ of one-particle irreducible (1PI) correlation functions. One-particle irreducible diagrams cannot be broken into two disconnected diagrams {\it via} cutting a single internal line. The 1PI generating functional is given by the Legendre transformation of the generating functional for connected graphs, $W[J(\varphi)]$:
\begin{equation}
  \label{eq:3.1.1}
  \Gamma[\varphi] \equiv W[J(\varphi)] - \int d^{4}x \; \varphi J.
\end{equation}
The mean field $\varphi$ is defined in presence of a source $J$,
\begin{equation}
  \label{eq:3.1.2}
  \varphi = \frac{\delta W}{\delta J} = \langle \phi (x)\rangle_{J}. 
\end{equation}
The generating functional of connected graphs, $W[J]$, is defined as
\begin{equation}
  \label{eq:3.1.3}
  \exp{(i W[J])} = \int \mathcal{D}\phi \; \exp{\{i\int d^{4}x\; \mathcal{L}(\phi)+J\phi\}}. 
\end{equation}
A general $n$-point correlation function is then defined as
\begin{equation}
  \label{eq:3.1.4}
  \langle \phi(x_{1})\dots\phi(x_{n})\rangle_{J} = (-i)^{n}\frac{\delta^{n}i W[J]}{\delta J(x_{1})\dots\delta J(x_{n})}.
\end{equation}
The generating functional $\Gamma[\varphi]$ contains all physical predictions of the theory \cite{Peskin:1995ev}. Its stationary point $\left.\frac{\delta\Gamma}{\delta\varphi}\right|_{J}=0$ gives the vacuum expectation value of the field $\phi$. The second derivative of $\Gamma$ gives the inverse propagator, whose zeroes give the masses of the particles in the theory. Higher derivatives give the 1PI amplitudes that can be used to compute S-matrix elements. 

A convenient way to compute $\Gamma[\varphi]$ uses the background field method \cite{Abbott:1981ke,Abbott:1980hw}. It is equivalent to the saddle-point approximation for non-gauge fields. Starting from the definition of $W[J]$ in Eq.~\eqref{eq:3.1.3} above, we expand $\phi$ around its classical solution \cite{Peskin:1995ev}, $\phi = \varphi + \eta$. Here, $\varphi$ is defined as in Eq.~\eqref{eq:3.1.2} and $\eta$ are the quantum fluctuations of the field. Equation~\eqref{eq:3.1.3} now takes the form
\begin{align}
\begin{aligned}
  \label{eq:3.1.5}
  \exp{(i W[J])} = \int \mathcal{D}\eta \; &\exp{\Big\{i\int d^{4}x\; \left(\mathcal{L}(\varphi)+J\varphi\right) + i\int d^{4}x\; \eta(x) \left(\frac{\delta \mathcal{L}}{\delta\phi(x)}+J\right)_{\phi=\varphi}} \\
 &+ \frac{i}{2}\int d^{4}x\, d^{4}y\; \eta(x)\eta(y)\left.\frac{\delta^{2}\mathcal{L}}{\delta\phi(x)\delta\phi(y)}\right|_{\phi=\varphi}+\dots \Big\}.
\end{aligned}
\end{align}
The integration of the first term gives a constant factor. The second term vanishes for the tree-level approximation of $\varphi$ in presence of the source $J$, upon using the classical equations of motion. However, $\varphi$ of Eq.~\eqref{eq:3.1.2} is defined at all orders in perturbation theory, spoiling the cancellation. Nevertheless, we can write $J(x) = J_{1}(x)+\delta J(x)$, such that $\left(\delta \mathcal{L}/\delta\phi(x)+J_{1}\right)_{\phi=\varphi}$ vanishes exactly \cite{Peskin:1995ev}. The difference, $\delta J(x)$, will start to contribute at the loop level, similar to a counterterm, see \cite{Peskin:1995ev}. The third term of Eq.~\eqref{eq:3.1.5} can be evaluated as a Gaussian, yielding
\begin{equation}
  \label{eq:3.1.6}
  \exp{(i W[J])} = \exp{\left\{i\int d^{4}x\; \left(\mathcal{L}(\varphi)+J\varphi\right)\right\} }\; \left(\text{det }\left.-\frac{\delta^{2}\mathcal{L}}{\delta\phi\delta\phi}\right|_{\phi=\varphi}\right)^{\mp\tfrac{1}{2}}(1+\dots).
\end{equation}
The sign of the exponent of the determinant is $(-)$ for bosonic and $(+)$ for fermionic fields. From this, we find explicitly for Eq.~\eqref{eq:3.1.1}:
\begin{equation}
  \label{eq:3.1.7}
  \Gamma[\varphi] = S[\varphi] \mp \frac{1}{2} \log{\text{det }\left[-\frac{\delta^{2}\mathcal{L}}{\delta\phi\delta\phi}\right]_{\phi=\varphi}} + \dots 
\end{equation}
Comparing this to a diagrammatic way of obtaining $\Gamma[\varphi]$, we see that the first term, the action $S[\varphi]$, comes from tree-level contributions. The second term comes from one-loop diagrams. The dots collect terms of higher order. 

Our goal will now be to construct $\Gamma[\varphi]$ of the UV theory and restrict it to cases where only light degrees of freedom with low momenta appear as external states. To be more precise, consider the case of two types of fields: light fields $\ell$, coupled to a source $j$, and heavy fields $H$, coupled to a source $J$. The 1PI generating functional of this theory is $\Gamma[l,h]$, where $l = \langle \ell\rangle_{jJ}$ and $h= \langle H \rangle_{jJ}$. Having no external $H$ fields is equivalent to setting $J=0$, as we will never vary $\Gamma$ with respect to $J$. Since $\delta\Gamma/\delta h = -J$, the case of vanishing $J$ is equivalent to evaluating $\Gamma[l,h]$ at the point $h=\bar{h}(l)$ where
\begin{equation}
  \label{eq:3.1.8}
  \left.\frac{\delta\Gamma[l,h]}{\delta h} = 0 \right|_{h=\bar{h}}.
\end{equation}
In the low-energy limit, we are also not interested in the high-frequency components of $\ell$, as they are also never produced. Therefore, we also require $\delta\Gamma[l,h]/\delta l = 0$, for the high-frequency components. Let $\gamma[l]$ be the generating functional that satisfies these conditions. Since there is no explicit heavy field in the description any more, we say we have ``integrated out'' the heavy degree of freedom from the theory. The functional $\gamma[l]$ is the one-light-particle-irreducible generating functional. It generates 1PI graphs for low energetic fields $\ell$. 

The tree-level approximation to $\gamma[l]$ is now given by $S[l,\bar{h}(l)]$, where $\bar{h}$ is given by Eq.~\eqref{eq:3.1.8}. This means we solve the equations of motion of $H$ (and also the high-energy modes of $\ell$) in terms of the low-energy modes of $\ell$. 

The one-loop result gets two contributions. First, the functional form of $\Gamma[l,h]$ changes when the one-loop terms are included. Second, these imply a redefinition of the stationary point in Eq.~\eqref{eq:3.1.8}. 
\begin{align}
\begin{aligned}
  \label{eq:3.1.9}
  \gamma[l] &= \Gamma_{\text{tree}}[l,\bar{h}_{\text{tree}} + \bar{h}_{\text{1-loop}}] + \Gamma_{\text{1-loop}}[l,\bar{h}_{\text{tree}}]+\dots\\
 &= \Gamma[l,\bar{h}_{\text{tree}}]+ \bar{h}_{\text{1-loop}}\cdot\left.\frac{\delta\Gamma}{\delta h}\right|_{h=\bar{h}_{\text{tree}}} + \Gamma_{\text{1-loop}}[l,\bar{h}_{\text{tree}}]+\dots
\end{aligned}
\end{align}
However, the second term of the second line vanishes at the considered order \cite{Burgess:2007pt}. Summarizing this, we write the low-energy generating functional as
\begin{equation}
  \label{eq:3.1.10}
  \gamma[l] = S[l,\bar{h}_{\text{tree}}(l)] + \Gamma_{\text{1-loop}}[l,\bar{h}_{\text{tree}}(l)]+\dots
\end{equation}

In this definition of $\gamma[l]$, we see that we need to solve the equations of motion of the heavy field in a first approximation. This amounts to solving 
\begin{equation}
  \label{eq:3.1.11}
  (\Box + M^{2})^{-1}\approx \frac{1}{M^{2}} \left(1-\frac{\Box}{M^{2}}+\frac{\Box^{2}}{M^{4}}-\dots\right).
\end{equation}
From this expansion we see some aspects of the low-energy EFT.
\begin{itemize}
\item The non-local interactions involving heavy fields of the full theory become local interactions in the EFT. This is connected to the uncertainty principle, the high energies needed to produce the heavy fields are only ``available'' for very short times, $\Delta t \sim 1/ \Delta E$, making them local.
\item The effects of the heavy field come with factors of $1/M^{2}$. In theories where the couplings in the equations of motion do not grow in the limit $M\rightarrow\infty$, the heavy fields decouple as stated by Appelquist and Carazzone \cite{Appelquist:1974tg}. In theories in which the couplings grow with $M$, for example because of spontaneous symmetry breaking or mixing effects, the decoupling does not take place. Reinserting the solution of the equation of motion in the Lagrangian generates non-renormalizable interactions without $1/M$ suppression. We see this in detail in Section~\ref{ch:SM+S}. 
\item Symmetries of the light fields in the full theory are still symmetries of the effective Lagrangian. 
\end{itemize}

Orthogonally to the functional approach we just presented, we can also integrate out the heavy field by diagrammatic methods. In this approach, we consider the amplitude of a given process explicitly in the UV theory. Then, we expand in $1/M^{2}$ and match to the amplitude of the same process in the low-energy EFT. If we do this for all processes, we also arrive at Eq.~\eqref{eq:3.1.10}. Otherwise, we are restricted to the given subset of processes. The aspects of the low-energy EFT discussed above also hold if we integrate out the field diagrammatically. 

The procedure of integrating out a heavy field can also be applied for several different mass scales consecutively \cite{Georgi:1994qn,Henning:2014wua}. Starting at a high scale $\Lambda_{1}$, we evolve the parameters to the scale $\Lambda_{2}<\Lambda_{1}$ of the heaviest particle, using the renormalization group equations (RGE). This particle is then integrated out, either {\it via} Eq.~\eqref{eq:3.1.10} or {\it via} the diagrammatic method. The effective theory of the remaining fields is further evolved using the corresponding RGE until the next threshold $\Lambda_{3}<\Lambda_{2}$ is reached and particles with masses $\Lambda_{3}$ are integrated out.

If the action is expanded in terms of a small parameter, applying the equations of motion of a field in $\gamma$ does not change the observables at a given order in the small parameter. This can be seen from a field redefinition $\phi(x) \rightarrow\tilde\phi = \phi(x) -\varepsilon^{n} f(x)$. The action $S$ becomes
\begin{equation}
  \label{eq:3.1.12}
  S[\tilde\phi] = S[\phi] - \varepsilon^{n}f(x) \frac{\delta S}{\delta \phi} + \mathcal{O}(\varepsilon^{n+1})
\end{equation}
An appropriate choice of $f(x)$ at the order $\varepsilon^{n}$ corresponds to applying the equations of motion in $S$ and further corrections come at order $\varepsilon^{n+1}$. In general, canonical field redefinitions in the action do not change scattering matrix elements \cite{Haag:1958vt,Donoghue:1992dd}.
\subsection{The Bottom-Up Approach}
\label{ch:EFT.bottom-up}
In the bottom-up approach, the UV theory is either unknown, or it is known but it is impossible to find its low-energy description in top-down approach. The latter is the case for QCD, where we have different degrees of freedom at high energies (quarks and gluons) and at low energies (pions, kaons, {\it etc.}). The application of bottom-up EFTs in cases where the UV is unknown is very convenient, as no commitment to a specific model and therefore only a few assumptions are made. Instead, the model-independent bottom-up approach focusses on what we know and what we see at the current experimental scale.\\

From the discussion of the last preceding section, we see that the effects of the high-energy physics are encoded in a series of operators that are composed of the low-energy fields and ordered in a systematic expansion \cite{Ovrut:1979pk,Ovrut:1980eq}. This tells us what we need to build the bottom-up effective field theory: The particle content at the given energy scale, the symmetries that these particles obey, and a power counting that defines a consistent expansion. The coefficients of the operators, called Wilson coefficients, can be specified for a given model, see \cite{Henning:2014wua}. In a model-independent analysis, they are free parameters to be determined by experiment. 
The first ingredient of the bottom-up EFT, the particle content, is rather easily found: We need to specify which degrees of freedom are present and propagating at the chosen energy scale. 

For the symmetries, two different assumptions can be made. Either, we can assume the low-energy symmetry also holds in the UV, as usually is the case for gauge symmetries, or we can assume that the new-physics sector breaks the symmetry. The higher order operators will therefore also violate the symmetry at some point. $\mathcal{CP}$-symmetry is an example for the second kind. In any case, the underlying assumptions regarding the symmetries should be spelled out clearly. 

The power counting gives the expected (natural) size of the Wilson coefficient of an effective operator. Additional symmetries of the UV can suppress some coefficients below that size. From the general discussion of decoupling and non-decoupling EFTs we see that there are two different types of power counting. In a decoupling EFT, the leading-order Lagrangian is renormalizable and the effects form the UV are suppressed by $1/\Lambda$. The expansion is therefore given by canonical dimensions. Higher order operators have a larger canonical dimension and are suppressed by higher powers of $1/\Lambda$, as the energy dimension of the product of operator and coefficient must always be equal to four. The scale of suppression, $\Lambda$, is the same for all operators. We identify it with the lowest-lying scale of new physics. If, in a particular UV-model, the operator is generated by effects from a higher scale $\Lambda_{2}>\Lambda$, the bottom-up analysis can still be done in terms of $\Lambda$ alone, without loss of generality. The Wilson coefficient of the corresponding operator is then of $\mathcal{O}(\Lambda/\Lambda_{2})$.

In non-decoupling EFTs, the leading-order Lagrangian usually contains operators of canonical dimension larger than four, making it non-renormalizable already at leading order. Therefore, an expansion in canonical dimensions cannot consistently be done. Instead, the renormalization procedure gives a guideline for a consistent expansion: The one-loop diagrams built from leading-order vertices need to be renormalized. Counterterms that are needed, but not included in the leading-order Lagrangian, will be included at next-to-leading order. This makes the theory renormalizable order by order in a loop expansion. In this expansion, the cutoff of the theory, $\Lambda$, is identified with $4\pi v$ \cite{Manohar:1983md,Georgi:1985kw,Georgi:1992dw}, where $v$ is the low-energy scale. This identification puts one-loop diagrams of leading-order Lagrangian parametrically at the same order as the next-to-leading order tree-level diagrams, $v^{2}/\Lambda^{2} = 1/16\pi^{2}$. This defines a consistent power counting for a non-decoupling effective field theory. 

We conclude from this discussion that the assumptions on the new physics consistently define the Lagrangian at leading order and the power counting. Given a set of assumptions, we cannot simply choose a leading-order Lagrangian or a power counting at our will. They are always connected, as the power counting is homogenous for the leading-order Lagrangian in order to have an unambiguously defined expansion.
\section{A Toy Example}
After this formal introduction we would like to illustrate the concept of top-down and bottom-up effective field theories in the light of an explicit example. Consider the following Lagrangian of a light scalar $\ell$ and a heavy scalar $H$ \cite{Burgess:2007pt}.
\begin{align}
\begin{aligned}
  \label{eq:3.2.1}
  \mathcal{L}&= \frac{1}{2}\partial_{\mu}\ell \partial^{\mu}\ell + \frac{1}{2}\partial_{\mu}H \partial^{\mu}H - \frac{1}{2} m^{2}\ell^{2} - \frac{1}{2}M^{2}H^{2}\\
  &- \frac{g_{\ell}}{4!}\ell^{4} - \frac{g_{H}}{4!}H^{4} - \frac{g_{\ell H}}{4}\ell^{2}H^{2} - \frac{\tilde{m}}{2}\ell^{2}H - \frac{\tilde{g}_{H}}{3!}M H^{3}
\end{aligned}
\end{align}
For experiments at energies $E\sim (m,\tilde{m}) \ll M$ we can formulate an effective, low-energy Lagrangian --- a top-down effective field theory. Integrating out the heavy scalar is in a first approximation given by the first term of Eq.~\eqref{eq:3.1.10}. We therefore need the equations of motion:
\begin{subequations}
  \label{eq:3.2.2}
\begin{equation}
\label{eq:3.2.2a}
  (\Box + m^{2})\ell= - \frac{g_{\ell}}{3!}\ell^{3} - \frac{g_{\ell H}}{2}H^{2}\ell - \tilde{m} \ell H
\end{equation}
\begin{equation}
\label{eq:3.2.2b}
  (\Box + M^{2})H  = - \frac{g_{H}}{3!}H^{3} - \frac{g_{\ell H}}{2}H\ell^{2} - \frac{\tilde{m}}{2}\ell^{2} - \frac{\tilde{g}_{H}}{2}M H^{2}
\end{equation}
\end{subequations}
The second equation can be solved for $H$, order by order in $1/M^{2}$. We find
\begin{equation}
\label{eq:3.2.3}
H = -\frac{\tilde{m}}{2M^{2}}\ell^{2} + \frac{g_{\ell H}\tilde{m}}{4M^{4}}\ell^{4} - \frac{\tilde{m}}{2M^{4}}\Box(\ell^{2}) + \mathcal{O}(\frac{1}{M^{6}}),
\end{equation}
by using Eq.~\eqref{eq:3.1.11}. Inserting this back into Eq.~\eqref{eq:3.2.1}, we find
\begin{equation}
  \label{eq:3.2.4}
  \mathcal{L_{\text{eff}}}= \frac{1}{2}\partial_{\mu}\ell \partial^{\mu}\ell - \frac{1}{2} m^{2}\ell^{2} -\frac{g_{\ell}}{4!}\ell^{4}+ \frac{\tilde{m}^{2}}{8M^{2}}\ell^{4} - \frac{g_{\ell H}\tilde{m}^{2}}{16M^{4}}\ell^{6} - \frac{\tilde{m}^{2}}{8M^{4}}\ell^{2}\Box(\ell^{2}).
\end{equation}
The last term, however, is not independent. It can be rewritten using integration by parts as well as the equations of motion of $\ell$.
\begin{equation}
  \label{eq:3.2.5}
  \ell^{2}\Box (\ell^{2}) = -4 \ell^{2} \partial_{\mu}\ell \partial^{\mu}\ell = \frac{4}{3}\ell^{3}\Box\ell
\end{equation}
Finally, we obtain the effective low-energy Lagrangian for the light fields $\ell$ \cite{Burgess:2007pt}, up to terms of $\mathcal{O}(1/M^{6})$:
\begin{equation}
  \label{eq:3.2.6}
  \mathcal{L_{\text{eff}}}= \frac{1}{2}\partial_{\mu}\ell \partial^{\mu}\ell - \frac{1}{2} m^{2}\ell^{2} -\frac{1}{4!}\left(g_{\ell}-3\frac{\tilde{m}^{2}}{M^{2}}-4\frac{m^{2}\tilde{m}^{2}}{M^{4}}\right)\ell^{4} + \frac{\tilde{m}^{2}}{M^{4}}\left(\frac{g_{\ell}}{36}-\frac{g_{\ell H}}{16}\right)\ell^{6} 
\end{equation}
Let us now analyze the same scenario in the bottom-up picture. At very low energies, we have a renormalizable theory of only $\ell$ fields. Their dynamics are described by
\begin{equation}
  \label{eq:3.2.7}
  \mathcal{L} = \frac{1}{2}\partial_{\mu}\ell \partial^{\mu}\ell - \frac{1}{2} m^{2}\ell^{2} -\frac{g_{\ell}}{4!}\ell^{4},
\end{equation}
which gives the equation of motion for $\ell$:
\begin{equation}
  \label{eq:3.2.8}
  \Box \ell = -m^{2} \ell - \frac{g_{\ell}}{3!}\ell^{3}
\end{equation}
The indirect effects of the heavy field $H$ are encoded in effective operators. Since the Lagrangian in Eq.~\eqref{eq:3.2.7} is renormalizable, we have a decoupling EFT with dimensional power counting. The leading new physics effects arise at the level of dimension-six operators, as dimension-five operators are forbidden by the $Z_{2}$ symmetry of the field $\ell$. The building blocks of the operators are just derivatives and the light field $\ell$. We find the following list of operators at this order\footnote{Derivatives of higher powers of the fields can always be reduced to a combination of operators where the derivatives act on a single field $\ell$.}:
\begin{equation}
    \label{3.2.9}
    \ell^{6}, \qquad (\partial_{\mu}\ell)(\partial^{\mu}\ell)\ell^{2}, \qquad (\Box\ell)\ell^{3}, \qquad (\Box\ell) (\Box\ell)
\end{equation}
Using integration by parts and the equations of motion in Eq.~\eqref{eq:3.2.8}, we find only one independent operator. The next-to-leading order (NLO) Lagrangian is then given by
\begin{equation}
  \label{eq:3.2.10}
  \mathcal{L}_{\text{NLO}} = \frac{c_{6}}{\Lambda^{2}}\ell^{6}.
\end{equation}
This operator, together with corrections of $\mathcal{O}(1/\Lambda^{2})$ of the leading order $\ell^{4}$ interaction describe the leading effects of new physics on the interactions of the scalar $\ell$. 

By explicitly comparing the effective Lagrangians of the top-down and the bottom-up approach, we find the following matching conditions. First, we identify the cutoff scale $\Lambda$, with the mass of the heavy scalar, $M$. When the experimental energies reach this scale, the heavy particle is produced explicitly and the effective description breaks down. Further, we find an explicit formula for the Wilson coefficient $c_{6}$ as well as for the shift of $g_{\ell}$ due to $\mathcal{O}(1/M^{2})$ effects. 
\begin{equation}
  \label{eq:3.2.11}
  c_{6} = \frac{\tilde{m}^{2}}{M^{2}}\left(\frac{g_{\ell}}{36}-\frac{g_{\ell H}}{16}\right), \qquad \qquad \Delta g_{\ell} = 3\frac{\tilde{m}^{2}}{M^{2}}+4\frac{m^{2}\tilde{m}^{2}}{M^{4}}
\end{equation}
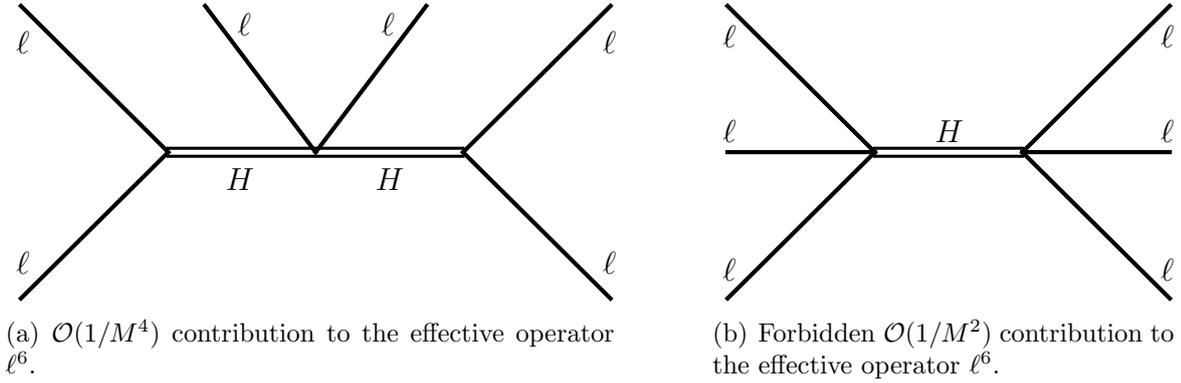
\begin{figure}[t]
\begin{center}
\subfigure[$\mathcal{O}(1/M^{4})$ contribution to the effective operator $\ell^{6}$. \label{fig:EFT.ex1}]{
\begin{overpic}[height=4cm]{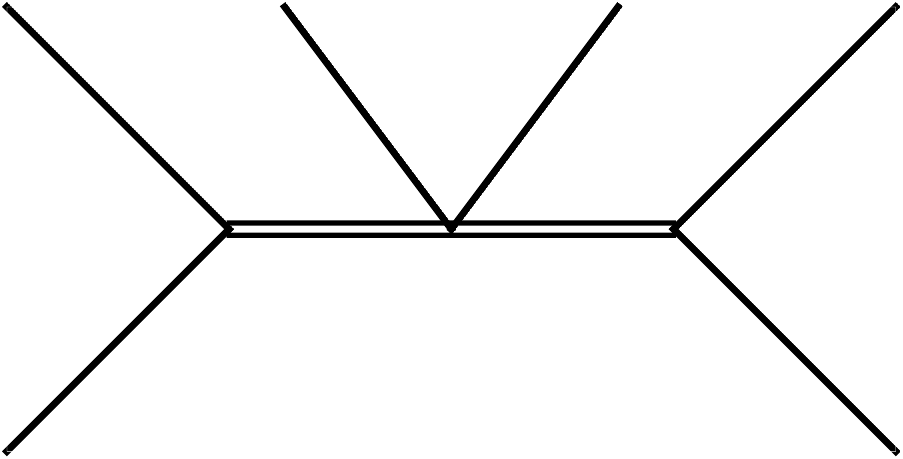}
\put (0,5){$\ell$}
\put (0,42){$\ell$}
\put (37,45){$\ell$}
\put (61,45){$\ell$}
\put (98,42){$\ell$}
\put (98,5){$\ell$}
\put (35,19){$H$}
\put (60,19){$H$}
\end{overpic}}\hfill
\subfigure[Forbidden $\mathcal{O}(1/M^{2})$ contribution to the effective operator $\ell^{6}$. \label{fig:EFT.ex2}]{
\begin{overpic}[height=4cm]{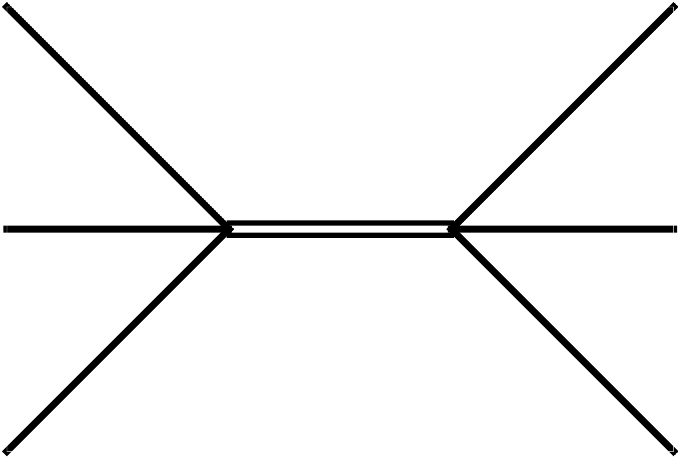}
\put (0,5){$\ell$}
\put (0,36){$\ell$}
\put (0,57){$\ell$}
\put (97,5){$\ell$}
\put (97,36){$\ell$}
\put (97,57){$\ell$}
\put (47,36){$H$}
\end{overpic}}\hfill
\end{center}
\caption[Diagrammatic illustration of some aspects of the toy model.]{Diagrammatic illustration of some aspects of the example.}
\label{fig:EFT.example}
\end{figure}
We can also understand the contributions to the effective Lagrangian diagrammatically. Integrating out a heavy field at tree-level corresponds to drawing tree-level diagrams with internal heavy fields and then expanding the propagator in terms of $1/M^{2}$. The contribution of $c_{6}$ that is proportional to $g_{\ell H}$ comes, for example, from the diagram in Fig. \ref{fig:EFT.ex1}. The contribution that is proportional to $g_{\ell}$ comes from a tree-level two-to-two diagram {\it via} a term proportional to $\Box\ell$ and applying the equations of motion of Eq.~\eqref{eq:3.2.8}. The diagrammatic approach also shows why there is no contribution to $c_{6}$ that comes with a single suppression in $1/M^{2}$. In order to have such a contribution, we need a diagram with six external $\ell$ fields and a single heavy field propagator. Such a diagram is given in Fig. \ref{fig:EFT.ex2}. Since the involved vertices violate the $Z_{2}$ symmetry of the theory, no such contribution can exist. However, the latter information comes from the specific UV model considered here and goes beyond a general bottom-up analysis. 
\section{Chiral Perturbation Theory}
\label{ch:chiPT}
Due to the strong dynamics of QCD, different approaches to perturbation theory of quarks and gluons needed to be developed. Below the mass of the $\rho$-meson, the resonance region, the theory can be described in terms of the eight pseudoscalar mesons $\pi, K $ and $\eta$, as well as symmetry relations among them \cite{Weinberg:1978kz,Gasser:1983yg,Gasser:1984gg,Leutwyler:1993iq,Pich:1995bw}. These eight mesons can be identified as an octet of light pseudo-Goldstone bosons, coming from the spontaneous and explicit breaking of chiral symmetry. The pseudo-Goldstone nature ensures that they are naturally light compared to the other resonances of the hadronic spectrum. With the information about the particle spectrum and the symmetries, we can build an EFT --- known as chiral perturbation theory.

The Lagrangian of QCD,
\begin{equation}
  \label{eq:3.3.2}
  \mathcal{L}_{\text{QCD}} = -\frac{1}{4}G^{a}_{\mu\nu}G^{\mu\nu}_{a} + i \bar{q}_{L}\gamma^{\mu} D_{\mu} q_{L} +  i \bar{q}_{R}\gamma^{\mu} D_{\mu} q_{R}
\end{equation}
exhibits a global $SU(3)_{L}\times SU(3)_{R}$ chiral symmetry of the three lightest quarks, as discussed in Section \ref{ch:SMsym}. The quark condensate in the vacuum spontaneously breaks \cite{'tHooft:1979bh} this symmetry down to the vectorial subgroup, $SU(3)_{L+R}$. Since we only consider the Goldstone fields and not the massive excitations, the Goldstone symmetry will be non-linearly realized. We write the eight Goldstone bosons, associated with the breaking of this symmetry, as in the non-linear sigma model, in terms of the matrix
\begin{equation}
  \label{eq:3.3.3}
  U= \exp{\left(i\frac{\Phi}{F_{\pi}}\right)},
\end{equation}
where
\begin{equation}
  \label{eq:3.3.4}
  \Phi=\sum_{a=1}^{8}\lambda_{a}\varphi_{a}=\sqrt{2}\begin{pmatrix}\frac{\pi^{0}}{\sqrt{2}} + \frac{\eta_{8}}{\sqrt{6}} & \pi^{+}& K^{+} \\ \pi^{-} & -\frac{\pi^{0}}{\sqrt{2}} + \frac{\eta_{8}}{\sqrt{6}} & K^{0}\\K^{-} & \bar{K}^{0} & -2 \frac{\eta_{8}}{\sqrt{6}}\end{pmatrix}
\end{equation}
and $F_{\pi}$ is the pion decay constant $F_{\pi}\approx 92~\text{MeV}$. Under the $SU(3)_{L}\times SU(3)_{R}$ symmetry, the $U$ transforms as
\begin{equation}
  \label{eq:3.3.4a}
  U \rightarrow g_{L} U g_{R}^{\dagger}, \qquad \text{where } g_{L,R} \in SU(3)_{L,R}. 
\end{equation}
Goldstone bosons with a non-linearly realized symmetry can always be brought to the exponential representation, as was shown in \cite{Coleman:1969sm,Callan:1969sn,Feruglio:1992wf}. The $\lambda_{a}$ in Eq.~\eqref{eq:3.3.4} are the Gell-Mann matrices. They are related to the $T_{a}$, the (traceless) generators of $SU(3)$, {\it via} $T_{a} = \tfrac{1}{2} \lambda_{a}$ and normalized to $\langle T_{a} T_{b} \rangle = \frac{1}{2} \delta_{ab}$. 
In general, the Goldstones can couple to external (classical) fields. These can include left-handed ($l_{\mu}$) or right-handed ($r_{\mu}$) vector currents or (pseudo) scalar fields ($\hat{p}$)$\hat{s}$. All of these are hermitean and matrices in flavor space. 

Weinberg showed in \cite{Weinberg:1978kz} that the effective expansion in chiral perturbation theory is equivalent to an expansion in momenta. The leading-order Lagrangian is of order $p^{2}$ and can be written as:
\begin{equation}
  \label{eq:3.3.7}
  \mathcal{L}_{\text{LO}} = \frac{F_{\pi}^{2}}{4} \langle D_{\mu}U^{\dagger} D^{\mu}U \rangle + \frac{F_{\pi}^{2}}{4} \langle U^{\dagger} \chi + \chi^{\dagger}U\rangle,
\end{equation}
where $\chi = 2 B_{0} (\hat{s}+i \hat{p})$. The covariant derivative of $U$ is given by
\begin{equation}
  \label{eq:3.3.8}
  D_{\mu} U = \partial_{\mu} U -i l_{\mu}U + i U r_{\mu}. 
\end{equation}
Including the $U(1)$ gauge symmetry of QED, the Lagrangian becomes \cite{Urech:1994hd,Knecht:1999ag,Neufeld:1995mu,Meissner:1997fa,Knecht:1997jw,Ecker:2000zr}
\begin{equation}
  \label{eq:3.3.9}
  \mathcal{L}_{\text{LO}} = -\frac{1}{4}F_{\mu\nu}F^{\mu\nu}+\frac{F_{\pi}^{2}}{4} \langle D_{\mu}U^{\dagger} D^{\mu}U \rangle + \frac{F_{\pi}^{2}}{4} \langle U^{\dagger} \chi + \chi^{\dagger}U\rangle +C\langle Q U Q U^{\dagger}\rangle,
\end{equation}
where gauge fixing terms have been omitted. $F_{\mu\nu}$ is the usual abelian field strength tensor. $Q$ is the charge matrix of the quarks, $Q = \tfrac{e}{3}\,\text{diag}(2,-1,-1) =\tfrac{e}{2}\left(\lambda_{3}+\frac{1}{\sqrt{3}}\lambda_{8}\right)$. The covariant derivative of $U$ changes to
\begin{equation}
  \label{eq:3.3.10}
  D_{\mu} U = \partial_{\mu} U -i QA_{\mu}U + i U Q A_{\mu}.
\end{equation}
For simplicity, we set the external vector currents to zero. We also neglect the external pseudoscalar current $\hat{p}$. The scalar $\hat{s}$ reduces to the mass matrix of the quarks, $\hat{s} = \text{diag}(m_{u},m_{d},m_{s})$. The last term in Eq.~\eqref{eq:3.3.9} is a potential for the Goldstones that is induced radiatively by photon loops. Its naive scale $\Lambda^{2}$ gets suppressed by a loop factor of $1/16\pi^{2}$ \cite{Das:1967it}. Terms with $Q^{4}$ scale only as $(\log{\Lambda})/16\pi^{2}$ and are therefore suppressed. Electromagnetism and the quark masses break the chiral symmetry of Eq.~\eqref{eq:3.3.4a} explicitly. The corresponding spurions $Q$ and $\chi$ transform as
\begin{equation}
\label{3.3.11}
\chi \rightarrow g_{L} \chi g^\dagger_{R} ,\qquad  Q_{L}\rightarrow g_{L} Q_{L} g^\dagger_{L},\qquad  Q_{R}\rightarrow g_{R} Q_{R} g^\dagger_{R},  
\end{equation} 
with the identification $Q_L=Q_R=Q$.

The leading-order Lagrangian, Eq.~\eqref{eq:3.3.9}, is non-renormalizable, as the presence of the exponential $U$ induces operators at arbitrarily high canonical dimension. The power counting of the EFT is therefore given by the loop counting of non-decoupling EFTs. 

An $L$-loop diagram with $B$ external Goldstones, $X$ external gauge fields and $\lambda$ external $\chi$ fields scales as \cite{Buchalla:2013eza}

\begin{align}
  \label{eq:3.3.12}
  \begin{aligned}
    D&\sim \frac{F_{\pi}^{2}}{\Lambda^{2L}}\;  \left(\frac{\varphi}{F_{\pi}}\right)^{B} \; \left(\frac{F_{\mu\nu}}{F_{\pi}}\right)^{X} \; \left(\chi \right)^{\lambda} \; \left(F_{\pi}Q\right)^{\tau+2\delta} \; \left(\frac{CQ^{2}}{F_{\pi}^{2}}\right)^{\rho} \;p^{d_{p}},
  \end{aligned}
\end{align}
where $\tau(\delta)$ gives the number of vertices involving one (two) gauge field(s) and $\rho$ gives the number of vertices coming from the Goldstone boson potential. We find this formula using topological identities \cite{Buchalla:2012qq,Buchalla:2013rka,Buchalla:2013eza,mastersthesis} of Feynman diagrams, such as the conservation of ends (of lines) and the Euler characteristic of planar graphs. The power of momentum $d_{p}$ in Eq.~\eqref{eq:3.3.12} scales as
\begin{equation}
  \label{eq:3.3.13}
  d_{p}= 2L+2 - X - 2\lambda -\tau - 2\delta - 2\rho.
\end{equation}
It gives the superficial degree of divergence, indicating when a loop diagram might becomes divergent and requires a counterterm. It is bounded from above, giving a finite number of counterterms at a given loop order. Since the counterterms are needed to renormalize the theory at the given (loop) order, we conclude that the NLO basis should at least contain these operators. Finding the NLO operators therefore amounts to identifying the counterterms. For consistency, we expect the coefficients of the  NLO operators to be of the same size as the counterterms \cite{Manohar:1983md,Georgi:1985kw,Georgi:1992dw}, $\mathcal{O}(F_{\pi}^{2}/\Lambda^{2}) = \mathcal{O}(1/16\pi^{2})$. This links the cutoff $\Lambda$ to the scale of low-energy physics, the decay constant $F_{\pi}$. From Eq.~\eqref{eq:3.3.13}, we can identify the classes of one-loop counterterms. When working in dimensional regularization, we see that $d_{p}$ yields also the number of derivatives of the operator. We find the following Lorentz-invariant classes of counterterms ($d;X,\lambda,2\rho+\tau+2\delta$):
\begin{equation}
  \label{eq:3.3.14}
  \begin{array}{llrl}
    (4;0,0,0): & UD^{4}        & (2;0,1,0): & UD^{2}\chi\\
    (2;0,0,2): & UD^{2}Q^{2} \hspace{2cm}& (0;0,1,2): & U\chi Q^{2}\\
    (0;0,0,4): & UQ^{4}        & (2;1,0,1): & D^{2}UFQ\\
    (0;0,2,0): & U\chi^{2}    & (0;2,0,2): & UF^{2}Q^{2}\\
  \end{array}
\end{equation}
Here, the labels correspond to the fields being present in the operator to the indicated power, $F$ stands for the field strength tensor instead of the field $A$ itself. The only exception holds for $U$, where no limit on the power is present. Due to the structure of the interactions, there needs to be a $Q$ with every $F$ (or $A$), whereas additional powers of $Q$ are always possible, due to internal photon lines. Without QED and for vanishing quark masses, the power counting formula above reduces to $d_{p} = 2L+2$, reflecting the statement that chiral perturbation theory is an expansion in derivatives or equivalently in momenta \cite{Weinberg:1978kz}. \\

The classes of counterterms in Eq.~\eqref{eq:3.3.14} can also be found using the concept of chiral dimensions \cite{Buchalla:2013eza}, which is equivalent to the loop counting. Rewriting Eq.~\eqref{eq:3.3.13}, we find
\begin{equation}
  \label{eq:3.3.15}
  2L+2 =d_{p}+ X + 2\lambda +\tau + 2\delta + 2\rho.
\end{equation}
As the order of the expansion is given by the number of loops, $L$, the left-hand side encodes the (chiral) order of the considered operator. The right-hand side tells us how the different objects in the operator contribute to the chiral order. We assign the following chiral dimensions \cite{Buchalla:2013eza}:
\begin{equation}
\label{eq:3.3.16}
  \begin{array}{lll}
    [U]_{\chi} = 0 \hspace{1.5cm} & [D_{\mu}]_{\chi} = 1 \hspace{1.5cm} & [\chi]_{\chi} = 2 \\
    {[A_{\mu}]_{\chi}=0}\hspace{1.5cm} &[Q]_{\chi} = 1 \hspace{1.5cm}&[\text{const.}]_{\chi}  = 0 \\
  \end{array}
\end{equation} 
These assignments have been used before \cite{Urech:1994hd,Nyffeler:1999ap,Hirn:2004ze,Hirn:2005fr,Hirn:2005sj}. However, they were not derived from a diagrammatical power counting, but rather from a homogenous counting of the leading-order Lagrangian. We see that $[\mathcal{L}_{\text{LO}}]_{\chi}=2$, using Eq.~\eqref{eq:3.3.16}. Operators of chiral order four give the classes of NLO operators. This condition yields the same Lorentz-invariant classes as in Eq.~\eqref{eq:3.3.14}. Within these classes, we construct the operators. In order to arrive at a minimal, non-redundant set, we use the equations of motion, integration by parts, and $SU(3)$ relations \cite{Borodulin:1995xd}. Further, we assume that the operators are even under $\mathcal{C}$ and $\mathcal{P}$ symmetry, as both QED and QCD conserve these discrete symmetries. We find the following operators \cite{Buchalla:2013eza}:\\
\renewcommand{\arraystretch}{1.5}
$UD^{4}:$
\begin{align}
\label{eq:3.3.17}
 \langle D_\mu U^\dagger D^\mu U\rangle^2,\qquad \langle D_\mu U^\dagger D_\nu U\rangle\, \langle D^\mu U^\dagger D^\nu U\rangle,\qquad\langle D_\mu U^\dagger D^\mu U D_\nu U^\dagger D^\nu U\rangle
\end{align}
$UD^{2}Q^{2}:$
\begin{align}
\label{eq:3.3.18}
 &\langle UD_{\mu}U^{\dagger}Q_{R}\rangle \langle UD^{\mu}U^{\dagger}Q_{R}\rangle + \langle U^{\dagger}D_{\mu}U Q_{L}\rangle \langle U^{\dagger}D^{\mu}U Q_{L}\rangle, \nonumber \\
 &\langle UD^{\mu}U^{\dagger}Q_{R}\rangle \langle U^{\dagger}D_{\mu}U Q_{L}\rangle, \qquad \langle D_{\mu}U^{\dagger}D^{\mu}U Q_{L}^{2}\rangle + \langle D_{\mu}UD^{\mu}U^{\dagger}Q_{R}^{2}\rangle, \nonumber \\
 &\langle D_{\mu}U D^{\mu}U^{\dagger}\rangle \langle Q_{L}U^{\dagger}Q_{R}U\rangle,  \qquad \langle D^{\mu}UD_{\mu}U^{\dagger} (UQ_{L}U^{\dagger}Q_{R}+ Q_{R}UQ_{L}U^{\dagger})\rangle, 
\end{align}
$UQ^{4}:$
\begin{align}
\label{eq:3.3.19}
\langle U^{\dagger}Q_{R}UQ_{L}\rangle \langle U^{\dagger}Q_{R}UQ_{L}\rangle,
\end{align}
$U\chi^{2}:$
\begin{align}
\label{eq:3.3.20}
& \langle \chi^{\dagger}U \chi^{\dagger}U + \chi U^{\dagger}\chi U^{\dagger}\rangle, \qquad \langle U\chi^{\dagger} + \chi U^{\dagger}\rangle \langle U\chi^{\dagger} + \chi U^{\dagger}\rangle, \nonumber \\
 & \langle U\chi^{\dagger} - \chi U^{\dagger}\rangle \langle U\chi^{\dagger} - \chi U^{\dagger}\rangle, \qquad \langle \chi \chi^{\dagger}\rangle,
\end{align}
$UD^{2}\chi:$
\begin{align}
\label{eq:3.3.21}
\langle D_{\mu}UD^{\mu}U^{\dagger}\rangle \langle U\chi^{\dagger} + \chi U^{\dagger}\rangle, \qquad \langle D_{\mu}UD^{\mu}U^{\dagger} (U\chi^{\dagger} + \chi U^{\dagger})\rangle,
\end{align}
$U\chi Q^{2}:$
\begin{align}
\label{eq:3.3.22}
\langle U\chi^{\dagger}+\chi U^{\dagger}\rangle \langle Q_{L}U^{\dagger}Q_{R}U\rangle,\qquad \langle (U\chi^{\dagger}+\chi U^{\dagger} + \chi^{\dagger}U + U^{\dagger}\chi)Q^{2}\rangle, \nonumber \\
   \langle (U\chi^{\dagger}\pm\chi U^{\dagger})Q_{R}UQ_{L}U^{\dagger}\rangle + \langle (U^{\dagger} \chi\pm\chi^{\dagger}U)Q_{L}U^{\dagger}Q_{R}U\rangle,
\end{align}
$D^{2}UFQ:$\\
There is no independent operator in this class.\\ \\
$UF^{2}Q^{2}:$
\begin{align}
\label{eq:3.3.24}
\langle Q_{L}U^{\dagger}Q_{R}U\rangle F_{\mu\nu}F^{\mu\nu}.
\end{align}
\renewcommand{\arraystretch}{1}
This list of operators is consistent with the results originally obtained in \cite{Urech:1994hd}. Operators that renormalize the operators of the leading-order Lagrangian have not been listed here. The Wilson coefficients of the operators, also called low-energy constants (LECs), can be obtained from experiment. Once they are determined, the leading and next-to-leading order Lagrangian can be used to make predictions of further experiments. 
\part{The Standard Model as Effective Field Theory}
  \chapter*{The Standard Model as Effective Field Theory}
\thispagestyle{fancyplain}
\markboth{The Standard Model as Effective Field Theory}{The Standard Model as Effective Field Theory}
Depending on the current situation at the experiments, different strategies for the analyses are pursued when the experimental results are investigated. At the Large Electron-Positron Collider (LEP), the underlying model was assumed to be the SM with only the Higgs missing. The experimental observations were used to fit the SM parameters and to put constraints on the Higgs mass. The paradigm shifted at the LHC. A Higgs-like particle was found and the SM seems to be complete. However, extensions of the SM are still anticipated, as we discussed in Section~\ref{ch:SM.openQ}. The search for these extensions can, for example, rely on proposing explicit UV models. This approach, however, requires many explicit assumptions and the list of available models is infinite. 

We saw in Chapter \ref{ch:EFT} that we can use bottom-up effective field theories to study theories with an unknown UV completion in a model-independent way, only relying on a few, very general assumptions. Using EFTs for these analyses is further justified, because the experimental collaborations did not find any new particles. This indicates a mass gap to the UV theory, which is an essential ingredient of any EFT. Thus, we use the Lagrangian that describes the current experimental observables as the leading-order Lagrangian of a bottom-up effective field theory. Since we do not know the couplings of the Higgs-like scalar precisely, we can make different assumptions about its nature. If we assume it is the SM-Higgs, the SM Lagrangian defines the leading-order Lagrangian of the EFT. The new physics decouples and the resulting bottom-up EFT is called SM-EFT. We discuss it in Chapter \ref{ch:SMEFT}. If we assume that the Higgs-like scalar comes from a strongly-coupled UV or has large mixings with other scalars, it does not decouple. In this case, it is more appropriate to use the electroweak chiral Lagrangian as bottom-up EFT. The leading-order Lagrangian is then more general than the SM. We discuss this in Chapter \ref{ch:ewXL}.

\chapter{The Decoupling EFT --- the SM-EFT}
\label{ch:SMEFT}
\thispagestyle{fancyplain}
Under the assumption that the new physics is decoupling, the scalar particle found at the LHC is written conveniently as part of the Higgs doublet, and the electroweak symmetry is realized linearly. This is analogous to the linear sigma model of Section~\ref{ch:2.2}, where also the complete multiplet (the $\pi_{i}$ and the $\sigma$) was included in the theory. Even though the underlying assumption is about the dynamics of the new physics and not the realization of the symmetry, the bottom-up EFT that we construct here is called linear EFT. We use the names linear EFT and decoupling EFT interchangeably. The renormalizable SM defines the leading-order Lagrangian of the EFT, $\mathcal{L}_{\text{LO}} \equiv \mathcal{L}_{\text{SM}}^{d=4}$ of Eq.~\eqref{eq:SMunbroken}. For that reason, the effective theory is also called Standard Model EFT, or SM-EFT. Because of the renormalizability of the SM, we have a decoupling EFT with the power counting given by canonical dimensions. This is equivalent to assuming that the new physics decouples completely from the SM. We use the full Lagrangian given by
\begin{equation}
  \label{eq:4.SM-EFT}
  \mathcal{L}_{\text{SM-EFT}} = \mathcal{L}_{\text{SM}}^{d=4} + \sum\limits_{d=5}^{\infty} \sum\limits_{i} \left(\frac{1}{\Lambda}\right)^{d-4} \tilde{c}^{(d)}_{i} \mathcal{O}_{i}^{(d)}.
\end{equation}
At each order in $(v/\Lambda)$, we write down all Lorentz- and gauge-invariant operators. The approximate symmetries that we discussed in Section \ref{ch:SMsym} might be violated by these operators. This can lead to a further suppression of the operator's coefficient, depending on the specific assumptions employed. Integrating by parts, applying Fierz identities, and using the equations of motion reduces the set of operators to a minimal set, {\it i.e.} a basis. However, the choice of a basis is not unique. Which particular basis we choose is irrelevant for physical observables, but some bases are more convenient for certain applications. One possible choice reduces the number of derivatives in the operators as much as possible. In the next section, we discuss the first orders of the expansion in Eq.~\eqref{eq:4.SM-EFT}. In general, we observe that the number of operators at each order increases substantially \cite{Henning:2015alf}.
\section{Higher Order Operators}
 \subsection{Dimension-Five Operators}
At dimension five, we can write down only one operator structure and its hermitean conjugate \cite{Weinberg:1979sa,Grzadkowski:2010es}. Weinberg discussed this operator first \cite{Weinberg:1979sa}, hence, the operator is sometimes called ``Weinberg operator''.
\begin{equation}
  \label{eq:4.dim5}
  \mathcal{O}_{\nu\nu} = (\widetilde{\phi}^\dagger\ell_{r})^{T}C (\widetilde{\phi}^\dagger\ell_{s}) + \text{h.c.}
\end{equation}
For three generations, the Wilson coefficient $\tilde{c}_{\nu\nu}^{rs}$ is a $3\times 3$ matrix carrying 12 independent parameters \cite{Henning:2015alf}. After symmetry breaking, $\mathcal{O}_{\nu\nu}$ generates Majorana masses for the left-handed neutrinos, $m_{\nu}^{rs}= \tilde{c}_{\nu\nu}^{rs} v^{2}/2\Lambda$. Compared to the electroweak scale $v$, the masses are suppressed by $v/\Lambda$, explaining why neutrinos are lighter than the other fields of the SM. 

An explicit model that generates this operator is given by adding heavy, right-handed neutrinos to the SM. Their quantum numbers allow a Yukawa interaction as well as a Majorana mass term. If the latter is very large, the right-handed neutrinos can be integrated out. The resulting low-energy EFT is given to first order by $\mathcal{O}_{\nu\nu}$ \cite{Willenbrock:2004hu}. This mechanism, generating very light left-handed neutrinos by introducing very heavy right-handed neutrinos, is called seesaw mechanism \cite{Kayser:1989iu}. 

Even though the operator is only suppressed by a single power of the new-physics scale $\Lambda$, the experimental bounds on neutrino masses indicate a strong suppression by $\Lambda \sim 10^{14}~\text{GeV}$ \cite{Gripaios:2015gxa}. Symmetry arguments support this suppression. Since $\mathcal{O}_{\nu\nu}$ generates a Majorana mass term for the left-handed neutrinos, it violates lepton number $L$ and also the $B-L$ symmetry. As the latter is a symmetry of the SM, we expect that $B-L$-violating effects are strongly suppressed. 
 \subsection{Dimension-Six Operators}
Operators of mass dimension six are suppressed by two powers of the new-physics scale. However, they are usually expected to be less suppressed than the dimension-five operator discussed above, as it is possible to write down operators that respect all accidental symmetries of the SM as well. In fact, the effects of many new-physics models arise to first approximation at the level of dimension-six operators \cite{Henning:2014wua}, making it a very popular EFT approach for LHC searches of beyond-the-Standard-Model physics. The LHC Higgs Cross Section Working Group \cite{LHCHXSWGwebsite} devoted a significant part of the CERN Higgs Yellow Report 4 to EFT analyses, using the dimension-six operators to a large extent.

\begin{table}[t] 
\centering
\renewcommand{\arraystretch}{1.5}
\begin{tabular}{||c|c||c|c||c|c||} 
\hline \hline
\multicolumn{2}{||c||}{$X^3$} & 
\multicolumn{2}{|c||}{$\phi^6$~ and~ $\phi^4 D^2$} &
\multicolumn{2}{|c||}{$\psi^2\phi^3$}\\
\hline
$\mathcal{O}_G$                & $f^{ABC} G_\mu^{A\nu} G_\nu^{B\rho} G_\rho^{C\mu} $ &  
$\mathcal{O}_\phi$       & $(\phi^\dagger\phi)^3$ &
$\mathcal{O}_{e\phi}$           & $(\phi^\dagger \phi)(\bar \ell_p e_r \phi)$\\
$\mathcal{O}_{\widetilde G}$          & $f^{ABC} \widetilde G_\mu^{A\nu} G_\nu^{B\rho} G_\rho^{C\mu} $ &   
$\mathcal{O}_{\phi\Box}$ & $(\phi^\dagger \phi)\raisebox{-.5mm}{$\Box$}(\phi^\dagger \phi)$ &
$\mathcal{O}_{u\phi}$           & $(\phi^\dagger \phi)(\bar q_p u_r \widetilde{\phi})$\\
$\mathcal{O}_W$                & $\varepsilon^{IJK} W_\mu^{I\nu} W_\nu^{J\rho} W_\rho^{K\mu}$ &    
$\mathcal{O}_{\phi D}$   & $\left(\phi^\dagger D^\mu\phi\right)^\star \left(\phi^\dagger D_\mu\phi\right)$ &
$\mathcal{O}_{d\phi}$           & $(\phi^\dagger \phi)(\bar q_p d_r \phi)$\\
$\mathcal{O}_{\widetilde W}$          & $\varepsilon^{IJK} \widetilde W_\mu^{I\nu} W_\nu^{J\rho} W_\rho^{K\mu}$ &&&&\\    
\hline \hline
\multicolumn{2}{||c||}{$X^2\phi^2$} &
\multicolumn{2}{|c||}{$\psi^2 X\phi$} &
\multicolumn{2}{|c||}{$\psi^2\phi^2 D$}\\ 
\hline
$\mathcal{O}_{\phi G}$     & $\phi^\dagger \phi\, G^A_{\mu\nu} G^{A\mu\nu}$ & 
$\mathcal{O}_{eW}$               & $(\bar \ell_p \sigma^{\mu\nu} e_r) \tau^I \phi W_{\mu\nu}^I$ &
$\mathcal{O}_{\phi \ell}^{(1)}$      & $(\phi^{\dagger} i \, \overleftrightarrow{D_{\mu}} \phi)(\bar \ell_p \gamma^\mu \ell_r)$\\
$\mathcal{O}_{\phi\widetilde G}$         & $\phi^\dagger \phi\, \widetilde G^A_{\mu\nu} G^{A\mu\nu}$ &  
$\mathcal{O}_{eB}$        & $(\bar \ell_p \sigma^{\mu\nu} e_r) \phi B_{\mu\nu}$ &
$\mathcal{O}_{\phi \ell}^{(3)}$      & $(\phi^{\dagger} i \, \overleftrightarrow{D_{\mu}^{I}} \phi)(\bar \ell_p \tau^I \gamma^\mu \ell_r)$\\
$\mathcal{O}_{\phi W}$     & $\phi^\dagger \phi\, W^I_{\mu\nu} W^{I\mu\nu}$ & 
$\mathcal{O}_{uG}$        & $(\bar q_p \sigma^{\mu\nu} T^A u_r) \widetilde{\phi}\, G_{\mu\nu}^A$ &
$\mathcal{O}_{\phi e}$            & $(\phi^{\dagger} i \, \overleftrightarrow{D_{\mu}} \phi)(\bar e_p \gamma^\mu e_r)$\\
$\mathcal{O}_{\phi\widetilde W}$         & $\phi^\dagger \phi\, \widetilde W^I_{\mu\nu} W^{I\mu\nu}$ &
$\mathcal{O}_{uW}$               & $(\bar q_p \sigma^{\mu\nu} u_r) \tau^I \widetilde{\phi}\, W_{\mu\nu}^I$ &
$\mathcal{O}_{\phi q}^{(1)}$      & $(\phi^{\dagger} i \, \overleftrightarrow{D_{\mu}} \phi)(\bar q_p \gamma^\mu q_r)$\\
$\mathcal{O}_{\phi B}$     & $ \phi^\dagger \phi\, B_{\mu\nu} B^{\mu\nu}$ &
$\mathcal{O}_{uB}$        & $(\bar q_p \sigma^{\mu\nu} u_r) \widetilde{\phi}\, B_{\mu\nu}$&
$\mathcal{O}_{\phi q}^{(3)}$      & $(\phi^{\dagger} i \, \overleftrightarrow{D_{\mu}^{I}} \phi)(\bar q_p \tau^I \gamma^\mu q_r)$\\
$\mathcal{O}_{\phi\widetilde B}$         & $\phi^\dagger \phi\, \widetilde B_{\mu\nu} B^{\mu\nu}$ &
$\mathcal{O}_{dG}$        & $(\bar q_p \sigma^{\mu\nu} T^A d_r) \phi\, G_{\mu\nu}^A$ & 
$\mathcal{O}_{\phi u}$            & $(\phi^{\dagger} i \, \overleftrightarrow{D_{\mu}} \phi)(\bar u_p \gamma^\mu u_r)$\\
$\mathcal{O}_{\phi WB}$     & $ \phi^\dagger \tau^I \phi\, W^I_{\mu\nu} B^{\mu\nu}$ &
$\mathcal{O}_{dW}$               & $(\bar q_p \sigma^{\mu\nu} d_r) \tau^I \phi\, W_{\mu\nu}^I$ &
$\mathcal{O}_{\phi d}$            & $(\phi^{\dagger} i \, \overleftrightarrow{D_{\mu}} \phi)(\bar d_p \gamma^\mu d_r)$\\
$\mathcal{O}_{\phi\widetilde WB}$ & $\phi^\dagger \tau^I \phi\, \widetilde W^I_{\mu\nu} B^{\mu\nu}$ &
$\mathcal{O}_{dB}$        & $(\bar q_p \sigma^{\mu\nu} d_r) \phi\, B_{\mu\nu}$ &
$\mathcal{O}_{\phi u d}$   & $i(\widetilde{\phi}^\dagger D_\mu \phi)(\bar u_p \gamma^\mu d_r)$\\
\hline \hline
\end{tabular}
\caption[Dimension-six operators without four-fermion ones.]{Dimension-six operators without four-fermion operators, taken from \cite{Grzadkowski:2010es}. Indices of chirality are suppressed. \label{tab:SMEFTdim6.1}}
\end{table}

\begin{table}[t]
\centering
\renewcommand{\arraystretch}{1.5}
\makebox[\textwidth][r]{
\begin{tabular}{||c|c||c|c||c|c||}
\hline\hline
\multicolumn{2}{||c||}{$(\bar LL)(\bar LL)$} & 
\multicolumn{2}{|c||}{$(\bar RR)(\bar RR)$} &
\multicolumn{2}{|c||}{$(\bar LL)(\bar RR)$}\\
\hline
$\mathcal{O}_{\ell\ell}$        & $(\bar \ell_p \gamma_\mu \ell_r)(\bar \ell_s \gamma^\mu \ell_t)$ &
$\mathcal{O}_{ee}$               & $(\bar e_p \gamma_\mu e_r)(\bar e_s \gamma^\mu e_t)$ &
$\mathcal{O}_{\ell e}$               & $(\bar \ell_p \gamma_\mu \ell_r)(\bar e_s \gamma^\mu e_t)$ \\
$\mathcal{O}_{qq}^{(1)}$  & $(\bar q_p \gamma_\mu q_r)(\bar q_s \gamma^\mu q_t)$ &
$\mathcal{O}_{uu}$        & $(\bar u_p \gamma_\mu u_r)(\bar u_s \gamma^\mu u_t)$ &
$\mathcal{O}_{\ell u}$               & $(\bar \ell_p \gamma_\mu \ell_r)(\bar u_s \gamma^\mu u_t)$ \\
$\mathcal{O}_{qq}^{(3)}$  & $(\bar q_p \gamma_\mu \tau^I q_r)(\bar q_s \gamma^\mu \tau^I q_t)$ &
$\mathcal{O}_{dd}$        & $(\bar d_p \gamma_\mu d_r)(\bar d_s \gamma^\mu d_t)$ &
$\mathcal{O}_{\ell d}$               & $(\bar \ell_p \gamma_\mu \ell_r)(\bar d_s \gamma^\mu d_t)$ \\
$\mathcal{O}_{\ell q}^{(1)}$                & $(\bar \ell_p \gamma_\mu \ell_r)(\bar q_s \gamma^\mu q_t)$ &
$\mathcal{O}_{eu}$                      & $(\bar e_p \gamma_\mu e_r)(\bar u_s \gamma^\mu u_t)$ &
$\mathcal{O}_{qe}$               & $(\bar q_p \gamma_\mu q_r)(\bar e_s \gamma^\mu e_t)$ \\
$\mathcal{O}_{\ell q}^{(3)}$                & $(\bar \ell_p \gamma_\mu \tau^I \ell_r)(\bar q_s \gamma^\mu \tau^I q_t)$ &
$\mathcal{O}_{ed}$                      & $(\bar e_p \gamma_\mu e_r)(\bar d_s\gamma^\mu d_t)$ &
$\mathcal{O}_{qu}^{(1)}$         & $(\bar q_p \gamma_\mu q_r)(\bar u_s \gamma^\mu u_t)$ \\ 
&& 
$\mathcal{O}_{ud}^{(1)}$                & $(\bar u_p \gamma_\mu u_r)(\bar d_s \gamma^\mu d_t)$ &
$\mathcal{O}_{qu}^{(8)}$         & $(\bar q_p \gamma_\mu T^A q_r)(\bar u_s \gamma^\mu T^A u_t)$ \\ 
&& 
$\mathcal{O}_{ud}^{(8)}$                & $(\bar u_p \gamma_\mu T^A u_r)(\bar d_s \gamma^\mu T^A d_t)$ &
$\mathcal{O}_{qd}^{(1)}$ & $(\bar q_p \gamma_\mu q_r)(\bar d_s \gamma^\mu d_t)$ \\
&&&&
$\mathcal{O}_{qd}^{(8)}$ & $(\bar q_p \gamma_\mu T^A q_r)(\bar d_s \gamma^\mu T^A d_t)$\\
\hline\hline
\multicolumn{2}{||c||}{$(\bar LR)(\bar RL)$ and $(\bar LR)(\bar LR)$} &
\multicolumn{4}{|c||}{$B$-violating}\\\hline
$\mathcal{O}_{\ell edq}$ & $(\bar \ell_p^j e_r)(\bar d_s q_t^j)$ &
$\mathcal{O}_{duq}$ & \multicolumn{3}{|c||}{$\varepsilon^{\alpha\beta\gamma} \varepsilon_{jk} 
 \left[ (d^\alpha_p)^T C u^\beta_r \right]\left[(q^{\gamma j}_s)^T C \ell^k_t\right]$}\\
$\mathcal{O}_{quqd}^{(1)}$ & $(\bar q_p^j u_r) \varepsilon_{jk} (\bar q_s^k d_t)$ &
$\mathcal{O}_{qqu}$ & \multicolumn{3}{|c||}{$\varepsilon^{\alpha\beta\gamma} \varepsilon_{jk} 
  \left[ (q^{\alpha j}_p)^T C q^{\beta k}_r \right]\left[(u^\gamma_s)^T C e_t\right]$}\\
$\mathcal{O}_{quqd}^{(8)}$ & $(\bar q_p^j T^A u_r) \varepsilon_{jk} (\bar q_s^k T^A d_t)$ &
$\mathcal{O}_{qqq}^{(1)}$ & \multicolumn{3}{|c||}{$\varepsilon^{\alpha\beta\gamma} \varepsilon_{jk} \varepsilon_{mn} 
  \left[ (q^{\alpha j}_p)^T C q^{\beta k}_r \right]\left[(q^{\gamma m}_s)^T C \ell^n_t\right]$}\\
$\mathcal{O}_{\ell equ}^{(1)}$ & $(\bar \ell_p^j e_r) \varepsilon_{jk} (\bar q_s^k u_t)$ &
$\mathcal{O}_{qqq}^{(3)}$ & \multicolumn{3}{|c||}{$\varepsilon^{\alpha\beta\gamma} (\tau^I \varepsilon)_{jk} (\tau^I \varepsilon)_{mn} 
  \left[ (q^{\alpha j}_p)^T C q^{\beta k}_r \right]\left[(q^{\gamma m}_s)^T C \ell^n_t\right]$}\\
$\mathcal{O}_{\ell equ}^{(3)}$ & $(\bar \ell_p^j \sigma_{\mu\nu} e_r) \varepsilon_{jk} (\bar q_s^k \sigma^{\mu\nu} u_t)$ &
$\mathcal{O}_{duu}$ & \multicolumn{3}{|c||}{$\varepsilon^{\alpha\beta\gamma} 
  \left[ (d^\alpha_p)^T C u^\beta_r \right]\left[(u^\gamma_s)^T C e_t\right]$}\\
\hline\hline
\end{tabular}}
\caption[Four-fermion operators.]{Four-fermion operators, taken from \cite{Grzadkowski:2010es}. Indices of chirality are suppressed. \label{tab:SMEFTdim6.2}}
\end{table}
Historically, the subset of the four-fermion interactions was considered first, thereby focussing on $B$- or $L$-violating \cite{Weinberg:1979sa,Wilczek:1979hc,Weldon:1980gi,Abbott:1980zj} as well as $B$- and $L$-conserving \cite{Ruckl:1983hz,Eichten:1983hw,Burges:1983zg} operators. This was motivated by the fact that a gauge theory in the UV introduces current-current interactions of four-fermion type, similar to Fermi's theory of weak interactions \cite{Fermi:1934hr}. Later, attempts to find more complete lists were made, the first one by Buchm{\"u}ller and Wyler \cite{Buchmuller:1985jz}. Other suggestions followed \cite{Hagiwara:1993ck,Giudice:2007fh}, but it was only in 2010, when Grzadkowski, Iskrzynski, Misiak, and Rosiek \cite{Grzadkowski:2010es} presented a complete and non-redundant set of dimension-six operators. Based on the geographical location of this collaboration, the basis is sometimes called Warsaw basis. It also followed the guideline of reducing the number of derivatives as much as possible. Table \ref{tab:SMEFTdim6.1} and \ref{tab:SMEFTdim6.2} list the operators of the Warsaw basis, excluding hermitean conjugates.

For one generation of fermions, there are 76 operators (including hermitean conjugates) that conserve Baryon number $B$ and eight that do not conserve $B$. For three generations, the number increases to a total of 2499 independet parameters \cite{Alonso:2013hga,Henning:2015alf} for $B$- and $L$-conserving operators. Usually, additional assumptions are made in order to reduce this large number to a manageable set. In the flavor sector for example, minimal flavor violation (MFV) assumes that ``the dynamics of flavor violation is completely determined by the structure of the ordinary Yukawa couplings'' \cite{D'Ambrosio:2002ex}. Assuming a weakly-coupled, renormalizable UV-completion introduces a further suppression by explicit loop factors to some of the operators \cite{Arzt:1994gp,Einhorn:2013kja}. On top, assumptions about the approximate symmetries such as $\mathcal{CP}$, custodial, and $B/L$ can be made. Presently, most of the data analyses use only subsets of the operators. However, global analyses have also been made \cite{Han:2004az,Berthier:2015oma,Berthier:2015gja,Falkowski:2015jaa}.

Apart from the Warsaw basis \cite{Grzadkowski:2010es}, other bases were proposed. Even though the physics content does not depend on this choice, some computations are more conveniently performed by means of a different basis. The most prominent other bases are the HISZ \cite{Hagiwara:1993ck} and the SILH \cite{Giudice:2007fh,Contino:2013kra} basis. For an easy translation between the different bases, a computer code called {\tt Rosetta} \cite{Falkowski:2015wza} was developed. 
 \subsection{Dimension Seven and Above }
Operators beyond dimension six have been considered only recently in a systematic way \cite{Lehman:2014jma,Lehman:2015coa}. Since the number of possible gauge-invariant combinations of SM fields grows rapidly with the operator dimension, the conventional method of listing all possible operator structures by hand becomes tedious. Lehman and Martin developed a general approach to construct these operators in a systematic way, based on Hilbert-Series techniques \cite{Lehman:2015via}. The authors of \cite{Henning:2015alf} showed explicitly that there are 30 (1542) independent operators at dimension seven and 993 (44807) operators at dimension eight for one (three) generations of fermions. In addition, general statements connecting $B$ and $L$ with the dimension of the operators were shown, {\it e.g.} all odd-dimensional operators violate $B-L$ \cite{deGouvea:2014lva,Kobach:2016ami}. 

In general, operators of dimension larger than six are only sub-leading with respect to the dimension-six operators and can be neglected at the current phase of LHC physics. Exceptions can arise in some cases, for example coming from special symmetry constructions \cite{Contino:2016jqw}. However, we should keep in mind that it is the square of the amplitude that enters the observables. Effects from dimension-six operators squared are of the same order as the interference effects of dimension-eight operators with the SM, making the dimension-eight terms important for future precision analyses. 
\section{Loop Corrections and Renormalization Within the SM-EFT}
\label{ch:SM-EFT.renorm}
Introducing operators with dimension higher than four changes the structure of the theory. When considered in loops, these operators change the renormalization group equations (RGE) of the SM parameters. Further, non-SM operator structures arise. However, the theory remains renormalizable order by order in the effective expansion. One-loop diagrams with a single dimension-six insertion are renormalized within the dimension-six operators, although field-redefinitions and equations of motion are necessary to express all operators in the original basis. 

At first, subsets of operators were considered in specific processes \cite{Grojean:2013kd,Elias-Miro:2013mua}. Later, the complete set of dimension-six operators in the Warsaw basis was renormalized at one loop \cite{Jenkins:2013zja,Jenkins:2013wua,Alonso:2013hga,Alonso:2014zka}. The result can be expressed in terms of the RGE of the Wilson coefficients:
\begin{equation}
  \label{eq:4.1}
  \mu \frac{d}{d\mu} \tilde{c}_{i} = \gamma_{ij} \tilde{c}_{j},
\end{equation}
where $\gamma_{ij}$ is the anomalous dimension matrix of the Wilson coefficients $\tilde{c}_{i}$. This result has several important consequences. First, the RG equations give the scale dependence of the Wilson coefficients. This dependence becomes important when measurements at low energy (usually the electroweak scale, $v$) are compared to predictions of UV models (usually at a scale $\Lambda\gg v$). In order to match the EFT and the model correctly, the Wilson coefficients have to be evolved using the renormalization-group equations, as discussed in Section \ref{ch:EFT.top-down}. Also, a precise determination of Wilson coefficients from future experimental data, like the high-luminosity phase of the LHC, relies on the one-loop improved computation \cite{Berthier:2015oma}. 

Some example processes have been computed using the dimension-six basis, explicitly taking the one-loop corrections into account \cite{Ghezzi:2015vva,Hartmann:2015oia,Hartmann:2015aia,Bylund:2016phk,David:2015waa}. However, the size of the RG effects is expected to be $(1/16\pi^{2}) \cdot (v^{2}/\Lambda^{2})$, which is well below the current precision of the LHC. Logarithms of the form $\log{\Lambda^{2}/m_{h}^{2}}$ might enhance the coefficients, but these effects amount only to one order of magnitude or less for most of the present applications \cite{Alonso:2013hga}. 

The renormalization further introduces a mixing between the dimension-six operators that occurs when the set is evolved to a different energy using the RGE in Eq.~\eqref{eq:4.1}.  This mixing yields the second important insight. It explicitly shows that the operator set found in \cite{Grzadkowski:2010es} is complete and non-redundant. However, the mixing also alters the pattern of Wilson coefficients that a UV model introduces. This was studied for the class of universal theories in \cite{Wells:2015cre}. 

Further, the authors of \cite{Alonso:2014rga} found that the structure of the anomalous dimension matrix $\gamma_{ij}$ is approximately holomorphic. This has been interpreted in favor of supersymmetry as UV completion. 
  \chapter{The Electroweak Chiral Lagrangian}
\label{ch:ewXL}
\thispagestyle{fancyplain}
\section{The Construction of the Leading-Order Lagrangian}
\label{ch:ewXL.LO}
We assume now that the scalar found at the LHC is not the SM-Higgs. Instead, we assume that it belongs to a new-physics sector that is non-decoupling. Thus, the leading-order Lagrangian is not given by the SM. However, we cannot remove the Higgs doublet completely, since the $W^{\pm}$ and $Z$ are massive and we therefore need three Goldstone bosons. Instead, we write the scalar doublet in terms of the physical $h$\footnote{We refer to $h$ as the Higgs, even though it is only a scalar and not necessarily the SM-Higgs. Whenever we mean the latter, we call it ``SM-Higgs'' explicitly.} and the three Goldstone bosons as
\begin{equation}
  \label{eq:5.1}
  \phi = \frac{v+h}{\sqrt{2}} \; U \, \binom{0}{1} = \sqrt{2} \Phi \binom{0}{1}, \qquad \text{where } \;U = \exp{\left\{ i \frac{2 \, T_{a} \varphi_{a}}{v}\right\}}.
\end{equation}
The Goldstone bosons, $\varphi_{a}$, are written in the exponential representation with $T_{a}$ being the generators of $SU(2)$. Without the physical Higgs, $h$, this construction is similar to the non-linear sigma model discussed in Section \ref{ch:nlsm}. The structure of the Lagrangian therefore depends only on the pattern of spontaneous global symmetry breaking, $SU(2)_{L}\times SU(2)_{R}\rightarrow SU(2)_{V}$. It does not distinguish between cases of dynamical symmetry breaking \cite{Appelquist:1980vg,Chivukula:2000mb} (as in chiral perturbation theory with the identification $F_{\pi}=v$) or a scalar obtaining a vacuum expectation value (as in the SM). The $SU(2)_{L}\times SU(2)_{R}$ symmetry is realized non-linearly on $\varphi_{a}$, giving the resulting framework the name ``non-linear'' Lagrangian. Other representations of the Goldstones are physically equivalent \cite{Donoghue:1992dd,Haag:1958vt,Coleman:1969sm,Callan:1969sn,Feruglio:1992wf}, as we discussed in the context of the non-linear sigma model in Section~\ref{ch:nlsm}. The exponential representation, however, has the advantage that the symmetry acts linearly on $U$:
\begin{equation}
  \label{eq:5.2}
  U \rightarrow g_{L} U g_{R}^{\dagger}, \qquad \text{where } g_{L,R} \in SU(2)_{L,R}
\end{equation}
The Lagrangian describing the electroweak interactions and containing the three Goldstone bosons but not the Higgs was discussed in the context of heavy Higgs models \cite{Appelquist:1980vg,Longhitano:1980tm,Dobado:1990zh,Herrero:1993nc}. It reads
\begin{align}
\begin{aligned}
  \label{eq:5.3}
  \mathcal{L}^{\text{heavy}}_{\text{Higgs}} = & -\frac{1}{2} \langle G_{\mu\nu}G^{\mu\nu}\rangle -\frac{1}{2}\langle W_{\mu\nu}W^{\mu\nu}\rangle -\frac{1}{4} B_{\mu\nu}B^{\mu\nu}\\
&+i\bar{q}_{L}\slashed{D}q_{L} +i\bar{\ell}_{L}\slashed{D}\ell_{L} +i\bar{u}_{R}\slashed{D}u_{R} +i\bar{d}_{R}\slashed{D}d_{R} +i\bar{e}_{R}\slashed{D}e_{R} \\
&+\frac{v^2}{4}\ \langle D_\mu U^\dagger D^\mu U\rangle  - \frac{v}{\sqrt{2}} \left[ \bar{q}_{L} Y_u U P_{+}q_{R} + \bar{q}_{L} Y_d U P_{-}q_{R} + \bar{\ell}_{L}  Y_e U P_{-}\ell_{R} + \text{ h.c.}\right],
\end{aligned}
\end{align}
with
\begin{equation}
\begin{array}{c}
  \label{eq:5.4}
  D_{\mu} U = \partial_{\mu}U + ig W_{\mu}U - i g' B_{\mu}U T_{3},\\ \\
  P_{\pm} = \frac{1}{2} \pm T_{3}, \qquad P_{12} = T_{1}+ i T_{2}, \qquad P_{21} = T_{1}-i T_{2}.
\end{array}
\end{equation}
The doublets $q_{R}=(u_{R},d_{R})^{T}$ and $\ell_{R}=(0,e_{R})^{T}$ combine the right-handed singlets for a clearer notation. The projectors $P_{12}$ and $P_{21}$ will be useful later. 

We now include the scalar particle that was found at the LHC \cite{Chatrchyan:2012xdj,Aad:2012tfa,CMS:2015kwa,ATLAS-CONF-2015-044}. Assuming it has the same quantum numbers as the physical SM-Higgs makes it a singlet under $SU(2)_{L}\times SU(2)_{R}$: $h\rightarrow h$. In the SM, $h$ would have couplings to $\mathcal{L}^{\text{heavy}}_{\text{Higgs}}$ that are fixed by Eq.~\eqref{eq:SMunbroken} and Eq.~\eqref{eq:5.1}. Since, by assumption, we relax the connection between the Goldstone bosons and $h$, we allow for general $\mathcal{O}(1)$ couplings of the Higgs to $\mathcal{L}^{\text{heavy}}_{\text{Higgs}}$ \cite{Contino:2010mh,Feruglio:1992wf,Koulovassilopoulos:1993pw,Burgess:1999ha,Grinstein:2007iv,Azatov:2012bz}. Additionally, we allow scenarios in which the Higgs is a pseudo-Nambu-Goldstone boson of a larger symmetry in the UV ({\it e.g.} composite Higgs models, see Section~\ref{ch:MCHM} and \cite{Agashe:2004rs,Contino:2006qr}). Similarly to the $\varphi_{a}$ in the exponential, $U$, we allow arbitrary powers of $h$ in the Lagrangian. Since $h$ is a singlet, the interactions with $h$ are general polynomials in $h$. The scale at which the larger symmetry is broken and the Higgs is generated is given by $f$. For the rest of this chapter, however, we assume that this scale is close to the electroweak scale, $f\simeq v$. We relax this assumption in Chapter \ref{ch:relation}.\\

The resulting Lagrangian is given by \cite{mastersthesis,Buchalla:2013rka,Buchalla:2013eza,Contino:2010rs,Contino:2010mh}
\begin{align}
\begin{aligned}
  \label{eq:LO}
  \mathcal{L}_{\text{LO}} =& -\frac{1}{2} \langle G_{\mu\nu}G^{\mu\nu}\rangle -\frac{1}{2}\langle W_{\mu\nu}W^{\mu\nu}\rangle -\frac{1}{4} B_{\mu\nu}B^{\mu\nu} \\
  &+i\bar{q}_{L}\slashed{D}q_{L} +i\bar{\ell}_{L}\slashed{D}\ell_{L} +i\bar{u}_{R}\slashed{D}u_{R} +i\bar{d}_{R}\slashed{D}d_{R} +i\bar{e}_{R}\slashed{D}e_{R} \\
  &+\frac{v^2}{4}\ \langle D_\mu U^\dagger D^\mu U\rangle \left( 1+F_U(h)\right) +\frac{1}{2} \partial_\mu h \partial^\mu h - V(h)\\
  &- \frac{v}{\sqrt{2}} \left[ \bar{q}_{L} \left( Y_u +\sum\limits^\infty_{n=1} Y^{(n)}_u \left(\frac{h}{v}\right)^n \right) U P_{+}q_{R} + \bar{q}_{L} \left( Y_d +  \sum\limits^\infty_{n=1} Y^{(n)}_d \left(\frac{h}{v}\right)^n \right) U P_{-}q_{R} \right.\\
  &\left. + \bar{\ell}_{L}  \left( Y_e +\sum\limits^\infty_{n=1} Y^{(n)}_e \left(\frac{h}{v}\right)^n \right) U P_{-}\ell_{R} + \text{ h.c.}\right]
\end{aligned}
\end{align}
Even though we motivated the Lagrangian phenomenologically, we can also construct it systematically unter the following assumptions:
\paragraph{Particles:} We assume SM particle content and we include three Goldstone bosons for the longitudinal components of the massive $W^{\pm}$ and $Z$. We include the Higgs as scalar singlet, not connected by symmetry to the Goldstone bosons. We do not assume any other light particle. Heavy particles, at or above the cutoff $\Lambda=4\pi f\simeq 4\pi v$, are integrated out. The latter assumption requires that the transverse gauge bosons and fermions of the SM are weakly coupled to the electroweak-symmetry-breaking sector, {\it i.e.} the Goldstones and the Higgs. A strong coupling to this sector will lead to a mass of the order of $\Lambda$, in contradiction to the assumption above. We assume that other particles of a new-physics sector that is not connected to electroweak symmetry breaking have masses of $\mathcal{O}(\Lambda)$ or above. We discuss the case with particles at an intermediate scale below the cutoff in Section~\ref{ch:scale.f}. The assumption of a weak coupling to new physics removes operators of the form $X_{\mu\nu}X^{\mu\nu}\,F(h)$ from the leading order Lagrangian. We further eliminate operators of the form $(\partial_{\mu}h) (\partial^{\mu}h)\,F(h)$ and $\bar{\Psi}\slashed{D}\Psi\,F(h)$ using field redefinitions without loss of generality \cite{mastersthesis,Buchalla:2013rka}. 
\paragraph{Symmetries:} We assume $SU(3)_{C}\times SU(2)_{L}\times U(1)_{Y}$ gauge invariance and conservation of $B$ and $L$. We further assume that the new physics conserves $\mathcal{CP}$, custodial and flavor symmetry at leading order. This removes the operator $\mathcal{O}_{\beta}$ (see Eq.~\eqref{eq:5.NLO.1}) from the leading-order Lagrangian. We discuss this below, in Section \ref{ch:Tparam}. The assumptions on the (approximate) global symmetries are not needed necessarily, they can also be relaxed in a more general approach. Currently, they are phenomenologically motivated. 
\paragraph{Power counting:} The power counting of this EFT is given by the loop expansion of non-decoupling EFTs. We discuss it in detail in Section~\ref{ch:5.pc} below.\\

The Lagrangian in Eq.~\eqref{eq:LO} generalizes the Higgs couplings of the SM, while the gauge and fermion sector remains untouched. The framework therefore allows us to test the SM-Higgs hypothesis in a systematic way. New physics that modifies the gauge or fermion sector will be subdominant. The description contains the SM in the limit
\begin{align}
  \begin{aligned}
    \label{eq:5.5}
F_{U}(h) &=2 \frac{h}{v} + \frac{h^{2}}{v^{2}}, \\
V(h)&=\frac{\lambda}{4}v^{2} h^{2}+ \frac{\lambda}{4}v h^{3}+ \frac{\lambda}{16}h^{4},\\
Y_{\Psi}&=Y^{(1)}_{\Psi},\qquad \text{ and }\qquad Y^{(n\geq 2)}_{\Psi}=0.
  \end{aligned}
\end{align}
With the generalized couplings, the Lagrangian in this expansion is also called ``the electroweak chiral Lagrangian with a light Higgs'' (ew$\chi\mathcal{L}$). 
\section{The Power Counting}
\label{ch:5.pc}
For arbitrary couplings in $\mathcal{L}_{\text{LO}}$ that do not coincide with Eq.~\eqref{eq:5.5}, the Lagrangian is non-renormalizable. It is therefore a non-decoupling EFT, very similar to chiral perturbation theory that we discussed in Section~\ref{ch:chiPT}. The power counting is thus given by a loop expansion, making the theory renormalizable order by order in the EFT. Loops of leading order generate divergencies that require counterterms, which are included at NLO. We expect the counterterms to be of the same size as the loop diagrams, such that we can write down the loop factors explicitly. The effective expansion of the Lagrangian is given by \cite{mastersthesis}
\begin{equation}
  \label{eq:5.6}
  \mathcal{L}_{\text{ew}\chi} = \mathcal{L}_{\text{LO}}+ \sum\limits_{L=1}^{\infty} \sum\limits_{i}\left(\frac{1}{16\pi^{2}}\right)^{L} c^{(L)}_{i}\mathcal{O}^{(L)}_{i}
\end{equation}
For this expansion, we identify $\Lambda = 4\pi v \approx M$, which comes from naive dimensional analysis (NDA) \cite{Manohar:1983md,Georgi:1992dw}. Here, $\Lambda$ is the cutoff of the EFT and $M$ is the mass of a heavy particle, for example a resonance. Differences of $\mathcal{O}(1)$ between these scales are encoded in the Wilson coefficients. The expansion parameter of the EFT in Eq.~\eqref{eq:5.6} is therefore $v^{2}/M^{2}\approx v^{2}/\Lambda^{2} = 1/16\pi^{2}$.

An $L$-loop diagram with $B$ external Goldstone bosons, $H$ external Higgs fields, $F^{1(2)}_{L/R}$ external left-/right-handed (anti)fermions, and $X$ external gauge fields, written in field-strength tensors $X_{\mu\nu}$, scales as \cite{mastersthesis,Buchalla:2013rka,Buchalla:2013eza}
\begin{align}
  \begin{aligned}
    \label{eq:5.7}
     \mathcal{D}&\sim \left(\frac{(v Y )^{\rho}(vg)^{\alpha+2\vartheta +\gamma+2 \delta+\tau}}{\Lambda^{2L}}\right) \; \left(\frac{\varphi}{v}\right)^{B} \; \left(\frac{h}{v}\right)^{H} \;\bar{\Psi}^{F^{1}_{L}}_{L}\;\Psi^{F^{2}_{L}}_{L}\;\bar{\Psi}^{F^{1}_{R}}_{R}\;\Psi^{F^{2}_{R}}_{R} \;\left(\frac{X_{\mu\nu}}{v}\right)^{X} \\  
&\times (\lambda v^{2})^{\nu}\;\frac{p^{d_{p}}}{v^{F_{L}+F_{R}-2}}.
  \end{aligned}
\end{align}
Here, $\rho$ is the number of Yukawa vertices ($Y$); $\alpha (\vartheta)$ the number of triple (quartic) gauge interactions with gauge coupling $g$; $\gamma$ the number of gauge-fermion interactions; $\nu$ the number of vertices from the Higgs potential with coupling $\lambda$; and $\tau(\delta)$ the number of vertices with Goldstone bosons, Higgs fields and one (two) gauge fields. The superficial degree of divergence is
\begin{equation}
  \label{eq:5.8}
  d_{p}=2L+2-X-\tfrac{1}{2}(F_{L}+F_{R})-\rho-2\delta-\tau-2\nu-\alpha-2\vartheta-\gamma.
\end{equation}
Again, it is bounded from above and the number of Goldstone bosons does not contribute. Also, the number of Higgs legs does not contribute. As in chiral perturbation theory, we can rearrange Eq.~\eqref{eq:5.8} and find
\begin{equation}
  \label{eq:5.9}
  2L+2 = d_{p}+X+\tfrac{1}{2}(F_{L}+F_{R})+\rho+2\delta+\tau+2\nu+\alpha+2\vartheta+\gamma\equiv \chi.
\end{equation}
We define the chiral dimensions as \cite{Buchalla:2013eza,Nyffeler:1999ap,Hirn:2004ze,Hirn:2005fr,Hirn:2005sj}
\begin{equation}
\label{eq:5.10}
  \begin{array}{c}
    [U]_{\chi} \;=\; [h]_{\chi}\;=\;[X_{\mu}]_{\chi}\;=\;0 \\ \\
    {[D_{\mu}]_{\chi} }\;=\; [\bar{\Psi}\Psi]_{\chi} \;=\;[g,g',g_{s}]_{\chi} \;=\;[Y_{\Psi}]_{\chi} \;=\; 1 \\ \\
    {[\lambda]_{\chi}} \; = \; 2
  \end{array}
\end{equation} 
We further define the chiral order $\chi$ as the sum of the chiral dimensions of all components of a term in the Lagrangian. The leading-order Lagrangian is homogeneously at chiral order two, $ [\mathcal{L}_{\text{LO}}]_{\chi} =2$.

While this construction considers only leading-order vertices within the diagrams so far, we can also generalize it to contain vertices of higher-order operators. Inserting a local vertex of loop order $L_{i}$ (chiral order $2L_{i}+2$) $n_{i}$ times in the diagram, modifies the power counting to
\begin{equation}
  \label{eq:5.11}
  2L+2 +2\sum\limits_{i}n_{i}L_{i}= d_{p}+X+\tfrac{1}{2}(F_{L}+F_{R})+\rho+2\delta+\tau+2\nu+\alpha+2\vartheta+\gamma\equiv \chi.
\end{equation}
An operator of chiral order four is therefore either a one-loop ($L=1$) graph of leading-order vertices ($L_{i}=0$), or a tree-level diagram ($L=0$) with one next-to-leading order insertion ($n_{i}=L_{i}=1$). 

Before we apply the chiral dimensions in the construction of the next-to-leading order operators in the next section, we discuss some further implications. If a symmetry is explicitly broken by a weak interaction (like gauge or Yukawa), the corresponding spurion will come with a weak coupling and therefore with a chiral dimension. This is important for the operator $\mathcal{O}_{\beta}$, which we discuss in Section \ref{ch:Tparam}.

There are other operators that are naively of chiral order two, {\it i.e.} leading order. These operators are $(\bar{\Psi}\Psi)^{2}$ and $X_{\mu\nu}\bar{\Psi}\sigma^{\mu\nu}\Psi$. The assumption of a weak coupling between the SM fields and the new physics introduces powers of couplings for the operators, increasing their chiral order to at least four. Phenomenology supports these assumptions further. Neither four-fermion interactions nor large dipole moments are observed in nature. 

Assigning a chiral dimension to weak couplings is not equivalent to an expansion in this weak coupling. Instead, the loop counting requires this assignment. An additional expansion in weak couplings is possible on top of the EFT expansion, if the coupling is sufficiently small. 

Since the counting of chiral dimensions is based on the superficial degree of divergence, we might identify some operators as needed counterterms even though an explicit computation reveals that the diagram is finite. We nevertheless include these operators at next-to-leading order in the EFT, as they can receive finite contributions at the same order, coming from integrating out heavy particles. This justifies the identification $4\pi v \lesssim M$ in the expansion above. 

The assignment of chiral dimensions in Eq.~\eqref{eq:5.10} practically yields a suppression of $1/f$ for the strongly-coupled fields, $h$ and $\varphi$; $1/\Lambda$ for $D_{\mu}$ and $g X$; and $1/(f\sqrt{\Lambda})$ for $\Psi$. This is similar to the NDA power counting of \cite{Manohar:1983md,Georgi:1992dw,Gavela:2016bzc}. However, the authors of \cite{Manohar:1983md,Georgi:1992dw,Gavela:2016bzc} do not assign a counting to the weak couplings, which yields an incorrect scaling of some operators \cite{Buchalla:2013eza,Buchalla:2016sop}. 
\section[The Operators at NLO]{The Operators at Next-to-Leading Order}
\label{ch:chiralNLO}
Operators of chiral order four define the next-to-leading order of the electroweak chiral Lagrangian. With $g$, a generic gauge coupling; $Y$, a Yukawa; and $w$, any of the two; the classes of operators are \cite{mastersthesis,Buchalla:2013rka}: $Uh D^{4},$ $ g^{2}X^{2}Uh,$ $ gXD^{2}Uh, $ $ w^{2}\Psi^{2} UhD,$ $  Y \Psi^{2} UhD^{2},$ $ gXUh\Psi^{2},$ $w^{2}\Psi^{4}Uh,$ and the classes of the leading order with two more powers of $w$. Subsets of the NLO operators have also been discussed in \cite{Bagger:1993zf,Wang:2006im,Azatov:2012bz,Alonso:2012px,Alonso:2012pz,Brivio:2013pma,Gavela:2014vra,Alonso:2014wta,Brivio:2014pfa,Hierro:2015nna,Brivio:2015kia}.

For a convenient construction of the operators, we define building blocks that transform as the adjoint of the left-handed $SU(2)_{L}$ symmetry.
\begin{equation}
  \label{eq:5.12}
  L_{\mu} \equiv i U D_{\mu}U^{\dagger}\qquad \text{ and } \qquad \tau_{L} \equiv UT_{3}U^{\dagger}
\end{equation}
The operator $\tau_{L}$ breaks custodial symmetry explicitly, as it is only invariant under $SU(2)_{L}\times U(1)_{Y}$ instead of $SU(2)_{L}\times SU(2)_{R}$. In fact, $\tau_{L}$ is the only possible spurion of the breaking of custodial symmetry \cite{Buchalla:2014eca}. We prove this as follows.

A generic spurion $s$ has an even number of $SU(2)_{L}$ indices, as all invariants in the Lagrangian are built from $U,h,$ gauge fields, and fermion bilinears that all carry an even number of $SU(2)_{L}$ indices. Without loss of generality, we can write the spurion as $2\times 2$ matrix, $s_{ab} = c_{0}\delta_{ab} + c_{j}\sigma^{j}_{ab}$, with complex coefficients $c_{i}$. The spurion must have one of the following transformation properties under $SU(2)_{L}\times SU(2)_{R}$:
\begin{equation}
  \label{eq:5.13}
  \begin{array}{c}
  a) \, s \rightarrow s \qquad   b) \, s \rightarrow g_{L} \,s\, g_{R}^{\dagger} \qquad c) \, s \rightarrow g_{R} \,s\, g_{L}^{\dagger} \\
  d) \, s \rightarrow g_{L} \,s\, g_{L}^{\dagger} \qquad e) \, s \rightarrow g_{R} \,s\, g_{R}^{\dagger}
  \end{array}
  \end{equation}
Keeping the spurion at a constant value breaks the symmetry in the desired way. However, gauge symmetries cannot be broken by any spurion and must remain exact. The trivial invariant $a)$ does not break custodial symmetry. A spurion that transforms as $b)$ or $c)$ violates the $SU(2)_{L}$ gauge symmetry. The scenario $d)$ also breaks this gauge symmetry, unless $s\sim \mathbf{1}$, the trivial case. Scenario $e)$ preserves gauge invariance if $s\sim \mathbf{1}$ or $s\sim T^{3}_{R}$, because only the $U(1)_{Y}$ subgroup of $SU(2)_{R}$ is gauged. This gives $T^{3}_{R}$ and therefore $\tau_{L}$ as the only, non-trivial spurion for the breaking of custodial symmetry, in contrast to the claims in \cite{Contino:2013kra}.

We list the next-to-leading operators in the following sections explicitly in Landau gauge. In this gauge, the Faddeev-Popov ghost Lagrangian coincides with the corresponding Lagrangian of the SM \cite{Appelquist:1980vg,Herrero:1993nc}. As there are no direct couplings of the ghosts to the Goldstone bosons, the non-renormalizability of the Goldstone sector does not affect diagrams with external ghosts. Therefore, there are no ghost fields needed in the counterterms. 

We write the polynomials in $h$ as $F_{i}(h)$, starting at linear order in the field,
\begin{equation}
  \label{eq:5.14}
  F_{i}(h)\equiv \sum\limits_{n=1}^\infty f_{i,n}\left(\frac{h}{v}\right)^n.
\end{equation}
We reduce the list of operators to a minimal set, {\it i.e.} a basis. This basis generalizes the complete Higgs-less basis of \cite{Buchalla:2012qq}. We list the equations of motion and other relations for the reduction in Appendix~\ref{ch:appB}. The authors of \cite{Nyffeler:1999ap,Grojean:2006nn} also discussed some of these relations.
\subsection[NLO Operators of the LO Classes With Two More Weak Couplings]{NLO Operators of the Leading-Order Classes With Two More Weak Couplings}
\label{ch:Tparam}
Almost all operators in this group will renormalize the operators of the leading-order Lagrangian. We do not list those operators again. The only structure we find that was previously not present is
\begin{equation}
  \label{eq:5.NLO.1}
  \mathcal{O}_{\beta} = g'^{2}v^{2}\langle L_{\mu}\tau_{L}\rangle \langle L^{\mu}\tau_{L}\rangle \, (1+ F_{\beta}(h)).
\end{equation}
This operator is related to the electroweak T-parameter \cite{Peskin:1990zt,Peskin:1991sw}. It breaks custodial symmetry. If we assume that the new physics conserves this symmetry, it can only be broken by SM effects, {\it i.e.} hypercharge or Yukawa couplings. The spurion $\tau_{L}$ then comes together with $g'^{2}$ or $Y^{2}$. This gives the operator two more chiral dimensions, leading to a total chiral order of four and moving it from leading to next-to-leading order. 
\subsection{NLO Operators of the Class $Uh D^{4}$} 
The operators of this class generalize the four derivative operators of the Higgs-less Lagrangian \cite{Longhitano:1980iz,Longhitano:1980tm}. The $\mathcal{CP}$-even operators are
\begin{align}
  \begin{aligned}
    \label{eq:5.NLO.2}  
    \mathcal{O}_{D1} &= \langle L_\mu L^\mu \rangle^2 \ (1+F_{D1}(h)), \\
    \mathcal{O}_{D2} &= \langle L_\mu L_\nu\rangle \ \langle L^\mu L^\nu\rangle  \ (1+F_{D2}(h)), \\
    \mathcal{O}_{D3} &= \left(\langle \tau_L L_\mu\rangle \ \langle \tau_L L^\mu\rangle \right)^2 \ (1+F_{D3}(h)),\\ 
    \mathcal{O}_{D4} &= \langle \tau_L L_\mu \rangle \ \langle \tau_L L^\mu\rangle \ \langle L_\nu L^\nu \rangle \ (1+F_{D4}(h)),\\
    \mathcal{O}_{D5} &= \langle \tau_L L_\mu\rangle \  \langle \tau_L L_\nu\rangle \ \langle L^\mu L^\nu\rangle \ (1+F_{D5}(h)),
  \end{aligned} 
\end{align}
\begin{align} 
  \begin{aligned}
    \label{eq:5.NLO.3}
    \mathcal{O}_{D6} &= i \langle \tau_L L_\mu L_\nu \rangle \ \langle \tau_L L^\mu\rangle \ \frac{\partial^\nu h}{v} (1+F_{D6}(h)),
  \end{aligned} 
\end{align}
\begin{align} 
  \begin{aligned}
    \label{eq:5.NLO.4}
    \mathcal{O}_{D7} &= \langle L_\mu L^\mu \rangle \ \frac{\partial_\nu h\, \partial^\nu h}{v^2} (1+F_{D7}(h)),\\
    \mathcal{O}_{D8} &= \langle L_\mu L_\nu \rangle \ \frac{\partial^\mu h\, \partial^\nu h}{v^2} (1+F_{D8}(h)),\\
    \mathcal{O}_{D9} &= \langle \tau_L L_\mu \rangle \ \langle \tau_L L^\mu \rangle \ \frac{\partial_\nu h\, \partial^\nu h}{v^2} (1+F_{D9}(h)),\\
    \mathcal{O}_{D10} &= \langle \tau_L L_\mu \rangle \ \langle \tau_L L_\nu \rangle \ \frac{\partial^\mu h\, \partial^\nu h}{v^2} (1+F_{D10}(h)),
  \end{aligned} 
\end{align}
\begin{align}
  \begin{aligned}
    \label{eq:5.NLO.5}
    \mathcal{O}_{D11} &= \frac{(\partial_\mu h\, \partial^\mu h)^2}{v^4} (1+F_{D11}(h)).
  \end{aligned} 
\end{align}
The $\mathcal{CP}$-odd operators are
\begin{align}
  \begin{aligned}
    \label{eq:5.NLO.6}
    \mathcal{O}_{D12} &= \langle L_\mu L^\mu \rangle \ \langle \tau_L L_\nu \rangle \ \frac{\partial^\nu h}{v} (1+F_{D12}(h)), \\
    \mathcal{O}_{D13} &= \langle L_\mu L_\nu \rangle \ \langle \tau_L L^\mu\rangle \ \frac{\partial^\nu h}{v} (1+F_{D13}(h)), \\
    \mathcal{O}_{D14} &= \langle \tau_L L_\mu \rangle \  \langle \tau_L L^\mu \rangle \ \langle \tau_L L_\nu \rangle \ \frac{\partial^\nu h}{v} (1+F_{D14}(h)), \\
    \mathcal{O}_{D15} &= \langle\tau_L L_\mu \rangle \ \frac{\partial^\mu h\, \partial_\nu h\, \partial^\nu h}{v^3} (1+F_{D15}(h)).
  \end{aligned}
\end{align}
If the leading-order Lagrangian preserves custodial symmetry, only the operators $\mathcal{O}_{Di}$ with $i\in \{1,2,7,8,11\}$ are needed as counterterms. The other operators receive a further suppression by the weak couplings that accompany the spurions. They might still be generated with finite contributions at the same, sub-leading, order. 
\subsection{NLO Operators of the Class $ g^{2}X^{2}Uh$}
The $\mathcal{CP}$-even operators of this class are
\begin{align}
  \begin{aligned}
    \label{eq:5.NLO.7}
    \mathcal{O}_{Xh1}&=g^{\prime 2} B_{\mu\nu} B^{\mu\nu} \, F_{Xh1}(h), \\
    \mathcal{O}_{Xh2}&=g^2 \langle W_{\mu\nu} W^{\mu\nu}\rangle \, F_{Xh2}(h), \\
    \mathcal{O}_{Xh3}&=g^2_s \langle G_{\mu\nu} G^{\mu\nu}\rangle \, F_{Xh3}(h), \\
    \mathcal{O}_{XU1}&=g'gB_{\mu\nu}\langle W^{\mu\nu}\tau_L\rangle \, (1+F_{XU1}(h)), \\
    \mathcal{O}_{XU2}&=g^2 \langle W_{\mu\nu}\tau_L\rangle^2 \, (1+F_{XU2}(h)). \\
  \end{aligned}
\end{align}
There are also $\mathcal{CP}$-odd operators in this class. They are
\begin{align}
  \begin{aligned}
    \label{eq:5.NLO.8}
    \mathcal{O}_{Xh4}&=g^{\prime 2} \varepsilon_{\mu\nu\lambda\rho} B^{\mu\nu} B^{\lambda\rho} \, F_{Xh4}(h), \\
    \mathcal{O}_{Xh5}&=g^2 \varepsilon_{\mu\nu\lambda\rho} \langle W^{\mu\nu} W^{\lambda\rho}\rangle \, F_{Xh5}(h), \\
    \mathcal{O}_{Xh6}&=g^2_s \varepsilon_{\mu\nu\lambda\rho} \langle G^{\mu\nu} G^{\lambda\rho}\rangle \, F_{Xh6}(h), \\
    \mathcal{O}_{XU4} &=g'g\varepsilon_{\mu\nu\lambda\rho}\langle \tau_L W^{\mu\nu}\rangle B^{\lambda\rho}\, (1+F_{XU4}(h)), \\
    \mathcal{O}_{XU5}&=g^2\varepsilon_{\mu\nu\lambda\rho}\langle\tau_LW^{\mu\nu}\rangle \langle\tau_LW^{\lambda\rho}\rangle \, (1+F_{XU5}(h)).  \\
  \end{aligned}
\end{align}
Since all of the operators contain the gauge fields explicitly, they come with two powers of the corresponding gauge coupling. The operator $\mathcal{O}_{XU1}$ has an explicit factor of $g'$, hence there is no need for an additional weak coupling to accompany the spurion in case of weak custodial symmetry breaking. As the custodial symmetry is respected by $SU(2)_{L}$, this argument does not hold for $\mathcal{O}_{XU2}$. The operators in this class generalize the operators discussed in \cite{Longhitano:1980tm,Appelquist:1993ka}. 
\subsection{NLO Operators of the Class $ gXD^{2}Uh $}
The comments on the previous class of next-to-leading order operators also apply here. Operators without explicit $B$ fields, but containing the spurion $\tau_{L}$ are further suppressed if custodial symmetry is only weakly broken. The $\mathcal{CP}$-even operators are
\begin{align}
  \begin{aligned}
    \label{eq:5.NLO.9}
    \mathcal{O}_{XU3}&=g\varepsilon_{\mu\nu\lambda\rho} \langle W^{\mu\nu}L^{\lambda}\rangle\langle\tau_L L^{\rho}\rangle \, (1+F_{XU3}(h)), \\
    \mathcal{O}_{XU7}&=ig'B_{\mu\nu}\langle\tau_L[L^{\mu},L^{\nu}] \rangle \, F_{XU7}(h), \\
    \mathcal{O}_{XU8}&=ig\langle W_{\mu\nu}[L^{\mu},L^{\nu}] \rangle \, F_{XU8}(h), \\
    \mathcal{O}_{XU9}&=ig\langle W_{\mu\nu}\tau_L\rangle \langle \tau_L[L^{\mu},L^{\nu}]\rangle \, F_{XU9}(h).
  \end{aligned}
\end{align}
The $\mathcal{CP}$-odd operators are
\begin{align}
  \begin{aligned}
    \label{eq:5.NLO.8.1}
    \mathcal{O}_{XU6}&=g\langle W_{\mu\nu}L^{\mu}\rangle\langle \tau_LL^{\nu}\rangle \, (1+F_{XU6}(h)), \\
    \mathcal{O}_{XU10} &=ig'\varepsilon_{\mu\nu\lambda\rho}B^{\mu\nu}\langle\tau_L[L^{\lambda},L^{\rho}] \rangle \, F_{XU10}(h), \\
    \mathcal{O}_{XU11} &=ig\varepsilon_{\mu\nu\lambda\rho}\langle W^{\mu\nu}[L^{\lambda},L^{\rho}] \rangle \, F_{XU11}(h), \\
    \mathcal{O}_{XU12}&=ig\varepsilon_{\mu\nu\lambda\rho}\langle W^{\mu\nu}\tau_L\rangle \langle \tau_L[L^{\lambda},L^{\rho}] \rangle \, F_{XU12}(h).
  \end{aligned}
\end{align}
The list of operators in this section reduces to the list of \cite{Longhitano:1980tm,Appelquist:1993ka} in the Higgs-less case.
\subsection{NLO Operators of the Class $ w^{2}\Psi^{2} UhD$}
Lorentz-invariance allows only vector currents for the fermion bilinears in this class. The weak couplings $w$ in front of the operators can either be Yukawa or gauge couplings. Since the chirality of the fermions is conserved in vector currents, an even number of the chirality-changing Yukawa couplings is needed. The operators are:
\begin{align}
  \begin{aligned}
    \label{eq:5.NLO.9.1}
    \mathcal{O}_{\psi V1}&=-w^{2}(\bar{q}_{L}\gamma^\mu q_{L})\ \langle  \tau_L L_\mu \rangle \, (1+F_{\psi V1}(h)), \\
    \mathcal{O}_{\psi V2}&=-w^{2}(\bar{q}_{L}\gamma^\mu \tau_L q_{L})\ \langle \tau_L L_\mu \rangle \, (1+F_{\psi V2}(h)), \\
    \mathcal{O}_{\psi V3}&=-w^{2}(\bar{q}_{L}\gamma^\mu U P_{12} U^\dagger q_{L})\ \langle  L_\mu U P_{21}U^\dagger\rangle \, (1+F_{\psi V3}(h)),  \\
    \mathcal{O}_{\psi V4}&=-w^{2}(\bar{u}_{R}\gamma^\mu u _{R})\ \langle  \tau_L L_\mu \rangle \, (1+F_{\psi V4}(h)),  \\
    \mathcal{O}_{\psi V5}&=-w^{2}(\bar{d}_{R}\gamma^\mu d _{R})\ \langle \tau_L L_\mu \rangle \, (1+F_{\psi V5}(h)), \\
    \mathcal{O}_{\psi V6}&=-w^{2}(\bar{u}_{R}\gamma^\mu d _{R})\ \langle  L_\mu UP_{21} U^\dagger\rangle \,  (1+F_{\psi V6}(h)),  \\
    \mathcal{O}_{\psi V7}&=-w^{2}(\bar{\ell}_L\gamma^\mu \ell_{L})\ \langle \tau_L L_\mu \rangle \, (1+F_{\psi V7}(h)), \\ 
    \mathcal{O}_{\psi V8}&=-w^{2}(\bar{\ell}_L\gamma^\mu \tau_L  \ell_{L})\  \langle \tau_L L_\mu \rangle \, (1+F_{\psi V8}(h)),   \\
    \mathcal{O}_{\psi V9}&=-w^{2}(\bar{\ell}_L\gamma^\mu U P_{12} U^\dagger \ell_{L})\ \langle  L_\mu U P_{21} U^\dagger\rangle \, (1+F_{\psi V9}(h)),  \\
    \mathcal{O}_{\psi V10}&=-w^{2}(\bar{e}_{R}\gamma^\mu e _{R})\ \langle \tau_L L_\mu \rangle \, (1+F_{\psi V10}(h)),  \\
    \mathcal{O}^\dagger_{\psi V3}, \quad & \mathcal{O}^\dagger_{\psi V6}, \quad \mathcal{O}^\dagger_{\psi V9}. 
  \end{aligned}
\end{align}
These operators generalize the Higgs-less operators discussed in \cite{Appelquist:1984rr,Bagan:1998vu,ManzanoFlecha:2002cx}. We write the minus sign to be consistent with the operators in \cite{Buchalla:2012qq} in the limit $F_{i}\rightarrow 0$. 
\subsection{NLO Operators of the Class $  Y \Psi^{2} UhD^{2}$}
The fermion bilinears in this class can be either scalar or tensor currents. Only the scalar currents are needed as counterterms. However, the operators with tensor currents might still receive finite contributions. The operators with scalar currents are:
\begin{align}
  \begin{aligned}
    \label{eq:5.NLO.10}
    \mathcal{O}_{\psi S1} &= Y (\bar{q}_{L} U P_{+}q_{R}) \langle L_{\mu}L^{\mu}\rangle (1+F_{\psi S1}(h)),\\ 
    \mathcal{O}_{\psi S2} &= Y (\bar{q}_{L} U P_{-}q_{R}) \langle L_{\mu}L^{\mu}\rangle (1+F_{\psi S2}(h)),\\
    \mathcal{O}_{\psi S3} &= Y (\bar{q}_{L} U P_{+}q_{R}) \langle\tau_{L}L_{\mu}\rangle \langle\tau_{L}L^{\mu}\rangle (1+F_{\psi S3}(h)),\\
    \mathcal{O}_{\psi S4} &= Y (\bar{q}_{L} U P_{-}q_{R}) \langle\tau_{L}L_{\mu}\rangle \langle\tau_{L}L^{\mu}\rangle (1+F_{\psi S4}(h)),\\
    \mathcal{O}_{\psi S5} &= Y (\bar{q}_{L} U P_{12}q_{R})\langle\tau_{L}L_{\mu}\rangle \langle U P_{21}U^{\dagger}L^{\mu}\rangle (1+F_{\psi S5}(h)),\\ 
    \mathcal{O}_{\psi S6} &= Y (\bar{q}_{L} U P_{21}q_{R})\langle\tau_{L}L_{\mu}\rangle \langle U P_{12}U^{\dagger}L^{\mu}\rangle (1+F_{\psi S6}(h)), \\
    \mathcal{O}_{\psi S7} &= Y(\bar{\ell}_{L} U P_{-}\ell_{R}) \langle L_{\mu}L^{\mu}\rangle (1+F_{\psi S7}(h)),\\
    \mathcal{O}_{\psi S8} &= Y (\bar{\ell}_{L} U P_{-}\ell_{R}) \langle\tau_{L}L_{\mu}\rangle \langle\tau_{L}L^{\mu}\rangle (1+F_{\psi S8}(h)),\\
    \mathcal{O}_{\psi S9} &= Y (\bar{\ell}_{L} U P_{12}\ell_{R}) \langle\tau_{L}L_{\mu}\rangle \langle U P_{21}U^{\dagger}L^{\mu}\rangle (1+F_{\psi S9}(h)),\\ 
    \mathcal{O}_{\psi S10} &= Y (\bar{q}_{L} U P_{+}q_{R}) \langle\tau_{L}L_{\mu}\rangle \left(\partial^{\mu}\tfrac{h}{v}\right) (1+F_{\psi S10}(h)),\\ 
    \mathcal{O}_{\psi S11} &= Y (\bar{q}_{L} U P_{-}q_{R}) \langle\tau_{L}L_{\mu}\rangle \left(\partial^\mu \tfrac{h}{v}\right) (1+F_{\psi S11}(h)), \\
    \mathcal{O}_{\psi S12} &= Y (\bar{q}_{L} U P_{12}q_{R}) \langle U P_{21}U^{\dagger}L_{\mu}\rangle\left(\partial^{\mu}\tfrac{h}{v}\right) (1+F_{\psi S12}(h)),\\
    \mathcal{O}_{\psi S13} &= Y (\bar{q}_{L} U P_{21}q_{R}) \langle U P_{12}U^{\dagger}L_{\mu}\rangle\left(\partial^{\mu}\tfrac{h}{v}\right) (1+F_{\psi S13}(h)),\\ 
    \mathcal{O}_{\psi S14} &= Y (\bar{q}_{L} U P_{+}q_{R}) \left(\partial_{\mu}\tfrac{h}{v}\right)\left(\partial^{\mu}\tfrac{h}{v}\right)(1+F_{\psi S14}(h)), \\
    \mathcal{O}_{\psi S15} &= Y (\bar{q}_{L} U P_{-}q_{R}) \left(\partial_{\mu}\tfrac{h}{v}\right)\left(\partial^{\mu}\tfrac{h}{v}\right) (1+F_{\psi S15}(h)),\\ 
    \mathcal{O}_{\psi S16} &= Y (\bar{\ell}_{L} U P_{-}\ell_{R}) \langle\tau_{L}L_{\mu}\rangle \left(\partial^{\mu}\tfrac{h}{v}\right) (1+F_{\psi S16}(h)),\\
    \mathcal{O}_{\psi S17} &= Y (\bar{\ell}_{L} U P_{12}\ell_{R}) \langle U P_{21}U^{\dagger}L_{\mu}\rangle\left(\partial^{\mu}\tfrac{h}{v}\right) (1+F_{\psi S17}(h)),\\
    \mathcal{O}_{\psi S18} &= Y (\bar{\ell}_{L} U P_{-}\ell_{R}) \left(\partial_{\mu}\tfrac{h}{v}\right)\left(\partial^{\mu}\tfrac{h}{v}\right) (1+F_{\psi S18}(h)).
  \end{aligned}
\end{align}
The operators with a tensor current are:
\begin{align}
  \begin{aligned}
    \label{eq:5.NLO.11}
    \mathcal{O}_{\psi T1} &= Y (\bar{q}_{L} \sigma_{\mu\nu}U P_{+}q_{R})\langle \tau_{L} L_{\mu}L_{\nu}\rangle (1+F_{\psi T1}(h)), \\
    \mathcal{O}_{\psi T2} &= Y (\bar{q}_{L} \sigma_{\mu\nu}U P_{-}q_{R})\langle \tau_{L} L_{\mu}L_{\nu}\rangle (1+F_{\psi T2}(h)),\\
    \mathcal{O}_{\psi T3} &= Y (\bar{q}_{L} \sigma_{\mu\nu}U P_{12}q_{R})\langle\tau_{L}L^{\mu}\rangle \langle U P_{21}U^{\dagger}L^{\nu}\rangle (1+F_{\psi T3}(h)), \\
    \mathcal{O}_{\psi T4} &= Y (\bar{q}_{L} \sigma_{\mu\nu}U P_{21}q_{R})\langle\tau_{L}L^{\mu}\rangle \langle U P_{12}U^{\dagger}L^{\nu}\rangle (1+F_{\psi T4}(h)),\\
    \mathcal{O}_{\psi T5} &= Y (\bar{\ell}_{L} \sigma_{\mu\nu}U P_{12}\ell_{R}) \langle\tau_{L}L^{\mu}\rangle \langle U P_{21}U^{\dagger}L^{\nu}\rangle (1+F_{\psi T5}(h)),\\
    \mathcal{O}_{\psi T6} &= Y (\bar{\ell}_{L} \sigma_{\mu\nu}U P_{-}\ell_{R}) \langle \tau_{L} L_{\mu}L_{\nu}\rangle (1+F_{\psi T6}(h)), \\
    \mathcal{O}_{\psi T7} &= Y (\bar{q}_{L} \sigma_{\mu\nu}U P_{+}q_{R}) \langle\tau_{L}L^{\mu}\rangle \left(\partial^{\nu}\tfrac{h}{v}\right) (1+F_{\psi T7}(h)), \\
    \mathcal{O}_{\psi T8} &= Y (\bar{q}_{L} \sigma_{\mu\nu}U P_{-}q_{R})\langle\tau_{L}L^{\mu}\rangle \left(\partial^{\nu}\tfrac{h}{v}\right) (1+F_{\psi T8}(h)), \\
    \mathcal{O}_{\psi T9} &= Y (\bar{q}_{L}\sigma_{\mu\nu} U P_{21}q_{R})\langle U P_{12}U^{\dagger}L^{\mu}\rangle\left(\partial^{\nu}\tfrac{h}{v}\right) (1+F_{\psi T9}(h)),\\
    \mathcal{O}_{\psi T10} &= Y (\bar{q}_{L}\sigma_{\mu\nu} U P_{12}q_{R})\langle U P_{21}U^{\dagger}L^{\mu}\rangle\left(\partial^{\nu}\tfrac{h}{v}\right) (1+F_{\psi T10}(h)),\\
    \mathcal{O}_{\psi T11} &= Y (\bar{\ell}_{L} \sigma_{\mu\nu}U P_{-}\ell_{R}) \langle\tau_{L}L^{\mu}\rangle \left(\partial^{\nu}\tfrac{h}{v}\right) (1+F_{\psi T11}(h)),\\
    \mathcal{O}_{\psi T12} &= Y (\bar{\ell}_{L}\sigma_{\mu\nu} U P_{12}\ell_{R}) \langle U P_{21}U^{\dagger}L^{\mu}\rangle\left(\partial^{\nu}\tfrac{h}{v}\right) (1+F_{\psi T12}(h)).
  \end{aligned}
\end{align}
The hermitean conjugates of all the operators are also in the basis, even though we do not list them. The Yukawa coupling in front of the operator is needed, as the operators involve fermions of both chiralities. Some of the operators were discussed in the Higgs-less case in \cite{Peccei:1989kr}.
\subsection{NLO Operators of the Class $ gXUh\Psi^{2}$}
\label{ch:5.NLO.dipole}
The operators of this class are not required as counterterms because the one-loop diagrams yielding these structures are finite. However, these operators might still be generated with finite coefficients. The operators are:
\begin{align}
  \begin{aligned}
    \label{eq:5.NLO.12}
    \mathcal{O}_{\psi X1} &= g' (\bar{q}_{L} \sigma_{\mu\nu}U P_{+}q_{R}) B^{\mu\nu}(1+F_{\psi X1}(h)),\\ 
    \mathcal{O}_{\psi X2} &= g' (\bar{q}_{L} \sigma_{\mu\nu}U P_{-}q_{R}) B^{\mu\nu} (1+F_{\psi X2}(h)),\\ 
    \mathcal{O}_{\psi X3} &= g (\bar{q}_{L} \sigma_{\mu\nu}U P_{+}q_{R}) \langle \tau_{L} W^{\mu\nu}\rangle (1+F_{\psi X3}(h)),\\ 
    \mathcal{O}_{\psi X4} &= g (\bar{q}_{L} \sigma_{\mu\nu}U P_{-} q_{R}) \langle \tau_{L} W^{\mu\nu}\rangle (1+F_{\psi X4}(h)),\\
    \mathcal{O}_{\psi X5} &= g (\bar{q}_{L} \sigma_{\mu\nu}U P_{12}q_{R})\langle U P_{21} U^{\dagger} W^{\mu\nu}\rangle (1+F_{\psi X5}(h)),\\
    \mathcal{O}_{\psi X6} &= g (\bar{q}_{L} \sigma_{\mu\nu}U P_{21}q_{R}) \langle U P_{12} U^{\dagger} W^{\mu\nu}\rangle (1+F_{\psi X6}(h)),\\
    \mathcal{O}_{\psi X7} &= g_{s}(\bar{q}_{L} \sigma_{\mu\nu}G^{\mu\nu}U P_{+}q_{R}) (1+F_{\psi X7}(h)),\\
    \mathcal{O}_{\psi X8} &= g_{s}(\bar{q}_{L} \sigma_{\mu\nu}G^{\mu\nu}U P_{-}q_{R}) (1+F_{\psi X8}(h)),\\
    \mathcal{O}_{\psi X9} &= g' (\bar{\ell}_{L} \sigma_{\mu\nu}U P_{-}\ell_{R}) B^{\mu\nu} (1+F_{\psi X9}(h)),\\
    \mathcal{O}_{\psi X10} &= g (\bar{\ell}_{L} \sigma_{\mu\nu}U P_{-}\ell_{R}) \langle \tau_{L} W^{\mu\nu}\rangle (1+F_{\psi X10}(h)),\\
    \mathcal{O}_{\psi X11} &= g (\bar{\ell}_{L} \sigma_{\mu\nu}U P_{12}\ell_{R}) \langle U P_{21} U^{\dagger} W^{\mu\nu}\rangle (1+F_{\psi X11}(h)). 
\end{aligned}
\end{align}
The hermitean conjugates of these operators are also independent operators of the basis. 
\subsection{NLO Operators of the Class $w^{2}\Psi^{4}Uh$}
The four-fermion operators can be further grouped according to their chirality structure. 
The $\bar LL\bar LL$ operators are
\begin{align*}
  \begin{aligned}
    \mathcal{O}_{LL1}&= w^{2} (\bar{q}_{L}\gamma^\mu q_{L})\, (\bar{q}_{L}\gamma_\mu q_{L}) (1+F_{LL1}(h)),\\
    \mathcal{O}_{LL2}&= w^{2} (\bar{q}_{L}\gamma^\mu T^a q_{L})\, (\bar{q}_{L}\gamma_\mu T^a q_{L}) (1+F_{LL2}(h)),\\
    \mathcal{O}_{LL3}&= w^{2} (\bar{q}_{L}\gamma^\mu q_{L})\, (\bar \ell_{L}\gamma_\mu \ell_{L}) (1+F_{LL3}(h)),\\   
    \mathcal{O}_{LL4}&= w^{2} (\bar{q}_{L}\gamma^\mu T^a q_{L})\, (\bar \ell_{L}\gamma_\mu T^a \ell_{L}) (1+F_{LL4}(h)),\\            
    \mathcal{O}_{LL5}&= w^{2} (\bar \ell_{L}\gamma^\mu \ell_{L})\, (\bar \ell_{L}\gamma_\mu \ell_{L}) (1+F_{LL5}(h)),\\
    \mathcal{O}_{LL6}&= w^{2} (\bar{q}_{L}\gamma^\mu UT_3U^\dagger q_{L})\, (\bar{q}_{L}\gamma_\mu UT_3U^\dagger q_{L}) (1+F_{LL6}(h)),\\
    \mathcal{O}_{LL7}&= w^{2} (\bar{q}_{L}\gamma^\mu UT_3U^\dagger q_{L})\, (\bar{q}_{L}\gamma_\mu q_{L}) (1+F_{LL7}(h)),\\
    \mathcal{O}_{LL8}&= w^{2} (\bar{q}_{L,\alpha}\gamma^\mu UT_3U^\dagger q_{L,\beta})\, (\bar{q}_{L,\beta}\gamma_\mu UT_3 U^\dagger q_{L,\alpha}) (1+F_{LL8}(h)),\\
    \mathcal{O}_{LL9}&= w^{2} (\bar{q}_{L,\alpha}\gamma^\mu  UT_3U^\dagger q_{L,\beta})\,  (\bar{q}_{L,\beta}\gamma_\mu  q_{L,\alpha}) (1+F_{LL9}(h)),\\
    \mathcal{O}_{LL10}&= w^{2}  (\bar{q}_{L}\gamma^\mu UT_3U^\dagger q_{L})\, (\bar \ell_{L}\gamma_\mu UT_3U^\dagger \ell_{L}) (1+F_{LL10}(h)),
 \end{aligned}
\end{align*}
\begin{align}
  \begin{aligned}
    \label{eq:5.NLO.13} 
    \mathcal{O}_{LL11}&= w^{2} (\bar{q}_{L}\gamma^\mu UT_3U^\dagger q_{L})\, (\bar \ell_{L}\gamma_\mu \ell_{L}) (1+F_{LL11}(h)),\\
    \mathcal{O}_{LL12}&= w^{2} (\bar{q}_{L}\gamma^\mu q_{L})\, (\bar \ell_{L}\gamma_\mu UT_3U^\dagger \ell_{L}) (1+F_{LL12}(h)),\\
    \mathcal{O}_{LL13}&= w^{2} (\bar{q}_{L}\gamma^\mu  UT_3U^\dagger \ell_{L})\, (\bar \ell_{L}\gamma_\mu  UT_3U^\dagger q_{L}) (1+F_{LL13}(h)),\\  
    \mathcal{O}_{LL14}&= w^{2} (\bar{q}_{L}\gamma^\mu  UT_3U^\dagger \ell_{L})\, (\bar \ell_{L}\gamma_\mu  q_{L}) (1+F_{LL14}(h)),\\
    \mathcal{O}_{LL15}&= w^{2}  (\bar \ell_{L}\gamma^\mu UT_3U^\dagger \ell_{L})\, (\bar \ell_{L}\gamma_\mu UT_3U^\dagger \ell_{L}) (1+F_{LL15}(h)),\\ 
    \mathcal{O}_{LL16}&= w^{2} (\bar \ell_{L}\gamma^\mu UT_3U^\dagger \ell_{L})\, (\bar \ell_{L}\gamma_\mu \ell_{L}) (1+F_{LL16}(h)).
  \end{aligned}
\end{align}
The $\bar RR\bar RR$ operators are
\begin{align}
  \begin{aligned}
    \label{eq:5.NLO.14}
    \mathcal{O}_{RR1}&= w^{2} (\bar{u}_{R}\gamma^\mu u _{R})\, (\bar{u}_{R}\gamma_\mu u_{R}) (1+F_{RR1}(h)),\\
    \mathcal{O}_{RR2}&= w^{2} (\bar{d}_{R}\gamma^\mu d _{R})\, (\bar{d}_{R}\gamma_\mu d_{R}) (1+F_{RR2}(h)),\\            
    \mathcal{O}_{RR3}&= w^{2} (\bar{u}_{R}\gamma^\mu u _{R})\, (\bar{d}_{R}\gamma_\mu d_{R}) (1+F_{RR3}(h)),\\            
    \mathcal{O}_{RR4}&= w^{2} (\bar{u}_{R}\gamma^\mu T^A u _{R})\, (\bar{d}_{R}\gamma_\mu T^A d_{R}) (1+F_{RR4}(h)),\\               
    \mathcal{O}_{RR5}&= w^{2} (\bar{u}_{R}\gamma^\mu u _{R})\, (\bar{e}_{R}\gamma_\mu e_{R}) (1+F_{RR5}(h)),\\  
    \mathcal{O}_{RR6}&= w^{2} (\bar{d}_{R}\gamma^\mu d _{R})\, (\bar{e}_{R}\gamma_\mu e_{R}) (1+F_{RR6}(h)),\\    
    \mathcal{O}_{RR7}&= w^{2} (\bar{e}_{R}\gamma^\mu e _{R})\, (\bar{e}_{R}\gamma_\mu e_{R}) (1+F_{RR7}(h)). 
  \end{aligned}
\end{align}
The $\bar LL\bar RR$ operators are
\begin{align}
  \begin{aligned}
    \label{eq:5.NLO.15}
    \mathcal{O}_{LR1}&= w^{2} (\bar{q}_{L}\gamma^\mu q_{L})\, (\bar{u}_{R}\gamma_\mu u_{R}) (1+F_{LR1}(h)),\\           
    \mathcal{O}_{LR2}&= w^{2} (\bar{q}_{L}\gamma^\mu T^A q_{L})\, (\bar{u}\gamma_\mu T^A u_{R}) (1+F_{LR2}(h)),\\             
    \mathcal{O}_{LR3}&= w^{2} (\bar{q}_{L}\gamma^\mu q_{L})\, (\bar{d}_{R}\gamma_\mu d_{R}) (1+F_{LR3}(h)),\\     
    \mathcal{O}_{LR4}&= w^{2} (\bar{q}_{L}\gamma^\mu T^A q_{L})\, (\bar{d}_{R}\gamma_\mu T^A d_{R}) (1+F_{LR4}(h)),\\            
    \mathcal{O}_{LR5}&= w^{2} (\bar{u}_{R}\gamma^\mu u_{R})\, (\bar \ell_{L}\gamma_\mu \ell_{L}) (1+F_{LR5}(h)),\\
    \mathcal{O}_{LR6}&= w^{2} (\bar{d}_{R}\gamma^\mu d_{R})\, (\bar \ell_{L}\gamma_\mu \ell_{L}) (1+F_{LR6}(h)),\\
    \mathcal{O}_{LR7}&= w^{2} (\bar{q}_{L}\gamma^\mu q_{L})\, (\bar{e}_{R}\gamma_\mu e_{R}) (1+F_{LR7}(h)),\\           
    \mathcal{O}_{LR8}&= w^{2} (\bar \ell_{L}\gamma^\mu \ell_{L})\, (\bar{e}_{R}\gamma_\mu e_{R}) (1+F_{LR8}(h)),\\
    \mathcal{O}_{LR9}&= w^{2} (\bar{q}_{L}\gamma^\mu \ell_{L})\, (\bar{e}_{R}\gamma_\mu d_{R}) (1+F_{LR9}(h)),\\               
    \mathcal{O}_{LR10}&= w^{2}  (\bar{q}_{L}\gamma^\mu UT_3U^\dagger q_{L})\, (\bar{u}_{R}\gamma_\mu u_{R}) (1+F_{LR10}(h)),\\              
    \mathcal{O}_{LR11}&= w^{2} (\bar{q}_{L}\gamma^\mu T^A UT_3U^\dagger q_{L})\, (\bar{u}_{R}\gamma_\mu T^A u_{R}) (1+F_{LR11}(h)),\\
    \mathcal{O}_{LR12}&= w^{2} (\bar{q}_{L}\gamma^\mu UT_3U^\dagger q_{L})\, (\bar{d}_{R}\gamma_\mu d_{R}) (1+F_{LR12}(h)),\\
    \mathcal{O}_{LR13}&= w^{2} (\bar{q}_{L}\gamma^\mu T^A UT_3U^\dagger q_{L})\, (\bar{d}_{R}\gamma_\mu T^A d_{R}) (1+F_{LR13}(h)),\\
    \mathcal{O}_{LR14}&= w^{2} (\bar{u}_{R}\gamma^\mu u_{R})\, (\bar \ell_{L}\gamma_\mu UT_3U^\dagger \ell_{L}) (1+F_{LR14}(h)),\\  
    \mathcal{O}_{LR15}&= w^{2} (\bar{d}_{R}\gamma^\mu d_{R})\, (\bar \ell_{L}\gamma_\mu UT_3U^\dagger \ell_{L}) (1+F_{LR15}(h)),\\
    \mathcal{O}_{LR16}&= w^{2} (\bar{q}_{L}\gamma^\mu UT_3U^\dagger q_{L})\, (\bar{e}_{R}\gamma_\mu e_{R}) (1+F_{LR16}(h)),\\   
    \mathcal{O}_{LR17}&= w^{2} (\bar \ell_{L}\gamma^\mu UT_3U^\dagger \ell_{L})\, (\bar{e}_{R}\gamma_\mu e_{R}) (1+F_{LR17}(h)),\\
    \mathcal{O}_{LR18}&= w^{2} (\bar{q}_{L}\gamma^\mu UT_3U^\dagger \ell_{L})\, (\bar{e}_{R}\gamma_\mu d_{R}) (1+F_{LR18}(h)). 
  \end{aligned}
\end{align}
The $\bar LR\bar LR$ operators are
\begin{align}
  \begin{aligned}
    \label{eq:5.NLO.16}
    \mathcal{O}_{ST1}&= w^{2} \varepsilon_{ij}\, (\bar{q}_{L}^i u_{R})\, (\bar{q}_{L}^j d_{R}) (1+F_{ST1}(h)),\\
    \mathcal{O}_{ST2}&= w^{2} \varepsilon_{ij}\, (\bar{q}_{L}^i T^A u_{R})\, (\bar{q}_{L}^j T^A d_{R}) (1+F_{ST2}(h)),\\
    \mathcal{O}_{ST3}&= w^{2} \varepsilon_{ij}\, (\bar{q}_{L}^i u_{R})\, (\bar \ell_{L}^j e_{R}) (1+F_{ST3}(h)),\\
    \mathcal{O}_{ST4}&= w^{2} \varepsilon_{ij}\, (\bar{q}_{L}^i \sigma^{\mu\nu} u_{R})\, (\bar \ell_{L}^j \sigma_{\mu\nu}  e_{R}) (1+F_{ST4}(h)),\\ 
    \mathcal{O}_{ST5}&= w^{2} (\bar{q}_{L} UP_+ q_{R})\, (\bar{q}_{L} UP_- q_{R}) (1+F_{ST5}(h)),\\
    \mathcal{O}_{ST6}&= w^{2} (\bar{q}_{L} UP_{21} q_{R})\, (\bar{q}_{L} UP_{12} q_{R}) (1+F_{ST6}(h)),\\
    \mathcal{O}_{ST7}&= w^{2} (\bar{q}_{L} UP_+ T^A q_{R})\, (\bar{q}_{L} UP_- T^A q_{R}) (1+F_{ST7}(h)),\\
    \mathcal{O}_{ST8}&= w^{2} (\bar{q}_{L} UP_{21} T^A q_{R})\, (\bar{q}_{L} UP_{12} T^A q_{R}) (1+F_{ST8}(h)),\\
    \mathcal{O}_{ST9}&= w^{2} (\bar{q}_{L} UP_+ q_{R})\, (\bar \ell_{L} UP_- \ell_{R}) (1+F_{ST9}(h)),\\
    \mathcal{O}_{ST10}&= w^{2} (\bar{q}_{L} UP_{21} q_{R})\, (\bar \ell_{L} UP_{12} \ell_{R}) (1+F_{ST10}(h)),\\
    \mathcal{O}_{ST11}&= w^{2} (\bar{q}_{L}\sigma^{\mu\nu} UP_+ q_{R})\, (\bar \ell_{L}\sigma_{\mu\nu} UP_- \ell_{R}) (1+F_{ST11}(h)),\\
    \mathcal{O}_{ST12}&= w^{2} (\bar{q}_{L}\sigma^{\mu\nu} UP_{21} q_{R})\, (\bar \ell_{L}\sigma_{\mu\nu} UP_{12} \ell_{R}) (1+F_{ST12}(h)). 
  \end{aligned}
\end{align}
The operators in which the total hypercharge of the fermions is not zero ($\mathsf{Y}(\psi^4)= \pm 1$), but compensated by the hypercharge of $U$, are
\begin{align}
  \begin{aligned}
    \label{eq:5.NLO.17}                                             
    \mathcal{O}_{FY1}&= w^{2} (\bar{q}_{L} UP_+ q_{R})\, (\bar{q}_{L} UP_+ q_{R}) (1+F_{FY1}(h)),\\
    \mathcal{O}_{FY2}&= w^{2} (\bar{q}_{L} UP_+ T^A q_{R})\, (\bar{q}_{L} UP_+ T^A q_{R}) (1+F_{FY2}(h)),\\
    \mathcal{O}_{FY3}&= w^{2} (\bar{q}_{L} UP_- q_{R})\, (\bar{q}_{L} UP_- q_{R}) (1+F_{FY3}(h)),\\
    \mathcal{O}_{FY4}&= w^{2} (\bar{q}_{L} UP_- T^A q_{R})\, (\bar{q}_{L} UP_- T^A q_{R}) (1+F_{FY4}(h)),\\
    \mathcal{O}_{FY5}&= w^{2} (\bar{q}_{L} UP_- q_{R})\, (\bar q_{R} P_+ U^\dagger q_{L}) (1+F_{FY5}(h)),\\
    \mathcal{O}_{FY6}&= w^{2} (\bar{q}_{L} UP_- T^A q_{R})\, (\bar q_{R} P_+ U^\dagger T^A q_{L}) (1+F_{FY6}(h)),\\                                     
    \mathcal{O}_{FY7}&= w^{2} (\bar{q}_{L} UP_- q_{R})\, (\bar \ell_{L} UP_- \ell_{R}) (1+F_{FY7}(h)),\\
    \mathcal{O}_{FY8}&= w^{2} (\bar{q}_{L}\sigma^{\mu\nu} UP_- q_{R})\, (\bar \ell_{L}\sigma_{\mu\nu} UP_- \ell_{R}) (1+F_{FY8}(h)),\\
    \mathcal{O}_{FY9}&= w^{2} (\bar \ell_{L} UP_- \ell_{R})\, (\bar q_{R} P_+ U^\dagger q_{L}) (1+F_{FY9}(h)),\\
    \mathcal{O}_{FY10}&= w^{2} (\bar \ell_{L} UP_- \ell_{R})\, (\bar \ell_{L} U P_- \ell_{R}) (1+F_{FY10}(h)),\\
    \mathcal{O}_{FY11}&= w^{2} (\bar \ell_{L} UP_- q_{R})\, (\bar q_{R} P_+ U^\dagger \ell_{L}) (1+F_{FY11}(h)).
  \end{aligned}
\end{align}
These operators trivially extend the list of operators in the Higgs-less basis \cite{Buchalla:2012qq}, as there are no derivatives in this class. Some of them were also discussed in \cite{Appelquist:1984rr}. The operators without the matrix $U$ have also been discussed in the dimension-six basis of \cite{Grzadkowski:2010es}. The weak couplings can either be gauge or Yukawa couplings. Some of the operators of this class are not needed as counterterms. They can, however, be generated {\it via} tree-level exchange of heavy resonances. Therefore, we keep them in the list of operators. 
\section{One-Loop Renormalization}
The consistent power counting of the electroweak chiral Lagrangian ensures the renormalizability, order by order in the effective expansion. Many authors studied the renormalization of the chiral Lagrangian, in general and in the context of specific processes. Usually, the scattering of longitudinally polarized gauge bosons is analyzed. The considered processes include: $W_{L}W_{L}\rightarrow Z_{L}Z_{L}$ analyzed in \cite{Espriu:2013fia}; $V_{L}V_{L}\rightarrow V_{L}V_{L}$, $V_{L}V_{L}\leftrightarrow hh$, and $hh\rightarrow hh$ for $V\in \{Z,W^{\pm}\}$ in \cite{Delgado:2013hxa}; $\gamma\gamma\rightarrow W_{L}W_{L}$ and $\gamma\gamma\rightarrow Z_{L}Z_{L}$ in \cite{Delgado:2014jda}; and $V_{L}V_{L} \rightarrow \bar{t}t$ in \cite{Castillo:2016erh}. Using the Goldstone boson equivalence theorem \cite{Cornwall:1974km,Vayonakis:1976vz} simplifies the computation of the scattering of longitudinal components of gauge bosons. The theorem states that the longitudinal component of the gauge field is described by the Goldstone boson at energies $E$ above the gauge boson mass $m_{V}$. Corrections arise at the order $\mathcal{O}(m_{V}/E)$. The list of processes given previously therefore yields the one-loop renormalization of the Goldstone boson sector. In these computations, the authors found agreement between the counterterms listed in Section \ref{ch:chiralNLO} and the counterterms needed for the renormalization.\\

Recently, the authors of \cite{Gavela:2014uta} considered a subset of the complete electroweak chiral Lagrangian with a light Higgs for a diagrammatical, one-loop renormalization. In \cite{Guo:2015isa}, Guo, Ruiz-Femen\'ia, and Sanz-Cillero considered the complete chiral Lagrangian with external gauge fields and fermions. In particular, they computed all divergent contributions of one-loop diagrams with Goldstone and Higgs fields in the loop, using the background field method \cite{Abbott:1981ke,Abbott:1980hw} and the heat-kernel \cite{Donoghue:1992dd}. A geometric approach to the scalar sector confirms this computation in \cite{Alonso:2015fsp}.  
Sanz-Cillero {\it et al.} found a generic matrix element to be \cite{Sanz-Cillero:2015bia}:
\begin{equation}
  \label{eq:5.15}
  \mathcal{M} \sim \underbrace{ \frac{p^2}{v^2}  }_{\text{LO (tree)}} \, + \, \bigg(\, \underbrace{ c_{k,n}^r }_{\text{ NLO (tree) }} \quad -\quad \underbrace{   \frac{ \Gamma_{k,n} }{16\pi^2}\ln\frac{p}{\mu} \quad +\quad ...  }_{\text{NLO (1-loop)}    }\quad \bigg)\,\,\, \frac{p^4}{v^4}\, \,\, +\,\,\, \mathcal{O}(p^6).
\end{equation}
Here, $p$ is the momentum of the process, $c_{k,}^r$ is the renormalized next-to-leading order coupling of the operator $k$ with $n$ Higgs legs, and $\Gamma_{k,n}$ encodes the running of the Wilson coefficients. We see in Eq.~\eqref{eq:5.15} that one-loop diagrams of the leading order Lagrangian contribute at the same order as tree-level diagrams with one single next-to-leading order vertex. The NLO operators renormalize the one-loop divergences, as stated by the power counting. 

Further, we observe that there is no mixing between the operators at the one-loop order. The one-loop diagrams with one NLO operator insertion, which introduce a mixing, are of chiral order six and therefore further suppressed in the electroweak chiral Lagrangian. This is in contrast to the SM-EFT, where the dimension-six operators mix at the one-loop level.

The running of the Wilson coefficients can be found from Eq.~\eqref{eq:5.15} and is given by \cite{Guo:2015isa}
\begin{equation}
  \label{eq:5.16}
  \mu \frac{d}{d\mu} c_{k,}^r = - \frac{\Gamma_{k,n}}{16\pi^{2}}.
\end{equation}
The effects of the running are important when we match the EFT to a UV model at a very high scale and want to use the EFT at a lower scale. We discussed this already in the context of the renormalization within SM-EFT in Section \ref{ch:SM-EFT.renorm}. However, these effects are loop suppressed and therefore well below the current experimental precision. Hence, we do not include them in our analysis. When the experimental precision reaches the sub-percent level, these corrections start to become important. The explicit computation of \cite{Guo:2015isa} also confirmed the structure of some of the counterterms we discussed in Section \ref{ch:chiralNLO}. Our NLO basis is therefore complete within the subset of one-loop diagrams considered in \cite{Guo:2015isa}. 

However, the complete renormalization of the electroweak chiral Lagrangian including the light Higgs is still missing. It will provide the complete running of the Wilson coefficients. Further, the complete renormalization gives the full list of counterterms, which is a subset of the operators that are based on the superficial degree of divergence. Nevertheless, finite contributions to the other operators of our list are always possible. Therefore, Eqs.~\eqref{eq:5.NLO.1} -- \eqref{eq:5.NLO.17} give the full list of NLO operators. 
  \chapter[Relation Between the SM-EFT and the $ew\chi\mathcal{L}$]{Relation Between the SM-EFT and the Electroweak Chiral Lagrangian}
\label{ch:relation}
\thispagestyle{fancyplain}
\section{General Considerations --- the Double Expansion}
We will now combine the two different expansions, which we discussed in Chapters~\ref{ch:SMEFT} and~\ref{ch:ewXL}, to obtain a phenomenologically interesting scenario. In particular, we assume that the Higgs sector is governed by a loop expansion of a non-decoupling EFT. The scale of new physics, $f$, that we assumed to be close to the electroweak scale $v$ can now be much higher. The new-physics sector that is non-decoupling at the scale $f$  decouples from the electroweak scale in the limit $f\rightarrow \infty$. In this case, we recover the SM. However, there are UV models in which we cannot take this limit, for example when the Higgs is a dilaton \cite{Goldberger:2008zz}. 

Our theory contains now three scales: the electroweak scale $v$, the scale of new physics $f$, and the cutoff $\Lambda \equiv 4 \pi f$. From these scales, we define two expansion parameters,
\begin{equation}
  \label{eq:6.1}
  \frac{f^{2}}{\Lambda^{2}} \equiv \frac{1}{16\pi^{2}} \qquad \text{ and } \qquad \xi \equiv \frac{v^{2}}{f^{2}}.
\end{equation}
In general, the EFT is a double expansion in both of these parameters. 
\begin{figure}[h!]
    \centering
    \includegraphics[width=0.55\textwidth,trim= 0 25 75 20,clip]{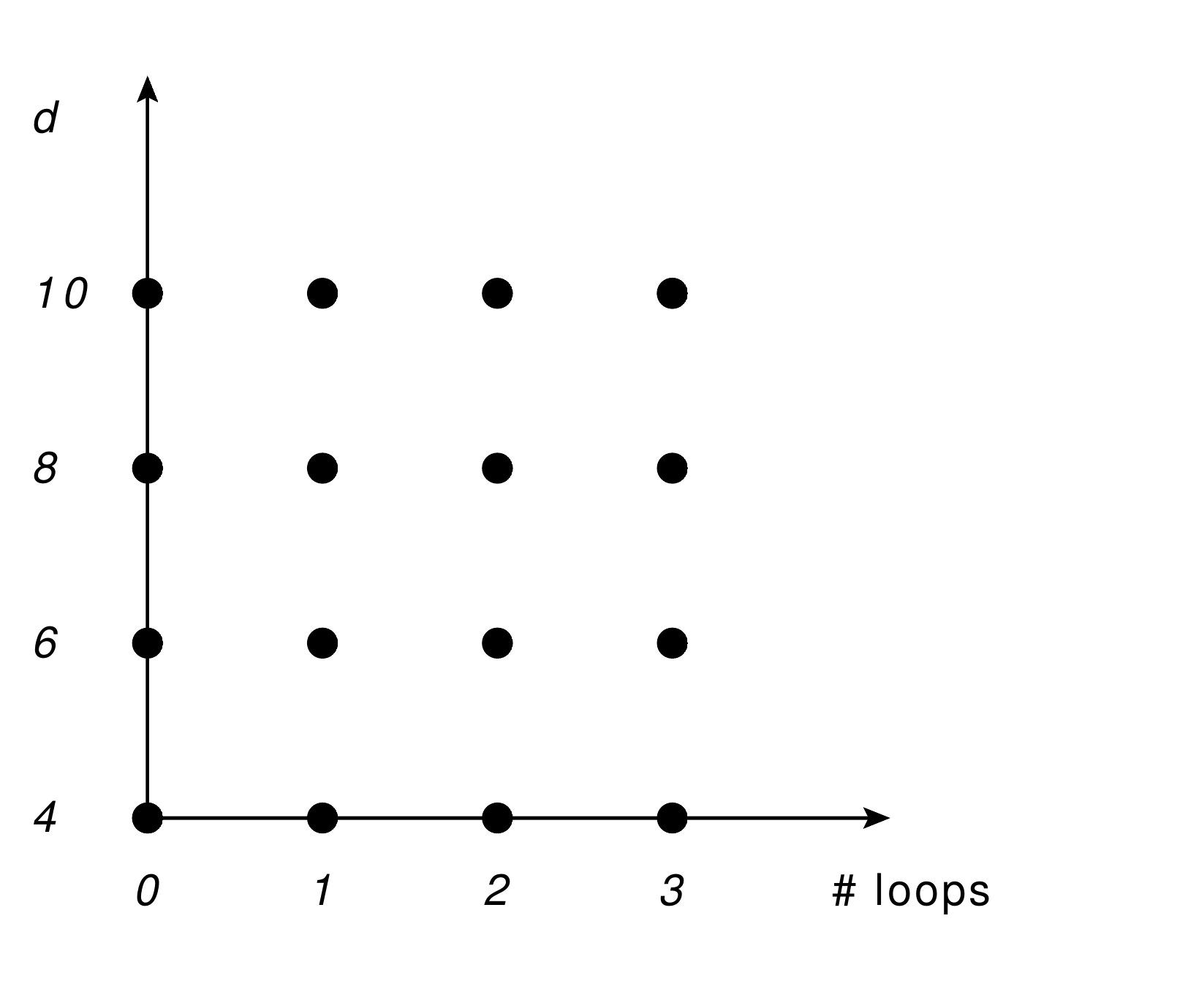}
    \caption[A visualization of the EFT's space of operators.]{A visualization of the EFT's space of operators \cite{Buchalla:2014eca}.}
    \label{fig:double.exp}
  \end{figure}

Figure~\ref{fig:double.exp} visualizes the possible expansions. Each black dot in the diagram represents (classes of) operators or terms in a physical amplitude. They are grouped according to the loop order at which they arise and the canonical dimension they have. The renormalizable SM of Eq.~\eqref{eq:SMunbroken}, for example, is given by the dot in the lower left corner. The loop expansion along the abscissa is equivalent to the expansion in chiral dimensions that we introduced in Eq.~\eqref{eq:5.10}. The expansion in canonical dimensions along the ordinate is equivalent to the expansion in $\xi$. Every power of $\xi$ introduces two powers of the new-physics scale $f$ in the denominator. An operator with canonical dimension $d$ therefore scales as $\xi^{(d-4)/2}$. 
\begin{figure}[h]
\begin{center}
\subfigure[SM-EFT: Expansion in canonical dimensions. \label{fig:double.lin}]{
\begin{overpic}[width = 0.47\textwidth]{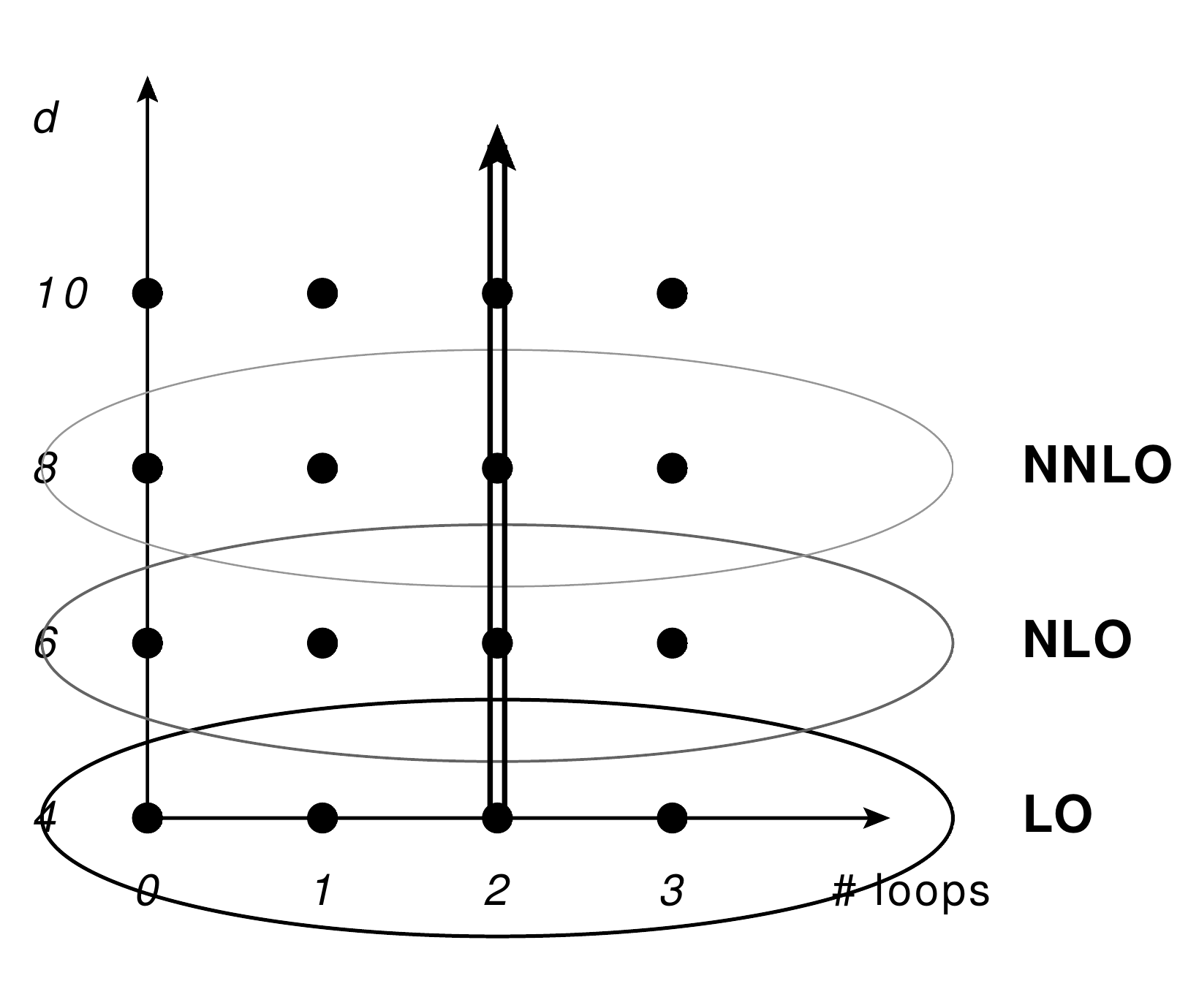}
\end{overpic}}\hfill
\subfigure[$ew\chi\mathcal{L}$: Expansion in loops. \label{fig:double.chiral}]{
\begin{overpic}[width = 0.47\textwidth]{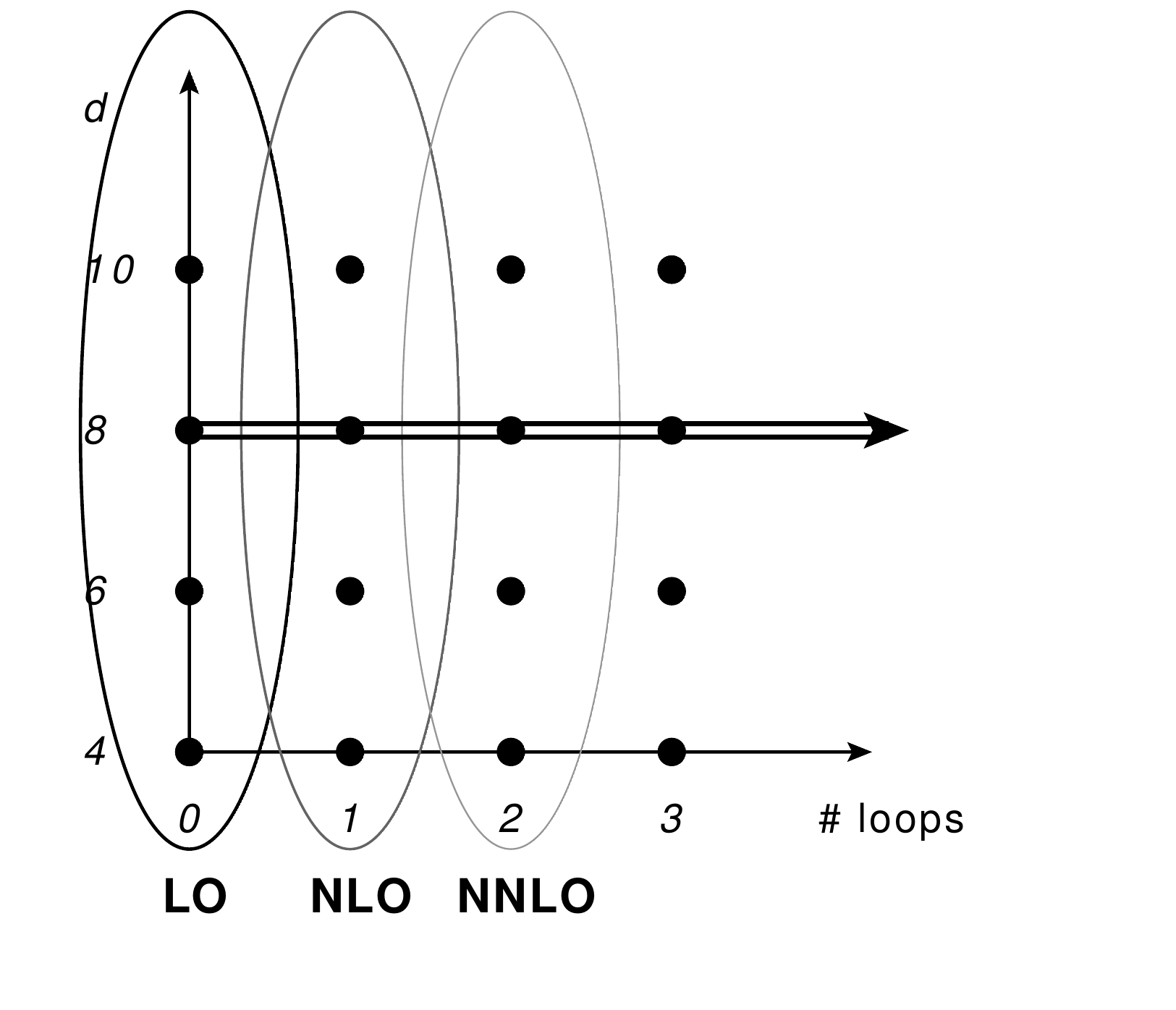}
\end{overpic}}
\end{center}
\caption[Different expansions of the EFT.]{Different expansions of the EFT.}
\label{fig:double.cases}
\end{figure}

The SM-EFT, which we discussed in Chapter~\ref{ch:SMEFT}, is an expansion in canonical dimensions. This corresponds to an expansion row by row from the bottom to the top of the diagram, see Fig.~\ref{fig:double.lin}. The electroweak chiral Lagrangian is an expansion in loops. As we show in Fig.~\ref{fig:double.chiral}, it corresponds to an expansion column by column, from left to right in the diagram. When we consider the EFT to all orders, both expansions cover the full theory space, {\it i.e.} all dots in the diagram of Fig.~\ref{fig:double.exp}. The two different expansions simply organize the operators in a different way. When we restrict our analysis to a given order in the EFT, the organisation of the operators becomes important. Operators that contribute in one expansion at a given order are not necessarily of the same order in the other expansion. 
\begin{figure}[h!]
    \centering
    \includegraphics[width=0.55\textwidth,trim= 0 40 75 20,clip]{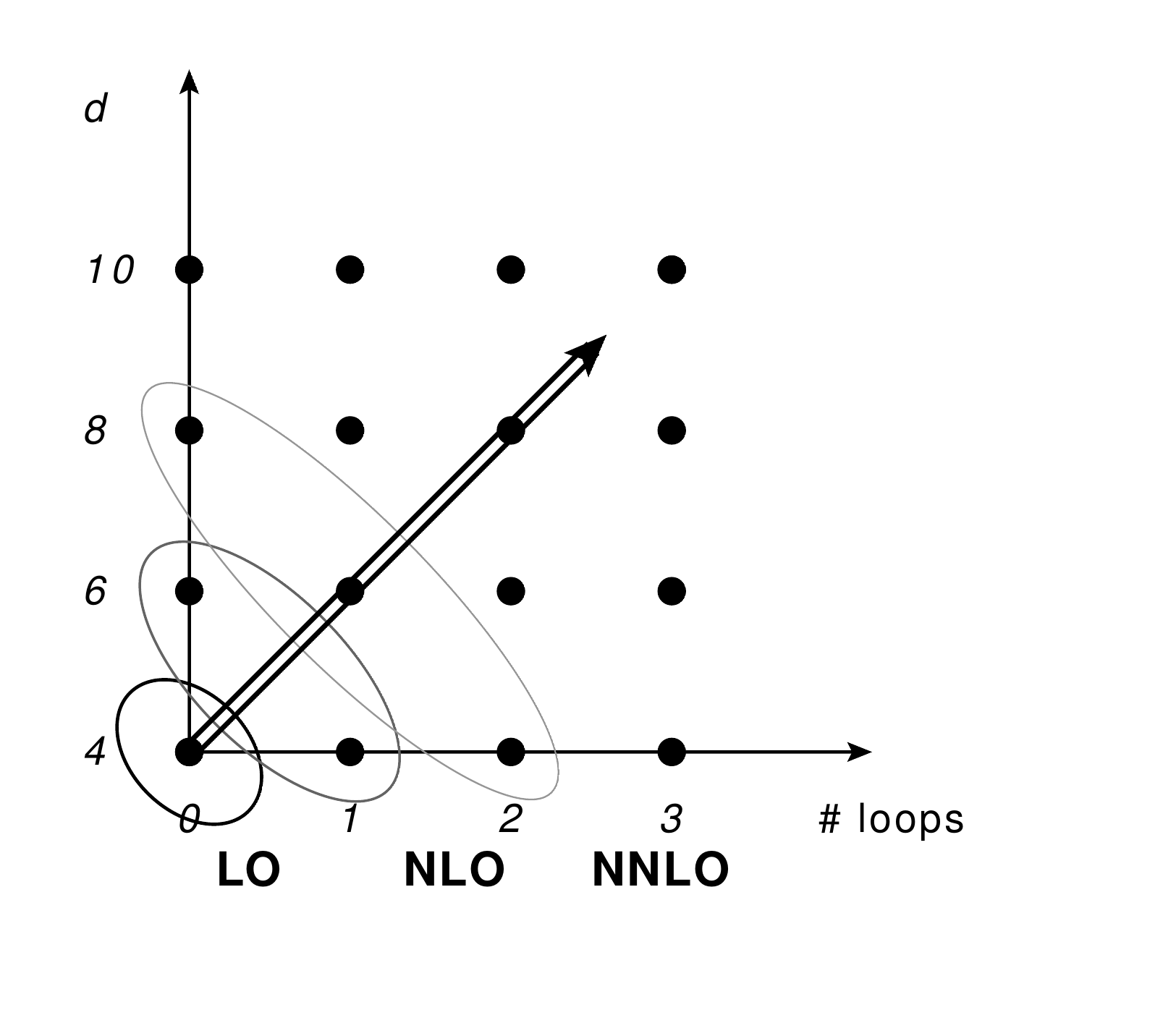}
    \caption[Double expansion in loops and canonical dimensions.]{The double expansion in loops (equivalent to an expansion in chiral dimensions) and canonical dimensions (equivalent to an expansion in $\xi$). }
    \label{fig:double.double}
  \end{figure}

The double expansion in the parameters of Eq.~\eqref{eq:6.1} goes through the diagram of Fig.~\ref{fig:double.exp} in a skewed way. The ``angle'' depends on the size of $\xi$. The expansion is at ``$45^{\circ}$'' if $\xi$ is of the order of the loop factor. This is illustrated in Fig.~\ref{fig:double.double}. For $\xi$ of order one, $\xi=\mathcal{O}(1)$, we approach the electroweak chiral Lagrangian and the angle ``flattens''. In the limit $\xi\rightarrow 1$, we reach the expansion of Fig.~\ref{fig:double.chiral}. For small $\xi$, the expansion becomes ``steeper'' and we approach the expansion in canonical dimensions of Fig.~\ref{fig:double.lin} in the limit $\xi\rightarrow 0$. Which of the scenarios is the most appropriate to describe nature therefore depends on the scale $f$. 
\section{Physics at the Scale $f$}
\label{ch:scale.f}
\begin{figure}[!t]
  \begin{center} 
    \begin{tikzpicture}
      \draw[thick] {(5em,0em) rectangle (15em,5em)};
      \draw (10em,4em) node[anchor = base] {No states at $f$};
      \draw[-,thick] (5em,3.75em) -- (15em,3.75em);
      \draw (10em,2em) node[anchor = base] {Double expansion};
      \draw (10em,0.5em) node[anchor = base] {Fig. \ref{fig:double.double}};
      \draw[thick] {(25em,0em) rectangle (35em,5em)};
      \draw (30em,4em) node[anchor = base] {States at $f$};
      \draw[-,thick] (25em,3.75em) -- (35em,3.75em);
      \draw (30em,2em) node[anchor = base] {SM-EFT + };
      \draw (30em,0.5em) node[anchor = base] {Double expansion};
      \draw[->,thick] (10em,7.5em) -- (10em,5em);
      \draw[->,thick] (30em,7.5em) -- (30em,5em);
      \draw[-,thick] (10em,7.5em) -- (30em,7.5em);
      \draw[-,thick] (20em,10em) -- (20em,7.5em);
      \draw[thick] {(15em,10em) rectangle (25em,15em)};
      \draw (20em,14em) node[anchor = base] {$v<f<3$ TeV};
      \draw (20em,10.5em) node[anchor = base] {Are there states at $f$?};
      \draw[-,thick] (15em,13.75em) -- (25em,13.75em);
      \draw[thick] {(30em,10em) rectangle (40em,15em)};
      \draw (35em,14em) node[anchor = base] {$f\gg 3$ TeV};
      \draw[-,thick] (30em,13.75em) -- (40em,13.75em);
      \draw (35em,12em) node[anchor = base] {SM-EFT};
      \draw (35em,10.5em) node[anchor = base] {Fig. \ref{fig:double.lin}};
      \draw[->,thick] (20em,17.5em) -- (20em,15em);
      \draw[->,thick] (35em,17.5em) -- (35em,15em);
      \draw[-,thick] (20em,17.5em) -- (35em,17.5em);
      \draw[-,thick] (30em,17.5em) -- (30em,20em);
      \draw[thick] {(5em,20em) rectangle (15em,25em)};
      \draw (10em,24em) node[anchor = base] {$f=v$};
      \draw[-,thick] (5em,23.75em) -- (15em,23.75em);
      \draw (10em,22em) node[anchor = base] {$ew\chi\mathcal{L}$};
      \draw (10em,20.5em) node[anchor = base] {Fig. \ref{fig:double.chiral}};
      \draw[thick] {(25em,20em) rectangle (35em,25em)};
      \draw (30em,24em) node[anchor = base] {$f>v$};
      \draw (30em,20.5em) node[anchor = base] {Where exactly is $f$?};
      \draw[-,thick] (25em,23.75em) -- (35em,23.75em);
      \draw[<->,thick] (15em,22.5em) -- (25em,22.5em);
      \draw[-,thick] (20em,25em) -- (20em,22.5em);
    \end{tikzpicture}
  \end{center}
  \caption{Different assumptions on the scale $f$.}
  \label{fig:6.scale.f}
\end{figure}
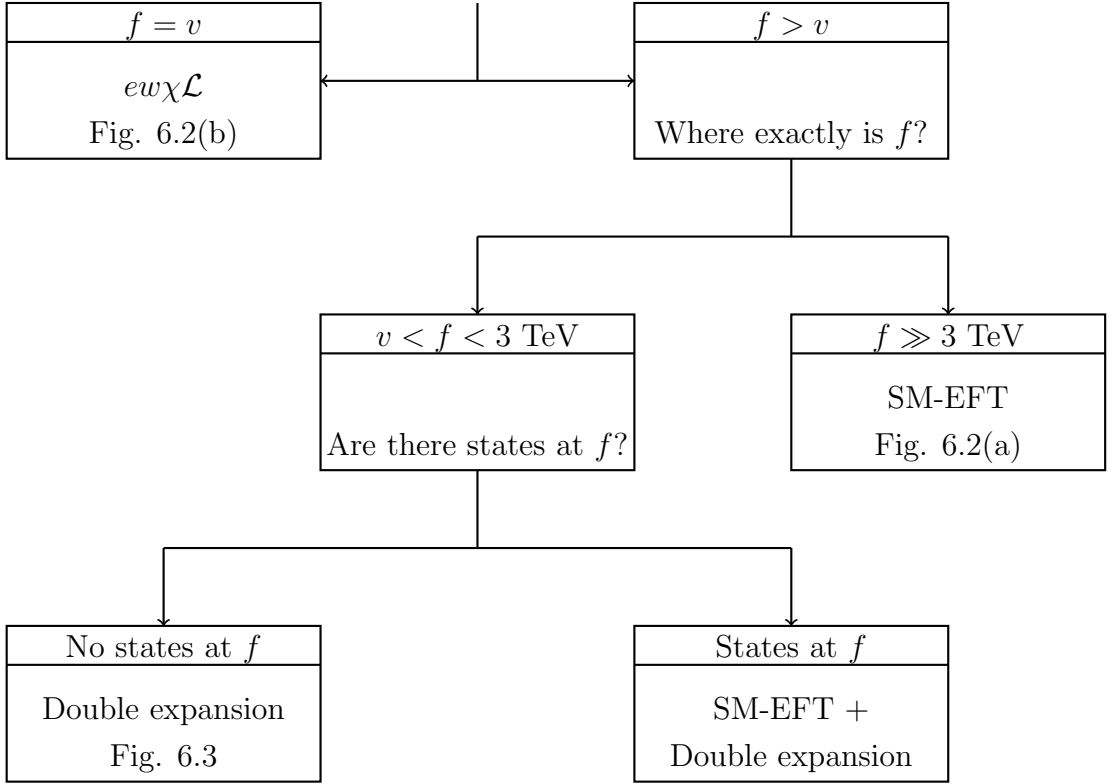

The distinction between the scales $f$ and $v$ gives us many different scenarios that we summarize in Fig.~\ref{fig:6.scale.f}. The case in which $f$ is at the weak scale $v$ gives us the electroweak chiral Lagrangian, which we discussed in Chapter~\ref{ch:ewXL}. If the scale $f$ is above the scale $v$, we can distinguish two scenarios. In one case, we have $f\gg 3~\text{TeV}$ and therefore $\xi \ll 1/16\pi^{2}$. The expansion in $\xi$ and therefore in canonical dimensions is now more appropriate. In the other case, $f$ lies between $v$ and $3~\text{TeV}$ and the expansion depends on whether there are states at the scale $f$ or not. Without those states, we have the electroweak chiral Lagrangian with an additional expansion in $\xi$, {\it i.e.} the double expansion of Fig.~\ref{fig:double.double}. We will discuss this in more detail in Section~\ref{ch:6.xiexpand}. Similar contributions arise if there are states at $f$ that contribute to the low-energy EFT only when integrated out at loop level, for example additional fermions. If there are states at the scale $f$ that can be integrated out at tree level, the situation is more complicated. Integrating out those states introduces effective operators that are suppressed by $1/f^{2}$ and not connected to the loop counting of the non-decoupling sector. This adds a dimensional expansion in $1/f^{2}$ to the already existing double expansion. We assume in this case that $f$ is close to the TeV scale or above, as otherwise the experimental collaborations should have seen the first effects already. If the fluctuation at $750~\text{GeV}$ in the di-photon channel \cite{CMS:2015dxe,ATLAS:2015diphoton,Khachatryan:2016hje,Aaboud:2016tru} turns out to be a true signal of new physics, it will be in this category. 
\section{The Electroweak Chiral Lagrangian Expanded in Small $\xi$}
\label{ch:6.xiexpand}
We now assume that we can expand the EFT in terms of the two parameters in Eq.~\eqref{eq:6.1} and that there are no states at the scale $f$ that we can integrate out at tree level. Any new particle in the theory therefore contributes at the order $1/M^{2}\approx 1/16\pi^{2}f^{2} = 1/\Lambda^{2}$. Numerical differences between these scales are encoded in the Wilson coefficients. Schematically, the expansion is given in Fig.~\ref{fig:double.double}. 

For each operator, we identify its loop order $L$ by chiral dimensions, and its canonical dimension $d$. The Wilson coefficient is then of the order $(1/f^{d-4})\cdot (1/16\pi^{2})^{L}$. The authors of \cite{Giudice:2007fh} discussed the dimension-six operators that describe the low-energy effects of a strongly-interacting light Higgs (SILH). They give the Lagrangian of the dimension-six contributions as:
\begin{align*}
\begin{aligned}
\mathcal{L}_{\text{SILH}} &=\frac{c_H}{2f^2}\partial^{\mu}(\phi^{\dagger}\phi)\partial_{\mu}(\phi^{\dagger}\phi)-\frac{c_6\lambda}{f^2}(\phi^{\dagger}\phi)^3+\left(\frac{c_y y_f}{f^2}\phi^{\dagger}\phi{\bar{f}}_L\phi f_R+{\text{h.c.}}\right) \\
&+\frac{c_T}{2f^2}(\phi^{\dagger}\!\!\stackrel{\longleftrightarrow}{D^{\mu}}\!\!\phi) (\phi^{\dagger}\!\!\stackrel{\longleftrightarrow}{D_{\mu}}\!\!\phi)
\end{aligned}
\end{align*}
\begin{align}
\begin{aligned}
\label{eq:6.2}
&+ ig\frac{c_W}{2M^2} (\phi^{\dagger}\!\!\stackrel{\longleftrightarrow}{D_{\mu}^a}\!\!\phi)(D^{\nu}W_{\mu\nu})^a +ig'\frac{c_B}{2M^2} (\phi^{\dagger}\!\!\stackrel{\longleftrightarrow}{D_{\mu}}\!\!\phi) (\partial^{\nu}B_{\mu\nu})\\
&+ig\frac{c_{HW}}{(4\pi f)^2}D^{\mu}\phi^{\dagger}W_{\mu\nu}{D^{\nu}}\phi+ig'\frac{c_{HB}}{(4\pi f)^2}D^{\mu}\phi^{\dagger}{D^{\nu}}\phi B_{\mu\nu}\\
&+g'^{2}\frac{c_{\gamma}}{(4\pi f)^2}\frac{g^2}{g_V^2}\phi^{\dagger}\phi B_{\mu\nu}B^{\mu\nu} +g_s^2\frac{c_g}{(4\pi f)^2}\frac{y_t^2}{g_V^2}\phi^{\dagger}\phi G_{\mu\nu}^aG^{\mu\nu a}\\
&-g^2\frac{c_{2W}}{2(g_V M)^2}D_{\mu}W^{\mu\nu a}D^{\rho}W_{\rho\nu}^a -g'^{2}\frac{c_{2B}}{2(g_V M)^2}\partial_{\mu}B^{\mu\nu}\partial^{\rho}B_{\rho\nu}\\
&-g_s^2\frac{c_{2G}}{2(g_V M)^2}D_{\mu}G^{\mu\nu A}D^{\rho}G_{\rho\nu}^A\\
&+g^3\frac{c_{3W}}{(4\pi M)^2}\langle W^{\mu\nu}W_{\nu\rho}W_{\mu}^{\rho}\rangle +g_s^3\frac{c_{3G}}{(4\pi M)^2} \langle G^{\mu\nu}G_{\nu\rho}G_{\mu}^{\rho}\rangle
\end{aligned}
\end{align}
From the point of view of a bottom-up EFT, we would say that Eq.~\eqref{eq:6.2} does not give a consistent low-energy description of a strongly-coupled Higgs \cite{Buchalla:2014eca}. First, the approximation of keeping only dimension-six operators is justified only if $\xi$ is small and we are close to the decoupling limit. Second, fermionic terms are not discussed in Eq.~\eqref{eq:6.2}. They have been included later to promote $\mathcal{L}_{\text{SILH}}$ to be a complete dimension-six basis \cite{Contino:2013kra}. Third, in a bottom-up EFT we would not distinguish operators that are tree-level generated by resonances ($\mathcal{O}(1/M^{2})$) from operators that are generated at one-loop level at the scale $f$ ($\mathcal{O}(1/16\pi^{2}f^{2})$). Such a distinction depends on the particular UV model that is realized in nature \cite{Arzt:1994gp,Einhorn:2013kja,Jenkins:2013fya}. For the purpose of power counting, we identify the scales $4\pi f$ and $M$. This identification is called naive dimensional analysis (NDA) \cite{Manohar:1983md,Georgi:1992dw} . With all of this information, we can relate the operators in Eq.~\eqref{eq:6.2} to the dots of Fig.~\ref{fig:double.exp}. 

The first line of Eq.~\eqref{eq:6.2}, suppressed by $1/f^{2}$, corresponds to dimension-six operators that are part of the leading-order electroweak chiral Lagrangian. The T-parameter in the second line is written with a $1/f^{2}$ suppression as well. Depending on what we assume about custodial symmetry, we would introduce an additional suppression of this operator, see Section~\ref{ch:Tparam}. The last line of Eq.~\eqref{eq:6.2} has chiral order six, as three powers of the field strength tensors would imply also three powers of gauge couplings by our assumptions of Section~\ref{ch:ewXL.LO}. We therefore expect these operators at the order $\mathcal{O}(\xi/(16\pi^{2})^{2})$. The remaining operators are of chiral order four and of the order $\mathcal{O}(\xi/(16\pi^{2}))$. To summarize, $\mathcal{L}_{\text{SILH}}$ contains operators at $\mathcal{O}(\xi)$, some of the operators at $\mathcal{O}(\xi/(16\pi^{2}))$ and some operators of $\mathcal{O}(\xi/(16\pi^{2})^{2})$. 

Phenomenology allows values of $\xi = \mathcal{O}(0.1)$ \cite{ATLAS-CONF-2015-044,CMS:2015kwa}. This implies that terms of $\mathcal{O}(\xi^{2})$ could be of the same size as the $\mathcal{O}(1/16\pi^{2})$ terms. The operators of $\mathcal{O}(\xi/(16\pi^{2})^{2})$ are negligible compared to the other operators. If we keep the operators of $\mathcal{O}(1/16\pi^{2})$, we have to introduce at least the operators at $\mathcal{O}(\xi^{2})$. The Lagrangian that is expanded consistently in $\xi$ and the chiral order $\chi$ is then given by
\begin{equation}
  \label{eq:6.3}
  \mathcal{L} = \mathcal{L}^{\xi^{0}}_{\chi=2}+\mathcal{L}^{\xi^{1}}_{\chi=2}+\mathcal{L}^{\xi^{2}}_{\chi=2}+\mathcal{L}^{\xi^{0}}_{\chi=4}+\mathcal{L}^{\xi^{1}}_{\chi=4} + \mathcal{O}(\xi^{3})+\mathcal{O}(\tfrac{\xi^{2}}{16\pi^{2}})+\mathcal{O}(\tfrac{1}{(16\pi^{2})^{2}})
\end{equation}
With the identification of Eq.~\eqref{eq:5.1}, we write the doublet $\phi$ in terms of the Goldstone matrix $U$ and the physical Higgs, $h$. The SM Lagrangian of Eq.~\eqref{eq:SMunbroken} then gives the first term
 of Eq.~\eqref{eq:6.3}. The second and third term corresponds to the operators of dimension-six and dimension-eight that are of leading order in the chiral expansion. They are 
\begin{align}
  \begin{aligned}
    \label{eq:6.4}
     \mathcal{L}^{\xi^{1}}_{\chi=2}&= \frac{\tilde{c}_{\phi}v^{4}}{8}\xi \left(1+\frac{h}{v}\right)^{6}-\frac{\tilde{c}_{\phi\Box}}{2}\xi \partial_{\mu}h\partial^{\mu}h \left(1+\frac{h}{v}\right)^{2}-v\xi\bar{\Psi}\tilde{Y}_{\Psi\phi}^{6}U P \Psi\left(1+\frac{h}{v}\right)^{3},\\     \mathcal{L}^{\xi^{2}}_{\chi=2}&= \frac{\tilde{c}_{V}v^{4}}{16}\xi^{2} \left(1+\frac{h}{v}\right)^{8}+\frac{\tilde{c}_{k}}{2}\xi^{2} \partial_{\mu}h\partial^{\mu}h \left(1+\frac{h}{v}\right)^{4}-v\xi^{2}\bar{\Psi}\tilde{Y}_{\Psi\phi}^{8}U P \Psi\left(1+\frac{h}{v}\right)^{5}.
   \end{aligned}
\end{align}
The indices of the dimension-six Wilson coefficients indicate the operators of the Warsaw basis, see Table~\ref{tab:SMEFTdim6.1}. The dimension-eight operators have a similar structure as the dimension-six operators, but an additional factor of $(\phi^{\dagger}\phi)/f^{2}$. Further, we assume that custodial symmetry is respected by the new physics.

The term $\mathcal{L}^{\xi^{0}}_{\chi=4}$ corresponds to one-loop corrections of the SM. We do not list these terms explicitly here, but we keep in mind that they are part of a consistent expansion. Finally, the term $\mathcal{L}^{\xi^{1}}_{\chi=4}$ corresponds to all operators of Table~\ref{tab:SMEFTdim6.1} and Table~\ref{tab:SMEFTdim6.2} that are of chiral order four. 

For later convenience, we rescale the Wilson coefficients as follows:
\begin{align}
  \begin{aligned}
    \label{eq:6.5}
    \tilde{c}_{\phi} &= -\frac{2}{3}\lambda  a_{1},\\
    \tilde{c}_{\phi\Box} &= -a_{2},\\
    \tilde{c}_{V}&= -\lambda b_{1},\\
    \tilde{c}_{k}&=b_{2}.\\
  \end{aligned}
\end{align}
We define $a_{1}$ and $b_{1}$ with an additional factor of the SM-Higgs self-coupling, $\lambda$, to ensure that the coefficients $a_{1}$ and $b_{1}$ stay numbers of $\mathcal{O}(1)$ without an internal $\xi$ dependence. The terms of Eq.~\eqref{eq:6.4} that are of chiral order two have to  match the leading order chiral Lagrangian in Eq.~\eqref{eq:LO}. In order to have a canonically normalized kinetic term, we have to redefine $h$ \cite{mastersthesis,Buchalla:2013rka},
\begin{align}
\begin{aligned}
  \label{eq:6.6}
  h\rightarrow h \Big\lbrace 1-\tfrac{\xi}{2} a_{2} &\left(1+\tfrac{h}{v}+\tfrac{h^2}{3 v^2}\right)+\xi ^2 a_{2}^2 \left(\tfrac{3}{8}+\tfrac{h}{v}+\tfrac{13}{12} \left(\tfrac{h}{v}\right)^2+\tfrac{13}{24} \left(\tfrac{h}{v}\right)^3+\tfrac{13}{120} \left(\tfrac{h}{v}\right)^4\right) \\ 
&  -\xi ^2 b_{2} \left(\tfrac{1}{2}+\tfrac{h}{v}+\left(\tfrac{h}{v}\right)^2 +\tfrac{1}{2} \left(\tfrac{h}{v}\right)^3+\tfrac{1}{10} \left(\tfrac{h}{v}\right)^4\right)\Big\rbrace .
\end{aligned}
\end{align}
The parameter $v$ should describe the physical vacuum expectation value. We find it by imposing that the linear term in the potential, after the redefinition above, vanishes. We find
\begin{align}
\begin{aligned}
  \label{eq:6.7}
  v&=\sqrt{ \frac{2 \left(3 \mu^{2} \xi ^2 a_{2}^2-4 \mu^{2} \xi  a_{2}-4 \mu^{2} \xi ^2 b_{2}+8 \mu^{2}\right)}{\lambda\left(8 + 8 a_{1} \xi -4 a_{2}\xi +\xi^{2}(3 a_{2}^{2}-4 a_{1}a_{2}+8 b_{1} -4 b_{2})\right) }}\\
& =\sqrt{\frac{2\mu^{2}}{\lambda}}\left(  1-\frac{a_{1}}{2}\xi + \frac{\xi^{2}}{2}\left(\frac{3a_{1}^{2}}{4}-b_{1}\right)+\mathcal{O}\left(\xi^3\right) \right).
\end{aligned}
\end{align}
The quadratic term of the potential gives the physical Higgs mass $m_{h}$. This condition, together with Eq.~\eqref{eq:6.7} enables us to express the bare quantities $\mu$ and $\lambda$ in terms of the physical quantities $v$ and $m_{h}$, and the Wilson coefficients:
\begin{align}
  \begin{aligned}
    \label{eq:6.8} 
  \mu^{2} &= \frac{m_{h}^2}{2}\left(1+\xi (a_{2}-a_{1})+\xi^{2}(2a_{1}^{2}-a_{1}a_{2}-2 b_{1}+b_{2}) +\mathcal{O}\left(\xi^3\right) \right)\\
\lambda &= \frac{m_{h}^2}{v^2}\left(1+\xi \left(a_{2}-2a_{1}\right)+\xi^2 \left(4 a_{1}^{2}-2 a_{1}a_{2}-3b_{1}+b_{2}\right)+\mathcal{O}\left(\xi^3\right)\right)
  \end{aligned}
\end{align}
Then, the Lagrangian has the form of $\mathcal{L}_{\text{LO}}$ in Eq.~\eqref{eq:LO}. We have explicitly
\begin{align}
  \begin{aligned}
    \label{eq:6.9} 
    V(h) &= \frac{1}{2} h^2 m_{h}^2+\frac{1}{2} m_{h}^2 v^2 \left[\left(1+\xi  \left(\tfrac{4}{3} a_{1}-\tfrac{3}{2} a_{2}\right)+\xi^2\left(-\tfrac{2}{3} a_{1} a_{2}+\tfrac{15}{8}a_{2}^2+4 b_{1}-\tfrac{5}{2} b_{2}-\tfrac{8}{3}a_{1}^{2}\right)\right) \left(\frac{h}{v}\right)^{3}\right. \\
    &+ \left(\tfrac{1}{4} +\xi  \left(2 a_{1}-\tfrac{25}{12} a_{2}\right)+\xi^2 \left(-4 a_{1} a_{2}+\tfrac{11}{2} a_{2}^2+8 b_{1}-\tfrac{21}{4} b_{2} -4a_{1}^{2}\right)\right) \left(\frac{h}{v}\right)^{4}\\
    &+ \left(\xi  (a_{1}-a_{2})+ \xi^2 \left(-\tfrac{37}{6}a_{1} a_{2}-2 a_{1}^{2}+\tfrac{13}{2} a_{2}^2+7 b_{1}-5 b_{2}\right)\right) \left(\frac{h}{v}\right)^{5}\\
    &+\left(\tfrac{\xi}{6} (a_{1}-a_{2})+\xi^2 \left(-\tfrac{25}{6} a_{1} a_{2}-\tfrac{1}{3} a_{1}^{2}+\tfrac{176}{45} a_{2}^2+\tfrac{7}{2} b_{1}-\tfrac{27}{10} b_{2}\right)\right) \left(\frac{h}{v}\right)^{6}\\
    &+\left. \xi^2 \left(-\tfrac{4}{3} a_{1} a_{2}+\tfrac{6}{5} a_{2}^2+b_{1}-\tfrac{4}{5} b_{2}\right)  \left(\frac{h}{v}\right)^{7}+\tfrac{\xi^2}{8} \left(-\tfrac{4}{3} a_{1} a_{2}+\tfrac{6}{5} a_{2}^2+b_{1}-\tfrac{4}{5} b_{2}\right) \left(\frac{h}{v}\right)^{8}\right],
  \end{aligned}
\end{align}
\begin{align}
  \begin{aligned}
    \label{eq:6.10}
    F_{U}(h) &=\left(2-a_{2} \xi +\xi^2 \left(\tfrac{3}{4} a_{2}^2-b_{2}\right)\right)\left(\frac{h}{v}\right)+ \left(1-2 a_{2} \xi+\xi ^2 \left(3 a_{2}^2-3 b_{2}\right)\right)\left(\frac{h}{v}\right)^{2}\\
    &+\left(-\xi\tfrac{4}{3} a_{2}+\xi^2 \left(\tfrac{14}{3} a_{2}^2-4 b_{2}\right)\right)\left(\frac{h}{v}\right)^{3}+ \left(-\xi \tfrac{a_{2}}{3}+\xi^2 \left(\tfrac{11}{3} a_{2}^2-3 b_{2}\right)\right)\left(\frac{h}{v}\right)^{4}\\
    &+\xi^2 \left(\tfrac{22}{15} a_{2}^2-\tfrac{6}{5} b_{2}\right)\left(\frac{h}{v}\right)^{5}+\tfrac{\xi^2}{6} \left(\tfrac{22}{15} a_{2}^2-\tfrac{6}{5} b_{2}\right)\left(\frac{h}{v}\right)^{6} ,
  \end{aligned}
\end{align}
and
\begin{align}
  \begin{aligned}
    \label{eq:6.11} 
    \sum_{n=1}^{5}Y_{\Psi}^{(n)}\left(\tfrac{h}{v}\right)^{n} &=\left(Y_{\Psi\phi}+\xi  \left(2 \tilde{Y}_{\Psi\phi}^{6}-\tfrac{a_{2}}{2} Y_{\Psi\phi}\right)+\xi^2 \left(\tfrac{3}{8} a_{2}^2 Y_{\Psi\phi}-a_{2} \tilde{Y}_{\Psi\phi}^{6}-\tfrac{b_{2}}{2} Y_{\Psi\phi}+4 \tilde{Y}_{\Psi\phi}^{8}\right)\right)\frac{h}{v} \\
    &+\left(\xi  \left(3 \tilde{Y}_{\Psi\phi}^{6}-\tfrac{a_{2}}{2} Y_{\Psi\phi}\right)+\xi^2 \left(a_{2}^2 Y_{\Psi\phi}-4 a_{2} \tilde{Y}_{\Psi\phi}^{6}-b_{2} Y_{\Psi\phi}+10 \tilde{Y}_{\Psi\phi}^{8}\right)\right)\left(\frac{h}{v}\right)^{2}\\
    &+\left(\tfrac{\xi}{3}  \left(3 \tilde{Y}_{\Psi\phi}^{6}-\tfrac{a_{2}}{2} Y_{\Psi\phi}\right)+\xi^2 \left(\tfrac{13}{12} a_{2}^2 Y_{\Psi\phi}-\tfrac{29}{6} a_{2} \tilde{Y}_{\Psi\phi}^{6}-b_{2} Y_{\Psi\phi}+10 \tilde{Y}_{\Psi\phi}^{8}\right)\right)\left(\frac{h}{v}\right)^{3}\\
    &+ \xi^2 \left(\tfrac{13}{24} a_{2}^2 Y_{\Psi\phi}-\tfrac{5}{2} a_{2} \tilde{Y}_{\Psi\phi}^{6}-\tfrac{b_{2}}{2} Y_{\Psi\phi}+5 \tilde{Y}_{\Psi\phi}^{8}\right)\left(\frac{h}{v}\right)^{4}\\
    &+ \tfrac{\xi^2}{5} \left(\tfrac{13}{24} a_{2}^2 Y_{\Psi\phi}-\tfrac{5}{2} a_{2}\tilde{Y}_{\Psi\phi}^{6}-\tfrac{b_{2}}{2} Y_{\Psi\phi}+5\tilde{Y}_{\Psi\phi}^{8}\right)\left(\frac{h}{v}\right)^{5}.
  \end{aligned}
\end{align}
Here, $Y_{\Psi\phi}$ are the Yukawa matrices of $\mathcal{L}_{\text{LO}}$, and $\tilde{Y}_{\Psi\phi}^{6}$ and $\tilde{Y}_{\Psi\phi}^{8}$ are the ones from the dimension-six and dimension-eight operators. If they are not diagonal, we have to rotate to the mass-eigenstate basis. However, we do not perform this step here explicitly. The lowest order term of $V(h)$ and $F_{U}(h)$ were also discussed in \cite{Contino:2013gna}.

In $\mathcal{L}^{\xi^{1}}_{\chi=4}$, we find
\begin{align}
  \begin{aligned}
    \label{eq:6.12}
    \mathcal{L}^{\xi^{1}}_{\chi=4} &=-\frac{c_{\beta}\xi}{16\pi^{2}}g'^{2} v^2\langle L_{\mu}\tau_L\rangle^2\left(1+\frac{h}{v}\right)^{4} -\frac{c_{Xh1}\xi}{4\cdot 16\pi^{2}} g'^{2}B_{\mu\nu}B^{\mu\nu} \left[1-\left(1+\frac{h}{v}\right)^2\right] \\
    &- \frac{c_{Xh2}\xi}{2\cdot 16\pi^{2}}g^{2}\langle W_{\mu\nu}W^{\mu\nu}\rangle \left[1- \left(1+\frac{h}{v}\right)^2\right]- \frac{c_{Xh3}\xi}{2\cdot 16\pi^{2}}g_{s}^{2}\langle G_{\mu\nu}G^{\mu\nu}\rangle \left[1-\left(1+\frac{h}{v}\right)^2\right]\\
    &+ \frac{c_{XU1}\xi}{16\pi^{2}} g g'\langle W_{\mu\nu} \tau_{L}\rangle B^{\mu\nu} \left(1+\frac{h}{v}\right)^{2} + \frac{c_{\psi V7}\xi}{16\pi^{2}} (\bar{\ell}_{L}\gamma^{\mu}\ell_{L}) \langle L_{\mu} \tau_{L}\rangle \left(1+\frac{h}{v}\right)^{2}\\
    &-w^{2} \frac{c_{\psi V1}\xi}{16\pi^{2}} (\bar{q}_{L}\gamma^{\mu}q_{L}) \langle L_{\mu} \tau_{L}\rangle \left(1+\frac{h}{v}\right)^{2}-w^{2} \frac{c_{\psi V10}\xi}{16\pi^{2}} (\bar{e}_{R}\gamma^{\mu}e_{R}) \langle L_{\mu} \tau_{L}\rangle \left(1+\frac{h}{v}\right)^{2}\\
    &-w^{2} \frac{c_{\psi V4}\xi}{16\pi^{2}} (\bar{u}_{R}\gamma^{\mu}u_{R}) \langle L_{\mu} \tau_{L}\rangle \left(1+\frac{h}{v}\right)^{2}-w^{2} \frac{c_{\psi V5}\xi}{16\pi^{2}} (\bar{d}_{R}\gamma^{\mu}d_{R}) \langle L_{\mu} \tau_{L}\rangle \left(1+\frac{h}{v}\right)^{2}\\
    &-w^{2} \frac{c_{\psi V6}\xi}{16\pi^{2}} (\bar{u}_{R}\gamma^{\mu}d_{R}) \langle P_{21}U^{\dagger}L_{\mu}U\rangle \left(1+\frac{h}{v}\right)^{2}+ \text{h.c.}\\
    &-w^{2}\frac{c_{\psi Vq}\xi}{16\pi^{2}} \mathcal{O}_{q} \left(1+\frac{h}{v}\right)^{2}-w^{2}\frac{c_{\psi Vl}\xi}{16\pi^{2}} \mathcal{O}_{\ell} \left(1+\frac{h}{v}\right)^{2} + \mathcal{L}_{\psi^{4}} + \mathcal{L}_{\psi^2 X} 
\end{aligned}
\end{align}
Here, $\mathcal{L}_{\psi^{4}}$ refers to all four-fermion operators without $U$-fields and $\mathcal{L}_{\psi^2 X} $ contains the operators of Section~\ref{ch:5.NLO.dipole}. Coming from dimension-six operators, the four-fermion operators contain no Higgs fields. The operators of $\mathcal{L}_{\psi^2 X} $ come with $F_{\psi Xi}(h) = h/v$. The coefficients of all of these operators are of the order $\mathcal{O}(\xi/16\pi^{2})$. Further, we use the notation
\begin{align}
\begin{aligned}
\label{eq:6.13}
\mathcal{O}_{q}&= 2(\bar{q}\tau_{L}\gamma^{\mu}q) \langle L_{\mu} \tau_{L}\rangle + (\bar{q}UP_{12}U^{\dagger}\gamma^{\mu}q) \langle P_{21}U^{\dagger}L_{\mu}U\rangle + (\bar{q}UP_{21}U^{\dagger}\gamma^{\mu}q) \langle P_{12}U^{\dagger}L_{\mu}U\rangle, \\
\mathcal{O}_{\ell} &= 2(\bar{l}\tau_{L}\gamma^{\mu}l) \langle L_{\mu} \tau_{L}\rangle + (\bar{l}UP_{12}U^{\dagger}\gamma^{\mu}l) \langle P_{21}U^{\dagger}L_{\mu}U\rangle + (\bar{l}UP_{21}U^{\dagger}\gamma^{\mu}l) \langle P_{12}U^{\dagger}L_{\mu}U\rangle.
\end{aligned}
\end{align} 
Therefore, only one linear combination of $\mathcal{O}_{\psi V2}$, $\mathcal{O}_{\psi V3}$, and $\mathcal{O}^\dagger_{\psi V3}$, namely $\mathcal{O}_{q}$, enters at leading order in $\xi$. The same is true for $\mathcal{O}_{\psi V8}$, $\mathcal{O}_{\psi V9}$, and $\mathcal{O}^\dagger_{\psi V9}$ that form $\mathcal{O}_{\ell}$. We also redefined the gauge fields, such that the kinetic terms are canonically normalized. This subtracts the Higgs-independent part of the three operators $\mathcal{O}_{Xh1}$, $\mathcal{O}_{Xh2}$, and $\mathcal{O}_{Xh3}$ in Eq.~\eqref{eq:6.12}. The redefinition does not change the structure of the covariant derivatives in the kinetic terms of the matter fields in $\mathcal{L}^{\xi^{0}}_{\chi=2}$, because the product of gauge coupling and gauge field is renormalized together. The renormalization of the field is therefore absorbed by the renormalization of the coupling. 
\section{Phenomenological Implications}
\label{ch:6.pheno}
The two different EFTs, the decoupling SM-EFT and the non-decoupling chiral Lagrangian, seem like different scenarios, realized in nature. The different assumptions on the Higgs lead to a different realization of the symmetry in the Goldstone sector. It is linearly realized in the SM-EFT and non-linearly realized in the electroweak chiral Lagrangian. Some research groups \cite{Contino:2013kra,Isidori:2013cga,Alonso:2014wta,Gonzalez-Fraile:2014cya} therefore tried to answer the question: ``Is the symmetry linearly or non-linearly realized?'' Since the observed Higgs behaves like the SM-Higgs within the experimental uncertainties, most groups conclude that it is appropriate to assume the linear realization and use the SM-EFT \cite{Passarino:2012cb,Einhorn:2013tja,Falkowski:2014tna,Ellis:2014dva,Ellis:2014jta,Ghezzi:2015vva,Hartmann:2015oia,Hartmann:2015aia,Berthier:2015oma,Berthier:2015gja,Corbett:2015ksa}. However, with the discussion of the preceding sections in mind, we see that the choice is not about how the symmetry is realized. Rather, it is about whether the new physics is decoupling from the SM or not. Completely non-decoupling scenarios, like Technicolor \cite{Weinberg:1975gm,Weinberg:1979bn,Susskind:1978ms}, identify $f=v$ and are ruled out by the experiments \cite{Agashe:2014kda}. The parameter $\xi$ is therefore definitely smaller than one, $\xi<1$. Hence, the question to ask is: ``How small is $\xi$?'' It is more appropriate to use the double expansion if $\xi$ has at least the size of the loop factor. If $\xi\lesssim 1/16\pi^{2}$, it is better to use the SM-EFT. The parameter $\xi$ interpolates continuously between the different scenarios. In the transition region, both expansions have advantages and disadvantages, so we cannot give a definite recommendation. However, the two limits $\xi\lesssim 1/16\pi^{2}$ and $\xi\gg 1/16\pi^{2}$, have some general implications for observables. If the experimental measurements show these features, we can decide which of the two expansions is more appropriate to apply. \\

Operators of dimension six introduce correlations between different couplings in the Lagrangian. We can see them for example in Eq.~\eqref{eq:6.10}. At the order $\mathcal{O}(\xi)$, the couplings between the Goldstones and one Higgs, and between the Goldstones and two Higgs fields are modified by the same parameter, $a_{2}$. They are therefore correlated. If the new physics is decoupling and $\xi$ is small, we measure these correlations. Operators of dimension eight, {\it i.e.}~effects of order $\mathcal{O}(\xi^{2})$, break the correlations \cite{Contino:2013gna}. Uncorrelated couplings would therefore hint at larger values of $\xi$ and the double expansion. However, if we measure couplings that seem to follow the predicted pattern of correlations, it is still possible that this comes from a very specific UV model instead of a small $\xi$ expansion. \\

A second general observation comes from the reordering of the terms in the different effective expansions. At next-to-leading order, the SM-EFT predicts many different effects in the Higgs and in the gauge-fermion sector to arise at the order $\mathcal{O}(v^{2}/\Lambda^{2})$. The double expansion, on the other hand, predicts larger effects in the Higgs sector (at $\mathcal{O}(\xi)$) than in the gauge-fermion sector (at $\mathcal{O}(\xi/16\pi^{2})$). Measuring sizable effects in the Higgs sector would therefore indicate that the scale of new physics, $f$, is not too far above the weak scale and the double expansion is appropriate. The enhancement of effects in the Higgs sector is larger for operators involving more Higgs fields. All terms in the polynomials $F_{i}(h)$ are of the same chiral order, but have increasing canonical dimension. Therefore, the operators are in different orders of the SM-EFT. However, even the observation of double Higgs production will be very challenging at the LHC \cite{Azatov:2015oxa,Behr:2015oqq}. Hence, for practical purposes one can only compare the size of the effects of single Higgs processes to the size of the effects in the gauge-fermion sector in order to distinguish the two different expansions. 

Experimentally, the gauge-fermion sector was measured at the one-percent level by the experiments at LEP \cite{ALEPH:2005ab,Barbieri:2004qk,ALEPH:2010aa}. Current limits on single-Higgs couplings are at the order of ten percent \cite{ATLAS-CONF-2015-044,CMS:2015kwa}, some of the Higgs couplings have not been measured at all. Thus, there is no reason to assume that the double expansion is not justified for the current phase of the LHC Higgs analysis.
\part{The Application of the Effective Theory}
  \chapter[Specific Models and Their Relation to the EFTs]{Specific Models and Their Relation to the Effective Field Theories}
\thispagestyle{fancyplain}
A different approach, complementary to the use of EFTs, is given by postulating specific models. They are designed to address current experimental anomalies or open problems of the SM. We would like to understand electroweak symmetry breaking and tackle the hierarchy problem of the SM-Higgs sector that we discussed in Section~\ref{ch:SM.openQ}. There are different ways to solve the hierarchy problem. The quantum corrections to the Higgs mass can be small because a symmetry leads to cancellations between the large contributions, and the high-energy cutoff is therefore well below the Planck-scale \cite{Hill:2002ap,Goldberger:2008zz}. In all cases, the Higgs sector is modified with respect to the SM.\\ 

A very prominent example is Supersymmetry (SUSY). In SUSY, the symmetry group of the SM is extended by a symmetry that relates the bosonic and fermionic fields \cite{Gervais:1971ji,Golfand:1971iw,Volkov:1973ix,Wess:1974tw,Martin:1997ns}. By definition, these fields, called superpartners, have the same quantum numbers and masses, but a different spin. The quantum corrections from bosonic and fermionic loops then have the same magnitude, but a different sign. Hence, they cancel. However, the particle spectrum we observe at the experiments is not supersymmetric. This implies that SUSY must be broken. A detailed supersymmetric model that is able to describe the current SM-like observations is the Minimal Supersymmetric Standard Model (MSSM) \cite{Nilles:1983ge,Haber:1984rc,Haber:1990aw}. Current experimental searches for the MSSM and other supersymmetric models have not found any new particles \cite[and references therein]{CMSSUSY,ATLASSUSY}. Instead, they only set lower bounds on the masses of the SM superpartners. Effects of SUSY are in general decoupling and therefore at low energies described by the SM-EFT \cite{Huo:2015nka}.

Postulating a confining gauge theory that breaks the electroweak symmetry dynamically also solves the hierarchy problem. A prototype model of this type is Technicolor (TC). Inspired by the fact that chiral symmetry breaking of QCD also breaks the electroweak symmetry, but predicts gauge boson masses that are three orders of magnitude too small, the authors of \cite{Weinberg:1975gm,Weinberg:1979bn,Susskind:1978ms} introduced a scaled-up version of QCD. 

In general, we have an $SU(N_{\text{TC}})$ gauge theory with confinement that breaks a global $SU(2)_{L}\times SU(2)_{R}$ symmetry spontaneously to $SU(2)_{V}$. The three Goldstone bosons become the longitudinal components of the $W^{\pm}$ and $Z$ bosons, as in the SM. However, in TC there is no physical Higgs particle. Instead, the resonances of $SU(N_{\text{TC}})$ restore the unitarity. TC models of this type are now excluded by experiments. First of all, CMS \cite{Chatrchyan:2012xdj} and ATLAS \cite{Aad:2012tfa} observed a Higgs-like scalar particle that is not part of these TC models. Further, TC predicts \cite{Hill:2002ap,Contino:2010rs} large values for the Peskin-Takeuchi S parameter \cite{Peskin:1990zt,Peskin:1991sw} that is measured to be small \cite{Agashe:2014kda}. In order to generate masses for the quarks, TC has to be extended. These extensions \cite{Dimopoulos:1979es}  predict also flavor-changing neutral currents (FCNCs) that are not observed. Therefore, variations of the TC idea have been proposed. Because of the strongly-coupled nature of the UV-completions, their effects are in general not decoupling at low energies. Thus, the electroweak chiral Lagrangian describes the low-energy effects of these types of models. 

We focus on this class of models in the following section. First, we introduce composite Higgs models --- a variation of TC. Then, we show how they are connected to the electroweak chiral Lagrangian. If the first hints of new physics that come from experiments are just deviations of signals from the SM prediction and no new particles are observed, it is crucial to know which class of models induces which pattern of deviations in observables. In addition, we show that the chiral counting of non-decoupling EFTs can also be applied in the context of composite Higgs models. In Section~\ref{ch:SM+S}, we illustrate how a simple, renormalizable UV-completion induces a decoupling or a non-decoupling EFT in different regions of the parameter space. This also shows how an explicit model is matched to the EFTs of Chapters~\ref{ch:SMEFT} and \ref{ch:ewXL}.

\section{Composite Higgs Models}
\label{ch:MCHM}
Composite Higgs models (CHMs) are a class of UV-completion that is inspired by TC. As in TC, we assume a new, strong interaction that breaks the electroweak symmetry dynamically. In addition, we assume a pseudo-Nambu-Goldstone boson (pNGB) in the spectrum that we identify with the Higgs. Kaplan and Georgi first proposed this idea in \cite{Kaplan:1983fs,Kaplan:1983sm,Georgi:1984ef,Georgi:1984af,Dugan:1984hq}. The pNGB nature of the Higgs explains why it is much lighter than the other resonances of the strong sector, similar to the pions in the spectrum of QCD. The Higgs, now being a composite object, does not receive large quantum corrections to its mass, as the virtual effects are cut-off at the compositeness scale. The mass is protected by Goldstone's symmetry. Fermions and gauge bosons of the SM are external to this strong sector and elementary in this picture. 

\begin{figure}[!t]
  \begin{center} 
    \begin{tikzpicture}
      \draw {(0em,0em) circle (6em)};
      \draw (0em,4em) node[anchor = base] {$\mathcal{G}$};
      \draw {(-2em,-2em) circle (2.5em)};
      \draw {(2em,-2em) circle (2.5em)};
      \draw (-2em,-1em) node[anchor = base] {$\mathcal{G}_{\text{gauge}}$};
      \draw (2em,-1em) node[anchor = base] {$\mathcal{H}$};
    \end{tikzpicture}
  \end{center}
  \caption[Pattern of symmetry breaking at the scale $f$ in composite Higgs models.]{Pattern of symmetry breaking at the scale $f$ in composite Higgs models, taken from \cite{Contino:2010rs}.}
  \label{fig:7.pattern.SSB}
\end{figure}
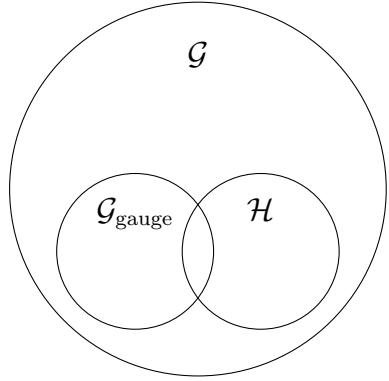
In general, we assume the following scenario \cite{Contino:2010rs}, which we also illustrate in Fig.~\ref{fig:7.pattern.SSB}: The strong sector has a global symmetry, $\mathcal{G}$, that is spontaneously broken to $\mathcal{H}$ at the scale $f$, giving $n = (\dim{\mathcal{G}}-\dim{\mathcal{H}})$ Goldstone bosons. A subgroup $\mathcal{G}_{\text{gauge}}\subset\mathcal{G}$ is gauged by external gauge fields. Its subgroup $\mathcal{G}_{\text{gauge}}\cap \mathcal{H}$ is the unbroken gauge group at the scale $f$, while the remainder $\mathcal{G}_{\text{br}} =  \mathcal{G}_{\text{gauge}}-(\mathcal{G}_{\text{gauge}}\cap \mathcal{H})$ is spontaneously broken. The spontaneous breaking of the gauge group requires $n_{\text{br}} = \dim{\mathcal{G}_{\text{br}}}$ of the $n$ Goldstone bosons. The external gauging of $\mathcal{G}_{\text{gauge}}$ also breaks the symmetry $\mathcal{G}$ explicitly. This explicit breaking induces a potential that breaks the electroweak symmetry at one-loop level, setting the electroweak scale, $v$, below the scale $f$. The massive $W^{\pm}$ and $Z$ bosons of the SM ``eat'' three of the $n-n_{\text{br}}$ remaining Goldstone bosons. The remaining $n-n_{\text{br}}-3$ Goldstone bosons get a mass of $\mathcal{O}(v)$ from the explicit breaking of $\mathcal{G}$, making them pNGBs. We identify the Higgs as one of them. 
\subsubsection{The Minimal Composite Higgs Model (MCHM)}
The authors of \cite{Agashe:2004rs,Contino:2006qr} proposed a minimal realization of the described pattern of symmetry breaking based on the coset $SO(5)/SO(4)$. In particular, they considered $\mathcal{G} = SO(5)\times U(1)_{X}$ and $\mathcal{H} = SO(4)\times U(1)_{X}$. The additional $U(1)_{X}$ is necessary to get the correct hypercharges of the SM particles. The strong dynamics is conformal at high energies and corresponds to a weakly-coupled, five-dimensional Anti-de Sitter (AdS) theory. In this picture, the Higgs is the fifth component of the five-dimensional gauge field. Agashe {\it et al.} computed the form factors of the strongly-coupled four-dimensional theory from the five-dimensional AdS theory in \cite{Agashe:2004rs}. They showed that the flavor problems of TC are solved and the contributions to the electroweak precision observables are small. However, this model is not renormalizable. Therefore, it is only an intermediate theory, not valid up to the Planck-scale. 

The breaking of $\mathcal{G}\rightarrow \mathcal{H}$ generates $n=4$ Goldstone bosons, transforming in the fundamental representation of $SO(4)$. The group $SO(4)$ is isomorphic to $SU(2)\times SU(2)$, which we identify with the global $SU(2)_{L}\times SU(2)_{R}$ of the Higgs sector in the SM. The four Goldstone bosons transform therefore also as a complex doublet of $SU(2)_{L}$. We identify it with the composite Higgs. The gauging of $\mathcal{G}_{\text{gauge}} =\mathcal{G}_{\text{SM}}= SU(2)_{L}\times U(1)_{Y}$ breaks $SO(5)$ explicitly and generates a potential for the Higgs at the one-loop level. The generated vacuum, however, tends to preserve $\mathcal{G}_{\text{SM}}$ \cite{Agashe:2004rs,Kaplan:1983fs,Kaplan:1983sm,Georgi:1984ef,Georgi:1984af,Dugan:1984hq}. Further contributions from fermion loops misalign the vacua and break the electroweak symmetry dynamically. Three of the Goldstone bosons become the longitudinal components of the massive gauge fields, the fourth one is the massive, Higgs-like scalar. Their masses are of the order of the electroweak scale. Fermions couple linearly to the operators of the strong sector and get their masses {\it via} mixing effects. This so-called ``partial compositeness'' \cite{Kaplan:1991dc} avoids the problems of TC with flavor-changing neutral currents \cite{Matsedonskyi:2012ym,DeSimone:2012fs,Carena:2014ria}. Different possibilities exist to group the SM fermions into $SO(5)$ multiplets. The authors of \cite{Agashe:2004rs} used the spinorial representation. Since this representation is four-dimensional, they called the model MCHM4. The authors of \cite{Contino:2006qr} considered the fundamental, five-dimensional representation and called the model MCHM5, as well as the anti-symmetric, ten-dimensional representation that they called MCHM10. 

In the following, we focus on the bosonic, $\mathcal{CP}$-even sector of these models. We do not specify the precise mechanism of the $SO(5)\rightarrow SO(4)$ breaking, instead we just parametrize the Goldstone bosons of the coset $SO(5)/SO(4)$. As in the chiral Lagrangians we discussed in Section~\ref{ch:chiPT} and Chapter~\ref{ch:ewXL}, we use the construction of Coleman {\it et al.} \cite{Coleman:1969sm,Callan:1969sn} to parametrize the Goldstones. We write the four Goldstones, $h_{\hat{a}}$ (with $\hat{a}=1,\dots,4$), in terms of the $SO(5)$ vector $\Sigma$. It is defined as
\begin{equation}
  \label{eq:7.1}
  \Sigma(h_{\hat a})={\mathcal U}\, \begin{pmatrix} 0_{4}\\1\end{pmatrix}, \qquad \text{where } \,{\mathcal U}=\exp(\sqrt{2}it^{\hat a} h_{\hat a}/f).
\end{equation}
Here, $t^{\hat a}$ are the broken generators spanning the coset. We list them in Appendix~\ref{ch:appA} explicitly. Expressed in terms of $h_{\hat{a}}$, we have 
\begin{equation}
  \label{eq:7.2}
  {\mathcal U} = \begin{pmatrix}[c|c]
\mathbf{1}-(1-c)\hat h \hat h^T & s \hat h \\
\hline
-s \hat h^T & c
\end{pmatrix},
\end{equation}
where $s=\sin{\left(|h|/f\right)}$, $c=\cos{\left(|h|/f\right)}$, $|h| = \sqrt{h_{\hat a}h_{\hat a}}$, and $\hat{h}_{\hat a}=h_{\hat a}/|h|$. This yields \cite{Agashe:2004rs,Contino:2006qr,Contino:2010rs}
\begin{equation}
  \label{eq:7.3}
  \Sigma(h_{\hat a})=\begin{pmatrix}\hat{h}_{\hat a} s \\ c\end{pmatrix}. 
\end{equation}
As already discussed, quantum effects generate a potential at one loop that breaks the $SO(5)$ symmetry explicitly. We parametrize this potential using the two $SO(5)$-breaking spurions that are consistent with SM gauge invariance, $\vec{n}=(0,0,0,0,1)^T$ and $t^{3}_{R}$, defined in Eq.~\eqref{eq:A.1}. The vector, $\vec{n}$, conserves custodial symmetry, while $t^{3}_{R}$ breaks it. We assume that this breaking comes from the effects of the SM only. This implies that $t^{3}_{R}$ comes with factors of weak couplings, $g'$ or $Y$. The spurions are related through $\vec{n} \vec{n}^T=1-4 t^{3}_{R} t^{3}_{R}$.\\

This low-energy description of the $SO(5)/SO(4)$ coset is a bottom-up, non-decoupling EFT. Its power counting is therefore given by chiral dimensions, as we discussed in Section~\ref{ch:chiPT} and Chapter~\ref{ch:ewXL}. At leading order, chiral order two, we have 
\begin{equation}
  \label{eq:7.4}
  \mathcal{L}_{\text{LO}}=\frac{f^2}{2}D_{\mu}\Sigma^TD^{\mu}\Sigma - \alpha\, \Sigma^T \vec{n} +4\beta\,\Sigma^T t_3^Rt_3^R \Sigma.
\end{equation}
The form of the potential depends on the representation of the fermions in $SO(5)$ \cite{Agashe:2004rs,Contino:2006qr}. Here, we choose the simplest form that leads to electroweak symmetry breaking, based on the MCHM4 \cite{Agashe:2004rs}. The coefficients of the potential, $\alpha$ and $\beta$, are generated at one-loop level and have therefore chiral order two. 

The isomorphism between $SO(4)$ and $SU(2)_{L}\times SU(2)_{R}$ allows us to relate the $SO(4)$ vector $h_{\hat{a}}$ to the $SU(2)_{L}\times SU(2)_{R}$ bi-doublet $\Phi$, defined in Eq.~\eqref{eq:2.1.6},
\begin{equation}
  \label{eq:7.5}
\sqrt{2}\Phi = (\widetilde{\phi},\phi) =\begin{pmatrix}h_4+ih_3&h_2+ih_1\\-(h_2-ih_1)&h_4-ih_3\end{pmatrix}=h_{\hat a}\lambda_{\hat a}\equiv |h| U.
\end{equation}
Here, we defined $\lambda_{\hat a}=(i{\vec{\sigma}},\mathbf{1})$, fulfilling the relation
\begin{equation}
  \label{eq:7.6}
  \lambda_{ij}^{\hat a}\lambda_{kl}^{\hat a\dagger}=2\delta_{il}\delta_{kj}.
\end{equation}
This gives
\begin{equation}
  \label{eq:7.7}
  t^a_{L,\hat a \hat b} \langle U\lambda^\dagger_{\hat b}\rangle = \langle T^{a}_{L} U\lambda^\dagger_{\hat a}\rangle,
\qquad
t^3_{R,\hat a\hat b} \langle U\lambda^\dagger_{\hat b}\rangle=-\langle U T^{3}_{R} \lambda^\dagger_{\hat a}\rangle,
\end{equation}
and
\begin{equation}
  \label{eq:7.8}
  \hat{h}_{\hat a}=\frac{1}{2}\langle U\lambda_{\hat a}^{\dagger}\rangle.
\end{equation}
We write the Lagrangian of Eq.~\eqref{eq:7.4} now in terms of the fields of the electroweak chiral Lagrangian of Chapter~\ref{ch:ewXL}. We find 
\begin{equation}
  \label{eq:7.9}
  \mathcal{L}_{\text{LO}}=  \frac{1}{2}\partial_{\mu}|h|\partial^{\mu}|h|+\frac{f^2}{4}\langle L_{\mu}L^{\mu}\rangle s^2 - \alpha\, c +\beta\, s^2.
\end{equation}
The potential exhibits spontaneous symmetry breaking for $\beta>0$ and $|\alpha| \leq 2\beta$. It generates the vacuum expectation value, $\langle |h| \rangle$, {\it via}
\begin{equation}
  \label{eq:7.10}
  \sin^2\frac{\langle |h| \rangle}{f} = 1-\left(\frac{\alpha}{2\beta}\right)^2.
\end{equation}
The mass of the physical scalar, $h\equiv |h| - \langle |h| \rangle$, is
\begin{equation}
  \label{eq:7.11}
  m^2_h=\frac{2\beta v^{2}}{f^4}.
\end{equation}
When we compare Eq.~\eqref{eq:7.9} to the Goldstone-kinetic term of $\mathcal{L}_{\text{LO}}$ in Eq.~\eqref{eq:LO}, we find \cite{Agashe:2004rs,Contino:2006qr,Contino:2010rs}
\begin{equation}
  \label{eq:7.12}
  f^{2}\, \sin^{2}{\left(\frac{\langle|h|\rangle + h}{f}\right)} = v^{2}\, \left(1+F_{U}(h)\right).
\end{equation}
This gives us a relation between $v$, $f$, and $\langle|h|\rangle$,
\begin{equation}
  \label{eq:7.13}
  \xi \equiv \frac{v^{2}}{f^{2}} = \sin^{2}{\left(\frac{\langle|h|\rangle}{f}\right)},
\end{equation}
and the expansion of $F_{U}(h)$ \cite{Contino:2010rs}:
\begin{equation}
  \label{eq:7.14}
  F_{U}(h) = 2 \sqrt{1-\xi} \,\left(\frac{h}{v}\right) + (1-2\xi)\,\left(\frac{h^{2}}{v^{2}}\right) - \tfrac{4}{3} \xi\sqrt{1-\xi}\,\left(\frac{h^{3}}{v^{3}}\right) + \dots
\end{equation}
The expression of the coefficients, $f_{U,n}$, of $F_{U}(h)=\sum f_{U,n} (h/v)^{n}$ to all orders is \cite{Buchalla:2013rka}
\begin{equation}
  \label{eq:7.14.b}
  f_{U,n} = \frac{2}{n!}\begin{cases}(1-2\xi) (-4\xi)^{\tfrac{n}{2}-1}, & \text{for }n\text{ even}\\
  \sqrt{1-\xi}(-4\xi)^{\tfrac{n-1}{2}}, & \text{for }n\text{ odd.}\end{cases}
\end{equation}
Also in this explicit model, $\xi$ controls the degree of decoupling. $W_{L}W_{L}$-scattering amplitudes violate perturbative unitarity at a scale $\Lambda\approx 4\pi\, v/\sqrt{\xi}$ \cite{Contino:2010rs}. The new physics decouples in the limit $\xi\rightarrow 0$ (for fixed $v$) and $F_{U}(h)$ approaches its SM form. The SM-Higgs unitarizes the amplitudes alone. A composite Higgs with generic $\xi\in(0,1)$ unitarizes the amplitudes only partly, the other resonances of the strong sector are needed for a complete unitarization. In the limit $\xi\rightarrow 1$, the Higgs does not contribute and only the resonances ensure the unitarization. This corresponds to the TC limit. \\

The authors of \cite{Contino:2011np} discussed the next-to-leading operators of the $SO(5)/SO(4)$ coset in detail. They defined the building blocks $d_{\mu}$ and $E_{\mu}$ through 
\begin{equation}
  \label{eq:7.15}
-i\mathcal{U}^\dagger D_{\mu} \mathcal{U} = d^{\hat a}_{\mu} t^{\hat a} + E^a_{\mu} t^a \equiv d_{\mu} + E_{\mu}.
\end{equation}
The unbroken generators, $t^{a}$, and the broken generators, $t^{\hat{a}}$, are given in Appendix~\ref{ch:appA}. $D_{\mu} = \partial_{\mu}+i A_{\mu}$ is the covariant derivative of the most general gauge field, $A_{\mu} = A_{\mu}^{\hat{a}}t^{\hat{a}}+A_{\mu}^{a}t^{a}$, with absorbed gauge coupling. Further, we consider the building blocks
\begin{equation}
  \label{eq:7.16}
\partial_{\mu} E_{\nu} - \partial_{\nu} E_{\mu} +i[E_{\mu},E_{\nu}]\equiv E_{\mu\nu}\equiv E^{L}_{\mu\nu} + E^{R}_{\mu\nu}
\end{equation}
and
\begin{equation}
  \label{eq:7.17}
f_{\mu\nu}=\mathcal{U}^\dagger F_{\mu\nu} \mathcal{U}\equiv f^{-}_{\mu\nu} + f^{L}_{\mu\nu} + f^{R}_{\mu\nu}.
\end{equation}
Here, the superscripts ``$L$'' and ``$R$'' refer to the operators that are multiplied with the $t^{a}_{L/R}$ generators, while $f^{-}_{\mu\nu}\equiv f^{-,\hat a}_{\mu\nu} t^{\hat a}$. The $\mathcal{CP}$-even, next-to-leading order operators are~\cite{Contino:2011np}
\begin{align}
  \begin{aligned}
    \label{eq:7.18}
    O_1 &= \langle d_{\mu} d^{\mu}\rangle^2,\\
    O_2 &= \langle d_{\mu} d_{\nu}\rangle \langle d^{\mu} d^{\nu}\rangle,\\
    O_3 &= \langle E^L_{\mu\nu} E^{L,\mu\nu}\rangle -\langle E^R_{\mu\nu} E^{R,\mu\nu}\rangle,\\
    O^+_4 &= \langle (f^L_{\mu\nu} + f^R_{\mu\nu}) i[d^{\mu}, d^{\nu}]\rangle,\\
    O^+_5 &= \langle (f^-_{\mu\nu})^2\rangle,\\
    O^-_4 &= \langle (f^L_{\mu\nu} - f^R_{\mu\nu}) i[d^{\mu}, d^{\nu}]\rangle,\\
    O^-_5 &= \langle (f^L_{\mu\nu})^2 - (f^R_{\mu\nu})^2\rangle.
  \end{aligned}  
\end{align}
With the identifications in Eqs.~\eqref{eq:7.7} and~\eqref{eq:7.8}, we can relate the operators to the NLO operators of the electroweak chiral Lagrangian, Eqs.~\eqref{eq:5.NLO.1} -- \eqref{eq:5.NLO.17}. We restrict the gauging to the SM gauge group and find
\begin{align*}
  \begin{aligned}
    O_1 &= \left(\frac{2}{f^2} \partial_\mu |h| \partial^\mu |h|+ s^2 \langle L_\mu L^\mu\rangle\right)^2 ,\\
    O_2 &= \left(\frac{2}{f^2} \partial_\mu |h| \partial_\nu |h| + s^2 \langle L_\mu L_\nu\rangle\right)^2 ,\\
    O^+_4 &= -s^2 \langle g D_\mu W^{\mu\nu} L_\nu -g'\partial_\mu B^{\mu\nu} \tau_L L_\nu +\frac{g^2}{2}(W_{\mu\nu})^2 +\frac{g'^{2}}{2}(B_{\mu\nu} T_3)^2 - g' g B_{\mu\nu} W^{\mu\nu}\tau_L\rangle,
  \end{aligned}
\end{align*}
\begin{align}
  \begin{aligned}
    \label{eq:7.19}
    O^+_5 &= s^2\langle g^2 (W_{\mu\nu})^2 + g'^{2}(B_{\mu\nu} T_3)^2 - 2  g' g B_{\mu\nu} W^{\mu\nu}\tau_L\rangle ,\\
    O^-_4 &=i\frac{c}{2}(s^2+2) \langle g W_{\mu\nu} [L^\mu,L^\nu] -g' B_{\mu\nu} \tau_L [L^\mu,L^\nu]\rangle \\  
    &+ 2c \langle g D_\mu W^{\mu\nu} L_\nu +g'\partial_\mu B^{\mu\nu} \tau_L L_\nu +\frac{g^2}{2}(W_{\mu\nu})^2 -\frac{g'^{2}}{2}(B_{\mu\nu} T_3)^2 \rangle ,\\
    O^-_5 &= 2 c\langle g^2 (W_{\mu\nu})^2 - g'^{2}(B_{\mu\nu} T_3)^2 \rangle.
  \end{aligned}
\end{align}
We note that the operator $O_{3}$ of Eq.~\eqref{eq:7.18} is redundant, $O_{3} = O^{-}_{5} - 2 O^{-}_{4}$. This was also mentioned in \cite{Alonso:2014wta}, where this and other cosets were considered. 

From the NLO operators of Section~\ref{ch:chiralNLO}, we find $\mathcal{O}_{D1,2,7,8,11}$, $\mathcal{O}_{Xh1,2}$, and $\mathcal{O}_{XU1,7,8}$. Some operators contain the terms $D_{\mu} W^{\mu\nu}$ and $\partial_{\mu} B^{\mu\nu}$ that are reducible when using the equations of motion. Further, we have to expand the trigonometric functions, $s$ and $c$, around $\langle |h|\rangle$ in order to find the explicit form of the $F_{i}(h)$.

When we rotate Eq.~\eqref{eq:7.19} to the physical basis using Eq.~\eqref{eq:2.2.16}, we find that no photon-photon-Higgs- and also no gluon-gluon-Higgs coupling is generated by the $SO(5)/SO(4)$ model. This was motivated in \cite{Giudice:2007fh} by a shift symmetry of the pNGB Higgs. It is true for the bosonic Lagrangian defined at the scale $f$, as we just derived. However, at the scale $v$, we have also integrated out the fermionic states of the scale $f$. This induces the operators $hG_{\mu\nu}G^{\mu\nu}$ and $hF_{\mu\nu}F^{\mu\nu}$ with coefficients of the order $\xi/16\pi^{2}$, {\it i.e.} at next-to-leading order. Additionally, explicit computations with new states at the scale $f$ confirm the appearance of those operators \cite{Falkowski:2007hz,Carena:2014ria}. 
\section{The Singlet Extension of the Standard Model}
\label{ch:SM+S}
We consider the SM extended with a real scalar singlet \cite{Schabinger:2005ei,Patt:2006fw,Bowen:2007ia,Englert:2011yb,Buttazzo:2015bka,Robens:2016xkb}. This model serves as a simple, renormalizable UV-completion. Even though this model is not favored by current experimental data \cite{Buttazzo:2015bka,Robens:2015gla}, it is a useful toy-model that illustrates the connection between a UV-completion and its low-energy EFTs \cite{Buchalla:2012qq,mastersthesis,Buchalla:2013rka,Buchalla:2016bse}. In particular, we will see that, depending on the region of parameter space, the heavy field either decouples at low energies or not. Therefore, the SM-EFT and the ew$\chi\mathcal{L}$ are both appropriate low-energy descriptions for different parametric limits.\\

We start with the Lagrangian of the SM, Eq.~\eqref{eq:SMunbroken}, and add a real scalar gauge singlet $S$. We further assume an additional $Z_{2}$ symmetry under which $S$ is odd and all the other fields are even. The full Lagrangian becomes
\begin{align}
  \begin{aligned}
    \label{eq:7.20}
  \mathcal{L}&=-\frac{1}{4} B_{\mu\nu}B^{\mu\nu} - \frac{1}{2}\langle W_{\mu\nu}W^{\mu\nu}\rangle - \frac{1}{2}\langle G_{\mu\nu}G^{\mu\nu}\rangle \\
  &+i\bar{q}_{L}^{i}\slashed{D}q_{L}^{i} +i\bar{\ell}_{L}^{i}\slashed{D}\ell_{L}^{i} +i\bar{u}_{R}^{i}\slashed{D}u_{R}^{i} +i\bar{d}_{R}^{i}\slashed{D}d_{R}^{i} +i\bar{e}_{R}^{i}\slashed{D}e_{R}^{i}\\
  &+(D^{\mu} \phi)^{\dagger} (D_{\mu} \phi)  +  \partial^{\mu}  S \partial_{\mu} S   - V(\phi,S)\\
  &-\bar{\ell}_{L}^{i}Y_{e}^{ij}\phi e_{R}^{j} - \bar{q}_{L}^{i}Y_{d}^{ij}\phi d_{R}^{j} - \bar{q}_{L}^{i}Y_{u}^{ij} (i\sigma_{2}\phi^{*}) u_{R}^{j} +\text{h.c.},
\end{aligned}
\end{align}
where
\begin{equation}
\label{eq:7.21}
V(\phi,S) =  -\frac{ \mu_{1}^{2}}{2} \phi^{\dagger} \phi  - \frac{\mu_{2}^{2}}{2} S^2  + \frac{\lambda_{1}}{4}  ( \phi^{\dagger} \phi )^2  + \frac{\lambda_{2}}{4}  S^4 + \frac{\lambda_{3}}{2}    \phi^{\dagger} \phi  S^2\,.
\end{equation}
The potential is bounded from below and has a stable minimum if \cite{Pruna:2013bma,Robens:2015gla}
\begin{equation}
\label{eq:7.22}
  \lambda_{1},  \;  \lambda_{2} >0\,, \qquad \text{and}\qquad   \lambda_{1}  \lambda_{2} - \lambda_{3}^{2} > 0   \,.
\end{equation}
Historically, this model is also called ``Higgs portal'' \cite{Schabinger:2005ei,Patt:2006fw,Englert:2011yb}, as it offers a renormalizable portal coupling to a sector that is otherwise not interacting with the SM. Models of this type are also appealing to describe dark matter sectors \cite{Burgess:2000yq,Davoudiasl:2004be}.
\subsection{The Physical Basis}
The potential in Eq.~\eqref{eq:7.21} has a non-trivial minimum, giving a vacuum expectation value to both, the Higgs and the scalar singlet:
\begin{equation}
  \label{eq:7.23}
  \phi  = \frac{v + h_{1}}{\sqrt{2}} \; U \, \binom{0}{1}\qquad \text{and} \qquad S = \frac{v_{s} + h_{2}}{\sqrt{2}}.
\end{equation}
Here, we wrote the Higgs doublet $\phi$ in terms of the Goldstone boson matrix $U$ and the Higgs field $h_{1}$, as before in Eq.~\eqref{eq:5.1}. The excitations of the vacuum, $h_{1}$ and $h_{2}$, mix kinetically in Eq.~\eqref{eq:7.21}. We obtain the physical states after the rotation
\begin{equation}
\label{eq:7.24}
\binom{h}{H}\; = \;\begin{pmatrix} \cos{\chi} & -\sin{\chi} \\ \sin{\chi} & \cos{\chi}\end{pmatrix}\, \cdot \,\binom{h_{1}}{h_{2}} \, ,
\end{equation}
with
\begin{equation}
\label{eq:7.25}
\tan(2 \chi) = \frac{  2  \lambda_{3}   v  v_{s} }{   \lambda_{2}  v_{s}^2 - \lambda_{1} v^2 } \,.
\end{equation}
We restrict the range of $\chi$, without loss of generality, to $\chi\in\left[-\pi/2,\pi/2\right]$. We see that the mixing vanishes in the limit $\lambda_{3}\rightarrow 0$, as expected. After this rotation, we find the masses of the two scalars,
\begin{equation}
\label{eq:7.26}
m_{h, H}^2 = \frac{1}{4} \left[   \lambda_{1} v^2 +    \lambda_{2}  v_{s}^2 \mp \sqrt{    (      \lambda_{1} v^2  -    \lambda_{2}  v_{s}^2   )^2 +  4  ( \lambda_{3}  v  v_{s}  )^2      } \right].
\end{equation}
We define $m_{h}\equiv m < M \equiv m_{H}$. The five parameters $\mu_{1}$, $\mu_{2}$, $\lambda_{1}$, $\lambda_{2}$, and $\lambda_{3}$ define the model in the interaction basis in Eqs.~\eqref{eq:7.20} and \eqref{eq:7.21}. We can relate them to five physical parameters $v$, $m$, $M$, $\chi$, and $f = \sqrt{v^{2}+v_{s}^{2}}$ using the relations~\eqref{eq:7.25}, \eqref{eq:7.26}, and 
\begin{equation}
  \label{eq:7.27}
  v^{2} = 2\; \frac{\mu_{1}^{2}\lambda_{2}-\mu_{2}^{2}\lambda_{3}}{\lambda_{1}  \lambda_{2} - \lambda_{3}^{2}}\, \qquad v_{s}^{2} = 2\; \frac{\mu_{2}^{2}\lambda_{1}-\mu_{1}^{2}\lambda_{3}}{\lambda_{1}  \lambda_{2} - \lambda_{3}^{2}}\,.
\end{equation}
We identify $m$ and $v$ with the Higgs mass and the electroweak vacuum expectation value. The three remaining parameters, $M$, $\chi$, and $f$, describe the dynamics beyond the SM. We use $\xi=v^{2}/f^{2}$, as before. For the reverse transformation, we use
\begin{align}
  \begin{aligned}
    \label{eq:7.28}
    \lambda_{1} &= \frac{2}{v^{2}}\left(m^{2} c^{2} +M^{2} s^{2}\right),\\
    \lambda_{2} & = \frac{2}{v^{2}} \frac{\xi}{1-\xi}\left(M^{2} c^{2} +m^{2} s^{2}\right),\\
    \lambda_{3} & = \frac{2}{v^{2}} \sqrt{\frac{\xi}{1-\xi}}\; c\, s\, (M^{2}-m^{2}),\\
    \mu_{1}^{2} &= \left(m^{2} c^{2} +M^{2} s^{2}\right)+ \sqrt{\frac{1-\xi}{\xi}}\;c\, s\,(M^{2}-m^{2}),\\
    \mu_{2}^{2} &= \left(M^{2} c^{2} +m^{2} s^{2}\right) + \sqrt{\frac{\xi}{1-\xi}}\; c\, s\, (M^{2}-m^{2}),
  \end{aligned}
\end{align}
where $c = \cos{\chi}$ and $s=\sin{\chi}$.

Expressed in the physical fields, $h$ and $H$, the scalar part of the Lagrangian of Eq.~\eqref{eq:7.20} is 
\begin{align}
  \begin{aligned}
    \label{eq:7.29}
    \mathcal{L} &= \frac{1}{2} \partial_{\mu} h \partial^{\mu} h + \frac{1}{2} \partial_{\mu} H \partial^{\mu} H - V(h,H) \\
    &+ \frac{v^2}{4} \langle D_{\mu} U^{\dagger} D^{\mu} U \rangle \left( 1 +  \frac{2 c }{v}  h +  \frac{2 s }{v}  H +  \frac{c^{2} }{v^2}  h^2  +  \frac{s^{2}}{v^2}  H^2 +  \frac{2\, s\, c }{v^2}  h H      \right) \\
    &- v \left(  \bar{q}_{L} Y_u U P_{+}q_{R} + \bar{q}_{L} Y_d U P_{-}q_{R} + \bar{\ell}_{L}  Y_e U P_{-}\ell_{R} + \text{ h.c.} \right)  \left[1 + \frac{c}{v} h + \frac{s}{v} H  \right],
  \end{aligned}
\end{align}
The potential is 
\begin{align}
  \begin{aligned}
    \label{eq:7.30}
    V(h,H) =&  \frac{1}{2} m^2 h^2 +   \frac{1}{2} M^2 H^2  - d_1 h^{3} - d_2  h^2 H - d_3 h H^2  - d_4 H^3 \\
    &- z_1 h^4 - z_2 h^3 H - z_3 h^2 H^2 -z_4 h H^3 - z_5 H^4 \,,
  \end{aligned}
\end{align}
where 
\begin{align*}
  \begin{aligned}
    d_{1} &= \frac{m^{2}}{2vv_{s}}\left(s^{3}v-c^{3}v_{s}\right) ,\\
    d_{2} &= -\frac{2m^{2}+M^{2}}{2vv_{s}}sc\left(sv+cv_{s}\right) ,\\
    d_{3} &= \frac{2M^{2}+m^{2}}{2vv_{s}}sc\left(cv-sv_{s}\right) ,\\
    d_{4} &= -\frac{M^{2}}{2vv_{s}}\left(c^{3}v+s^{3}v_{s}\right) ,
  \end{aligned}
\end{align*}
\begin{align}
  \begin{aligned}
    \label{eq:7.31}
    z_{1} &= -\frac{1}{8v^{2}v_{s}^{2}}\left[m^{2}\left(s^{3}v-c^{3}v_{s}\right)^{2}+M^{2}s^{2}c^{2}\left(sv+cv_{s}\right)^{2}\right],\\
    z_{2} &= \frac{sc}{2v^{2}v_{s}^{2}}(sv+cv_{s})\left[m^{2}\left(s^{3}v-c^{3}v_{s}\right)+M^{2}sc\left(cv-sv_{s}\right)\right],\\
    z_{3} &= -\frac{sc}{8v^{2}v_{s}^{2}}\left[m^{2}\left(6sc(sv+cv_{s})^{2}-2vv_{s}\right)+M^{2}\left(6sc(cv-sv_{s})^{2}+2vv_{s}\right)\right],\\
    z_{4} &= \frac{sc}{2v^{2}v_{s}^{2}}(cv-sv_{s})\left[M^{2}\left(c^{3}v+s^{3}v_{s}\right)+m^{2}sc \left(sv+cv_{s}\right)\right],\\
    z_{5} &= -\frac{1}{8v^{2}v_{s}^{2}}\left[M^{2}\left(c^{3}v+s^{3}v_{s}\right)^{2}+m^{2}s^{2}c^{2}\left(cv-sv_{s}\right)^{2}\right].
  \end{aligned}
\end{align}
So far, this describes the renormalizable SM singlet extension in the physical basis, without any approximation. However, we are interested in scenarios where $M\gg m$, such that we can integrate out the heavy field. This limit can naturally be realized with an approximate $SO(5)$ symmetry in the scalar sector, where the four real components of $\phi$ and the $S$ transform as a fundamental vector of $SO(5)$. We split the potential in Eq.~\eqref{eq:7.21} in a part that conserves $SO(5)$ and a part that breaks $SO(5)$ weakly, $V \equiv V_{0}+V_{1}$. In the exact $SO(5)$-limit, we have $\lambda_{1}=\lambda_{2}=\lambda_{3}=2M^{2}/f^{2}$ and $\mu_{1} = \mu_{2}=M$. Further, $\xi=\sin^{2}{\chi}$ and the light scalar becomes a Goldstone boson, {\it i.e.} $m=0$. Expressed in terms of $\Sigma^{2}\equiv \phi^{\dagger}\phi+S^{2} $, we have
\begin{equation}
  \label{eq:7.32}
  V_{0}= -\frac{\mu_{1}^{2}}{2} \Sigma^{2} + \frac{\lambda_{1}}{4}(\Sigma^{2})^{2}.
\end{equation}
Two of the five parameters of the model conserve the $SO(5)$ symmetry. We identify them with $M$ and $f$. The symmetry is weakly broken by the spurion $S^{2}$, which respects the $Z_{2}$ and an $SO(4)$ symmetry. The latter corresponds to  the custodial symmetry of the Higgs sector that we assume to hold. There are three renormalizable operators, which include the spurion $S^{2}$, that we write in $V_{1}$:
\begin{equation}
  \label{eq:7.33}
  V_{1}= \frac{\mu_{1}^{2}-\mu_{2}^{2}}{2} S^{2} + \frac{\lambda_{1}+\lambda_{2}-2\lambda_{3}}{4} S^{4} + \frac{\lambda_{3}-\lambda_{1}}{2} \Sigma^{2}S^{2}.
\end{equation}
This explicit breaking introduces a small mass for the light scalar, which becomes a pseudo-Nambu-Goldstone boson. The Goldstone symmetry keeps $m\ll M$, which is what we need for the effective expansion. We express the deviations from the $SO(5)$-limit in terms of 
\begin{equation}
  \label{eq:7.34}
  r \equiv \frac{m^{2}}{M^{2}} \qquad \text{and}\qquad \delta \equiv \frac{\omega}{\xi}-1,
\end{equation}
where $\omega \equiv \sin^{2}{\chi}$. These parameters are small for weak $SO(5)$ breaking, $r,\delta\ll 1$, which ensures $m\ll M$. The fifth parameter of the model (apart from $M$, $f$, $\delta$, and $r$) is $\xi$. It also breaks the $SO(5)$ symmetry, but is naturally of order unity, as the vacuum is degenerate in the strict $SO(5)$-limit. The small explicit breaking of $SO(5)$ through $\delta$ and $r$ lifts this degeneracy. We can also express the couplings $\lambda_{i}$ in the latter set of parameters:
\begin{align}
  \begin{aligned}
    \label{eq:7.36}
    \lambda_1 =&  \frac{2 M^2}{ f^2} \frac{  r + \omega (1-r)  }{  \xi }, \\ 
    \lambda_2 =& \frac{2 M^2}{ f^2} \frac{   1 - \omega (1-r) }{   1 - \xi }, \\
    \lambda_3 =& \frac{ 2 M^2 }{f^2}   (1-r)  \sqrt{     \frac{   \omega (1- \omega) }{  \xi (1- \xi) }    }.
  \end{aligned}
\end{align} 
When we expand the couplings in Eq.~\eqref{eq:7.33} to first order in the small parameters of Eq.~\eqref{eq:7.34}, we find
\begin{align}
  \begin{aligned}
    \label{eq:7.35}
    \mu_{1}^{2}-\mu_{2}^{2} & = M^{2}\frac{\delta}{2(1-\xi)},\\
    \lambda_{1}+\lambda_{2}-2\lambda_{3}&= \frac{2M^{2}}{f^{2}}\frac{r}{\xi(1-\xi)},\\
    \lambda_{1}-\lambda_{3} & = \frac{2M^{2}}{f^{2}} \left(\frac{r}{\xi}+\frac{\delta}{2(1-\xi)}\right).
  \end{aligned}
\end{align}
This shows that $V_{1}$ in Eq.~\eqref{eq:7.33} breaks $SO(5)$ weakly for $r,\delta\ll 1$. \\

\begin{figure}[t]
    \centering
    \includegraphics[width=0.75\textwidth]{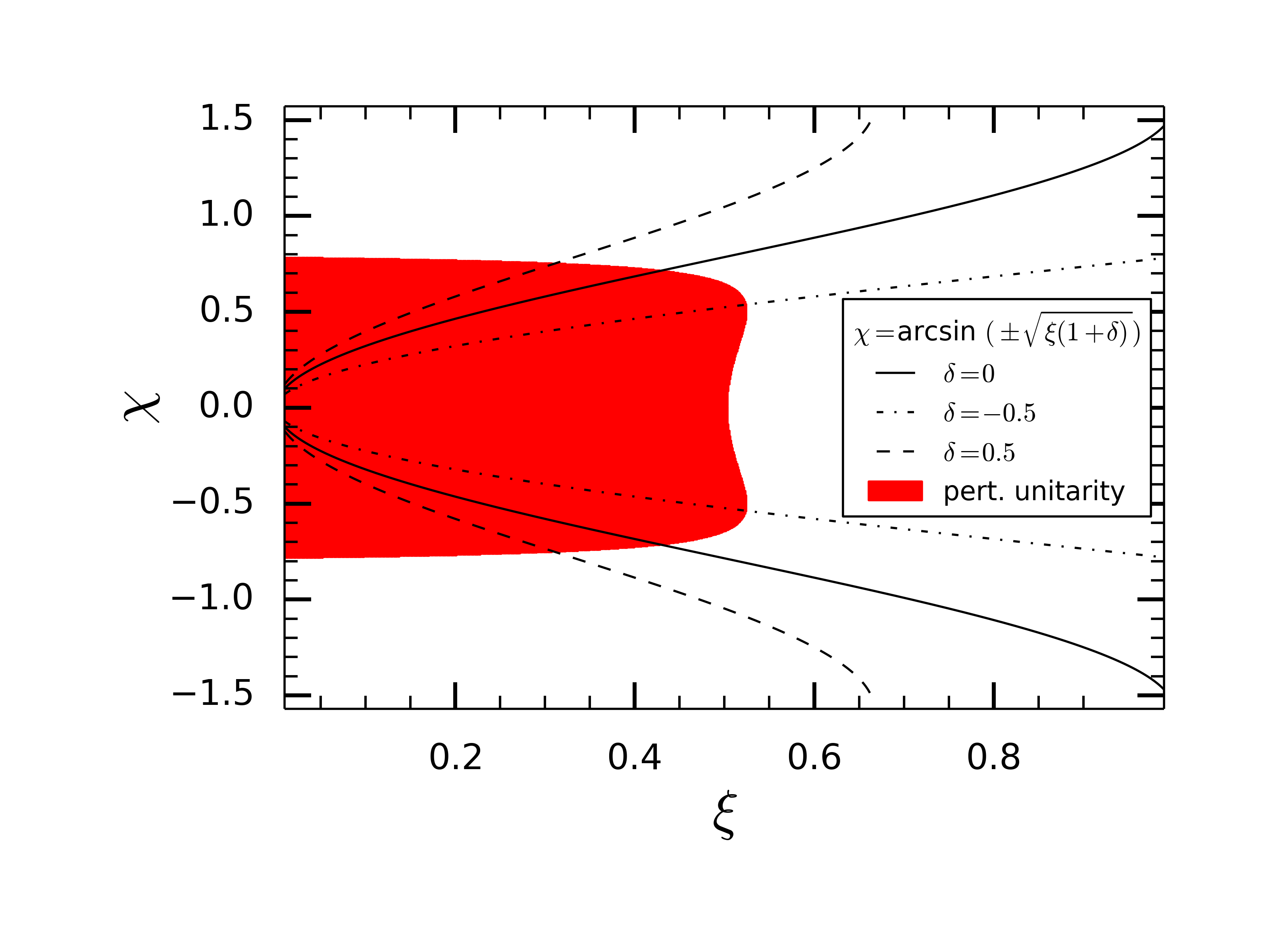}
    \caption[Parameter scan of the model for fixed $M=1~\text{TeV}$.]{Parameter scan of the model for fixed $M=1~\text{TeV}$. The red area is allowed by perturbative unitarity \cite{Pruna:2013bma}. The lines show the $SO(5)$-limit and small deviations from it.}
    \label{fig:param.scan}
  \end{figure}

Figure~\ref{fig:param.scan} shows a parameter scan of the model in the $\xi$-$\chi$-plane for fixed $M=1~\text{TeV}$. The red area defines the region of the parameter space that is allowed by perturbative unitarity. We use the relations for tree-level unitarity from \cite{Pruna:2013bma}. Additional lines show the $SO(5)$-limit and small deviations from it. 
\subsection{The Effective Descriptions}
We are interested in the limit of the model where $M\gg m$. In this limit, we can integrate out the heavy scalar and construct a low-energy EFT \cite{Buchalla:2016bse}. Similar setups are also discussed in \cite{Gorbahn:2015gxa,Chiang:2015ura ,Corbett:2015lfa,Brehmer:2015rna,Egana-Ugrinovic:2015vgy,Boggia:2016asg,Feruglio:2016zvt}. 

\begin{figure}[t]
\begin{center}
\subfigure[General case \label{fig:scalesC}]{
\begin{overpic}[width=4cm]{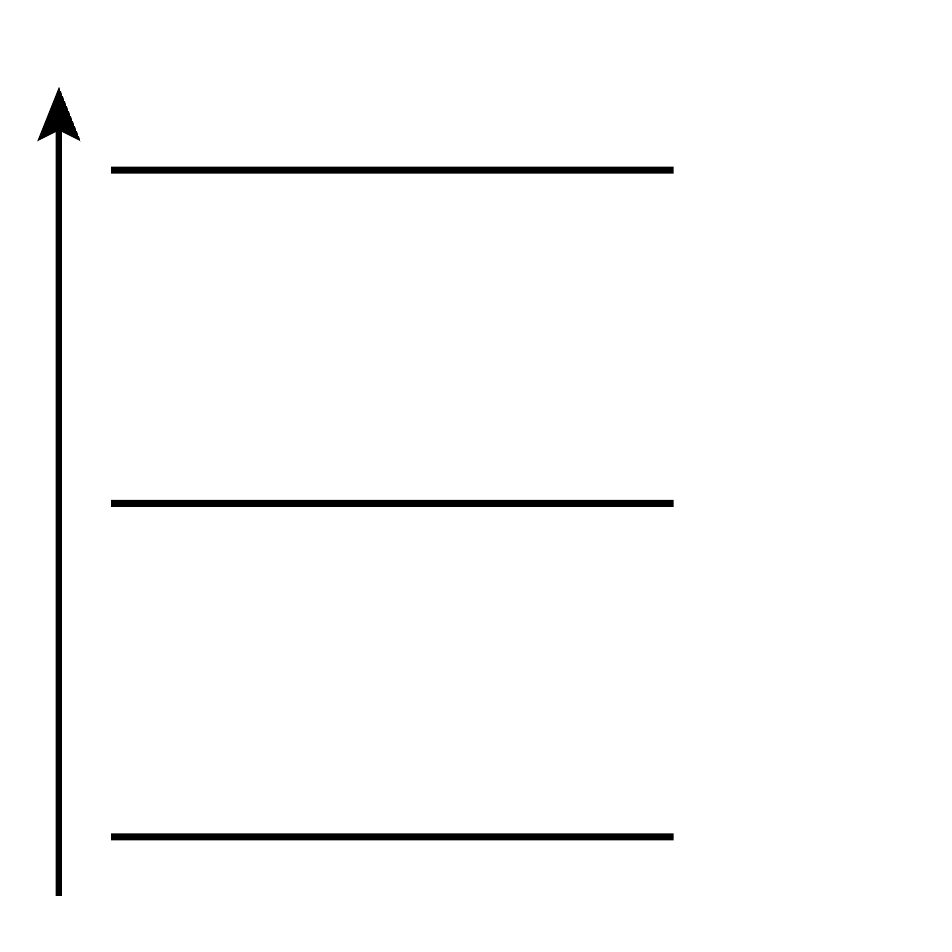}
\put (-10,90){$E$}
\put (80,80){$M$}
\put (95,60){$\Bigg\}\lambda$}
\put (80,45){$f$}
\put (95,25){$\Bigg\}\xi$}
\put (80,10){$v$}
\end{overpic}}\hfill
\subfigure[Case \ref{case.strong}), the strongly-coupled regime that introduces the non-decoupling EFT. \label{fig:scalesA}]{
\begin{overpic}[width=4cm]{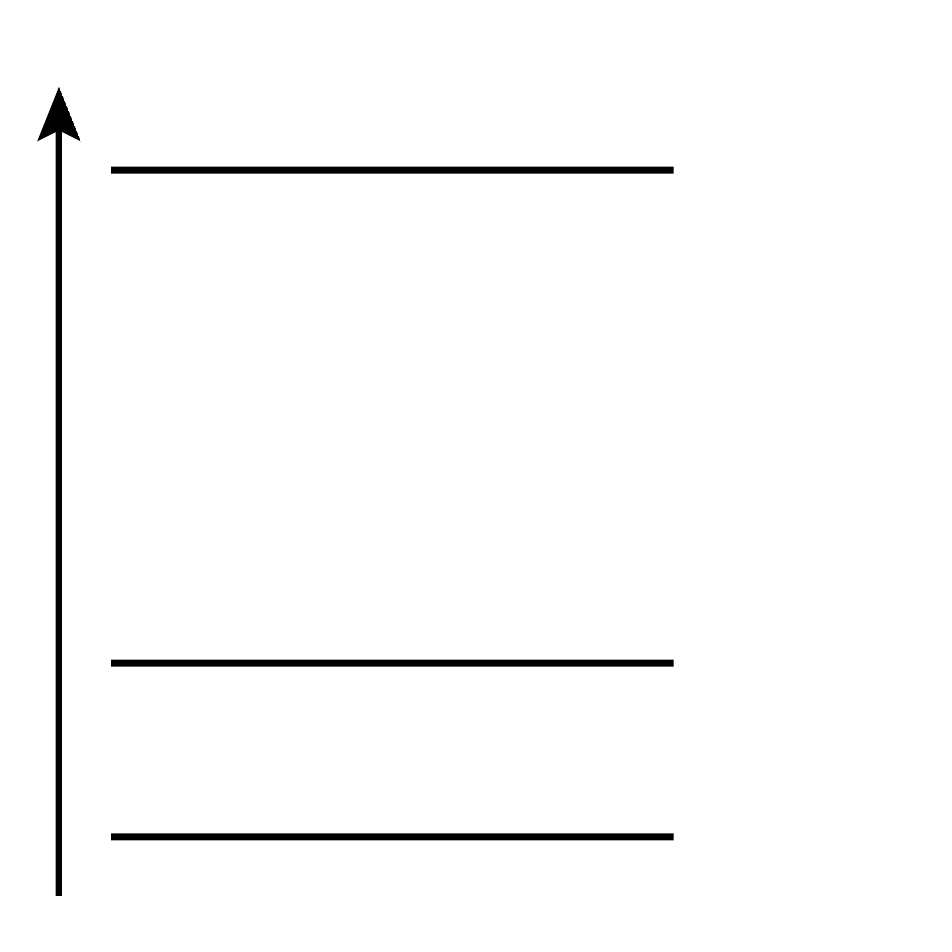}
\put (-10,90){$E$}
\put (80,80){$M$}
\put (80,30){$f$}
\put (80,10){$v$}
\end{overpic}}\hfill
\subfigure[Case \ref{case.weak}), the weakly-coupled regime that introduces the decoupling EFT. \label{fig:scalesB}]{
\begin{overpic}[width=4cm]{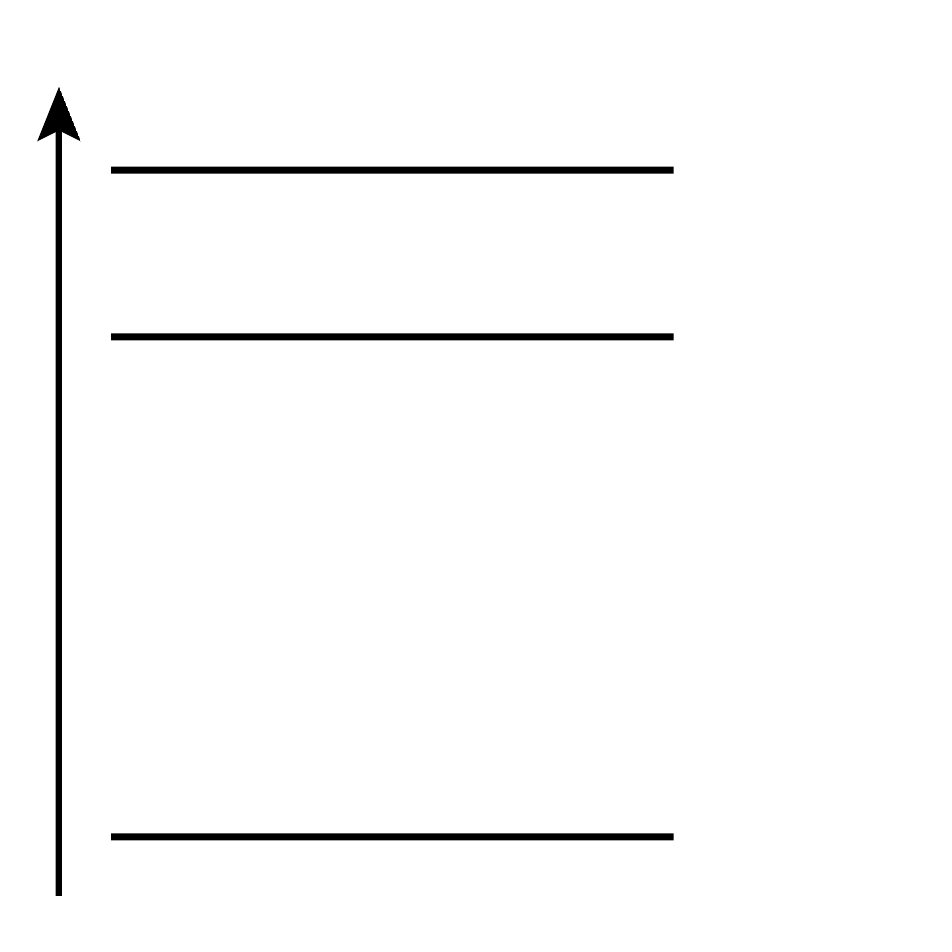}
\put (-10,90){$E$}
\put (80,80){$M$}
\put (80,65){$f$}
\put (80,10){$v$}
\end{overpic}}
\end{center}
\caption[Schematic picture of the different possible hierarchies.]{Schematic picture of the different possible hierarchies.}
\label{fig:7.scales}
\end{figure}
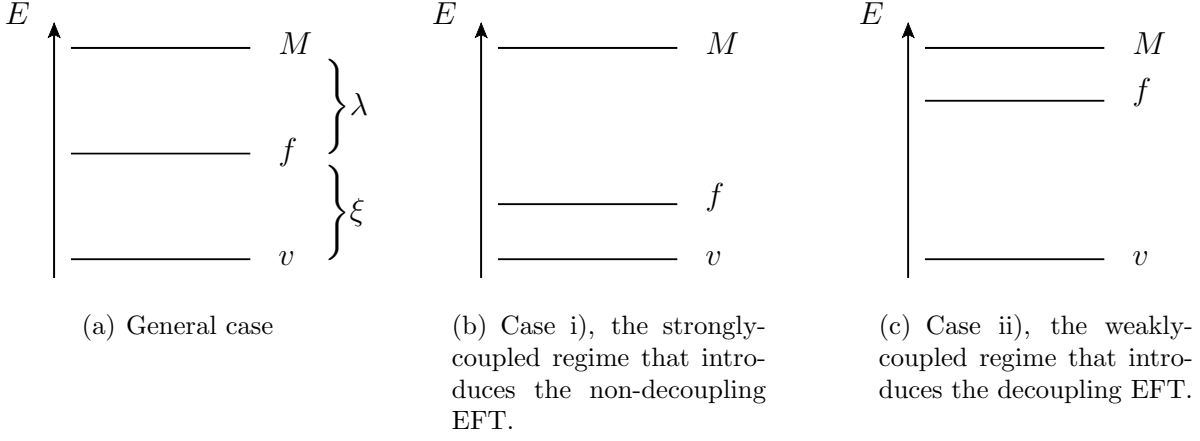

With $v=246~\text{GeV}$, $m=125~\text{GeV}$, and $M\gg m$, the parameter space is still not fully determined. The couplings $\lambda_{i}$ give the ratio of $M$ and $f$ close to the $SO(5)$-limit, see Eq.~\eqref{eq:7.36}. The ratio of $v$ to $f$ is parametrized by $\xi$. We illustrate the hierarchies of the scales in Fig.~\ref{fig:scalesC}. In particular, we distinguish two different cases: 
\begin{enumerate}[i)]
\item \label{case.strong}strongly-coupled regime:
  \begin{equation}
    \label{eq:7.37}
    |\lambda_{i}|\lesssim 32\pi^{2}\,, \qquad m \sim v \sim f \ll M \qquad \Rightarrow \quad \xi,\omega = \mathcal{O}(1)
  \end{equation}
  Figure~\ref{fig:scalesA} illustrates the hierarchy of scales in this case. \\
\item \label{case.weak}weakly-coupled regime:
  \begin{equation}
    \label{eq:7.38}
    \lambda_{i}=\mathcal{O}(1)\,, \qquad m \sim v \ll f \sim M \qquad \Rightarrow \quad \xi,\omega \ll 1
  \end{equation}
  Figure~\ref{fig:scalesB} illustrates the hierarchy of scales in this case.
\end{enumerate}
We assume that $|\lambda_{i}|$ stays below the nominal strong-coupling limit, $|\lambda_{i}|=32\pi^{2}$, which corresponds to $M=4\pi f$. Otherwise, the perturbative description in terms of a single particle, $H$, would not be valid anymore. Independent of the actual size of $|\lambda_{i}|$, we still assume that the differences between the couplings in Eq.~\eqref{eq:7.35} are small, such that $\delta,r \ll 1$. 

\subsubsection{Case \ref{case.strong}) --- the Non-Decoupling Limit}
In the first case, $f$ is of the order of $v$; $\xi=\mathcal{O}(1)$; and the model is strongly coupled. We write the Lagrangian of $H$, Eq.~\eqref{eq:7.29}, as
\begin{equation}
  \label{eq:7.39}
  \mathcal{L}=-\frac{1}{2} H (\Box+M^{2})H + J_{1}H + J_{2}H^{2}+J_{3}H^{3}+J_{4}H^{4}.
\end{equation}
The currents, $J_{i}$, are
\begin{align}
  \begin{aligned}
    \label{eq:7.40}
    J_{1} &= d_{2}h^{2}+z_{2}h^{3}+\frac{v^{2}}{2}\langle D_{\mu}U^{\dagger} D^{\mu}U\rangle \left(\frac{s}{v}+\frac{sc}{v^{2}}h\right)-s J_{\Psi},\\
    J_{2} &= d_{3}h + z_{3}h^{2}+\frac{s^{2}}{4}\langle D_{\mu}U^{\dagger} D^{\mu}U\rangle,\\
    J_{3} &= d_{4}+z_{4}h,\\
    J_{4} &= z_{5},
  \end{aligned}
\end{align}
with
\begin{equation}
  \label{eq:7.41}
  J_{\Psi}\equiv \bar{q}_{L} Y_u U P_{+}q_{R} + \bar{q}_{L} Y_d U P_{-}q_{R} + \bar{\ell}_{L}  Y_e U P_{-}\ell_{R} + \text{ h.c.}
\end{equation}
As we discussed in Section~\ref{ch:EFT.top-down}, we need to expand the equation of motion of $H$ in $1/M$. It is therefore necessary to make the $M$-dependence in Eq.~\eqref{eq:7.39} explicit and write
\begin{align}
  \begin{aligned}
    \label{eq:7.42}
    J_{i}&\equiv M^{2} J_{i}^{0} + \bar{J_{i}},\\
    d_{i}&\equiv M^{2} d_{i}^{0} + \bar{d_{i}},\\
    z_{i}&\equiv M^{2} z_{i}^{0} + \bar{z_{i}}.\\
  \end{aligned}
\end{align}
The Lagrangian becomes 
\begin{equation}
  \label{eq:7.42.1}
  \mathcal{L}\equiv\mathcal{L}^{0} M^{2} + \bar{\mathcal{L}}.
\end{equation}
We insert the expansion,
\begin{equation}
  \label{eq:7.43}
  H = H_{0}+H_{1}+H_{2}+\dots,
\end{equation}
with $H_{n}=\mathcal{O}(1/M^{2n})$, into the Lagrangian in Eq.~\eqref{eq:7.42.1} and derive the equation of motion,
\begin{equation}
  \label{eq:7.44}
\frac{\delta \mathcal{L}^{0}}{\delta H} M^{2} +\frac{\delta \bar{\mathcal{L}}}{\delta H}=  -\left(\Box+M^{2}-2J_{2}\right)H + J_{1}+3J_{3}H^{2}+4J_{4}H^{3}=0.
\end{equation}
We solve it order by order in $1/M^{2}$. At $\mathcal{O}(M^{2})$, the equation of motion is an algebraic equation that defines $H_{0}$,
\begin{equation}
  \label{eq:7.45}
 \left. \frac{\delta \mathcal{L}^{0}}{\delta H}\right|_{H_{0}} =  J_{1}^{0}+(-1+2J_{2}^{0})H_{0}+3J_{3}^{0}H_{0}^{2}+4J_{4}^{0}H_{0}^{3}=0.
\end{equation}
At $\mathcal{O}(1)$, the equation of motion is
\begin{equation}
  \label{eq:7.45.1}
  \left.\frac{\delta^{2}\mathcal{L}^{0}}{(\delta H)^{2}}\right|_{H_{0}}H_{1} + \left.\frac{\delta\bar{\mathcal{L}}}{\delta H}\right|_{H_{0}}=0.
\end{equation}
We find an equation for $H_{1}$ in terms of $H_{0}$,
\begin{equation}
  \label{eq:7.46}
  H_{1} = \frac{(-\Box+2\bar{J_{2}})H_{0}+\bar{J_{1}}+3\bar{J_{3}}H_{0}^{2}+4\bar{J_{4}}H_{0}^{3}}{M^{2}(1-2J_{2}^{0}-6J_{3}^{0}H_{0}-12J_{4}^{0}H_{0}^{2})} .
\end{equation}
Inserting Eq.~\eqref{eq:7.43} back into Eq.~\eqref{eq:7.42.1} gives the effective Lagrangian, organized by an expansion in $1/M^{2}$. To find $H_{0}$ from Eq.~\eqref{eq:7.45}, we rewrite $\mathcal{L}^{0}$ as
\begin{equation}
  \label{eq:7.47}
  \mathcal{L}^{0} = -\frac{1}{2}\left[H_{0}+\frac{1}{2v}\left(s(s\,H_{0}+c\,h)^{2}+\sqrt{\tfrac{\xi}{1-\xi}}c(c\,H_{0}-s\,h)^{2}\right)\right]^{2}.
\end{equation}
With the auxiliary field
\begin{equation}
  \label{eq:7.48}
R\equiv H_{0}+\frac{1}{2v}\left(s(s\,H_{0}+c\,h)^{2}+\sqrt{\tfrac{\xi}{1-\xi}}c(c\,H_{0}-s\,h)^{2}\right) ,
\end{equation}
the Lagrangian $\mathcal{L}^{0}$ becomes
\begin{equation}
  \label{eq:7.49}
  \mathcal{L}^{0} = -\frac{1}{2} R^{2}.
\end{equation}
The equation of motion at $\mathcal{O}(M^{2})$, Eq.~\eqref{eq:7.45}, is
\begin{equation}
    \label{eq:7.50}
    \left.\frac{\delta \mathcal{L}^{0}}{\delta H}\right|_{H_{0}} = \frac{\delta \mathcal{L}^{0}}{\delta R} \frac{\delta R}{\delta H_{0}} =0.
\end{equation}
There are two different solutions of 
\begin{equation}
  \label{eq:7.51}
      \frac{\delta \mathcal{L}^{0}}{\delta R}=-\left[H_{0}+\frac{1}{2v}\left(s(s\,H_{0}+c\,h)^{2}+\sqrt{\tfrac{\xi}{1-\xi}}c(c\,H_{0}-s\,h)^{2}\right)\right]=0,
\end{equation}
and one of
\begin{equation}
  \label{eq:7.52}
      \frac{\delta R}{\delta H_{0}}= \left[1+\frac{1}{2v}\left(2s^{2}(s\,H_{0}+c\,h)+2\sqrt{\tfrac{\xi}{1-\xi}}c^{2}(c\,H_{0}-s\,h)\right)\right] =0. 
\end{equation}
The Lagrangian with the correct solution inserted for $H_{0}$ should describe the effects when $H$ is integrated out at tree level. Diagrammatically, this corresponds to all tree-level diagrams with only internal $H$-lines that form the effective Lagrangian for $h$. This diagrammatic picture helps us to find the right expansion of $H_{0}$.\\

A general Feynman diagram with $I$ internal $H$ lines, $V_{n}$ vertices from the interaction terms $J_{n}H^{n}$ (with $n=1,\dots,4$), and $L$ loops fulfills the topological identities
\begin{align}
  \begin{aligned}
    \label{eq:7.53}
    2 I &= V_{1} + 2 V_{2} + 3 V_{3} + 4 V_{4},\\
    L &= I -(V_{1}+V_{2}+V_{3}+V_{4})+1.
  \end{aligned}
\end{align}
For tree-level diagrams, $L=0$, we find
\begin{equation}
  \label{eq:7.54}
  V_{1} = V_{3}+2 V_{4}+2.
\end{equation}
\begin{figure}[t]
  \begin{center}
    \subfigure[$V_{2}=V_{3}=V_{4}=0, \; V_{1}=2$ \label{fig:model.1}]{
      \begin{tikzpicture}
        \draw[-,thick] (5em,2em) -- (15em,2em);
        \draw {(4em,2em) circle (1em)};
        \draw (4em,2em) node[anchor = center] {$J_{1}$};
        \draw {(16em,2em) circle (1em)};
        \draw (16em,2em) node[anchor = center] {$J_{1}$};
      \end{tikzpicture}
    }\hfill
    \subfigure[$V_{2}=1,\; V_{3}=V_{4}=0, \; V_{1}=2$ \label{fig:model.2}]{
      \begin{tikzpicture}
        \draw[-,thick] (5em,2em) -- (9em,2em);
        \draw[-,thick] (11em,2em) -- (15em,2em);
        \draw {(4em,2em) circle (1em)};
        \draw (4em,2em) node[anchor = center] {$J_{1}$};
        \draw {(10em,2em) circle (1em)};
        \draw (10em,2em) node[anchor = center] {$J_{2}$};
        \draw {(16em,2em) circle (1em)};
        \draw (16em,2em) node[anchor = center] {$J_{1}$};
      \end{tikzpicture}
    }\\
    \subfigure[$V_{2}=V_{4}=0, \; V_{3}=1,\; V_{1}=3$\label{fig:model.3}]{
      \begin{tikzpicture}
        \draw[-,thick] (5em,4em) -- (9em,4em);
        \draw[-,thick] (11em,4em) -- (15em,4em);
        \draw[-,thick] (10em,5em) -- (10em,9em);
        \draw {(4em,4em) circle (1em)};
        \draw (4em,4em) node[anchor = center] {$J_{1}$};
        \draw {(10em,4em) circle (1em)};
        \draw (10em,4em) node[anchor = center] {$J_{3}$};
        \draw {(10em,10em) circle (1em)};
        \draw (10em,10em) node[anchor = center] {$J_{1}$};
        \draw {(16em,4em) circle (1em)};
        \draw (16em,4em) node[anchor = center] {$J_{1}$};
      \end{tikzpicture}
    }\hfill
    \subfigure[$V_{2}=V_{3}=0, \; V_{4}=1,\; V_{1}=4$ \label{fig:model.4}]{
      \begin{tikzpicture}
        \draw[-,thick] (5em,2em) -- (9em,2em);
        \draw[-,thick] (11em,2em) -- (15em,2em);
        \draw[-,thick] (10em,3em) -- (10em,6em);
        \draw[-,thick] (10em,1em) -- (10em, -2em); 
        \draw {(4em,2em) circle (1em)};
        \draw (4em,2em) node[anchor = center] {$J_{1}$};
        \draw {(10em,2em) circle (1em)};
        \draw (10em,2em) node[anchor = center] {$J_{4}$};
        \draw {(10em,-3em) circle (1em)};
        \draw (10em,-3em) node[anchor = center] {$J_{1}$};
        \draw {(10em,7em) circle (1em)};
        \draw (10em,7em) node[anchor = center] {$J_{1}$};
        \draw {(16em,2em) circle (1em)};
        \draw (16em,2em) node[anchor = center] {$J_{1}$};
      \end{tikzpicture}
    }
  \end{center}
  \caption[Illustration of the effective expansion of $\mathcal{L}$ in terms of $J_{i}$.]{Illustration of Eq.~\eqref{eq:7.54}, the expansion of the effective Lagrangian in terms of $J_{i}$.}
  \label{fig:model.organic}
\end{figure}
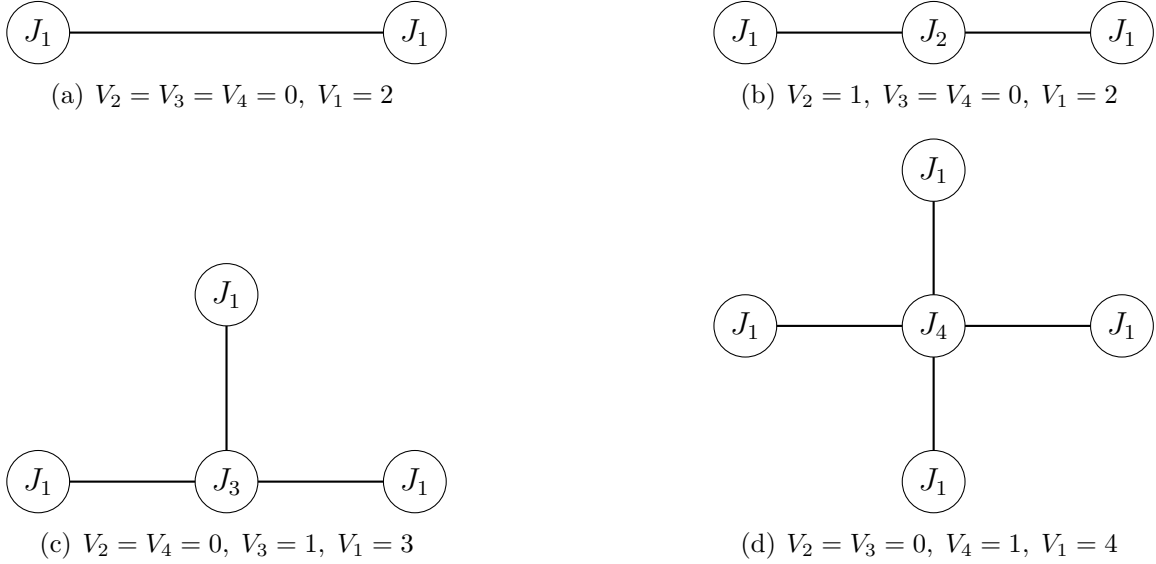

Figure~\ref{fig:model.organic} illustrates Eq.~\eqref{eq:7.54} for different values of $V_{i}$. Since $V_{3},V_{4}\geq 0$, we have $V_{1}\geq 2$. The effective, tree-level Lagrangian contains therefore terms with at least two powers of $J_{1}$. Thus, the expansion of $H_{0}$ contains terms with at least one power of $J_{1}$. Since Eq.~\eqref{eq:7.45} only depends on the currents $J_{i}^{0}$, which are functions of $h$ only, we conclude that the expansion of $H_{0}$ in terms of $h$ starts as $J_{1}^{0}$, {\it i.e.} at $\mathcal{O}(h^{2})$. This singles out one of the three solutions of Eq.~\eqref{eq:7.50},
\begin{align}
  \begin{aligned}
    \label{eq:7.55}
    H_{0} &= -\frac{v+(s^{2}c-c^{2}sW)h}{s^{3}+Wc^{3} } + \sqrt{\frac{(v+(s^{2}c-c^{2}sW)h)^{2}}{(s^{3}+Wc^{3})^{2}}-\frac{(sc^{2}+cs^{2}W)h^{2}}{s^{3}+Wc^{3}}}\\
    &= -\frac{v}{2} (c s (c+s W))\frac{h^{2}}{v^{2}}- \frac{v}{2}\left(c^2 s^2 (c W-s) (c+s W)\right) \frac{h^{3}}{v^{3}}\\
    &-\frac{v }{8}\left(c^2 s^2 (c+s W) \left(5 c^3 s W^2+W \left(c^4-8 c^2 s^2+s^4\right)+5 c s^3\right)\right)\frac{h^{4}}{v^{4}}+\dots,
  \end{aligned}
\end{align}
with $W = \sqrt{\xi/(1-\xi)}$. This, of course, also solves Eq.~\eqref{eq:7.45}. In particular, we see that integrating out $H$ in the strong coupling limit, {\it i.e.} when $\xi, \omega = \mathcal{O}(1)$, generates polynomials of $h$, and higher powers of $h$ are not suppressed by $1/M^{2}$. The expansion is therefore not in canonical dimensions, but rather in chiral dimensions, which is characteristic for the non-decoupling EFT. We discuss this in more detail below.

We now insert the expansion of $H$, Eq.~\eqref{eq:7.43}, back in Eq.~\eqref{eq:7.42.1} and expand the Lagrangian in $1/M^{2}$. Since the solution of $H_{0}$ in Eq.~\eqref{eq:7.55} corresponds to $\delta \mathcal{L}^{0}/\delta R=-R=0$, the Lagrangian at $\mathcal{O}(M^{2})$ cancels, see Eq.~\eqref{eq:7.49}. 

At $\mathcal{O}(1)$, we have the leading-order Lagrangian,
\begin{equation}
  \label{eq:7.55.1}
  \mathcal{L}_{\text{LO}}^{\text{SM+S}} = \left. \frac{\delta \mathcal{L}^{0}}{\delta H}\right|_{H_{0}} H_{1} + \left.\bar{\mathcal{L}}\right|_{H_{0}}.
\end{equation}
The first term, the contribution proportional to $H_{1}$, vanishes by the equation of motion, see Eq.~\eqref{eq:7.45}. We therefore have
\begin{align}
  \begin{aligned}
    \label{eq:7.56}
    \mathcal{L}_{\text{LO}}^{\text{SM+S}} &= \left.\bar{\mathcal{L}}\right|_{H_{0}}\\
    &=\frac{1}{2} (\partial h)^{2} - \frac{m^{2}}{2} h^{2} + d_{1}h^{3}+\bar{z_{1}}h^{4}+\frac{1}{2} (\partial H_{0})^{2} + \bar{J_{1}}H_{0} + \bar{J_{2}}H_{0}^{2}+ \bar{J_{3}}H_{0}^{3}+ \bar{J_{4}}H_{0}^{4}\\
    &+ \frac{v^2}{4} \langle D_{\mu} U^{\dagger} D^{\mu} U \rangle \left( 1 +  \frac{2 c }{v}  h +  \frac{c^{2} }{v^2}  h^2\right) - v J_{\Psi}\left(1+c\frac{h}{v}\right).
  \end{aligned}
\end{align}
The kinetic term of $h$ in Eq.~\eqref{eq:7.56} is not yet canonically normalized, 
\begin{equation}
  \label{eq:7.57}
  \mathcal{L}_{h,\text{kin}} = \frac{1}{2} (\partial h)^{2}+\frac{1}{2} (\partial H_{0})^{2} = \frac{1}{2} (\partial h)^{2} \left(1+ \left(\frac{\delta H_{0}(h)}{\delta h}\right)^{2}\right).
\end{equation}
The field redefinition \cite{Buchalla:2013rka,mastersthesis}
\begin{align}
  \begin{aligned}
    \label{eq:7.58}
    \tilde{h} &= \int_{0}^{h} \sqrt{1+\left(\frac{\delta H_{0}(s)}{\delta s}\right)^{2}} ds \\
    &= h\left(1+\frac{c^{2}s^{2}}{6v^{2}}\left(c+s W\right)^{2}h^{2}+\frac{3c^{3}s^{3}}{8v^{3}} (c+s W)^{2}(c W-s) h^{3}+\mathcal{O}(h^{4})\right)
  \end{aligned}
\end{align}
brings it to the canonically normalized form. The leading-order Lagrangian is then given by Eq.~\eqref{eq:LO}, with
\begin{align}
  \begin{aligned}
    \label{eq:7.59} 
    V(h) &= \frac{m^{2}}{2}v^{2} \left[\left(\frac{h^{2}}{v^{2}}\right)+\left(c^{3}-s^{3}W\right) \left(\frac{h^{3}}{v^{3}}\right)-\frac{1}{12}\left(19 s^{2}c^{2}(s W +c)^{2}-3(s^{4}W^{2}+c^{4})\right) \left(\frac{h^{4}}{v^{4}}\right)\right.\\
  &\left. -\frac{s^{2}c^{2}}{2}(s W +c)^{3}\left(3(1-2s^{2})-\frac{c-s W}{c+ sW}\right) \left(\frac{h^{5}}{v^{5}}\right) +\dots\right],
  \end{aligned}
\end{align}
\begin{align}
  \begin{aligned}
    \label{eq:7.60}
    F_{U}(h) &= 2c \left(\frac{h}{v}\right) + c\left(c^{3}-s^{3}W\right) \left(\frac{h^{2}}{v^{2}}\right)-\frac{4s^{2}c^{3}}{3}\left(c+W s\right)^{2} \left(\frac{h^{3}}{v^{3}}\right) +\dots,
  \end{aligned}
\end{align}
and
\begin{align}
  \begin{aligned}
    \label{eq:7.61}
    \sum_{n=1}^{\infty}Y_{\Psi}^{(n)}\left(\tfrac{h}{v}\right)^{n} &= Y_{\Psi}\left[c \left(\frac{h}{v}\right)-\frac{s^{2}c}{2}(W s + c) \left(\frac{h^{2}}{v^{2}}\right) \right.\\
    &\left.-\frac{s^{2}c^{2}}{6}(W s +c)(4W sc +1-4s^{2}) \left(\frac{h^{3}}{v^{3}}\right) +\dots\right].
  \end{aligned}
\end{align}

We see that the leading-order Lagrangian is consistently at chiral order two. Here, the Higgs mass counts with chiral dimension one, $[m]_{\chi}=1$. The approximate $SO(5)$ symmetry keeps $m$ small compared to $M$. The $SO(5)$-violating couplings, Eq.~\eqref{eq:7.35}, are therefore weak and carry chiral dimension. 

At $\mathcal{O}(1/M^{2})$, NLO in the EFT, the Lagrangian is
\begin{equation}
  \label{eq:7.61.1}
  \mathcal{L}_{\text{NLO}}^{\text{SM+S}} = \left. \frac{\delta^{2}\mathcal{L}^{0}}{(\delta H)^{2}}\right|_{H_{0}}\frac{H_{1}^{2}}{2 M^{2}} + \left. \frac{\delta\mathcal{L}^{0}}{\delta H}\right|_{H_{0}}\frac{H_{2}}{M^{2}}+\left. \frac{\delta\bar{\mathcal{L}}}{\delta H}\right|_{H_{0}}\frac{H_{1}}{M^{2}}.
\end{equation}
The second term vanishes again by using the equation of motion, Eq.~\eqref{eq:7.45}. We rewrite the last term using Eq.~\eqref{eq:7.45.1} and find
\begin{align}
  \begin{aligned}
    \label{eq:7.62}
    \mathcal{L}_{\text{NLO}}^{\text{SM+S}}&= - \frac{1}{2} \left. \frac{\delta^{2}\mathcal{L}^{0}}{(\delta H)^{2}}\right|_{H_{0}}\frac{H_{1}^{2}}{M^{2}}\\
    &=\frac{\left[(-\Box+2\bar{J_2})H_{0} + \bar{J_1} + 3\bar{J_3} H_{0}^{2} + 4\bar{J_{4}} H_{0}^{3}\right]^2}{2 M^2 (1- 2 J_{2}^{0}-6 J_{3}^{0} H_{0}-12 J_{4}^{0} H_{0}^{2})}.
  \end{aligned}
\end{align}
These terms are consistently at chiral order four, with the assignment from before. Equation~\eqref{eq:7.62} contains operators that modify the leading-order operators in Eqs. \eqref{eq:7.59}--\eqref{eq:7.61} and some of the NLO operators of Eqs.~\eqref{eq:5.NLO.1}--\eqref{eq:5.NLO.17}. In particular, the model generates
\begin{equation}
  \label{eq:7.63}
  \mathcal{O}_{D1}, \mathcal{O}_{D7}, \mathcal{O}_{D11};\quad \mathcal{O}_{\psi S1}, \mathcal{O}_{\psi S2}, \mathcal{O}_{\psi S7}, \mathcal{O}_{\psi S14}, \mathcal{O}_{\psi S15}, \mathcal{O}_{\psi S18},
\end{equation}
the hermitean conjugates of the $\mathcal{O}_{\psi Si}$ in \eqref{eq:7.63}, and four-fermion operators coming from the square of the Yukawa terms contained in $\bar{J_{1}}$. The latter are
\begin{align}
  \begin{aligned}
    \label{eq:7.64}
    \mathcal{O}_{FY1}, \mathcal{O}_{FY3}, \mathcal{O}_{FY5}, \mathcal{O}_{FY7}, \mathcal{O}_{FY9}, \mathcal{O}_{FY10}, \mathcal{O}_{ST5}, \mathcal{O}_{ST9}, \mathcal{O}_{LR1},\mathcal{O}_{LR2},\\ 
    \mathcal{O}_{LR3}, \mathcal{O}_{LR4}, \mathcal{O}_{LR8}, \mathcal{O}_{LR9},  \mathcal{O}_{LR10}, \mathcal{O}_{LR11}, \mathcal{O}_{LR12}, \mathcal{O}_{LR13}, \mathcal{O}_{LR17}, \mathcal{O}_{LR18}.
  \end{aligned}
\end{align}
This list is larger than the list that was discussed in \cite{Buchalla:2012qq,Buchalla:2013rka,mastersthesis}. \\

So far, we discussed only tree-level effects. One-loop effects introduce additional, important contributions. We discuss them briefly here and give more details in \cite{Buchalla:2016bse}. 

\renewcommand{\arraystretch}{2}
\begin{table}[!h]
    \centering
    \begin{tabular}[t]{|c|c|c||c|}
\hline
\multirow{2}{*}{Diagram} &\multicolumn{2}{|c||}{Factors of $M$} &\multirow{2}{2cm}{Total size (incl. loop factor)} \\
& from momentum integral & from couplings & \\
\hline
&&&\\
\begin{tikzpicture}\draw[-] (0em,0em) -- (4em,0em);\draw[thick] {(2em,0.75em) circle (0.75em)};\draw (2em,0.75em) node[anchor = center] {$H$};\draw (0em,0em) node[anchor = east] {$h$};\draw (4em,0em) node[anchor = west] {$h$}; \end{tikzpicture}&$\frac{p^{4}}{p^{2}}\rightarrow M^{2}$&$ M^{2}$&$\frac{M^{4}}{16\pi^{2}}$\\
&&&\\
\begin{tikzpicture}\draw[-] (0em,0em) -- (1.25em,0em);\draw[thick] {(2em,0em) circle (0.75em)};\draw[-] (2.75em,0em) -- (4em,0em);\draw (2em,0em) node[anchor = center] {$H$};\draw (0em,0em) node[anchor = east] {$h$};\draw (4em,0em) node[anchor = west] {$h$};\end{tikzpicture}&$\frac{p^{4}}{p^{4}}\rightarrow \log{M}$&$M^{4}$&$\frac{M^{4}}{16\pi^{2}}\log{M}$\\
&&&\\
\begin{tikzpicture}\coordinate (Ursprung) at (2em,0em);\draw[thick] {(Ursprung) circle (0.75em)};\coordinate (a) at ( $ (Ursprung) + (45:0.75em) $ );\coordinate (b) at ( $ (Ursprung) + (90:0.75em) $ );\coordinate (c) at ( $ (Ursprung) + (135:0.75em) $ );\coordinate (d) at ( $ (Ursprung) + (180:0.75em) $ );\coordinate (e) at ( $ (Ursprung) + (225:0.75em) $ );\coordinate (f) at ( $ (Ursprung) + (270:0.75em) $ );\coordinate (g) at ( $ (Ursprung) + (315:0.75em) $ );\coordinate (h) at ( $ (Ursprung) + (0:0.75em) $ );\draw[-] (a) -- ++(45:1em);\draw[-] (b) -- ++(90:1em);\draw[-] (c) -- ++(135:1em);\draw[-] (d) -- ++(180:1em);\draw[-] (e) -- ++(225:1em);\draw[-] (f) -- ++(270:1em);\draw[-] (g) -- ++(315:1em);\draw (Ursprung) node[anchor = center] {$H$};\draw ($ (a) + (45:0.75em)$) node[anchor = south west] {$h$};\draw ($ (b) + (90:0.75em)$) node[anchor = south] {$h$};\draw ($ (c) + (135:0.75em)$) node[anchor = south east] {$h$};\draw ($ (d) + (180:0.75em)$) node[anchor = east] {$h$};\draw ($ (e) + (225:0.75em)$) node[anchor = north east] {$h$};\draw ($ (f) + (270:0.75em)$) node[anchor = north] {$h$};\draw ($ (g) + (315:0.75em)$) node[anchor = north west] {$h$};\draw ($ (h) + (0:0.75em)$) node[anchor = west] {$\vdots$};\end{tikzpicture}&\raisebox{2em}{$\frac{p^{4}}{p^{2I}}\rightarrow M^{4-2I}$}&\raisebox{2em}{$ M^{2I}$}&\raisebox{2em}{$\frac{M^{4}}{16\pi^{2}}$}\\
\hline
    \end{tabular}
    \caption{Parametric size of various one-loop diagrams in case \ref{case.strong}).}
    \label{tab:scaling.non-decoupling}
  \end{table}
\renewcommand{\arraystretch}{1}

Corrections to the effective Lagrangian, defined in Eqs.~\eqref{eq:7.55.1} and \eqref{eq:7.61.1}, arise when we integrate out $H$ at the one-loop level, as we discussed in Eq.~\eqref{eq:3.1.10}. Using the superficial degree of divergence, we find the expected divergence of the one-loop diagrams \cite{Peskin:1995ev}. The vertices, defined in Eq.~\eqref{eq:7.40}, introduce additional factors of $M^{2}$. We list the parametric size of various diagrams in Table~\ref{tab:scaling.non-decoupling}. The naive size of the contributions is $M^{4}/16\pi^{2}$, which is larger than the leading-order effects, $\mathcal{O}(v^{4})$. Our model, however, has an approximate $SO(5)$ symmetry that conserves the hierarchy $m\ll M$ naturally. Quantum corrections to $m$ must therefore be proportional to the $SO(5)$-breaking parameters $r$ and $\delta$, defined in Eq.~\eqref{eq:7.34}. These parameters are small, such that $r M^{2}=\mathcal{O}(v^{2})$ and $\delta M^{2}=\mathcal{O}(v^{2})$, see \eqref{eq:7.35}. The contributions we discussed in Table~\ref{tab:scaling.non-decoupling} are then of $\mathcal{O}(v^{2}M^{2}/16\pi^{2})\lesssim \mathcal{O}(v^{2}f^{2})\approx \mathcal{O}(v^{4})$, which means they are parametrically of the size of $\mathcal{L}_{\text{LO}}$. The approximate $SO(5)$ symmetry is therefore necessary to have a well-defined model without fine tuning. We see again that $M$ has to stay below the nominal strong-coupling limit $4\pi f$, as otherwise the model is not calculable and contributions from all loop orders are equally important. 

At NLO of the EFT, $\mathcal{O}(v^{2}/M^{2})$, also the one-loop diagrams of the leading-order Lagrangian become important. We expect the latter effects at $\mathcal{O}(1/16\pi^{2})\approx \mathcal{O}(\xi/16\pi^{2})\lesssim \mathcal{O}(v^{2}/M^{2})$, {\it i.e.} comparable to the NLO effects in Eq.~\eqref{eq:7.62}. 

We see that the one-loop effects are, in general, only slightly suppressed compared to the tree-level contributions we discussed before. We conclude that tree-level gives the features of the model only qualitatively. Numerically, the couplings are affected by one-loop contributions. However, since we work below the nominal strong-coupling limit, the loop effects are smaller than the tree-level effects. 
\subsubsection{Case \ref{case.weak}) --- the Decoupling Limit}
In the second case, $f$ is of the order of $M$; $\xi\ll 1$; and the model is weakly coupled. Since $v_{s}\gg v$ in Eq.~\eqref{eq:7.23}, we only expand $S$ around its vacuum expectation value, $S = (v_{H} + H_{s})/\sqrt{2}$, and keep $\phi$ explicit. There is no mixing between the two scalars at this level. We find the Lagrangian
\begin{align}
  \begin{aligned}
    \label{eq:7.65}
    \mathcal{L} &= (D^{\mu} \phi)^{\dagger} (D_{\mu} \phi) +\left(\frac{ \mu_{1}^{2}}{2}-\frac{\lambda_{3}v_{H}^{2}}{4}\right) \phi^{\dagger} \phi  - \frac{\lambda_{1}}{4}  ( \phi^{\dagger} \phi )^2 \\
    &\qquad + \frac{1}{2} \partial^{\mu}  H_{s} \partial_{\mu} H_{s} -\frac{1}{2} M_{s}^{2}H_{s}^{2} \\
    & - \frac{v_{H}\lambda_{3}}{2} \phi^{\dagger} \phi H_{s} -  \frac{\lambda_{3}}{4} \phi^{\dagger} \phi H_{s}^{2} - \frac{v_{H}\lambda_{2}}{4} H_{s}^{3}- \frac{\lambda_{2}}{16} H_{s}^{4},
  \end{aligned}
\end{align}
with 
\begin{equation}
  \label{eq:7.66}
  M_{s}^{2}=\mu_{2}^{2} = \frac{v_{H}^{2}\lambda_{2}}{2}. 
\end{equation}
This Lagrangian is like the Lagrangian in Eq.~\eqref{eq:7.39}, with the identifications $H=H_{s}$, $M=M_{s}$, and 
\begin{align}
  \begin{aligned}
    \label{eq:7.66.1}
    J_{1} &= - \frac{v_{H}\lambda_{3}}{2} \phi^{\dagger} \phi ,\\
    J_{2} &= -  \frac{\lambda_{3}}{4} \phi^{\dagger} \phi,\\
    J_{3} &= - \frac{v_{H}\lambda_{2}}{4} ,\\
    J_{4} &= - \frac{\lambda_{2}}{16}.\\
  \end{aligned}
\end{align}
The difference to the non-decoupling case and Eq.~\eqref{eq:7.40} is that the $J_{i}$ here are not proportional to the heavy scale $M_{s}^{2}$. Instead, only $J_{1}$ and $J_{3}$ are proportional to a single power of $v_{H}$, which is of the order of $M_{s}$. We solve the equation of motion of $H_{s}$, 
\begin{equation}
  \label{eq:7.67}
  -\left(\Box+M_{s}^{2}-2 J_{2}\right)H_{s} + J_{1}+3J_{3}H_{s}^{2}+4J_{4}H_{s}^{3}=0,
\end{equation}
order by order in $1/M_{s}^{2}$. Keeping in mind that $v_{H}/M_{s} = \mathcal{O}(1)$ by Eq.~\eqref{eq:7.66}, we find
\begin{equation}
  \label{eq:7.68}
  H_{s} = -\frac{\lambda_{3}}{2}\frac{v_{H}}{M_{s}^{2}}(\phi^{\dagger}\phi) + \left(\Box+\frac{\lambda_{3}}{2}\phi^{\dagger}\phi\right)\frac{\lambda_{3}}{2}\frac{v_{H}}{M_{s}^{4}}(\phi^{\dagger}\phi)- \frac{3\lambda_{2}\lambda_{3}^{2}}{16}\frac{v_{H}^{3}}{M_{s}^{6}}(\phi^{\dagger}\phi)^{2} +\mathcal{O}\left(\frac{1}{M_{s}^{4}}\right).
\end{equation}
Since the unsuppressed part, $H_{s,0}$, vanishes, we do not generate arbitrary high powers of the light field $\phi$ without suppression. Instead, powers of $(\phi^{\dagger}\phi)$ are systematically suppressed by $1/M_{s}^{2}$. The expansion is therefore given by canonical dimensions. When we insert Eq.~\eqref{eq:7.68} into Eq.~\eqref{eq:7.65}, we find the low-energy Lagrangian 
\begin{align}
  \begin{aligned}
    \label{eq:7.69}
    \mathcal{L} &= (D^{\mu} \phi)^{\dagger} (D_{\mu} \phi) +\left(\frac{ \mu_{1}^{2}}{2}-\frac{\lambda_{3}v_{H}^{2}}{4}\right) \phi^{\dagger} \phi  - \left(\frac{\lambda_{1}}{4} -\frac{\lambda_{3}^{2}}{8}\frac{v_{H}^{2}}{M_{s}^{2}}\right) ( \phi^{\dagger} \phi )^2 \\
    &+\frac{1}{8}\frac{\lambda_{3}^{2}v_{H}^{2}}{M_{s}^{4}}\partial^{\mu} (\phi^{\dagger}\phi)  \partial_{\mu}(\phi^{\dagger}\phi) +\mathcal{O}\left(\frac{1}{M_{s}^{4}}\right),
  \end{aligned}
\end{align}
in agreement to \cite{Gorbahn:2015gxa}. We see that only one dimension-six operator, $\mathcal{O}_{\phi\Box}$, is generated (after integrating by parts). \\

At low energies, the electroweak symmetry is broken and $\phi$ acquires a vacuum expectation value. We write $\phi$ as in Eq.~\eqref{eq:5.1} and perform a field redefinition to bring the kinetic term of $h$ to its canonically normalized form. These steps are equivalent to the $\xi$-expansion of Section~\ref{ch:6.xiexpand}, with the special choice of parameters
\begin{align}
  \begin{aligned}
    \label{eq:7.70}
    \tilde{c}_{\phi}=\tilde{c}_{V}=&\tilde{c}_{k}=0, \qquad \tilde{Y}_{\Psi\phi}^{6}=\tilde{Y}_{\Psi\phi}^{8}=0,\\
    &-\frac{\tilde{c}_{\phi\Box}\xi}{2}= \frac{\lambda_{3}^{2}}{\lambda_{2}^{2}}\frac{v^{2}}{v_{H}^{2}}\equiv \alpha^{2}. 
\end{aligned}
\end{align}
We note that $\alpha$ corresponds to the mixing angle $\chi$, defined in Eq.~\eqref{eq:7.35}, to first order in $v/v_{H}$. The effective, low-energy Lagrangian of the scalar sector, up to and including terms of the order $\mathcal{O}(1/M_{s}^{2})$, is then given by $\mathcal{L}_{\text{LO}}$ in Eq.~\eqref{eq:LO} with 
\begin{align}
  \begin{aligned}
    \label{eq:7.71} 
    V(h) = \frac{m^{2}}{2}v^{2} &\left[\left(\frac{h^{2}}{v^{2}}\right)+\left(1-\frac{3}{2}\alpha^{2}\right) \left(\frac{h^{3}}{v^{3}}\right)+\left(\frac{1}{4}-\frac{25}{12}\alpha^{2}\right) \left(\frac{h^{4}}{v^{4}}\right)\right.\\
&\left. -\alpha^{2} \left(\frac{h^{5}}{v^{5}}\right) -\frac{\alpha^{2}}{6} \left(\frac{h^{6}}{v^{6}}\right)\right],
  \end{aligned}
\end{align}
\begin{align}
  \begin{aligned}
    \label{eq:7.72}
    F_{U}(h) &= 2(1-\alpha^{2}) \left(\frac{h}{v}\right) + \left(1-2\alpha^{2}\right) \left(\frac{h^{2}}{v^{2}}\right)-\frac{4}{3}\alpha^{2} \left(\frac{h^{3}}{v^{3}}\right) -\frac{\alpha^{2}}{3}\left(\frac{h^{4}}{v^{4}}\right),
  \end{aligned}
\end{align}
and
\begin{align}
  \begin{aligned}
    \label{eq:7.73}
    \sum_{n=1}^{\infty}Y_{\Psi}^{(n)}\left(\tfrac{h}{v}\right)^{n} &= Y_{\Psi}\left[\left(1-\frac{\alpha^{2}}{2} \right)\left(\frac{h}{v}\right)-\frac{\alpha^{2}}{2} \left(\frac{h^{2}}{v^{2}}\right) -\frac{\alpha^{2}}{6} \left(\frac{h^{3}}{v^{3}}\right)\right].
  \end{aligned}
\end{align}
As already discussed before, the effective Lagrangian is expanded in canonical dimensions. Therefore, the polynomial $F_{U}$ has at most four, $V$ at most six, and the Yukawa interaction at most three powers of $h$. Tree-level corrections to this Lagrangian arise at $\mathcal{O}(1/M_{s}^{4})$.\\

When we integrate out $H_{s}$ at the one-loop level, we encounter similar diagrams as in case \ref{case.strong}). However, their parametric size is different, as the couplings in Eq.~\eqref{eq:7.66.1} are not proportional to $M_{s}^{2}$. Contributions from $J_{3}$ are more important than contributions from $J_{4}$, as the former is proportional to $v_{H}\approx M_{s}$. We show the size of the diagrams in Table~\ref{tab:scaling.decoupling}. \\
\renewcommand{\arraystretch}{2}
\begin{table}[!h]
    \centering
    \begin{tabular}[t]{|c|c|c||c|}
\hline
\multirow{2}{*}{Diagram} &\multicolumn{2}{|c||}{Factors of $M_{s}$} &\multirow{2}{2cm}{Total size (incl. loop factor)} \\
& from momentum integral & from couplings & \\
\hline
&&&\\
\begin{tikzpicture}\draw[-] (0em,0em) -- (4em,0em);\draw[thick] {(2em,0.75em) circle (0.75em)};\draw (2em,0.75em) node[anchor = center] {$H$};\draw (0em,0em) node[anchor = east] {$h$};\draw (4em,0em) node[anchor = west] {$h$}; \end{tikzpicture}&$\frac{p^{4}}{p^{2}}\rightarrow M_{s}^{2}$&$ 1$&$\frac{M_{s}^{2}}{16\pi^{2}}$\\
&&&\\
\begin{tikzpicture}\draw[-] (0em,0em) -- (1.25em,0em);\draw[thick] {(2em,0em) circle (0.75em)};\draw[-] (2.75em,0em) -- (4em,0em);\draw (2em,0em) node[anchor = center] {$H$};\draw (0em,0em) node[anchor = east] {$h$};\draw (4em,0em) node[anchor = west] {$h$};\end{tikzpicture}&$\frac{p^{4}}{p^{4}}\rightarrow \log{M_{s}}$&$M_{s}^{2}$&$\frac{M_{s}^{2}}{16\pi^{2}}\log{M_{s}}$\\
&&&\\
\begin{tikzpicture}\coordinate (Ursprung) at (2em,0em);\draw[thick] {(Ursprung) circle (0.75em)};\coordinate (a) at ( $ (Ursprung) + (45:0.75em) $ );\coordinate (b) at ( $ (Ursprung) + (90:0.75em) $ );\coordinate (c) at ( $ (Ursprung) + (135:0.75em) $ );\coordinate (d) at ( $ (Ursprung) + (180:0.75em) $ );\coordinate (e) at ( $ (Ursprung) + (225:0.75em) $ );\coordinate (f) at ( $ (Ursprung) + (270:0.75em) $ );\coordinate (g) at ( $ (Ursprung) + (315:0.75em) $ );\coordinate (h) at ( $ (Ursprung) + (0:0.75em) $ );\draw[-] (a) -- ++(45:1em);\draw[-] (b) -- ++(90:1em);\draw[-] (c) -- ++(135:1em);\draw[-] (d) -- ++(180:1em);\draw[-] (e) -- ++(225:1em);\draw[-] (f) -- ++(270:1em);\draw[-] (g) -- ++(315:1em);\draw (Ursprung) node[anchor = center] {$H$};\draw ($ (a) + (45:0.75em)$) node[anchor = south west] {$h$};\draw ($ (b) + (90:0.75em)$) node[anchor = south] {$h$};\draw ($ (c) + (135:0.75em)$) node[anchor = south east] {$h$};\draw ($ (d) + (180:0.75em)$) node[anchor = east] {$h$};\draw ($ (e) + (225:0.75em)$) node[anchor = north east] {$h$};\draw ($ (f) + (270:0.75em)$) node[anchor = north] {$h$};\draw ($ (g) + (315:0.75em)$) node[anchor = north west] {$h$};\draw ($ (h) + (0:0.75em)$) node[anchor = west] {$\vdots$};\end{tikzpicture}&\raisebox{2em}{$\frac{p^{4}}{p^{2I}}\rightarrow M_{s}^{4-2I}$}&\raisebox{2em}{$ M_{s}^{I}$}&\raisebox{2em}{$\frac{M_{s}^{4-I}}{16\pi^{2}}$}\\
\hline
    \end{tabular}
    \caption{Parametric size of various one-loop diagrams in case \ref{case.weak}).}
    \label{tab:scaling.decoupling}
  \end{table}
\renewcommand{\arraystretch}{1}
We see that only the corrections to $m^{2}$ are of $\mathcal{O}(M_{s}^{2}/16\pi^{2})$ and the contributions to the potential are further suppressed. We absorb this contribution of $\mathcal{O}(M_{s}^{2}/16\pi^{2})$ in the renormalization of $m^{2}$. The loop corrections are therefore, in general, suppressed in the weakly-coupled regime. 
\subsubsection{The Relation Between the Two Effective Descriptions}
We now compare the two low-energy EFTs of case \ref{case.strong}) and \ref{case.weak}). In case \ref{case.strong}), $\xi$ and $\omega$ are of $\mathcal{O}(1)$ and the model is strongly coupled. The solution of the equation of motion of the heavy field starts with an unsuppressed term. This yields polynomials of the light field $h$ in the effective Lagrangian that are not suppressed by the heavy mass $M$. The expansion is therefore not given by canonical dimensions, but instead in chiral dimensions. 

In case \ref{case.weak}), $\xi$ and $\omega$ are small and the model is weakly coupled. The solution to the equation of motion of the heavy field starts at $\mathcal{O}(1/M)$. Therefore, higher powers of $H$ are suppressed by higher powers of $1/M$ and canonical dimensions define this expansion. \\

The low-energy limits of this explicit model illustrate the statements we make in Section~\ref{ch:6.pheno} about the phenomenological implications of the two expansions. 

\begin{figure}[t]
  \centering
  \includegraphics[width=0.75\textwidth]{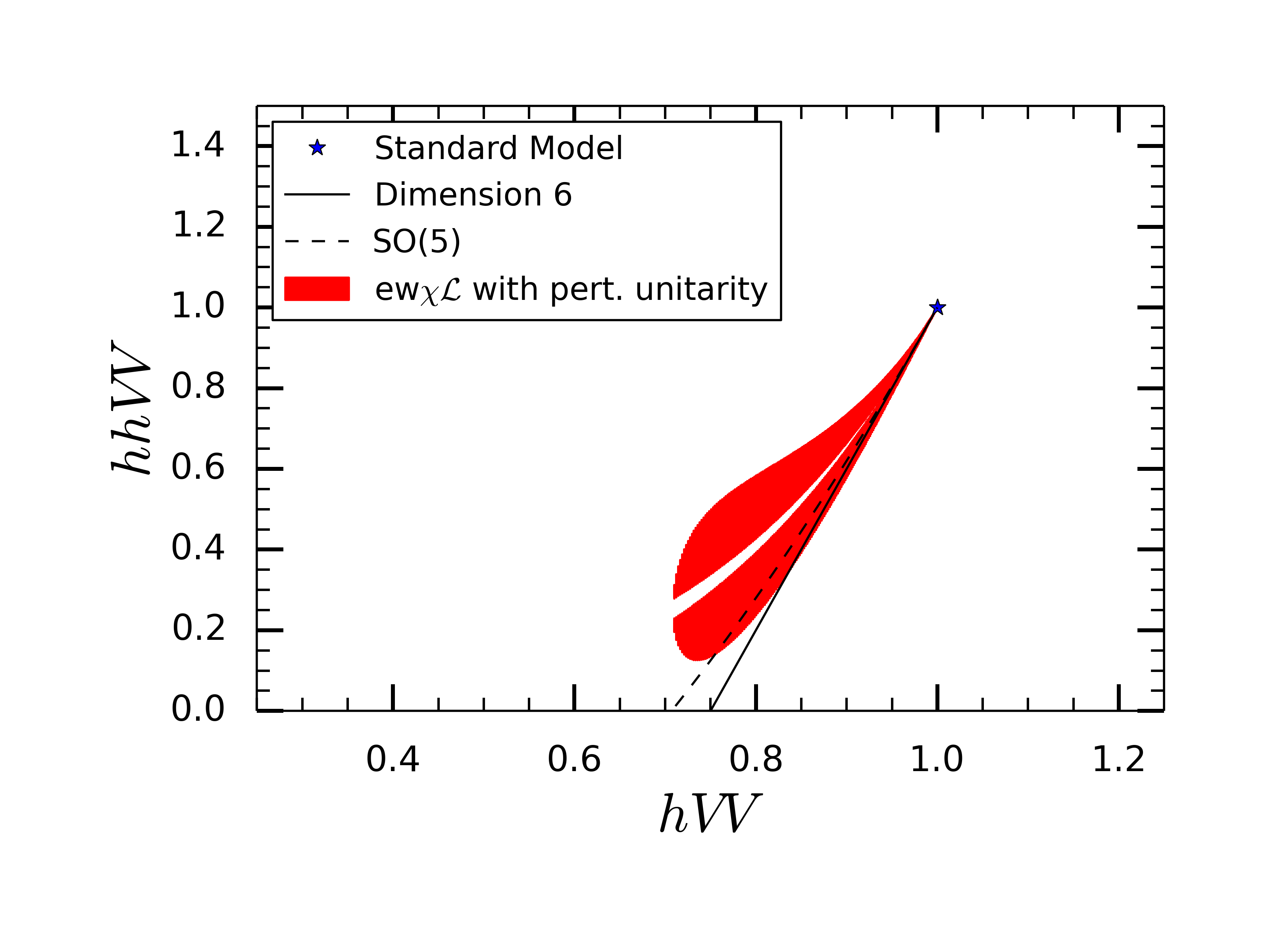}
  \caption[Allowed Higgs couplings to a pair of massive vector bosons in the considered scenarios.]{Allowed Higgs couplings to a pair of massive vector bosons in the considered scenarios. The red area is allowed by perturbative unitarity \cite{Pruna:2013bma}. The couplings are normalized to the SM values.}
  \label{fig:hVV.hhVV}
\end{figure}

In the weakly-coupled limit, there is a correlation between the coupling of a single Higgs to a pair of heavy vector bosons and the coupling of a pair of Higgs to a pair of heavy vectors. We see this in Eq.~\eqref{eq:7.72}. In the strongly-coupled limit, this correlation does not exist, see Eq.~\eqref{eq:7.60}. Figure~\ref{fig:hVV.hhVV} illustrates the allowed values of the couplings in the two EFTs, in the $SO(5)$-limit, and in the SM. We highlight the parameter region that is allowed by perturbative unitarity in red. We used the formulas of \cite{Pruna:2013bma} to find this region. The gap between the two red regions originates from the regions of parameter space in which $\chi \approx 0$. The Higgs couplings have to be close to their SM values in this case. 

We also observe in Eq.~\eqref{eq:7.60} that the SM-like coupling structures of $h$ to the other particles are modified by $\mathcal{O}(1)$ effects, whereas new, non-SM-like Lorentz structures arise at NLO and are therefore suppressed. In general, the polynomials in $h$ that appear at leading order show that vertices with a high number of Higgs particles are not suppressed in the strongly-coupled limit.\\

We introduced in Chapter~\ref{ch:relation} the parameter $\xi$ as the degree of decoupling. The singlet model also illustrates this. Close to the $SO(5)$-limit, $\xi$ is of the same size as $\omega$, see Eq.~\eqref{eq:7.34}. We therefore understand the mixing angle as the discriminating factor between the decoupling and the non-decoupling case. In other words, the transition from the strongly-coupled to the weakly-coupled regime requires $f\gg v$, which means $v_{s}\gg v$. Expanding the Lagrangian in $v/v_{s}$ is then equivalent to expanding in small $\chi$, see Eq.~\eqref{eq:7.25}. 

In the non-decoupling region, $\xi$ and therefore also $\omega$ are of order unity. Thus, the heavy mass eigenstate has a significant doublet component. This generates non-decoupling effects when we integrate out the heavy particle, as the doublet is not a reasonable approximation of the low-energy degree of freedom. In the weakly-coupled region, the mixing is small and the heavy particle mostly consists of the singlet. The corresponding low-energy EFT is well described by the doublet. The authors of \cite{Burgess:2014lza} also noted that large mixing effects induce a non-linear Lagrangian, {\it i.e.} a non-decoupling EFT. \\

After identifying the mixing angle as the discriminator between the two expansions, we investigate the transition between them. Starting from Eqs.~\eqref{eq:7.59}--\eqref{eq:7.61}, we expand $\mathcal{L}_{\text{LO}}^{SM+S}$ up to second order in $\chi$. At this order, we identify $\chi$ with $\alpha$ of Eq.~\eqref{eq:7.70}. The resulting expression for Eqs.~\eqref{eq:7.59}--\eqref{eq:7.61} agree with Eqs.~\eqref{eq:7.71}--\eqref{eq:7.73}. This shows that, in the limit of a small mixing angle, the decoupling EFT at dimension-six provides a correct description of the leading mixing effects. This is in contrast to the claims in \cite{Gorbahn:2015gxa,Boggia:2016asg}. When the mixing is larger, the operators at dimension-six, {\it i.e.} at $\mathcal{O}(v^{2}/v_{H}^{2})$, are not sufficient to describe all the effects properly and higher order operators have to be considered. This corresponds precisely to the resummation in $\xi$ that we discussed in Chapter~\ref{ch:relation}. For $v/v_{H}=\mathcal{O}(1)$, {\it i.e.} $\xi=\mathcal{O}(1)$, the non-decoupling description using the electroweak chiral Lagrangian is more appropriate.\\ 

Recently, the authors of \cite{Brehmer:2015rna} introduced a procedure called ``$v$-improved matching'', to improve the convergence between the decoupling EFT and the full model in cases where $\xi$ is not small. Instead of truncating the effective expansion at $\mathcal{O}(v^{2}/M^{2})$, they also considered contributions of $\mathcal{O}(v/M)^{d>2}$ in the Wilson coefficients of the dimension-six operators. In the singlet model, this results in finding $(2\cos{\chi})$ instead of $(2-\chi^{2})$ as first coefficient in $F_{U}(h)$.

This top-down inspired idea is connected to the non-decoupling EFT in the bottom-up approach. Adding more powers of $(\phi^{\dagger} \phi)/M^{2}\rightarrow v^{2}/M^{2}$ to an operator of the decoupling EFT, $\mathcal{O}_{d6}$, corresponds to adding terms of higher order in $\xi$ to the corresponding operator in the non-decoupling EFT. In the case at hand, adding more powers of $(\phi^{\dagger} \phi)$ to $\mathcal{O_{\phi\Box}} $ modifies the polynomial $\mathcal{F}_{h}(h)$ of the operator $(\partial^{\mu}h)(\partial_{\mu}h) \mathcal{F}_{h}(h)$. Since $\phi$ is expanded around $v$, every power of $(\phi^{\dagger} \phi)$ contributes to the first terms in $\mathcal{F}_{h}(h)$, effectively performing a resummation in $\xi$. When the kinetic term is brought to its canonical normalization using field redefinitions \cite{Buchalla:2013rka} these effects get shifted to all the Higgs couplings, already of the leading-order Lagrangian. Therefore, using the electroweak chiral Lagrangian for the bottom-up analysis would automatically include the effects of large $\xi$ and mixing angles. In this case, no information from the UV side is needed and the range of application would also cover cases with large mixing effects and rather strong coupling.

  \chapter{Effective Field Theories in Data Analysis}
\thispagestyle{fancyplain}
We will now discuss the process of data analysis, with emphasis on the use of effective field theories. We introduce the important ingredients for a fit to LHC-Higgs data and present the results of such a fit.
\section{General Aspects of Data Analysis}
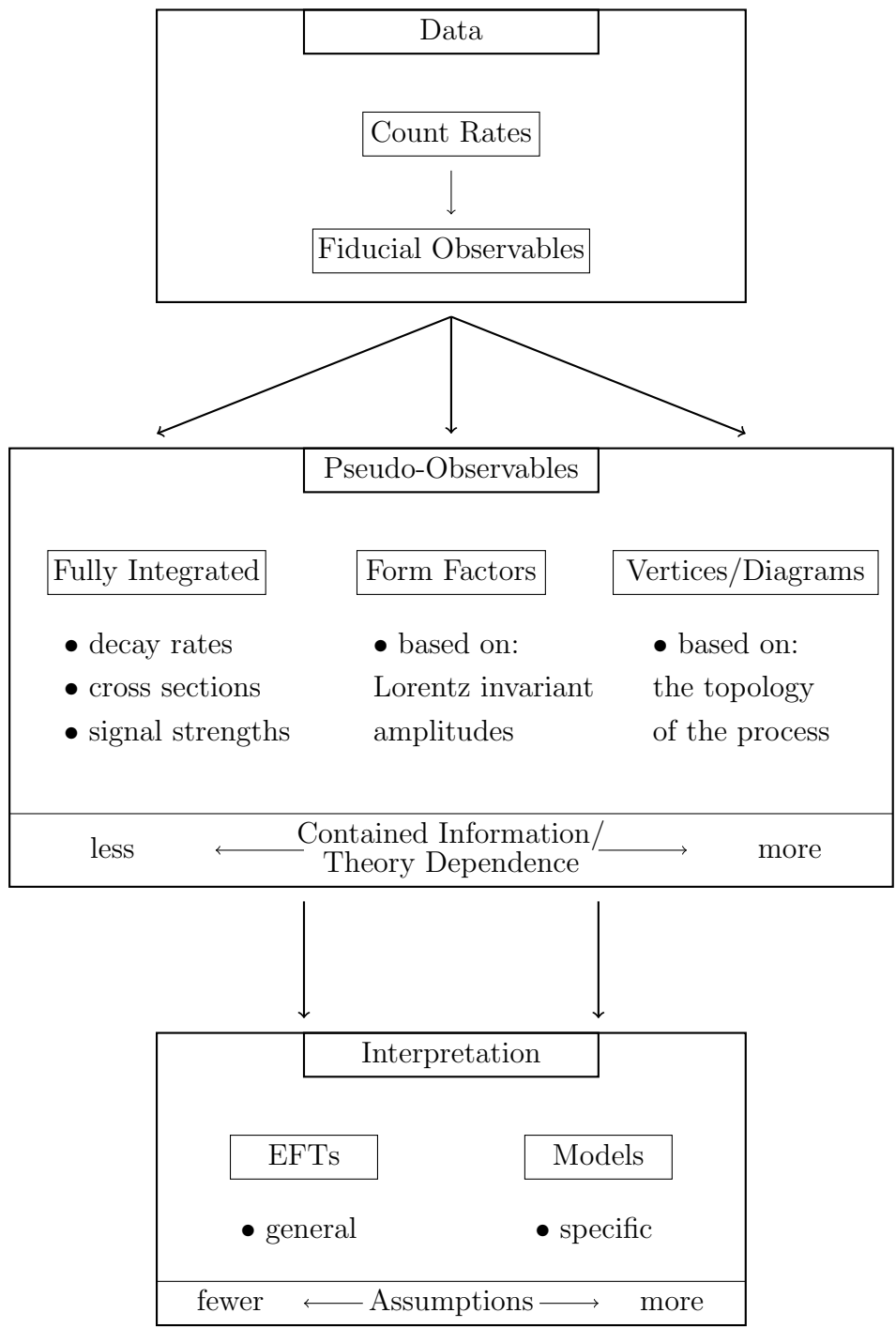
\begin{figure}[!h]
  \begin{center} 
    \begin{tikzpicture}
      \draw[thick] {(5em,0) rectangle (25em,10em)};
      \draw[thick] {(10em,8.5em) rectangle (20em,10em)};
      \draw (15em,9em) node[anchor = base] {Interpretation};
      \draw[thick] {(0em,15em) rectangle (30em,30em)};
      \draw[thick] {(10em,28.5em) rectangle (20em,30em)};
      \draw (15em,29em) node[anchor = base] {Pseudo-Observables};
      \draw[thick] {(5em,35em) rectangle (25em,45em)};
      \draw[thick] {(10em,43.5em) rectangle (20em,45em)};
      \draw (15em,44em) node[anchor = base] {Data};
      \draw (15em,48em) node[anchor=base] {};
      \draw {(12em,40em) rectangle (18em,41.5em)};
      \draw (15em,40.5em) node[anchor = base] {Count Rates};
      \draw[->] (15em,39.5em) -- (15em,38em);
      \draw {(10.3em,36em) rectangle (19.7em,37.5em)};
      \draw (15em,36.5em) node[anchor = base] {Fiducial Observables};
      \draw{(1.3em,25em) rectangle (8.7em,26.5em)};
      \draw (5em,25.5em) node[anchor = base] {Fully Integrated};
      \draw (1.5em,23em) node[anchor=base west] {$\bullet$ decay rates};     
      \draw (1.5em,21.5em) node[anchor=base west] {$\bullet$ cross sections};     
      \draw (1.5em,20em) node[anchor=base west] {$\bullet$ signal strengths};     
      \draw{(11.8em,25em) rectangle (18.2em,26.5em)};
      \draw (15em,25.5em) node[anchor = base] {Form Factors};
      \draw (12em,23em) node[anchor=base west] {$\bullet$ based on:};     
      \draw (12em,21.5em) node[anchor=base west] {Lorentz invariant};     
      \draw (12em,20em) node[anchor=base west] {amplitudes};     
      \draw{(20.5em,25em) rectangle (29.5em,26.5em)};
      \draw (25em,25.5em) node[anchor = base] {Vertices/Diagrams};
      \draw (21.5em,23em) node[anchor=base west] {$\bullet$ based on:};     
      \draw (21.5em,21.5em) node[anchor=base west] {the topology};     
      \draw (21.5em,20em) node[anchor=base west] {of the process};     
      \draw (0em,17.5em) -- (30em,17.5em);
      \draw[->] (10em,16.25em) -- (7em,16.25em);
      \draw (3.5em,16em) node[anchor = base] {less};  
      \draw[->] (20em,16.25em) -- (23em,16.25em);
      \draw (26.5em,16em) node[anchor = base] {more};
      \draw (15em,16.5em) node[anchor=base] {Contained Information/};
      \draw (15em,15.5em) node[anchor=base] {Theory Dependence};
      \draw{(7.5em,5em) rectangle (12.5em,6.5em)};
      \draw (10em,5.5em) node[anchor = base] {EFTs};
      \draw{(17.5em,5em) rectangle (22.5em,6.5em)};
      \draw (20em,5.5em) node[anchor = base] {Models};
      \draw (7.5em,3em) node[anchor=base west] {$\bullet$ general};
      \draw (17.5em,3em) node[anchor=base west] {$\bullet$ specific};
      \draw (5em,1.5em) -- (25em,1.5em);
      \draw (15em,0.5em) node[anchor = base] {Assumptions}; 
      \draw[->] (12em,0.75em ) -- (10em,0.75em); 
      \draw (7.5em,0.5em) node[anchor = base] {fewer};  
      \draw[->] (18em,0.75em ) -- (20em,0.75em); 
      \draw (22.5em,0.5em) node[anchor = base] {more};  
      \draw[thick,->] (15em,34.5em) -- (5em,30.5em);
      \draw[thick,->] (15em,34.5em) -- (15em,30.5em);
      \draw[thick,->] (15em,34.5em) -- (25em,30.5em);
      \draw[thick,->] (10em,14.5em) -- (10em,10.5em);
      \draw[thick,->] (20em,14.5em) -- (20em,10.5em);
    \end{tikzpicture}
  \end{center}
  \caption[Flow of information between data and interpretation.]{Flow of information between data (top) and interpretation (bottom).}
  \label{fig:8.flow}
\end{figure}

When data sets of large experimental collaborations are analyzed, the process is usually split in several steps between experimentalists and theorists. Generally speaking, the experimentalists transfer the primary quantities, {\it e.g.} count rates, from the detectors into quantities that are closer to theory, {\it e.g.} cross sections. The theorists then further analyze the result in light of a given framework, either a model or an effective field theory. The interfaces between experimentalists and theorists are given by ``pseudo-observables''. To be more precise, a pseudo-observable is defined to be ``any uniquely defined, QFT-consistent, expression giving one number'' \cite{PassarinoEFTNLO} that is implementable in any SM deformation. This general definition still gives a lot of freedom to the definition of actual pseudo-observables. Currently, the LHC Higgs Cross Section Working Group \cite{LHCHXSWGwebsite} discusses different sets of pseudo-observables for the Higgs analysis \cite{YellowReport4}.\\
Let us consider the schematic flow of information between data and interpretation in more detail in Fig.~\ref{fig:8.flow}.
The raw data, consisting of count rates in the detectors, are transformed into fiducial observables. These observables are defined in a fiducial volume of the detector, usually where the detection efficiency is the highest \cite{Gomez-Ambrosio:2015jea}. However, fiducial observables only have a limited use for direct analysis. Even though they have only a small amount of theory dependence \cite{HEFT2015Tackmann}, coming for example from the parton distribution functions of the protons, they come along with other difficulties. For instance, their definition requires a signal definition such that experimental efficiencies are close to production-mode independent \cite{HEFT2015Tackmann}. For this reason, the fiducial observables are then further processed to pseudo-observables. This process introduces an additional theory dependence, which should be kept at a minimum, as another motivation of pseudo-observables is their long-term use as experimental legacy. As an example for the theory dependence, consider the LEP pseudo-observable $\Gamma^{e}_{Z}$ \cite{Bardin:1999gt}. It represents the decay width of $Z$ to $e^{+}e^{-}$ and is given by the measured value and then corrected for initial state QED radiation \cite{PO:Passarino1}.\\ 
In general, the process of obtaining the pseudo-observables can be formulated as
\begin{equation}
\label{eq:8.13}
  \text{Fiducial Observable } = \text{ Pseudo-Observable } + \text{ Remainder}
\end{equation}
The ``Remainder''-term is small and well understood for the cases at hand. The degree of theory dependence can vary substantially when defining the pseudo-observables, see the center of Fig.~\ref{fig:8.flow}.\\
Very close to the experimental side, we have fully integrated pseudo-observables, such as $\Gamma_{Z}$ and $m_{Z}$, as well as $A_{fb}$ at LEP \cite{Bardin:1999gt}. The first results of the LHC Higgs analyses were presented in terms of signal strengths $\mu$. The advantage of these fully integrated quantities lies in their long-term usefulness. Since they are almost theory-independent, the analyses do not have to be redone if there are changes on the theory side, {\it e.g.} if higher order corrections become available. Their disadvantage is the reduced sensitivity to differential distributions of kinematic variables. However, with a limited amount of data, the distributions will have a large statistical error and the analyses can only be done using integrated quantities.\\ 
Form factors, on the other hand, provide more information on these distributions, as they parametrize Lorentz-invariant amplitudes. For Higgs production and decay a set of pseudo-observables was introduced in \cite{Gonzalez-Alonso:2014eva,Bordone:2015nqa,Greljo:2015sla}, relying on a pole decomposition of the amplitudes. Even closer to the theory side are pseudo-observables that are defined in terms of vertices and Feynman diagrams. However, one should be careful using this approach since not all couplings in a phenomenological Lagrangian are directly observable without further assumptions \cite{Trott:2014dma}. \\
The interpretation of the measurements, for any pseudo-observable, can now be done as an independent next step. Either a specific UV model or a model-independent EFT can be used for this. For the fit to Run-1 data presented here, we use the fully integrated signal strength, which we discuss in the following from an experimental and a theoretical point-of-view. 
 \subsection{The Experimentalists View on the Signal Strength --- Introducing the $\kappa$-Framework}
The expected number of events $n$ in Higgs analyses can be written as \cite{Bernon:2015hsa} $n = \mu n_{s} + n_{b}$, where $n_{s}$ is the number of expected events in the SM, $n_{b}$ is the expected background and $\mu$ is the signal strength. Since $\mu=1$ corresponds to the SM, we can write it as
\begin{equation}
  \label{eq:8.14}
  \mu = \frac{\sigma\cdot \epsilon \cdot A}{(\sigma\cdot \epsilon \cdot A)_{\text{SM}}}
\end{equation}
In this relation, $\sigma$ is the total cross section of the process and $\epsilon \cdot A$ is the product of efficiency and acceptance of the selection criteria \cite{Bernon:2015hsa}. Usually, two more assumptions are made to rewrite Eq.~\eqref{eq:8.14}. First, we assume that the total signal can be written as a sum of processes that exist for a SM Higgs boson, {\it i.e.} $\sigma = \sum\limits_{X,Y} \sigma(X) \cdot \text{Br }(h\rightarrow Y)$. Here, $X$ is the Higgs production channel of the process and $Y$ is its decay mode, see Section \ref{ch:SMprod.dec}. The second assumption we employ is that the product of efficiency and acceptance is equal to its SM value. We can now express the signal strength $\mu$ in terms of ``reduced efficiencies'' $\text{eff}(X,Y)$ \cite{Bernon:2015hsa}:
\begin{equation}
  \label{eq:8.15}
  \mu = \sum_{X,Y} \text{eff}(X,Y) \frac{\sigma(X) \cdot \text{Br }(h\rightarrow Y)}{(\sigma(X) \cdot \text{Br }(h\rightarrow Y))_{\text{SM}}}
\end{equation}
The experimental search is then carried out for different decay channels. Each of these analyses is further divided into several event categories, motivated by their different reduced efficiencies. For a global interpretation of all measured production and decay channels, we need not only the signal strength for each category, but also their reduced efficiencies. On top, for a reliable statistical analysis, we need to know all correlations between the channels. Since the errors cannot be assumed to be Gaussian in every case, it requires the full likelihood information to be known. Unfortunately, the experimental collaborations do not publish all these information for all channels. Therefore, we cannot use the signal strength of Eq.~\eqref{eq:8.15} to interpret Higgs data for all channels.

Another way to interpret experimental data uses the ``unfolded'' signal strengths in theory plane. These signal strengths are defined for each production mode $X$ and decay channel $Y$ as \cite{Bernon:2015hsa}
\begin{equation}
  \label{eq:8.16}
  \mu(X,Y) = \frac{\sigma(X) \cdot \text{Br }(h\rightarrow Y)}{(\sigma(X) \cdot \text{Br }(h\rightarrow Y))_{\text{SM}}}
\end{equation}
As they are provided by the experimental collaborations directly, they have the advantage that all different efficiencies are correctly taken into account. In addition, they are published in two-dimensional planes (see {\it e.g.} in \cite{ATLAS:2014yka}), so correlations can be read off in a Gaussian approximation. The unfolded signal strength $\mu(X,Y)$ of Eq.~\eqref{eq:8.16} can also be computed from a given Lagrangian, as we will see in the next subsection. Therefore, we choose the $\mu(X,Y)$ as pseudo-observables for our analysis at the end of this chapter.
\subsubsection{The $\kappa$-Framework}
For a first interpretation of the measurements, the LHC Higgs Cross Section Working Group introduced a set of pseudo-observables, the $\kappa$-framework \cite{LHCHiggsCrossSectionWorkingGroup:2012nn,Heinemeyer:2013tqa}. It is an interim framework to explore the coupling structure of the Higgs-like scalar, answering the question: {\it ``Do we observe the Standard Model or not?''} 

Assuming a production similar to the SM-Higgs with a narrow width, we write the signal strength in Eq.~\eqref{eq:8.16}. We further rewrite the branching ratios in terms of the individual decay rates, 
\begin{equation}
  \label{eq:8.17}
  \text{Br }(h\rightarrow Y) = \frac{\Gamma(h\rightarrow Y)}{\sum_{i}\Gamma(h\rightarrow i)}.
\end{equation}
The $\kappa_{i}$ are then defined as the ratio of the extracted production cross sections and decay widths with respect to their SM value:
\begin{equation}
  \label{eq:8.18}
  \kappa^{2}_{X} = \frac{\sigma(X\rightarrow h)}{\sigma(X\rightarrow h)_{\text{SM}}}, \hspace{2cm}\kappa^{2}_{Y} = \frac{\Gamma(h\rightarrow Y)}{\Gamma(h\rightarrow Y)_{\text{SM}}}.
\end{equation}
With this definition, we can divide the $\kappa_{i}$ into two categories. In the first category, we have the $\kappa_{i}$ that correspond to a simple rescaling of the SM couplings. It includes $\kappa_{W}, \kappa_{Z}$ and $\kappa_{f}$ for any fermion $f\in \{t,b,c,s,u,d,e,\mu,\tau\}$ in production as well as decay. The second category includes $\kappa_{i}$ that are, for SM topologies, not given by a single diagram alone. $\kappa_{\text{VBF}}, \kappa_{gg}, \kappa_{\gamma\gamma}$ and $\kappa_{Z\gamma}$ are in this category. 

The signal strength can be written as
\begin{equation}
  \label{eq:8.19}
  \mu(X,Y) = \frac{\kappa^{2}_{X}\kappa^{2}_{Y}}{\kappa^{2}_{h}},
\end{equation}
where $\kappa^{2}_{h} = \sum_{i}\Gamma(h\rightarrow i) / \sum_{i}\Gamma(h\rightarrow i)_{\text{SM}}$. Usually, different assumptions can be considered: In category one, we can define a single $\kappa_{V}$ for the couplings of Higgs to $W$ and $Z$ under the assumption of custodial symmetry. We can group the $\kappa_{i}$ of the Higgs coupled to a pair of fermions $\bar{f}f$ for up- and down-type quarks, or we assume a universal $\kappa_{f}$ for all fermions, {\it etc}. For the $\kappa_{i}$ in category two, we can either assume they depend only on the $\kappa_{i}$ of category one via the SM-topologies, or they are used as independent $\kappa_{i}$ to be fitted. In this way they are sensitive to Higgs admixtures of various types, new fermions in the loop induced processes, or decays to invisible channels. 

Recently, Passarino criticized this framework to be not suitable for LHC Run-2 analyses \cite{PO:Passarino2}. First, it only amounts to a rescaling of SM couplings by construction, so it will only test deviations in event rates rather than the event shapes ({\it i.e.} kinematic distributions). Second, only QCD corrections, being factorizable, can be taken into account. Electroweak corrections cannot be implemented. The third point of criticism of \cite{PO:Passarino2} states that the $\kappa$-framework is not QFT compatible, as it violates gauge symmetry and unitarity by ad-hoc variations of SM couplings. The authors of \cite{Henning:2014wua} made similar statements. We will see in the next section how the $\kappa$-framework is related to the electroweak chiral Lagrangian and how this counters the criticism above. 
 \subsection{The Theorists View on the Signal Strength --- Interpreting the $\kappa$-Framework}
We can also interpret the signal strengths $\mu(X,Y)$ of Eq.~\eqref{eq:8.16} model-independently within a well-defined set of assumptions in an effective field theory. For that purpose, we compute Eq.~\eqref{eq:8.16} using the electroweak chiral Lagrangian. We start from Eq.~\eqref{eq:5.6},
\begin{equation}
  \label{eq:8.20}
  \mathcal{L}_{\text{ew}\chi} = \mathcal{L}_{\text{LO}}+\mathcal{L}_{\text{NLO}} + \mathcal{O}\left(\frac{1}{(16\pi^{2})^{2}}\right),
\end{equation}
where $\mathcal{L}_{\text{LO}}$ is given by Eq.~\eqref{eq:LO} and the operators of $\mathcal{L}_{\text{NLO}}$ are given by Eqs.~\eqref{eq:5.NLO.1}--\eqref{eq:5.NLO.17}. For the available data of Run-1, we only need operators that have one single Higgs leg. The tree-level couplings to pairs of vectors and fermions are then given by $\mathcal{L}_{\text{LO}}$. If we further assume custodial symmetry to hold at leading order, the coupling of $h$ to a pair of vector bosons is only given by the first term of $F_{U}(h)$. We do not consider NLO contributions to $h\rightarrow WW,ZZ$, as they are suppressed with respect to the LO terms \cite{Buchalla:2013mpa}. The loop-induced processes $h\rightarrow gg,\gamma\gamma,Z\gamma$ are given by one-loop terms from $\mathcal{L}_{\text{LO}}$. However, amplitudes from local $\mathcal{L}_{\text{NLO}}$ operators are of the same relative size, $\mathcal{O}(\xi/16\pi^{2})$. We therefore keep them as well. The resulting Lagrangian that we can use to interpret Eq.~\eqref{eq:8.16} is given by \cite{Buchalla:2015wfa}
\begin{equation}
  \label{eq:8.fit}
\begin{array}{ll}
  \mathcal{L} &=2 c_{V} \left(m_{W}^{2}W_{\mu}^{+}W^{-\mu} +\dfrac{1}{2} m^2_Z Z_{\mu}Z^{\mu}\right) \dfrac{h}{v} - \sum_{f}c_{f} Y_{f} \bar{f} f h \\
 &+ \dfrac{e^{2}}{16\pi^{2}} c_{\gamma\gamma} F_{\mu\nu}F^{\mu\nu} \dfrac{h}{v}+ \dfrac{e g \sin{\theta_{w}}}{16\pi^{2}} c_{Z\gamma} Z_{\mu\nu}F^{\mu\nu} \dfrac{h}{v}+\dfrac{g_{s}^{2}}{16\pi^{2}} c_{gg}\langle G_{\mu\nu}G^{\mu\nu}\rangle\dfrac{h}{v}.
\end{array}
\end{equation}
Here, $Y_{f} $ is defined\footnote{Note that this definition differs by a factor of $\sqrt{2}$ from the one of Eq.~\eqref{eq:5.3}.} by the masses of the fermions, $Y_{f} = m_{f}/v$. Similar parametrizations were discussed before, using phenomenological motivations \cite{Carmi:2012yp,Espinosa:2012im,Giardino:2013bma,Ellis:2013lra,Einhorn:2013tja,Bernon:2014vta,Flament:2015wra}. In the normalization of Eq.~\eqref{eq:8.fit}, we expect the size of the Wilson coefficients $c_{i}$ to be
\begin{equation}
  \label{eq:8.21}
  c_{i} = \begin{cases} 1 + \mathcal{O}(\xi) & \text{for } i = V, t, b, \tau, c, s, \mu, u, d, e\\ \mathcal{O}(\xi) & \text{for } i = gg, \gamma\gamma, Z\gamma \end{cases}
\end{equation}
We compute the contributions to $\mu$ of Eq.~\eqref{eq:8.16}, using Eq.~\eqref{eq:8.fit}. For the different production modes $X \in \{  ggF, WH/ZH, VBF, ttH \}$ and decay channels $Y \in \{WW, ZZ, \gamma \gamma,Z\gamma,\bar{f}f\}$ we find 
\begin{align}
\begin{aligned}
  \label{eq:8.22}
\frac{\sigma(\text{VH})}{\sigma(\text{VH})_{\text{SM}}} &= c_V^2 \,, \qquad  \qquad \qquad \frac{\sigma(\text{VBF})}{\sigma(\text{VBF})_{\text{SM}}} = c_V^2 \,, \\
\frac{\sigma(\text{ttH})}{\sigma(\text{ttH})_{\text{SM}}} &=  c_{t}^2 \,,\qquad  \qquad \qquad \frac{\sigma(\text{ggF})}{\sigma(\text{ggF})_{\text{SM}}}\simeq    \frac{\Gamma(h \rightarrow gg)}{\Gamma(h \rightarrow gg)_{\text{SM}}} \,.
\end{aligned}
\end{align}
We express the ratio of the branching ratios as
\begin{equation}
  \label{eq:8.23}
  \frac{\text{Br }(h \rightarrow Y)}{\text{Br }(h\rightarrow Y)_{\text{SM}}} = \frac{\Gamma^{Y}/\Gamma^{Y}_{\text{SM}}}{\sum_{j} (\text{Br }(h\rightarrow j)_{\text{SM}} \times \Gamma^{j}/\Gamma^{j}_{\text{SM}})}.
\end{equation}
The tree-level decay rates for $h \rightarrow VV^*$ $(VV=W^+W^-, ZZ)$ and $h \rightarrow \bar{f}f $ get rescaled compared to the SM by $c_V^2$ and $c_f^2$, respectively. For the loop-induced decays, we have \cite{Gunion:1989we,Manohar:2006gz,Contino:2014aaa}
\begin{align*}
\begin{aligned}
  \frac{\Gamma(h \rightarrow \gamma \gamma)}{\Gamma(h \rightarrow \gamma \gamma)_{\text{SM}}} &=    \frac{  \left|  \sum\limits_{q}   \frac{4}{3} N_C  Q_q^2    c_{q} A_{1/2}(x_q) \eta^{q,\gamma\gamma}_{\text{QCD}}   + \sum\limits_{f=\tau,\mu,e}\frac{4}{3} c_{f}  A_{1/2}(x_{f})    + c_V A_{1}(x_W) + 2 c_{\gamma \gamma}  \right|^2    }{ \left| \sum\limits_{q}   \frac{4}{3} N_C  Q_q^2 A_{1/2}(x_q) \eta^{q,\gamma\gamma}_{\text{QCD}}   + \frac{4}{3}  A_{1/2}(x_{\tau})    +  A_{1}(x_W)   \right|^2    }   \,, 
\end{aligned}
\end{align*}
\begin{align}
\begin{aligned}
  \label{eq:8.24}
   \frac{\Gamma(h \rightarrow gg)}{\Gamma(h \rightarrow gg)_{\text{SM}}}  &=  \frac{   \left|  \sum\limits_{q}  \frac{1}{3} c_q A_{1/2}(x_q) \eta^{q,gg}_{\text{QCD}}   + \frac{1}{2}  c_{gg}      \right|^2     }{   \left| \sum\limits_{q}  \frac{1}{3}  A_{1/2}(x_q) \eta^{q,gg}_{\text{QCD}}    \right|^2    }   \,, \\
 \frac{\Gamma(h \rightarrow Z \gamma)}{\Gamma(h \rightarrow Z \gamma)_{\text{SM}}} &= \frac{\left| \sum\limits_{f} c_{f} N_C  Q_{f}A_{f}(x_{f},\lambda_{f}) \eta^{f,Z\gamma}_{\text{QCD}}+ c_{V}A_{W}(x_{W},\lambda_{W}) +c_{Z\gamma}\right|^{2}}{\left| \sum\limits_{f}  N_C  Q_{f}A_{f}(x_{f},\lambda_{f})+ A_{W}(x_{W},\lambda_{W}) \right|^{2}}
\end{aligned}
\end{align}
with $x_i =  4 m_i^2/m_h^2$, $\lambda_{i} = 4 m_i^2/m_Z^2$, $Q_{f}$ the electric charge of a fermion, and $T^{3}_{f}$ the third component of its weak isospin. The loop functions are 
\begin{align}
\begin{aligned}
    \label{eq:8.25}
    A_{1/2}(x) &=   \frac{3}{2}  x \left[    1 + (1-x) f(x) \right]  \,,  \qquad  \qquad A_{1}(x)  =  - [  2 + 3 x + 3 x (2-x) f(x)   ] \, , \\
    A_{f}(x_{f},\lambda_{f}) & = -2 \frac{T^{3}_{f}-2Q_{f}\sin^{2}{\theta_{w}}}{\sin{\theta_{w}}\cos{\theta_{w}}} \left[I_1\left(x_{f},\lambda_{f}\right)-I_2\left(x_{f},\lambda_{f}\right)\right]\, ,\\
    A_{W}(x_{W},\lambda_{W}) & = - \cot{\theta_{w}}\Big[4(3-\tan^{2}{\theta_{w}})I_{2}(x_{W},\lambda_{W})\\
&\qquad+\left((1+\tfrac{2}{x_{W}})\tan^{2}{\theta_{w}}-(5+\tfrac{2}{x_{W}})\right) I_{1}(x_{W},\lambda_{W})\Big] \, ,
\end{aligned}
\end{align}
with the definitions 
\begin{align}
\begin{aligned}
\label{eq:8.25a}
&I_1\left(\tau,\lambda\right)=\frac{\tau\lambda}{2\left(\tau-\lambda\right)}+\frac{\tau^2\lambda^2}{2\left(\tau-\lambda\right)^2}\left[f\left(\tau\right)-f\left(\lambda\right)\right]+\frac{\tau^2\lambda}{\left(\tau-\lambda\right)^2}\left[g\left(\tau\right)-g\left(\lambda\right)\right] \\ \text{and} \\
&I_2\left(\tau,\lambda\right)=-\frac{\tau\lambda}{2\left(\tau-\lambda\right)}\left[f\left(\tau\right)-f\left(\lambda\right)\right] \; .
\end{aligned}
\end{align}
The functions $f(x)$ and $g(x)$ read
\begin{align}
\begin{aligned}
\label{eq:8.26}
g\left(x\right)\; &=\; \begin{cases}
\sqrt{x-1}\, \arcsin \frac{1}{\sqrt{x}} & x\geq 1 \\
\frac{\sqrt{1-x}}{2}\left[\ln\frac{1+\sqrt{1-x}}{1-\sqrt{1-x}}-i\pi\right] \,\quad& x<1 \, \end{cases}\\
f(x)\; &=\; \begin{cases} \arcsin^2(1/\sqrt{x})\,  \quad & x\geq 1 \\[3pt]   - \dfrac{1}{4} \Big[\ln\Big( \frac{1+\sqrt{1-x}}{1-\sqrt{1-x}}\Big)- i\pi \Big]^2\,  & x<1. \end{cases} \, 
\end{aligned}
\end{align}
At this order, QCD corrections of $\mathcal{O}(\alpha_s)$ due to the exchange of hard gluons and quarks in production and decay can be taken into account. The QCD corrections factorize for tree-level amplitudes of production and decay and therefore cancel in the ratios. However, they do not cancel for $h \rightarrow \gamma \gamma, Z\gamma, gg$, where we included ${\eta^{t,gg}_{\text{QCD}}  = 1+11\alpha_{s}/4\pi}$ and $\eta^{t,\gamma\gamma}_{\text{QCD}} =\eta^{t,Z\gamma}_{\text{QCD}} = 1 -\alpha_{s}/\pi$ for the top-loop\cite{Manohar:2006gz,Contino:2013kra,Contino:2014aaa}. The effects of QCD corrections on other quark loops are negligibly small. We set $\eta^{f,Z\gamma}_{\text{QCD}}=0$ for $f\in\{e,\mu,\tau\}$. We further discuss the impact of QCD corrections on the final result at the end of Section \ref{sec:fit}. Non-factorizing electroweak corrections, which we expect to be small as well, are not taken into account.\\
By comparing Eqs.~\eqref{eq:8.22} and \eqref{eq:8.24} to Eq.~\eqref{eq:8.19}, we see that the $\kappa_{i}$ are equivalent to the Wilson coefficients $c_{i}$ under the following conditions \cite{Buchalla:2015wfa}:
\begin{itemize}
\item We directly have $c_{i}= \kappa_{i}$ for tree-level processes. With the assumption of custodial symmetry, we have $c_{V}=\kappa_{V}\equiv \kappa_{W}=\kappa_{Z}$. 
\item For loop-induced processes, the amplitude is given by the SM topologies scaled with $c_{i}$ plus one contact term. When using the $\kappa$-framework with independent $\kappa_{gg}$, $\kappa_{\gamma\gamma}$ and $\kappa_{Z\gamma}$, we have the same number of free parameters and therefore the two frameworks can be seen as equivalent.
\item The $\kappa_{h}$ of the total width is given by the SM topologies, but uses the modified couplings from above. This corresponds to the assumption of no invisible decay channels. 
\end{itemize}
The correspondence between the $\kappa_{i}$ and the $c_{i}$ is not one-to-one in point two above, since the $c_{i}$ are defined on the amplitude level while the $\kappa_{i}$ are defined at the level of the amplitude squared. Neglecting the small imaginary parts of the loop diagrams, we can find a linear translation between the two sets. Working directly within the framework of the electroweak chiral Lagrangian instead of the $\kappa_{i}$ brings several advantages. Since it is based on a consistent effective field theory, includig higher order electroweak and QCD corrections is well defined. Also, if experimental precision increases, the extension of the framework to kinematic distributions via higher order operators in the EFT is straightforward. However, if the Higgs is produced with large transverse momentum (``boosted'' Higgs), already the leading order effects introduce non-trivial kinematical distributions \cite{Grojean:2013nya,Schlaffer:2014osa,Buschmann:2014twa,Buschmann:2014sia}. On top, the interpretation of a non-SM result in terms of specific models is easier for the $c_{i}$, as the Wilson coefficients of models are often given in the literature. Having the correspondence of the $c_{i}$ and the $\kappa_{i}$ in mind, we see that the criticism of the $\kappa$-framework, especially regarding the consistency within quantum field theory given above \cite{PO:Passarino2,Henning:2014wua}, is not justified.\\

\begin{figure}[t]
\begin{center}
\subfigure[$\mathcal{O}\left(\frac{1}{16\pi^{2}}\right)$\label{fig:h2Zgamma.ferSM}]{
\begin{overpic}[width=0.3\textwidth]{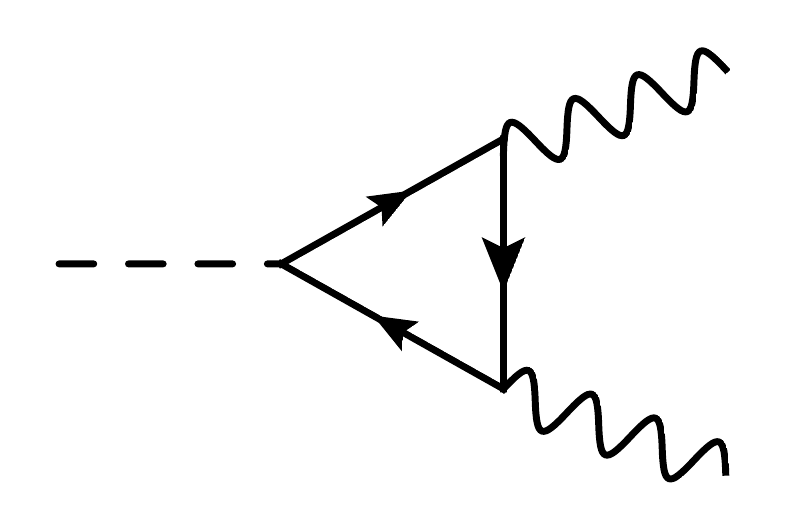}
\end{overpic}}\hfill
\subfigure[$\mathcal{O}\left(\frac{1}{16\pi^{2}}\right)$\label{fig:h2Zgamma.gauSM}]{
\begin{overpic}[width=0.3\textwidth]{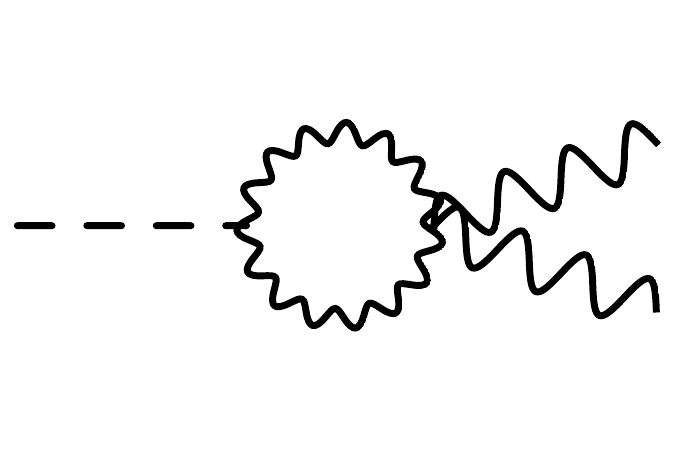}
\end{overpic}}\hfill
\subfigure[$\mathcal{O}\left(\frac{v^{2}}{\Lambda^{2}}\right)$\label{fig:h2Zgamma.treedim6}]{
\begin{overpic}[width=0.3\textwidth]{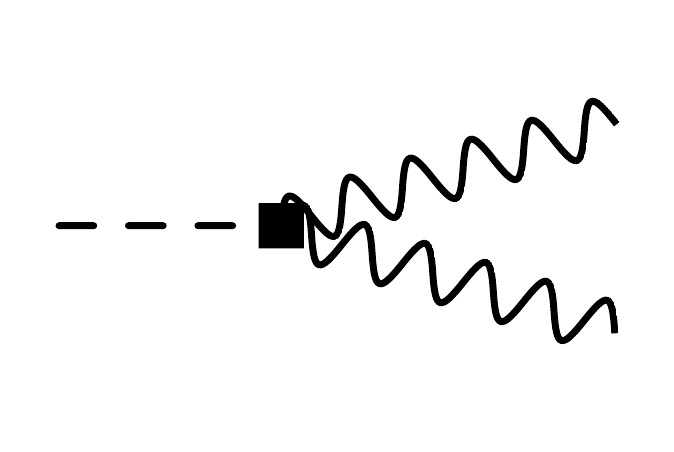}
\end{overpic}}\\
\subfigure[$\mathcal{O}\left(\frac{v^{2}}{16\pi^{2}\Lambda^{2}}\right)$\label{fig:h2Zgamma.ferdim61}]{
\begin{overpic}[width=0.3\textwidth]{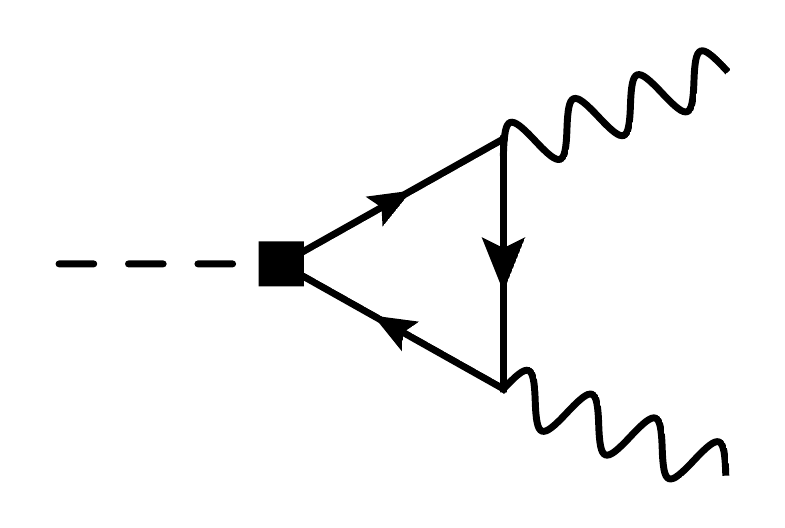}
\end{overpic}}\hfill
\subfigure[$\mathcal{O}\left(\frac{v^{2}}{16\pi^{2}\Lambda^{2}}\right)$\label{fig:h2Zgamma.ferdim62}]{
\begin{overpic}[width=0.3\textwidth]{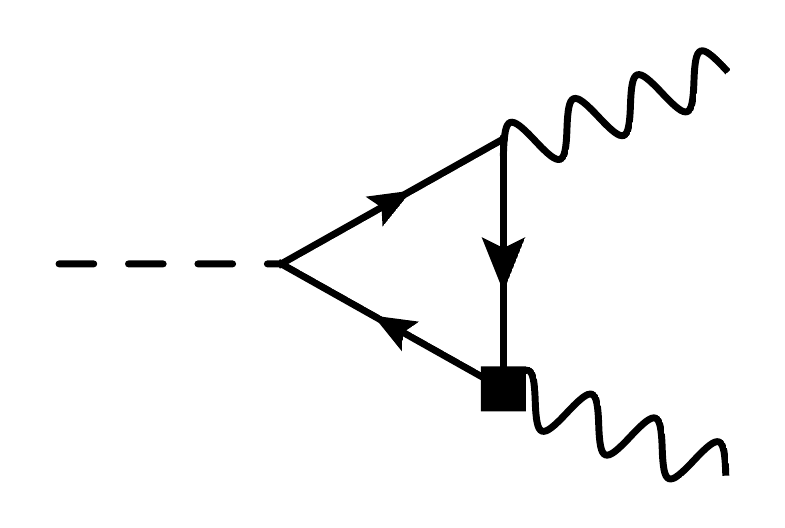}
\end{overpic}}\hfill
\subfigure[$\mathcal{O}\left(\frac{v^{2}}{16\pi^{2}\Lambda^{2}}\right)$\label{fig:h2Zgamma.gaudim6}]{
\begin{overpic}[width=0.3\textwidth]{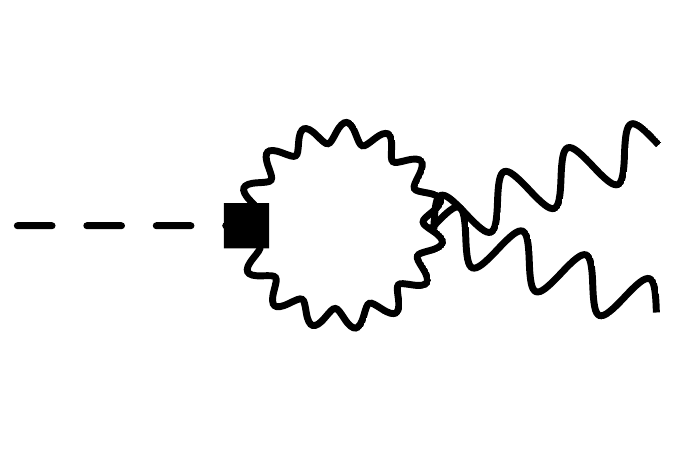}
\end{overpic}}
\hfill
\end{center}
\caption[Non-exhaustive list of contributions to $h\rightarrow Z\gamma$ in SM-EFT.]{Non-exhaustive list of contributions to the process $h\rightarrow Z\gamma$ in SM-EFT. The first line corresponds to leading order SM-EFT, the second line to next-to-leading order. Black squares dennote a dimension-six insertion.}
\label{fig:h2Zgamma.SMEFT}
\end{figure}

We can also connect the $\kappa$-framework to the effective field theory with dimension-six operators introduced in Chapter \ref{ch:SMEFT}. The observables in Eqs.~\eqref{eq:8.22} and \eqref{eq:8.24} can be computed using the Lagrangian in Eq.~\eqref{eq:4.SM-EFT} and the $\kappa_{i}$ are given as functions of the Wilson coefficients $\tilde{c}_{i}$ of the SM-EFT. However, since the power counting is different compared to the electroweak chiral Lagrangian, we do not find the equivalence that we found previously. The contributions to the decay of $h\rightarrow Z\gamma$ may serve as an example. At leading order in the SM-EFT, we have the SM contributions \cite{Kniehl:1990mq,CarloniCalame:2006vr,Bredenstein:2006rh} plus tree-level topologies of only one dimension-six insertion, see the Feynman diagrams in Figs. \ref{fig:h2Zgamma.ferSM} - \ref{fig:h2Zgamma.treedim6}. 
Since the SM loops are not modified, we only have one free parameter, the local $hZ\gamma$ coupling. At next-to-leading order in the SM-EFT \cite{Ghezzi:2015vva,Hartmann:2015oia,Hartmann:2015aia,David:2015waa} we also have one-loop diagrams with one insertion of dimension-six operators, see Figs. \ref{fig:h2Zgamma.ferdim61} -- \ref{fig:h2Zgamma.gaudim6}. This comes now with more free parameters than needed for the $\kappa$-framework, as the $Z\bar{f}f$ vertex is also modified at that order. In addition, there is a hierarchy between the contribution of Fig. \ref{fig:h2Zgamma.treedim6}, coming at $\mathcal{O}\left({v^{2}}/{\Lambda^{2}}\right)$, and the contributions of Figs. \ref{fig:h2Zgamma.ferdim61} -- \ref{fig:h2Zgamma.gaudim6}, coming at $\mathcal{O}\left({v^{2}}/{16\pi^{2}\Lambda^{2}}\right)$. Additional assumptions, such as a weakly coupled UV completion \cite{Arzt:1994gp,Einhorn:2013kja} can be used to bring all contributions to the same relative order, $\mathcal{O}({v^{2}}/{16\pi^{2}\Lambda^{2}})$, but the modified $Z\bar{f}f$ vertex still remains to be non-absorbable in the $\kappa_{i}$. 

Several different proposals exist in the literature to extend the $\kappa$-framework to also be sensitive to kinematical distributions. The inclusion of dimension-six operators at the one-loop level \cite{Ghezzi:2015vva,Hartmann:2015oia,Hartmann:2015aia}, as discussed before, is one example. However, as already said, it does not include the original $\kappa$-framework in a consistent limit. Another proposal uses the pole decomposition of amplitudes \cite{Gonzalez-Alonso:2014eva,Bordone:2015nqa,Greljo:2015sla}. With the justification of the $\kappa$-framework, based on a leading-order analysis using the electroweak chiral Lagrangian presented above, an extension to next-to-leading order \cite{Buchalla:2015qju} would serve as a natural generalization of the $\kappa$-framework. 
 \section{Foundations of Bayesian Data Analysis}
In this section, we introduce the basic concepts of Bayesian statistics for data analysis \cite{sivia2006data,D'Agostini:2003qr}. We will see that the Bayesian approach is very useful for the parameter estimation within effective field theories, as prior knowledge on the size of the Wilson coefficients from power counting can consistently be taken into account \cite{Schindler:2008fh,Wesolowski:2015fqa}.
 \subsection{Basics}
When analyzing experimental data, we usually face the following problem: We observe a certain effect, but we do not know the cause(s) of this effect. For example, when we measure a certain coupling, we would like to know whether the underlying theory is the SM or some model of new physics. This means that we have to infer causes by inductive reasoning rather than deductive reasoning, where the cause is known and we can derive the effects. It appears natural to assign probabilities, {\it i.e.} numbers between $0$ and $1$, to different causes. These probabilities, in the Bayesian picture, express our degree-of-belief in a certain event rather than the long-run-relative-frequency that the frequentist approach assigns to probabilities. The degree-of-belief always differs between different observers, as it depends on the prior state of knowledge. This can be seen in the simple example ``What are the chances for rain tonight?'' Our answer will differ, depending on our geographical location, the current appearance of the sky, whether we saw the weather forecast, {\it etc}. However, we would like to make these subjective statements more objective. Therefore, we will express the probability of an event $X$, with explicit background information $I$ given, as $\prob{X}{I}$. Probabilities obey by logical consistency the product rule:
\begin{align}
\begin{aligned}
  \label{eq:8.9}
\prob{X,Y}{I} &= \prob{X}{I} \cdot \prob{Y}{X,I}\\
&= \prob{Y}{I} \cdot \prob{X}{Y,I}\\
\end{aligned}
\end{align}
Here, the comma means a logical AND. The second line follows from the fact that the probability of $X$ and $Y$ being true at the same time is a commutative statement. The product rule can be visualized using Venn-diagrams, see Fig. \ref{fig:ProductRule}.
\definecolor{light-gray}{gray}{0.9}
\begin{figure}[h]
  \begin{center} \fbox{
    \begin{tikzpicture}
      \draw (0,0) ellipse (5em and 3em);
      \draw (0,0) node {$\prob{X}{I}$};
      \draw (7.5em,0) ellipse (5 em and 3em);
      \draw (7.5em,0) node {$\prob{Y}{I}$};
      \begin{scope}
        \clip {(0,0) ellipse (5em and 3em)};
        \fill[light-gray] {(7.5em,0) ellipse (5em and 3 em)};
      \end{scope}  
      \begin{scope}
        \draw (3.75em,-3em) -- (3.75em,0em);
        \draw (3.75em,-3em) node[anchor=north] {$\prob{X,Y}{I}$};
        \draw (12.5em,-3em) node[anchor=north] {I};
      \end{scope}
      \end{tikzpicture}}
  \end{center}
  \caption{A visualization of the product rule for probabilities.}
  \label{fig:ProductRule}
\end{figure}
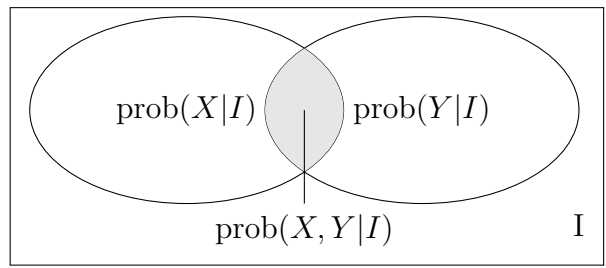 \\

By reordering Eq.~\eqref{eq:8.9}, we find {\it Bayes Theorem}, which will be of central importance throughout the remainder of this chapter. 
\begin{equation}
  \label{eq:8.BayesTheorem}
  \prob{X}{Y,I} = \frac{\prob{Y}{X,I} \cdot \prob{X}{I}}{\prob{Y}{I}}
\end{equation}
The significance of this Theorem will be more clear when we replace ($X \rightarrow$ hypothesis) and ($Y \rightarrow $ data) in Eq.~\eqref{eq:8.BayesTheorem}.
\begin{equation}
  \label{eq:8.10}
  \prob{\text{hypothesis}}{\text{data},I} = \frac{\prob{\text{data}}{\text{hypothesis},I} \cdot \prob{\text{hypothesis}}{I}}{\prob{\text{data}}{I}}
\end{equation}
We see that the quantity that we are interested in, $\prob{\text{hypothesis}}{\text{data},I}$ (the so-called {\it posterior probability density}), is given by the product of $\prob{\text{hypothesis}}{I}$ (called {\it prior}) times $\prob{\text{data}}{\text{hypothesis},I}$ (called {\it likelihood}) divided by $\prob{\text{data}}{I}$ (called {\it evidence}). For the purpose of the analysis presented here, we write Eq.~\eqref{eq:8.10} as
\begin{equation}
\label{eq:Bayesian}
\mathcal{P} \equiv posterior \sim prior \times likelihood
\end{equation}
We replace the equality with a proportionality, as the evidence does not play a role in parameter estimation. Instead, the evidence is useful for model-selection questions on which we do not focus our analysis. More details can be found in \cite{sivia2006data,D'Agostini:2003qr}. We can determine the constant of proportionality of Eq.~\eqref{eq:Bayesian} from the normalization condition that all probabilities have to fulfill
\begin{equation}
  \label{eq:8.11}
  \int \prob{A}{I} dA = 1.
\end{equation}
A very powerful tool that we can derive from the normalization condition is the {\it marginalization}:
\begin{equation}
  \label{eq:8.Marginalization}
  \prob{A}{I} = \int \prob{A,B}{I} dB
\end{equation}
This relation is also very important for our analysis. It states that we can sum over all mutually exclusive states $B$ and express the final probability independently of $B$. This allows us to include so-called {\it nuisance parameters} in our analysis --- parameters that are needed in the model, but are not of our primary interest. These can include higher order terms of the EFT, parameters of the prior, or even calibration constants of the experiment. Marginalization also allows us to reduce high-dimensional parameter spaces to one- and two-dimensional subspaces that we can visualize easily, similar to a geometric projection. The two-dimensional marginalization shows the correlation between the two parameters. The one-dimensional marginalized posterior pdf of a single parameter gives the result of the parameter estimation.\\
In the following section, we discuss the components of Eq.~\eqref{eq:Bayesian} in more detail.
\subsection{Prior Information}
\label{sec:prior}
The prior quantifies our initial state of knowledge before we analyze the data. In the case at hand we have an estimation of the size of the Wilson coefficients, based on the power counting: Deviations from the SM are expected at order $\mathcal{O}(\xi)$. The question is now, how we can express the qualitative statement of ``natural-sized coefficients'' in a quantitative way. The method of maximum entropy\cite{Jaynes:1957zza} provides a possible prescription to answer this question \cite{Schindler:2008fh}. It states that the prior probability should be chosen in a way that the entropy is maximized, while simultaneously all other necessary constraints are fulfilled. For example, if the prior should have a support only in a compact volume, $x\in [a_{\text{min}},a_{\text{max}}]$, the method of maximum entropy states that the prior should be uniform.
\begin{equation}
  \label{eq:8.12}
  \prob{X}{I} \sim \Theta(X-a_{\text{min}}) \Theta(a_{\text{max}}-X)
\end{equation}
Here, $\Theta(X)$ is the Heaviside step function. If the parameters are scattered around a mean value $\mu$ with a given variance $\sigma^{2}$, the prior should be a Gaussian. Qualitatively, we would interpret such a prior as: ``we are {\it confident} that the value is within $\mu\pm \sigma$'', and ``we are {\it almost certain} it is within $\mu\pm 3\sigma$'' \cite{D'Agostini:2003qr}. In general, the prior should reflect where we expect the posterior probability of the parameters to be concentrated, but also allow longer tails for possible surprises in the data. This introduces an uncertainty to the prior that can also consistently be taken into account using Bayes Theorem, Eq.~\eqref{eq:8.BayesTheorem}. Parameters of the prior, called hyperparameters, can then be understood as nuisance parameters and marginalized over. 

In a more abstract way, the prior can also determine the details of the analysis itself. For example, previous experiments measured the violation of custodial symmetry to be very small. We therefore choose our Lagrangian~\eqref{eq:8.fit} to respect custodial symmetry.

Appropriate use of the prior also avoids the problem of {\it overfitting} and {\it underfitting} \cite{Wesolowski:2015fqa}. Overfitting refers to a very fine-tuned result in an unnatural region of parameter space. A naturalness requirement on the parameter coming from the prior as well as a possible marginalization over higher-order terms reduces the risk of overfitting \cite{Wesolowski:2015fqa}. Underfitting on the other side refers to a fit where important features are not appropriately described by the underlying model, or EFT respectively. Also in this case a marginalization over higher-order contributions can reduce the risk of underfitting \cite{Wesolowski:2015fqa}. 
\subsection{Likelihood}
\label{sec:lilith}
The likelihood, $\prob{\text{data}}{\text{hypothesis},I}$, can be understood as the fraction of a large set of experiments (for fixed parameters) that gives a result in the considered interval of the observable \cite{Wesolowski:2015fqa}. In the past years, the experimental collaborations published the likelihood of the signal strength, Eq.~\eqref{eq:8.16}, in two-dimensional planes for each decay mode (as {\it e.g.} in \cite{ATLAS:2014yka}). For this, the production modes are combined to two effective production modes $\mu_{\text{ggF+ttH}}$ and $\mu_{\text{VBF+VH}}$. This combination is justified for the currently available data \cite{Bernon:2015hsa}. In some cases, the one-dimensional likelihood is published beyond the Gaussian approximation. 

For our analysis, we obtain the likelihood from the code {\tt Lilith} (``Light Likelihood Fit for the Higgs'') \cite{Bernon:2015hsa}. It gives the relevant likelihood for a given input-signal strength. The likelihood and correlations were extracted previously from the experimental publications and hard-coded into the program. This extraction uses the previously explained two-dimensional planes or the full one-dimensional likelihood, depending on what is published by the collaborations. In the two-dimensional case, a two-dimensional Gaussian is fitted to the $68\%$ confidence level (CL) interval and cross-checked for the $95\%$ CL interval \cite{Bernon:2015hsa}.
\subsection{The Posterior}
The posterior, $\mathcal{P}\equiv \prob{\text{hypothesis}}{\text{data},I}$, contains the full information about the parameters. This information is usually what we are interested in, but it is also crucial for propagation of errors and correlations if we want to use the extracted parameters further. If we can approximate the posterior by a multi-dimensional Gaussian, we can express the information in only a few numbers. These are the expectation value, $\langle c \rangle$; the standard deviation, $\sigma_{c}$, of a parameter $c$; and the covariance, $\cov{x}{y}$, of two parameters $x$ and $y$:
\begin{align}
  \begin{aligned}
    \label{eq:8.27}
\langle c_{i} \rangle &= \int c_{i}\,\mathcal{P}(c_{1},\dots,c_{i},\dots,c_{n}) d\vec{c}\\
\cov{x}{y} &= \langle x y \rangle - \langle x \rangle \langle y \rangle\\
\sigma_{c}&=\sqrt{\cov{c}{c}}.
  \end{aligned}
\end{align}
We can compute the correlation matrix $\rho$ from the covariance and summarize the correlations between all pairs of parameters,
\begin{equation}
  \label{eq:8.28}
  \rho_{x,y} = \frac{\cov{x}{y}}{\sigma_{x}\sigma_{y}}.
\end{equation}
We then understand the one sigma interval as ``the parameter $c$ lies to $68\%$ within $\langle c \rangle \pm \sigma_{c}$''. For non-Gaussian posteriors, {\it e.g.} with multiple maxima or asymmetries, these simple parametrizations cannot account for the full information and we have to use the complete posterior $\mathcal{P}$ instead.

The interplay of prior and likelihood leads to posteriors that are either prior or likelihood dominated. Being dominated by the prior means that the information of the likelihood does not constrain the parameter and the only information we can extract is the prior information we put in. In oder to avoid misleading conclusions, cross-checks with different priors should be made \cite{Wesolowski:2015fqa}. 

For a visualization of the multi-dimensional posterior, usually one- or two-dimensional marginalized posteriors are presented. The one-dimensional posterior contains the full information that can be inferred for a single parameter. The two-dimensional marginalized posterior shows the correlation between the two parameters. For a given set of $n$ parameters, we collect all $(n-1)n/2$ two-dimensional plots and all $n$ one-dimensional plots in a so-called corner plot \cite{Wesolowski:2015fqa,triangle.py}. 
\section{Numerical Methods for Finding the Posterior pdf}
When analyzing data with the effective Lagrangian, we are interested in the posterior probability density, see Eq.~\eqref{eq:Bayesian}. There are several ways how the posterior can be computed numerically. The first way is direct computation in a given, fixed grid in parameter space. The second way uses Markov Chain Monte Carlo techniques.
\subsection{Fixed Grid Computation of $\mathcal{P}$}
In the first way to evaluate Eq~\eqref{eq:Bayesian}, we provide a fixed grid in parameter space as input and determine the values of $\mathcal{P}$ explicitly. The resulting table of values contains the posterior for the entire parameter space considered. This approach has the advantage that we find all islands of non-vanishing posterior $\mathcal{P}$. However, if the posterior is centered in a small region, a large part of the computational power is wasted for regions that are of no interest. This is especially true for non-flat priors that have a large support, like for example Gaussian priors. In addition, for certain priors, the fixed grid comes with yet another disadvantage. Consider as an example the priors of Fig. \ref{fig:prior} below. For a fixed density of points, required by the desired resolution of the posterior, the flat prior needs the least number of points in the grid. The blue Gaussian for $a=1$ needs about twice as many points to be evaluated. As this applies to all $n$ dimensions of parameter space, the total number of points scales as $2^{n}$ and can easily be orders of magnitude larger, making it impossible to be used. A variable density of points increases the complexity of the code a lot. For that reason, we use a more convenient way of obtaining the posterior in our fit. We discuss it in the next section. 
\subsection{Markov Chain Monte Carlo}
Markov Chain Monte Carlo (MCMC) methods approximate the posterior pdf, $\mathcal{P}$, efficiently, especially in high-dimensional parameter spaces. In the following sections, we outline the basic principles of the Markov Chain Monte Carlo algorithm that we use to perform the fit to the LHC Higgs data. In contrast to the fixed grid computation, where we obtain the values of the posterior pdf as a table of values, we obtain a set of points that is distributed according to the posterior pdf $\mathcal{P}(x)$. We see the advantage of this in the following example \cite{D'Agostini:2003qr}. The expectation value of a function, $\langle f(x)\rangle$, is given by a computationally expensive integral of the form
\begin{equation}
  \label{eq:8.1}
  \langle f(x)\rangle = \int f(x) \mathcal{P}(x) dx.
\end{equation}
For a discrete set of points, $x_{i}$, we can replace the integral with a sum over all points. In addition, we collect the $x_{i}$ into bins $x_{b}$ of a given size. The probability of being in a given bin, $\mathcal{P}(x_{b})$, can be approximated via Bernoulli's Theorem to be $\mathcal{P}(x_{b})\approx m_{b} / N$, where $m_{b}$ is the number of $x_{i}$ in the bin $x_{b}$ and $N$ is the total number of points. The expectation value is then
\begin{equation}
  \label{eq:8.2}
  \langle f(x)\rangle \approx \sum_{\text{bins}} f(x_{b}) \frac{m_{b}}{N} = \frac{1}{N}\sum_{i} f(x_{i}).
\end{equation}
So if we have a set of points that is distributed according to $\mathcal{P}(x)$, we just have to average over all points instead of performing the integral of Eq.~\eqref{eq:8.1} in order to compute expectation values. A set of points that has this distribution is called Markov Chain. In the following we see different algorithms that construct such a Markov Chain.   
\subsubsection{Basics and the Metropolis-Hastings Algorithm}
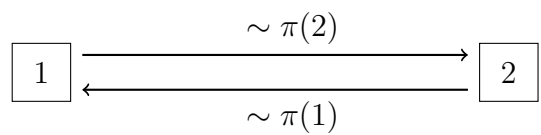
\begin{figure}[h]
  \begin{center}
    \begin{tikzpicture}
      \draw {(0,0) rectangle (0.05\textwidth,0.05\textwidth)} node (links) {};
      \draw (0.025\textwidth,0.025\textwidth)node {$1$};
      \draw {(0.4\textwidth,0) rectangle (0.45\textwidth,0.05\textwidth)} node (rechts) {};
      \draw (0.425\textwidth,0.025\textwidth)node {$2$};
      \draw[thick,->] (0.06\textwidth,0.04\textwidth) -- (0.39\textwidth,0.04\textwidth) node[anchor=east] {};
      \draw (0.24\textwidth,0.04\textwidth) node[anchor=south] {$\sim \pi(2)$};
      \draw[thick,->] (0.39\textwidth,0.01\textwidth) -- (0.06\textwidth,0.01\textwidth) node[anchor=west] {};
      \draw (0.24\textwidth,0.01\textwidth) node[anchor=north] {$\sim \pi(1)$};
      \end{tikzpicture}
  \end{center}
  \caption[The detailed balance condition.]{The detailed balance condition \cite{sivia2006data}.}
  \label{fig:8.DB}
\end{figure}

\begin{figure}[h]
  \begin{center}
    \begin{tikzpicture}
      \draw {(0,0) rectangle (20em,10em)};
      \draw {(0,8em) rectangle (20em,10em)};
      \draw (10em,8em) node[anchor=south]{Propose position: $\vec{p} = \vec{x_{i}} + r \vec{q}$};
      \draw (10em,5.75em) node[anchor=south]{Check: $\frac{\pi(p)}{\pi(x_{i})}> \text{rand }(0,1)$ ?};
      \draw {(0,8em) -- (10em,4em)};
      \draw {(20em,8em) -- (10em,4em)};
      \draw {(10em,4em) -- (10em,0em)};
      \draw (2em,4em) node[anchor=south west]{Yes:};
      \draw {(0em,4em) -- (20em,4em)};
      \draw (2em,1em) node[anchor=south west]{$x_{i+1} = p$};
      \draw (18em,4em) node[anchor=south east]{No:};
      \draw (18em,1em) node[anchor=south east]{$x_{i+1} = x_{i}$};
      \end{tikzpicture}
  \end{center}
  \caption[The Metropolis-Hastings algorithm.]{The Metropolis-Hastings algorithm \cite{Metropolis:1953,Hastings:1970}.}
  \label{fig:8.MH}
\end{figure}
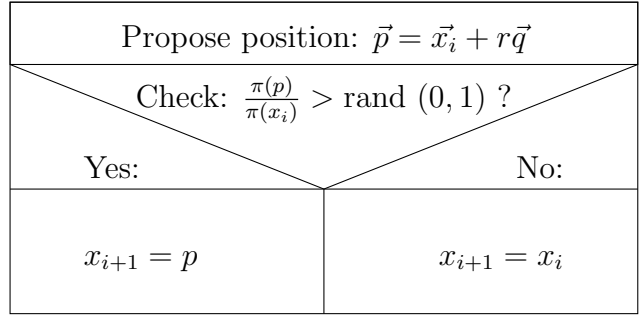

Let us start with a one-dimensional Markov Chain \cite{sivia2006data}: It is a series of points, where each point in the chain is given by an iteration based on the previous point. The distribution of all points in the chain resembles a previously given probability density $\pi(x)$ for a large number of steps. The iterations in the Markov Chain are given by a random walk in the following way. First, a random number generator proposes a potential next point. This point is then either accepted, {\it i.e.} the ``walker'' goes to this point, or rejected, {\it i.e.} the ``walker'' stays at its current position. The decision is based on the assigned probabilities $\pi$ of the original and the proposed point. In order to reach equilibrium between all possible transitions and therefore a stationary approximation to the probability density $\pi(x)$, we need to enforce a condition called {\it detailed balance} to our algorithm. Detailed balance means that the transition rate from a state $1$ to a state $2$ is the same as the transition rate from $2$ to $1$. This means that we cannot distinguish if we are going forwards or backwards within our Markov Chain. The condition of detailed balance is fulfilled if the acceptance rate of $1\rightarrow 2$ is proportional to $\pi(2)$ and the rate $2\rightarrow 1$ is proportional to $\pi(1)$, see Fig. \ref{fig:8.DB}.
An algorithm that generates Markov Chains fulfilling the detailed balance condition is the Metropolis-Hastings algorithm \cite{Metropolis:1953,Hastings:1970}. We explain it in Fig.~\ref{fig:8.MH}. 
A proposal position $p$ is generated from the current position $x_{i}$ in the chain, a random number $q\in (0,1)$ and a search radius $r$. If $\pi(p)>\pi(x)$, the proposal is always accepted, {\it i.e.} $x_{i+1} = p$. If $\pi(p)<\pi(x)$ the proposal is accepted with the probability $\pi(p)/\pi(x)$. If the proposal is rejected, the chain stays at its current position, {\it i.e.} $x_{i+1} = x_{i}$. As already indicated in the structogram in Fig.~\ref{fig:8.MH}, this can easily be extended to multi-dimensional probability densities. Since the algorithm only depends on the ratio $\pi(p)/\pi(x_{i})$, there is no need to find the overall normalization of the final probability density. This avoids another computationally expensive step in the analysis. However, the convergence of this algorithm is bad if the target density is highly anisotropic. For example, consider a two-dimensional Gaussian where the two widths differ by some orders of magnitude \cite{Goodman:2009MCMC}. In order to sample the probability distribution efficiently, the search radius should be of the order of the width. If it is too small, we need many steps to sample the full parameter space. If it is too large, we are not sensitive to the distribution. The problem gets even more complicated if the Gaussian with different widths is not aligned with the coordinate axes. In general, we need to tune the $N(N+1)/2$ parameters of the covariance matrix by hand to reach a reasonable convergence time for our Markov Chain. Using sampling algorithms that are affine invariant avoids this problem.
\subsubsection{The Affine Invariant Goodman-Weare Algorithm}
\begin{figure}[h]
  \begin{center}
    \begin{tikzpicture}
      \draw {(0,0) rectangle (30em,15em)};
      \draw {(2em,0) rectangle (30em, 13em)};
      \draw (2em,13em) node[anchor=south west]{For all walkers $X^{(i)}$ with $i \in [1,\dots, L]$ do:};
      \draw (2em,11em) node[anchor=south west]{Select random walker $X_{n}^{(j)}$ from the complementary ensemble};
      \draw {(2em,11em) -- (30em,11em)};
      \draw (2em,9em) node[anchor=south west]{Generate $Z$ from $g(z)$, as in Eq.~\eqref{eq:8.4}};
      \draw {(2em,9em) -- (30em,9em)};
      \draw (2em,7em) node[anchor=south west]{Propose position: $p= X_{n}^{(j)} + Z (X_{n}^{(i)}-X_{n}^{(j)})$};
      \draw {(2em,7em) -- (30em,7em)};
      \draw (16em,4.5em) node[anchor=south]{Check: $Z^{N-1}\frac{\pi(p)}{\pi(X_{n}^{(i)})}> \text{rand }(0,1)$ ?};
      \draw {(2em,7em) -- (16em,3em)};
      \draw {(30em,7em) -- (16em,3em)};
      \draw {(16em,3em) -- (16em,0em)};
      \draw {(2em,3em) -- (30em,3em)};
      \draw (4em,3em) node[anchor=south west]{Yes:};
      \draw (4em,0em) node[anchor=south west]{$X^{(i)}_{n+1} = p$};
      \draw (28em,3em) node[anchor=south east]{No:};
      \draw (28em,0em) node[anchor=south east]{$X^{(i)}_{n+1} = X^{(i)}_{n}$};
      \end{tikzpicture}
  \end{center}
  \caption[The Goodman-Weare algorithm.]{The Goodman-Weare algorithm \cite{Goodman:2009MCMC,2013PASP..125..306F}.}
  \label{fig:8.GW}
\end{figure}
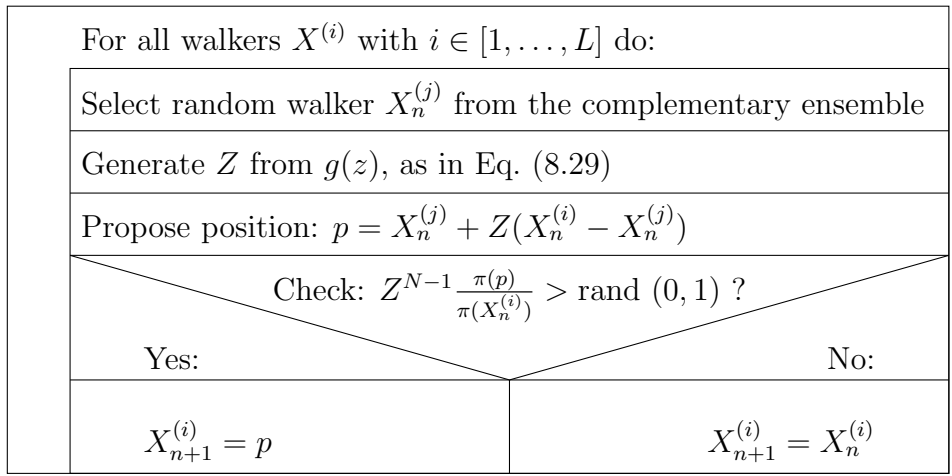

Affine invariant algorithms have the same performance if the parameter space is transformed with an affine transformation. Anisotropic and skewed distributions are sampled as efficient as isotropic distributions with such algorithms. Goodman and Weare proposed an algorithm that is affine invariant in \cite{Goodman:2009MCMC}. We explain the algorithm in Fig.~\ref{fig:8.GW}.
It samples a set of $L$ walkers at the same time. The proposed walk of a given walker $X^{(i)}$ is now based on the position of the other walkers at the same time step. In detail, the algorithm picks one of the other walkers randomly and performs a ``stretch move''\cite{2013PASP..125..306F}: The proposed new point lies on a line drawn between the two walkers.
\begin{equation}
  \label{eq:8.3}
  p= X_{n}^{(j)} + Z (X_{n}^{(i)}-X_{n}^{(j)})
\end{equation}
The random variable $Z$ is drawn from the distribution
\begin{equation}
  \label{eq:8.4}
  g(z) = \begin{cases}\frac{1}{\sqrt{z}} & \text{if } z\in[\frac{1}{a},a] \\ 0 & \text{otherwise} \end{cases}
\end{equation}
The parameter $a$ can be adjusted by hand, usually one assumes $a=2$. The proposal $p$ is accepted with a probability 
\begin{equation}
  \label{eq:8.5}
  q = \text{min }\left(1, Z^{N-1}\frac{\pi(p)}{\pi(X_{n}^{(i)})}\right),
\end{equation}
where $N$ is the dimension of the parameter space. Together, the Eqs.~\eqref{eq:8.3} -- \eqref{eq:8.5} ensure the detailed balance condition. The advantage of this affine invariant algorithm is that now only very few (basically just $a$ in Eq.~\eqref{eq:8.4}) parameters have to be tuned by hand. This is in contrast to the $N(N+1)/2$ parameters for the Metropolis-Hastings algorithm. 
\subsubsection{The Goodman-Weare Algorithm of {\tt emcee}}
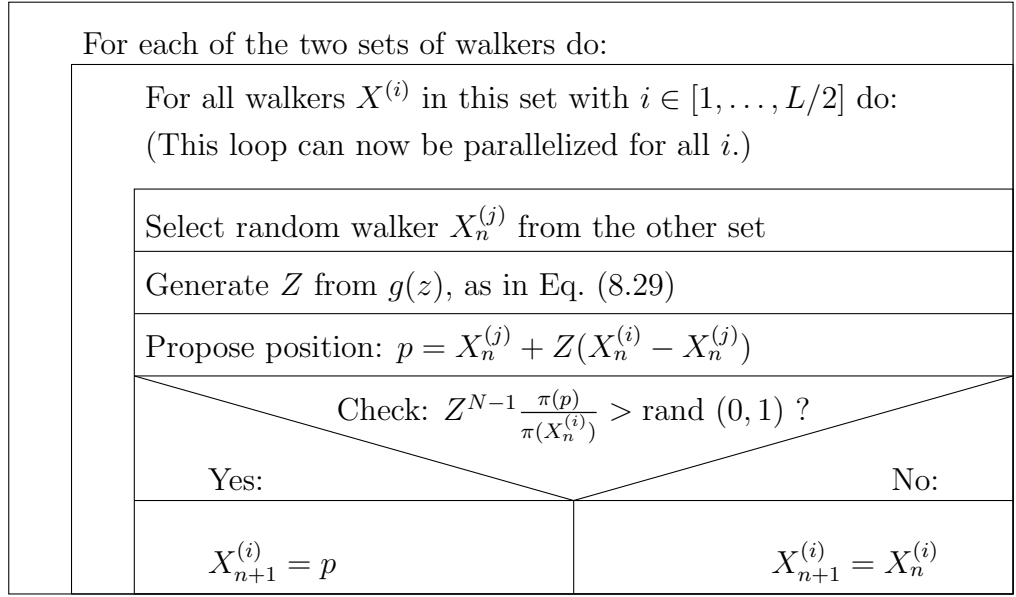
\begin{figure}[h]
  \begin{center}
    \begin{tikzpicture}
      \draw {(-2em,-2em) rectangle (30em,17em)};
      \draw(0,15em) node[anchor =south west]{For each of the two sets of walkers do:};
      \draw {(0,-2em) rectangle (30em,15em)};
      \draw {(2em,-2em) rectangle (30em, 11em)};
      \draw (2em,13em) node[anchor=south west]{For all walkers $X^{(i)}$ in this set with $i \in [1,\dots, L/2]$ do:};
      \draw (2em,11.5em) node[anchor=south west]{(This loop can now be parallelized for all $i$.)};
      \draw (2em,9em) node[anchor=south west]{Select random walker $X_{n}^{(j)}$ from the other set};
      \draw {(2em,9em) -- (30em,9em)};
      \draw (2em,7em) node[anchor=south west]{Generate $Z$ from $g(z)$, as in Eq.~\eqref{eq:8.4}};
      \draw {(2em,7em) -- (30em,7em)};
      \draw (2em,5em) node[anchor=south west]{Propose position: $p= X_{n}^{(j)} + Z (X_{n}^{(i)}-X_{n}^{(j)})$};
      \draw {(2em,5em) -- (30em,5em)};
      \draw (16em,2.5em) node[anchor=south]{Check: $Z^{N-1}\frac{\pi(p)}{\pi(X_{n}^{(i)})}> \text{rand }(0,1)$ ?};
      \draw {(2em,5em) -- (16em,1em)};
      \draw {(30em,5em) -- (16em,1em)};
      \draw {(16em,1em) -- (16em,-2em)};
      \draw {(2em,1em) -- (30em,1em)};
      \draw (4em,1em) node[anchor=south west]{Yes:};
      \draw (4em,-2em) node[anchor=south west]{$X^{(i)}_{n+1} = p$};
      \draw (28em,1em) node[anchor=south east]{No:};
      \draw (28em,-2em) node[anchor=south east]{$X^{(i)}_{n+1} = X^{(i)}_{n}$};
      \end{tikzpicture}
  \end{center}
  \caption[The algorithm of the program {\tt emcee}.]{The algorithm of the program {\tt emcee} \cite{2013PASP..125..306F}.}
  \label{fig:8.emcee}
\end{figure}

For our problem, the sampling of the posterior pdf for the parameters in Eq.~\eqref{eq:8.fit}, we use the {\tt python} code {\tt emcee}\cite{2013PASP..125..306F} that uses a variation of the Goodman-Weare algorithm \cite{Goodman:2009MCMC} called ``parallel stretch move''. It parallelizes the stretch move of the Goodman-Weare algorithm by simultaneously evolving the walkers based on the state of the other walkers. The computationally expensive loop through all walkers is now evaluated in parallel. However, this modification violates detailed balance \cite{2013PASP..125..306F}. In order to retain it, the set of walkers needs to be split into two disjoint sets. The walkers of set 1 are simultaneously evolved based on the state of the walkers in set 2 and {\it vice versa}. We give the resulting algorithm in the structogram in Fig.~\ref{fig:8.emcee}. 
\subsubsection{Testing the Quality of the Sampling}
There are several ways in which we can test the performance and convergence of an algorithm. The authors of \cite{2013PASP..125..306F} suggest the {\it autocorrelation time} $T_{\text{cor}}$ and {\it acceptance fraction} $a_{f}$ of the sample. The autocorrelation time measures after which time the correlation between two consecutive points in the chain vanishes. The correlation is defined as
\begin{equation}
  \label{eq:8.6}
  C_{f}(T) = \lim_{t \rightarrow \infty} \text{cov }[f(X(t+T)),f(X(t))]
\end{equation}
The autocorrelation time $T_{\text{cor}}$ is now defined as the time $T$ for which the correlation vanishes.
\begin{equation}
  \label{eq:8.7}
  \lim_{T\rightarrow T_{\text{cor}}} C_{f}(T) = 0
\end{equation}
It is an affine invariant measure of the performance and related to the number of evaluations of the posterior probability density required to produce independent samples of the target density. The autocorrelation time helps estimating for how many steps the Markov Chain Monte Carlo code should be evaluated. As a rule of thumb: 
\begin{equation}
  \label{eq:8.8}
  \text{Number of steps } \gg 10 \cdot T_{\text{cor}}
\end{equation}
The acceptance fraction $a_{f}$ is the ratio of accepted walks compared to all proposed walks. If it is too low  $(a_{f}\approx 0)$, the Markov Chain does not represent the target density, as almost all proposed steps are rejected and the chain contains only a few independent samples. A very high acceptance fraction $(a_{f}\approx 1)$ means that the chain performs a random walk without probing the target density. This does also not produce a representative sample. There seems to be no ideal acceptance fraction, but the authors of \cite{2013PASP..125..306F} recommend $0.2 \leq a_{f}\leq 0.5$ as a good range.
\section{Fit to LHC Higgs Data}
Many groups discussed fits of Higgs data \cite{Espinosa:2012ir,Azatov:2012bz,Azatov:2012rd,Falkowski:2013dza,Dumont:2013wma,Bernon:2014vta,deBlas:2014ula,Bergstrom:2014vla,Flament:2015wra,Corbett:2015ksa,Fichet:2015xla}. Here, we present the result of a fit to LHC data \cite{Buchalla:2015qju}, based on the electroweak chiral Lagrangian. For that purpose we first discuss our priors and the input we used to obtain the likelihood. Then, we present and discuss the posterior. 
\subsection{The Prior}
\begin{figure}[t]
    \centering
    \includegraphics[width=\textwidth]{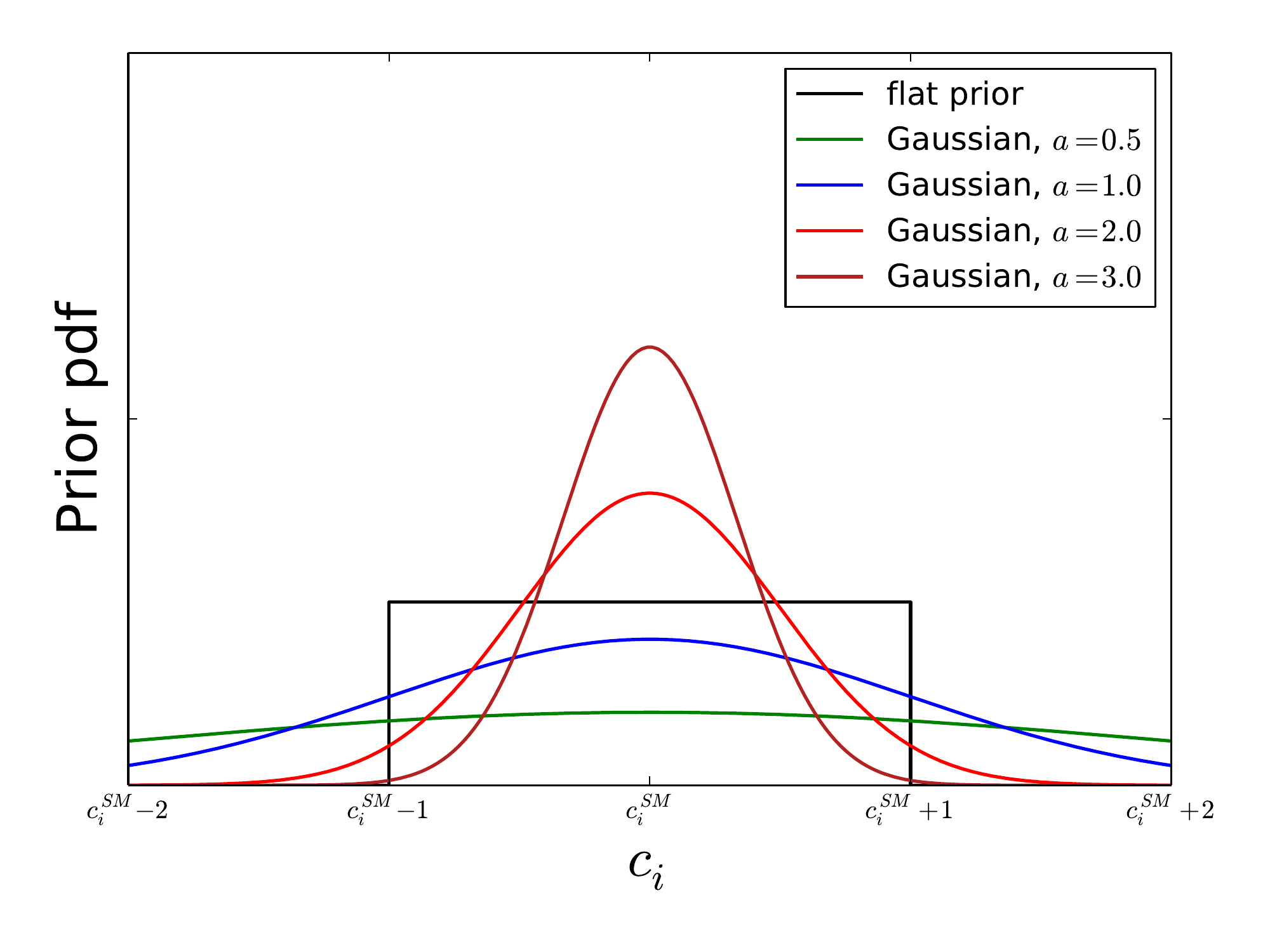}
    \caption[Different priors used for our analysis.]{The different priors used for our analysis.}
    \label{fig:prior}
  \end{figure}

Currently, only the decays to the fermions of the third generation as well as to $WW, ZZ $ and $\gamma\gamma$ are observed at the experiments. For $h\rightarrow \mu\mu$ and $h\rightarrow Z \gamma$, only upper bounds are reported. Since the contribution of the not measured couplings to the other observables is negligibly small, we do not fit them. This efficiently sets their prior to zero:
\begin{equation}
  \label{eq:8.29}
\prob{c_{i}}{I}=0, \hspace{2cm} \text{for } i\in \{e,\mu,u,d,c,s,Z\gamma\}
\end{equation}
This leaves a set of six parameters, $\{c_{V},c_{t},c_{b},c_{\tau},c_{\gamma\gamma},c_{gg}\}$ to be analyzed in the fit. We discuss the impact of extending the fit with $c_{\mu}$ and $c_{Z\gamma}$ later on. As already explained in Eq.~\eqref{eq:8.21}, we expect the size of the $c_{i}$ to be of the size $c_{i}^{\text{SM}}+\mathcal{O}(\xi)$. A first choice for a prior incorporating this information is a flat prior in the range $c_{i}^{\text{SM}}\pm 1$. The exact numerical value of the range is of course debatable, but the ``1'' is motivated by $\xi$ being of $\mathcal{O}(0.1)$ \cite{ATLAS-CONF-2015-044,CMS:2015kwa} and additional factors are of $\mathcal{O}(1)<10$. We therefore set the first prior to
\begin{equation}
  \label{eq:8.30}
  \prob{c_{i}}{I}_{\text{flat}}= \Theta(c_{i}-c_{i}^{\text{SM}}+1) \Theta(1+c_{i}^{\text{SM}}-c_{i}), \hspace{1cm} \text{for } i\in \{V,t,b,\tau,\gamma\gamma,gg\}
\end{equation}
In Section \ref{sec:prior}, we also discussed the use of Gaussian priors for cases, where the parameters are scattered around a mean value with a given variance. Since we assume the parameters are close to the SM with a certain width, related to $\xi$, we consequently use Gaussian priors. We quantified the width of the Gaussian in the following way: We assume that $c_{i}^{\text{SM}}\pm 1$ corresponds to $c_{i}^{\text{SM}}\pm a \sigma$ of the Gaussian. 
\begin{equation}
  \label{eq:8.31}
  \prob{c_{i}}{I}_{a}= \exp{\{-\sum\limits_{i}\frac{a^{2}(c_{i}^{\text{SM}}-c_{i})^{2}}{2}\}}, \hspace{1cm} \text{for } i\in \{V,t,b,\tau,\gamma\gamma,gg\}
\end{equation}
Figure \ref{fig:prior} visualizes the priors we use.
\subsection{The Likelihood {\it via} {\tt Lilith}}
We obtain the likelihood from the code {\tt Lilith-1.1.3}, as described in Section \ref{sec:lilith}. We take the branching ratios of the SM, needed in Eq.~\eqref{eq:8.23}, from \cite{Heinemeyer:2013tqa} for $m_{h}= 125~\text{GeV}$. We use them at the highest available order in QCD, see \cite{Dittmaier:2012vm}. For reducing the effects of higher order QCD corrections \cite{Contino:2014aaa} we use the pole masses $m_{t}^{\text{pole}}= 172.5~\text{GeV}$ and $m_{b}^{\text{pole}}= 4.75~\text{GeV}$ in the $gg\rightarrow h$ amplitude and the running masses\footnote{The running mass does not coincide with the $\overline{\text{MS}}$-mass \cite{Djouadi:1993ji,Spira:1995rr}.} $m_{t}= 188.03~\text{GeV}$ and $m_{b}= 3.44~\text{GeV}$ in the $h\rightarrow \gamma\gamma$ amplitude, see \cite{Contino:2014aaa}. We use the strong coupling constant $\alpha_{s}(m_{Z})=0.1185$. We take the experimental input from a modified version of the {\tt Lilith DB 15.09}, using the combined ATLAS and CMS result of Run-1 \cite{ATLAS-CONF-2015-044,CMS:2015kwa} plus ttH-data from \cite{Khachatryan:2015ila,Khachatryan:2014jba,Khachatryan:2014qaa,Aad:2015iha,Aad:2014eha,Aad:2015gra} that was not resolved in the combined analysis, as well as the VH dataset from the Tevatron \cite{Aaltonen:2013ioz}.
\subsection{The Posterior --- the Result}
\label{sec:fit}
\begin{figure}[t]
    \centering
    \includegraphics[width=\textwidth]{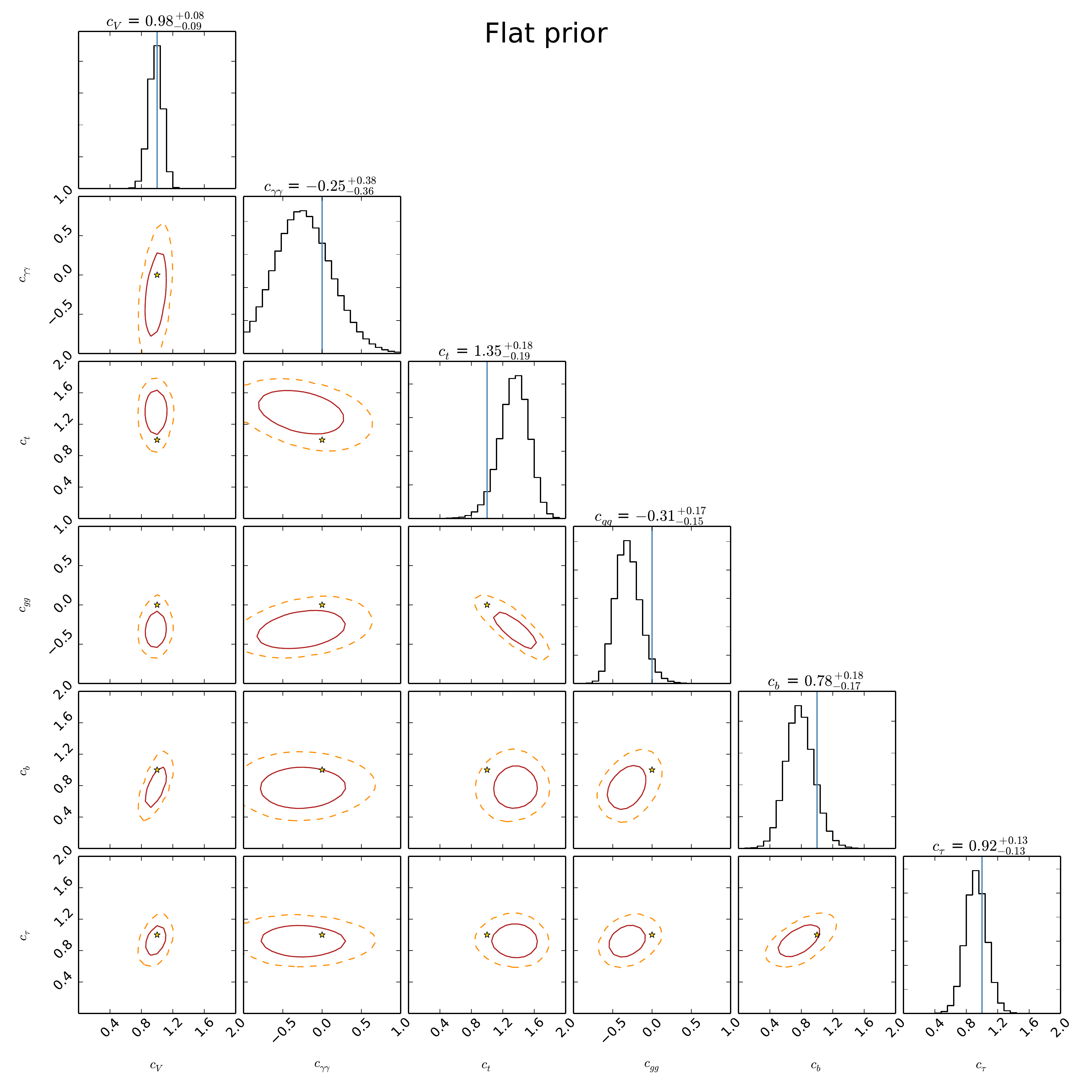}
    \caption[Cornerplot of the posterior for $\prob{c_{i}}{I}_{\text{flat}}$.]{Posterior probability density corner plot for the flat prior of Eq.~\eqref{eq:8.30}. The red solid (yellow dashed) contours indicate the $68\%$ ($95\%$) credibility intervals, the star is at the SM point. The mean values and their correlations, computed from Eqs.~\eqref{eq:8.27},\eqref{eq:8.28} are below. We discuss their relation to the parameters given in the plot in the main text.\\$ $\\ \mbox{\hspace{-2cm}$\begin{pmatrix} c_V \\  c_t  \\ c_b\\  c_\tau\\ c_{\gamma\gamma} \\c_{gg} \end{pmatrix}=\begin{pmatrix} 0.98  \pm 0.09\\ 1.34  \pm 0.19\\ 0.79  \pm 0.18\\0.92  \pm 0.14\\ -0.24  \pm 0.37\\ -0.30  \pm 0.16\end{pmatrix},\hspace{1.5cm} \rho = \begin{pmatrix} 1.0 &  0.01 &  0.67 &  0.37 &  0.41 &  0.10 \\ .  &  1.0  &  0.02 & -0.04 & -0.36 & -0.84 \\ .  &   .   &  1.0  &  0.58 &  0.02 &  0.37 \\ .  &   .   &   .   &  1.0  & -0.05 &  0.26 \\ .  &   .   &   .   &   .   &  1.0  &  0.31 \\ .  &   .   &   .   &   .   &   .   &  1.0\end{pmatrix}$}}
    \label{fig:8.res1}
  \end{figure}

\begin{figure}[t]
    \centering
    \includegraphics[width=\textwidth]{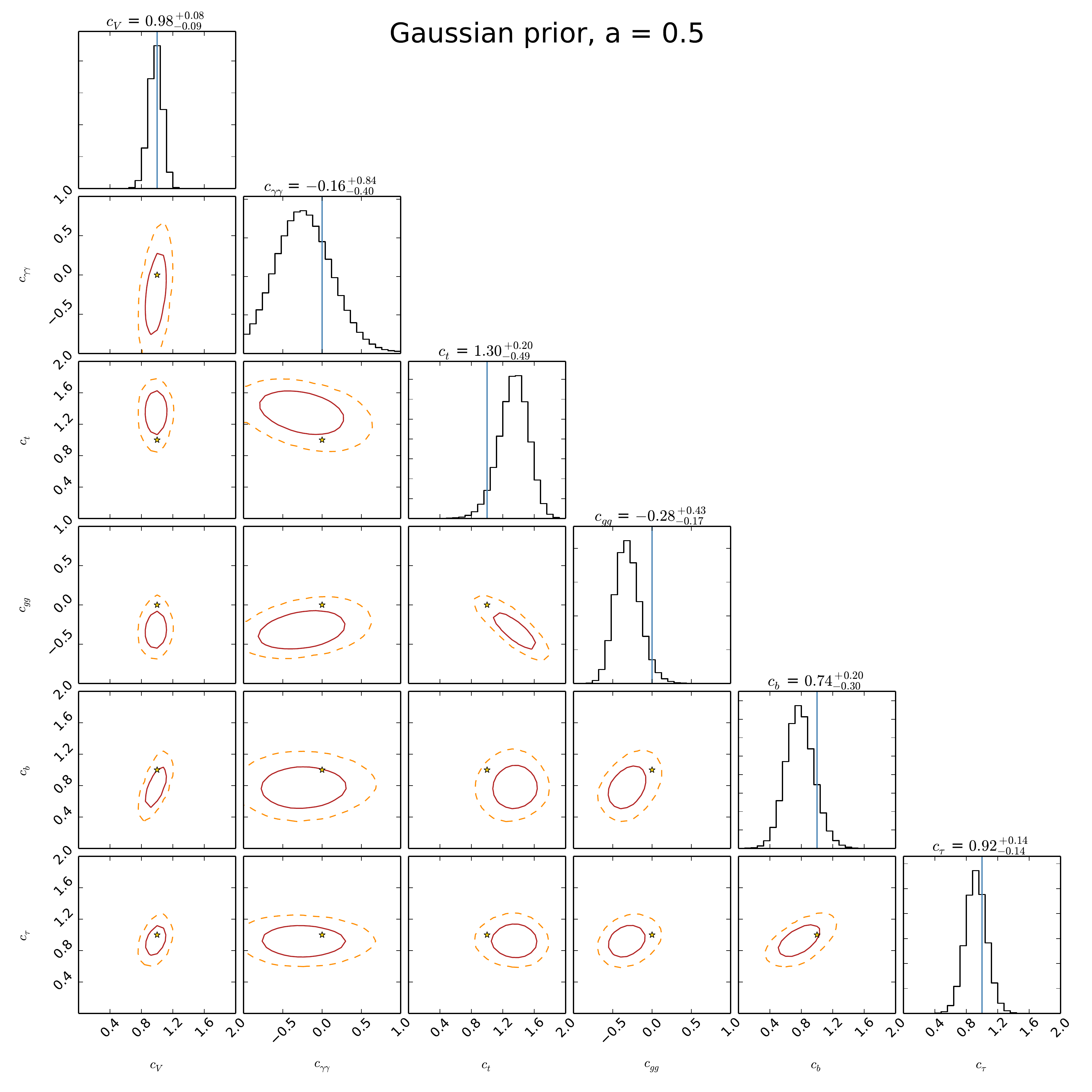}
    \caption[Cornerplot of the posterior for $\prob{c_{i}}{I}_{a = 0.5}$.]{Posterior probability density corner plot for the Gaussian prior of Eq.~\eqref{eq:8.31} with $a=0.5$. The red solid (yellow dashed) contours indicate the $68\%$ ($95\%$) credibility intervals, the star is at the SM point. The mean values and their correlations, computed from Eqs.~\eqref{eq:8.27},\eqref{eq:8.28} are below. We discuss their relation to the parameters given in the plot in the main text.\\$ $\\ \mbox{\hspace{-2cm}$\begin{pmatrix} c_V \\  c_t  \\ c_b\\  c_\tau\\ c_{\gamma\gamma} \\c_{gg} \end{pmatrix}=\begin{pmatrix} 0.98 \pm 0.09 \\ 0.93 \pm 0.96 \\ 0.57 \pm 0.56 \\ 0.85 \pm 0.38 \\ 0.1  \pm 0.88 \\ 0.0  \pm 0.73 \\ \end{pmatrix}\hspace{1.5cm} \rho = \begin{pmatrix} 1.0  &  0.05  &   0.18  &   0.13  &   0.13  &  -0.02 \\  .    &  1.0    & -0.03  & -0.06  & -0.92  &  -0.99 \\  .    &    .      &   1.0    &   0.0    &   0.04  &    0.09 \\  .    &    .      &     .      &   1.0    &   0.05  &    0.07 \\  .    &    .      &     .      &     .      &   1.0    &    0.91 \\  .    &    .      &     .      &     .      &     .      &    1.0\end{pmatrix}$}}
    \label{fig:8.res2}
  \end{figure}

\begin{figure}[t]
    \centering
    \includegraphics[width=\textwidth]{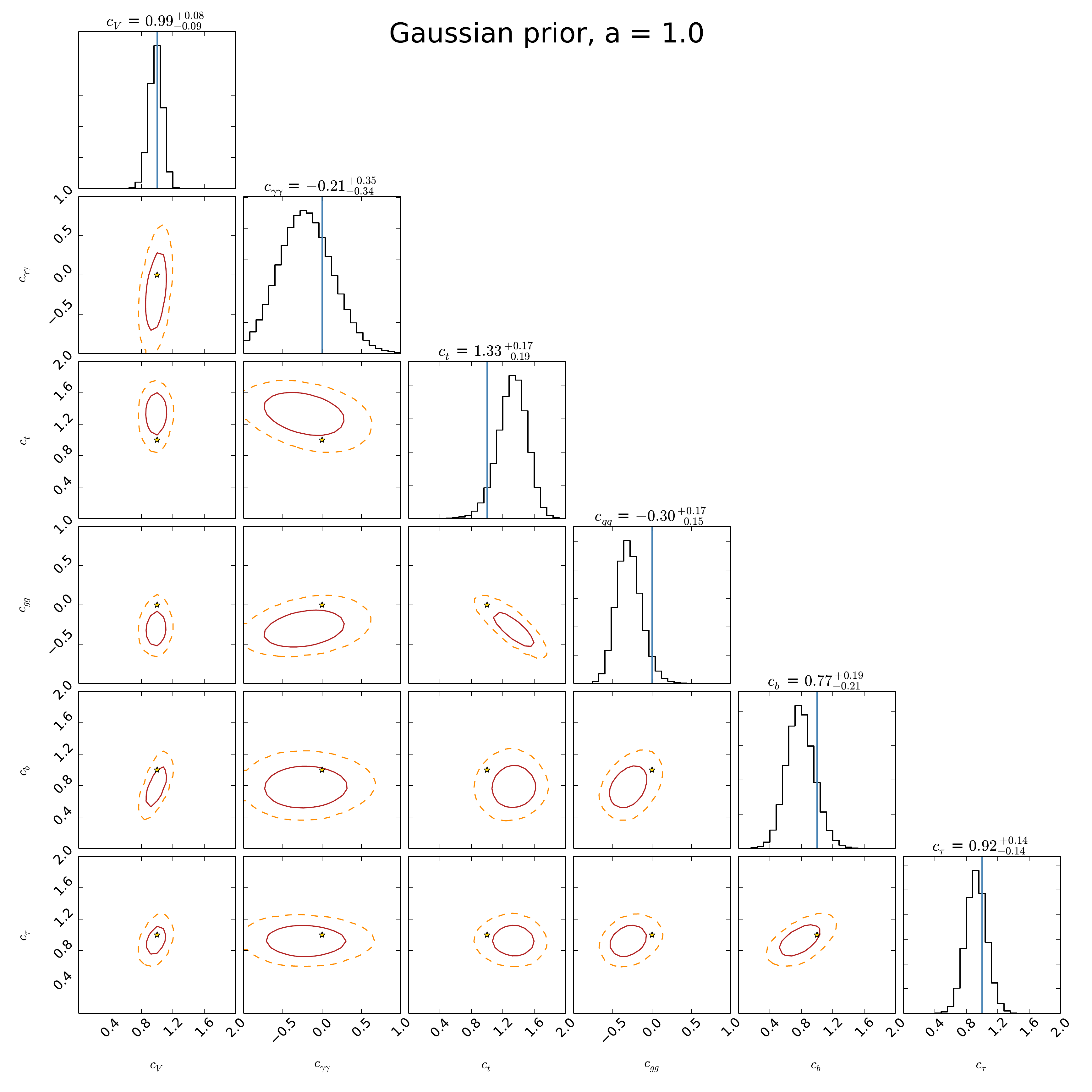}
    \caption[Cornerplot of the posterior for $\prob{c_{i}}{I}_{a = 1}$.]{Posterior probability density corner plot for the Gaussian prior of Eq.~\eqref{eq:8.31} with $a=1$. The red solid (yellow dashed) contours indicate the $68\%$ ($95\%$) credibility intervals, the star is at the SM point. The mean values and their correlations, computed from Eqs.~\eqref{eq:8.27},\eqref{eq:8.28} are below. We discuss their relation to the parameters given in the plot in the main text.\\$ $\\ \mbox{\hspace{-2cm}$\begin{pmatrix} c_V \\  c_t  \\ c_b\\  c_\tau\\ c_{\gamma\gamma} \\c_{gg} \end{pmatrix}=\begin{pmatrix}   0.98  \pm 0.08 \\  1.31  \pm 0.24 \\  0.67  \pm 0.44 \\  0.85  \pm 0.38 \\ -0.20  \pm 0.37 \\-0.28  \pm 0.20 \\\end{pmatrix}\hspace{1.5cm} \rho = \begin{pmatrix} 1.0  &  0.04  &  0.29  &  0.13  &   0.37  &  0.06 \\  .    &  1.0    &  0.02  &  0.0    & -0.45  &-0.89 \\  .    &    .      &  1.0    &  0.05  &   0.02  &  0.19 \\  .    &    .      &    .      &  1.0    & -0.01  &  0.07 \\  .    &    .      &    .      &    .      &   1.0    &  0.41 \\  .    &    .      &    .      &    .      &     .      &  1.0\end{pmatrix}$}}
    \label{fig:8.res3}
  \end{figure}

\begin{figure}[t]
    \centering
    \includegraphics[width=\textwidth]{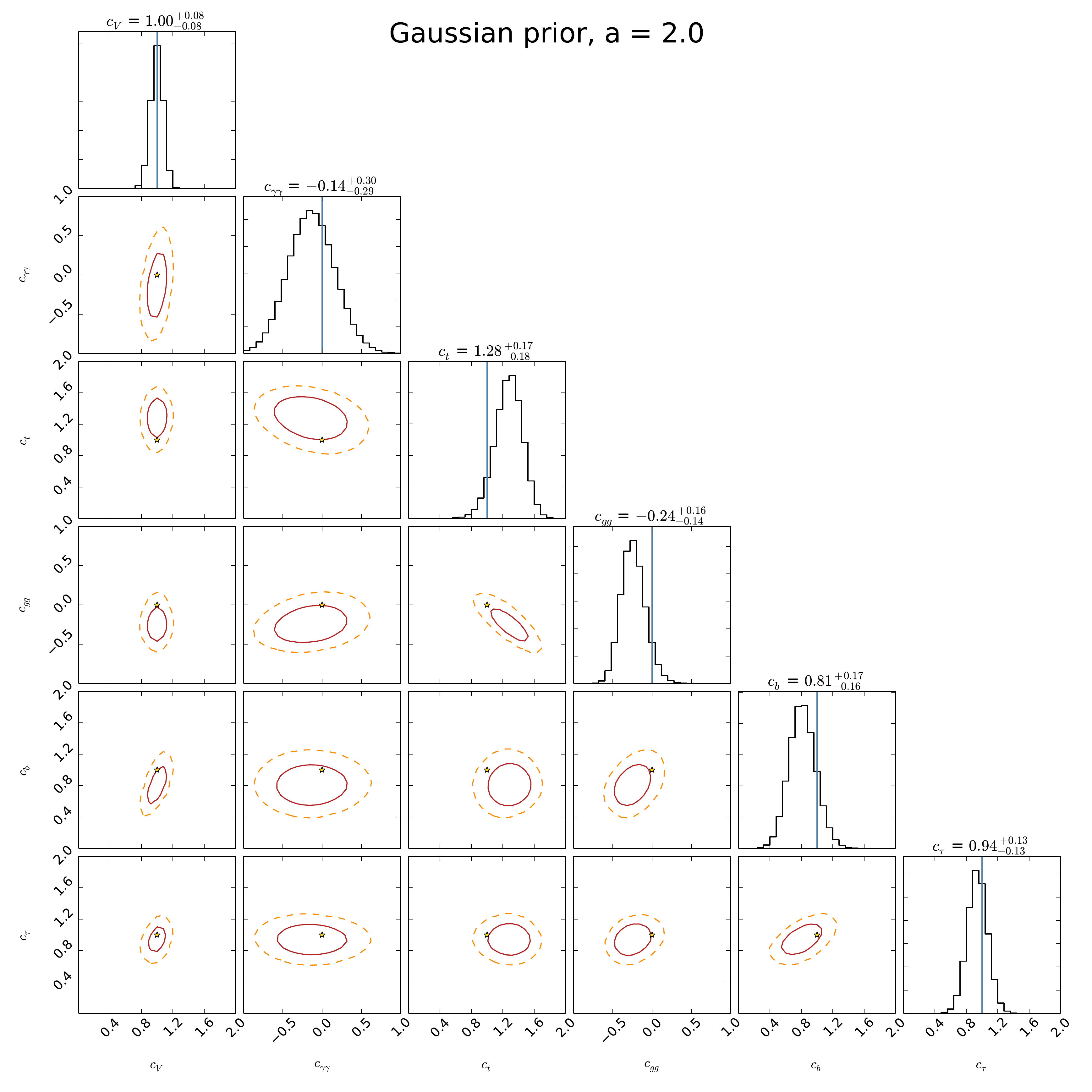}
    \caption[Cornerplot of the posterior for $\prob{c_{i}}{I}_{a = 2}$.]{Posterior probability density corner plot for the Gaussian prior of Eq.~\eqref{eq:8.31} with $a=2$. The red solid (yellow dashed) contours indicate the $68\%$ ($95\%$) credibility intervals, the star is at the SM point. The mean values and their correlations, computed from Eqs.~\eqref{eq:8.27},\eqref{eq:8.28} are below. We discuss their relation to the parameters given in the plot in the main text.\\$ $\\ \mbox{\hspace{-2cm}$\begin{pmatrix} c_V \\  c_t  \\ c_b\\  c_\tau\\ c_{\gamma\gamma} \\c_{gg} \end{pmatrix}=\begin{pmatrix}   1.0    \pm 0.08 \\  1.27  \pm 0.18 \\  0.82  \pm 0.17 \\  0.94  \pm 0.14 \\-0.13  \pm 0.30 \\-0.24  \pm 0.15 \\\end{pmatrix}\hspace{1.5cm} \rho = \begin{pmatrix} 1.0  &  0.07 &  0.65  &   0.31  &   0.35  &   0.04 \\  .    &  1.0   &  0.04  & -0.03  & -0.29  & -0.82 \\  .    &    .     &  1.0    &   0.50  &   0.02  &   0.37 \\  .    &    .     &    .      &   1.0    & -0.04  &   0.23 \\  .    &    .     &    .      &     .      &   1.0    &   0.24 \\  .    &    .     &    .      &     .      &     .      &   1.0\end{pmatrix}$}}
    \label{fig:8.res4}
  \end{figure}

\begin{figure}[t]
    \centering
    \includegraphics[width=\textwidth]{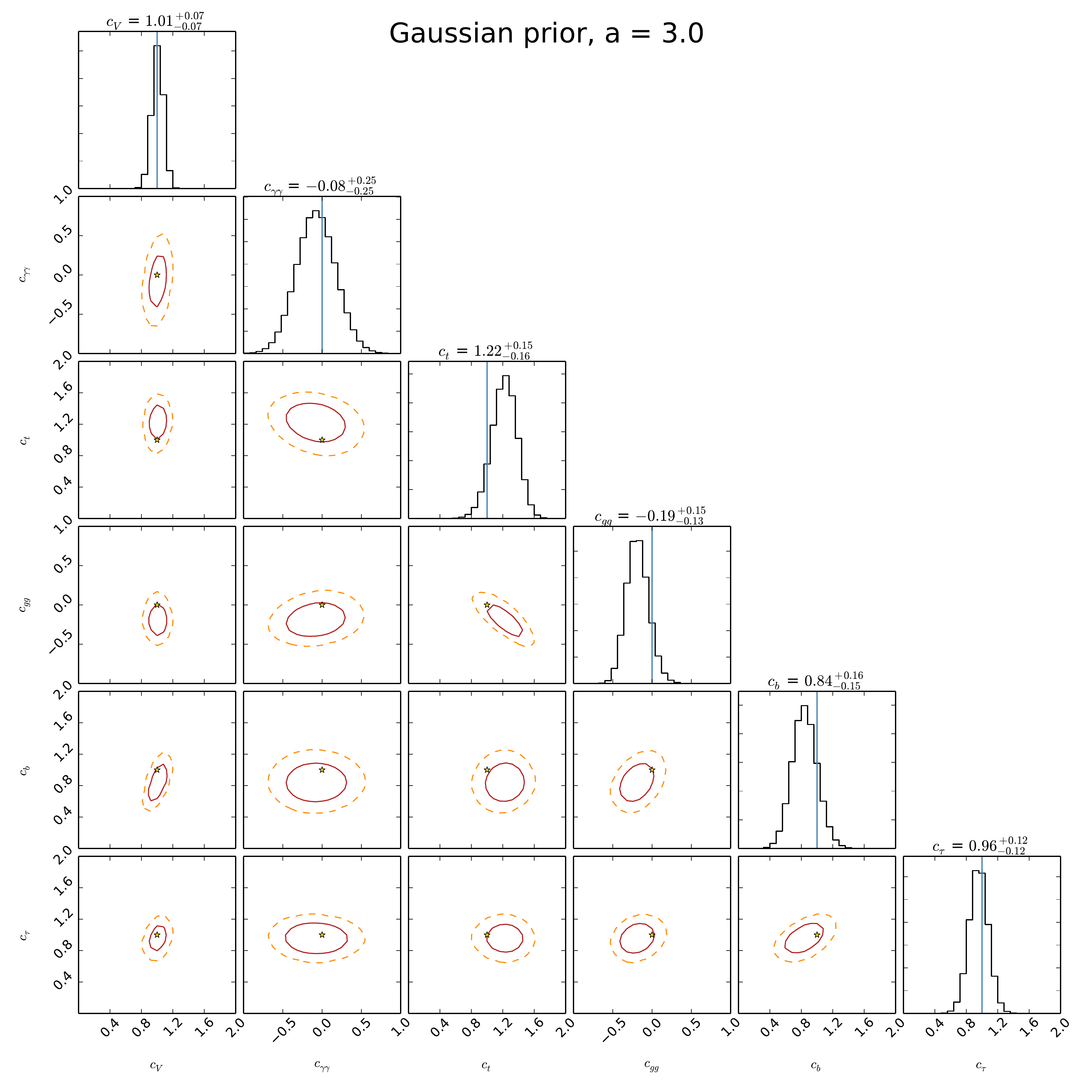}
    \caption[Cornerplot of the posterior for $\prob{c_{i}}{I}_{a = 3}$.]{Posterior probability density corner plot for the Gaussian prior of Eq.~\eqref{eq:8.31} with $a=3$. The red solid (yellow dashed) contours indicate the $68\%$ ($95\%$) credibility intervals, the star is at the SM point. The mean values and their correlations, computed from Eqs.~\eqref{eq:8.27},\eqref{eq:8.28} are below. We discuss their relation to the parameters given in the plot in the main text.\\$ $\\ \mbox{\hspace{-2cm}$\begin{pmatrix} c_V \\  c_t  \\ c_b\\  c_\tau\\ c_{\gamma\gamma} \\c_{gg} \end{pmatrix}=\begin{pmatrix}   1.01  \pm 0.07 \\  1.22  \pm 0.16 \\  0.85  \pm 0.16 \\  0.96  \pm 0.12 \\-0.07  \pm 0.25 \\-0.18  \pm 0.14 \\\end{pmatrix}\hspace{1.5cm} \rho = \begin{pmatrix} 1.0  &  0.1  &  0.61  &   0.3    &   0.31  & -0.02 \\  .    &  1.0  &  0.05  & -0.02  & -0.23  & -0.8   \\  .    &    .    &  1.0    &   0.51  &   0.0    &   0.38 \\  .    &    .    &    .      &   1.0    & -0.05  &   0.23 \\   .    &    .    &    .      &     .      &   1.0    &   0.19 \\  .    &    .    &    .      &     .      &     .      &   1.0   \\\end{pmatrix}$}}
    \label{fig:8.res5}
  \end{figure}

We find the posterior with the MCMC code {\tt emcee} \cite{2013PASP..125..306F} for 500 walkers and 50,000 steps each. We choose the initial points at random around the SM point and then ``burn them in'' for 10,000 additional steps. Figures \ref{fig:8.res1}--\ref{fig:8.res5} give the results, where we produced the corner plots using the code {\tt corner.py} \cite{Wesolowski:2015fqa,triangle.py}. We obtain the mean values and correlation matrices in the caption of the figures as discussed in Eqs.~\eqref{eq:8.27} and \eqref{eq:8.28}. The mean values and asymmetric errors given in the corner plot are based on the $68\%$ quantiles of the shown histograms of the final distribution. For symmetric, single-modal Gaussian distributions, they should agree with the numbers of the figure captions. 

The quality of the MCMC sampling is good, as the autocorrelation time is $T_{\text{cor}} = 80$ and the acceptance fraction is $a_{f}= 0.5$ for the flat prior. For the Gaussian priors, we have $60<T_{\text{cor}} < 100$ and $a_{f}\approx 0.4$.

We see that the data are compatible with the SM at one to two standard deviations, independent of the chosen prior. However, deviations of $\mathcal{O}(10\%)$ are still allowed, as the errors are large. This is somehow expected, as the analyses of the experimental collaborations \cite{ATLAS-CONF-2015-044,CMS:2015kwa} is based on the same data. In the results we see a particularly strong correlation between $c_{gg}$ and $c_{t}$. This comes from the fact that the bounds on $c_{t}$ from ttH production are still loose compared to ggF, where both $c_{gg}$ and $c_{t}$ enter, see Eq.~\eqref{eq:8.22}. The correlation between $c_{V}$ and $c_{b}$ is of experimental origin, as the decay $h\rightarrow \bar{b}b$ can only be measured in associated production with vector bosons (VH). In other production modes the QCD $\bar{b}b$ background is too large. 

We also see the influence of the different priors. For increasing $a$, the best fit point moves closer to the SM. For $a=0.5$, we see large differences between the mean values computed by {\tt corner.py} and with Eq.~\eqref{eq:8.27}. This is because the prior is non-vanishing at $c_{i} = - c_{i}^{\text{SM}}$. These points come with a high likelihood as Eqs.~\eqref{eq:8.22} and \eqref{eq:8.24} are invariant under a global flip of all signs. In this case, the posterior has two distinct maxima and the analysis in terms of Eqs.~\eqref{eq:8.27} and \eqref{eq:8.28} cannot give reliable results. However, from power counting we do not expect the $c_{i}$ to have such values. The Gaussian prior with $a=0.5$ is therefore probably poorly chosen. A more reliable choice of priors from this point of view is therefore $a=1$ or $a=2$. As already said, they give very similar results compared to each other and the flat prior, showing that the result is likelihood- and therefore data-dominated. 

Further cross-checks were made regarding the impact of QCD corrections. For that purpose, $\eta^{t}_{\text{QCD}}$ was set to one and its value at $\mathcal{O}(\alpha_{s})$ accuracy. The changes in the fit result are only at the percent level. Also, setting $\eta^{b}_{\text{QCD}} = 2$ changes the result only slightly. 

The best fit results can also be translated to the $\kappa$ framework and compared to the results published in \cite{ATLAS-CONF-2015-044}. This should be equal to the result of our fit using a flat prior. Unfortunately, \cite{ATLAS-CONF-2015-044} did not assume custodial symmetry. $\kappa_{Z}$ and $\kappa_{W}$ were fitted separately. This makes a direct comparison slightly more complicated. To see the difference between fitting the $c_{i}$ and fitting the $\kappa_{i}$, we perform the fit also in the parametrization of the $\kappa_{i}$. When neglecting the imaginary parts of the loop functions in Eq.~\eqref{eq:8.24}, we find a linear transformation between the two parameter sets. The obtained results agree within the error bars, but neglecting the imaginary parts introduces small differences. 

In the last paragraph, we address the stability of the fit upon extending it with the so far neglected parameters $c_{i}$, for $i\in \{e,\mu,u,d,c,s,Z\gamma\}$. For the parameters that are not constrained by data at all $(i\in \{e,u,d,c,s\})$ the posterior only consists of the prior. Putting an unreasonably large prior, or no prior at all, leads to overfitting \cite{Wesolowski:2015fqa}. The fit is driven to a very unnatural region of parameter space and the resulting marginalized posterior pdf of $c_{i}$ $(i\in \{V,t,b,\tau,\gamma\gamma,gg\})$ is different, {\it i.e.} the fit is unstable. A reasonable prior, based on power counting, avoids this and gives a posterior similar to the previously obtained one. For $c_{\mu}$ and $c_{Z\gamma}$, the situation is different. Data give an upper bound of a few times the SM value on the corresponding observables, effectively only allowing natural values for $c_{\mu}$ and $c_{Z\gamma}$. The marginalized posteriors for $c_{i}$ $(i\in \{V,t,b,\tau,\gamma\gamma,gg\})$ are close to the ones reported before and the fit is stable. However, in all cases, no additional information apart from the prior can be extracted for the unconstrained $c_{i}$. 
\part{Conclusion}
  \chapter{Conclusions and Outlook}
\thispagestyle{fancyplain}
The LHC has confirmed the SM to very good accuracy at its Run-1, from 2010 to 2013. So far, no direct signs for new physics have been observed. This absence of new particles up to the TeV-scale indicates a mass gap, which is an essential ingredient for effective field theories. Applied in the bottom-up approach, EFTs provide a systematic and model-independent framework to look for new physics. Therefore, they are a powerful tool for the current phase of the LHC. 

We discussed the EFTs for physics beyond the SM, emphasizing the systematics and the assumptions behind the expansions. Based on the underlying dynamics of the new physics, we distinguished two types of effective field theories: decoupling and non-decoupling EFTs. Motivated by the experimental situation, in which we have uncertainties in the Higgs couplings of ten percent or more on the one hand \cite{ATLAS-CONF-2015-044,CMS:2015kwa}, and the electroweak precision observables of the LEP experiment at the percent level on the other hand \cite{ALEPH:2005ab,Barbieri:2004qk,ALEPH:2010aa}, we studied the electroweak chiral Lagrangian as a non-decoupling EFT. The electroweak chiral Lagrangian describes large, {\it i.e.} leading order, effects in the Higgs couplings, whereas corrections to the gauge-fermion interactions arise at next-to-leading order. It is therefore the natural framework to test the SM-Higgs hypothesis effectively.\\

The power counting is essential for finding the complete basis of operators in bottom-up EFTs, which is crucial for a consistent analysis. We studied the power counting of non-decoupling EFTs and derived that they are expanded in chiral dimensions, a concept based on a loop expansion \cite{Buchalla:2013eza}. Using the chiral dimensions, we explicitly constructed the operators of the electroweak chiral Lagrangian at next-to-leading order \cite{mastersthesis,Buchalla:2013rka}. We also discussed how different assumptions on custodial symmetry assign weak couplings, and therefore chiral dimensions, to the spurions of custodial symmetry breaking \cite{Buchalla:2014eca}. We discussed how this reduces the number of operators in the basis at next-to-leading order.

Using the electroweak chiral Lagrangian, we gave a quantum field the\-o\-ret\-i\-cal justification of the $\kappa$-formalism \cite{Buchalla:2015wfa}. The $\kappa$-formalism was introduced as interim pa\-ra\-me\-triza\-tion of Higgs couplings \cite{LHCHiggsCrossSectionWorkingGroup:2012nn,Heinemeyer:2013tqa}. Having an EFT at the core of the analysis framework allows us not only to find out if we observe the SM-Higgs or not --- it also allows us to systematically interpret deviations from the SM. Further, we can consistently include radiative QCD corrections in the leading-order analysis. Also, the extension of the $\kappa$-formalism is straightforward and well-defined within the electroweak chiral Lagrangian \cite{Buchalla:2015wfa,Buchalla:2015qju}. Such an extension is currently discussed in the LHC Higgs cross section working group, as it is possible to measure new observables, like kinematical distributions, with an increasing amount of data in the near future. 

Chiral dimensions have a wide range of applications, as they describe the expansion of effective theories, where a strongly-coupled Goldstone sector is weakly coupled to external fermions and gauge fields. We showed the application of chiral dimensions in chiral perturbation theory coupled to QED \cite{Buchalla:2013eza,Buchalla:2016sop} and the minimal composite Higgs model based on the coset $SO(5)/SO(4)$ \cite{Buchalla:2014eca}. The extension of chiral perturbation theory and the electroweak chiral Lagrangian by resonances of the strong sector also uses chiral dimensions as a tool for power counting \cite{Ecker:1988te,Ecker:1989yg,Cirigliano:2006hb,Santos:2015mqa}. \\

We further discussed the power counting if the scale of the non-decoupling physics, $f$, is above the weak scale, $v$. The EFT becomes a double expansion in canonical dimensions ({\it i.e.} factors of $\xi = v^{2}/f^{2}$) and chiral dimensions ({\it i.e.} loop-factors of $1/16\pi^{2}$) \cite{Buchalla:2014eca} in this case. 

For moderate values of $\xi$, $1/16\pi^{2}\lesssim \xi\lesssim 1$, this is an appealing framework for the ongoing analyses at the LHC: Current phenomenology indicates that the scale of new physics is above the electroweak-scale, but non-decoupling models give a natural solution to the hierarchy problem. Using this information, we performed a Bayesian fit \cite{Buchalla:2015qju} of the leading EFT effects to the LHC-Higgs data of Run-1. The Bayesian framework is perfect for the EFT based analysis. We used the information from the power counting as prior information for the parameters. This reduces the risk of over- and underfitting. Based on the list of observed processes, the number of free parameters reduced to a manageable set of six. We found that the SM describes the data within one to two standard deviations. This corresponds to an uncertainty in the Higgs couplings of ten to twenty percent. Statistical uncertainties dominate in the data of Run-1. Future runs will increase the amount of available data, reducing the statistical uncertainties and therefore the uncertainties in the Higgs couplings. The electroweak chiral Lagrangian can also easily incorporate additional observables, like further decay channels. In this case, the number of free parameters increases slightly. 

The double expansion also helps to understand the relation between the decoupling SM-EFT and the non-decoupling electroweak chiral Lagrangian: In the limit of $\xi\lesssim 1/16\pi^{2}$, the expansion approaches the SM-EFT; for $\xi=\mathcal{O}(1)$, the electroweak chiral Lagrangian. We illustrated this explicitly using a renormalizable model, the SM extended with a scalar singlet \cite{Buchalla:2016bse}. Also in this toy-model, the parameter $\xi$ interpolates continuously between the different EFT regimes. Regions of parameter space with rather strong coupling induce the non-decoupling EFT, whereas regions with weak couplings induce the decoupling EFT. This observation is also true for other models that are analyzed at the LHC. 

In general, a double expansion of this type describes the EFT of a non-decoupling sector that decouples from the low-energy scale. This insight is not only applicable in the electroweak-symmetry-breaking sector of the SM, but also in other scenarios where decoupling and non-decoupling sectors interact. \\

Our analysis showed that using only the SM-EFT is not the best framework for the current analyses at the LHC. A consistent treatment to describe deviations from the SM within the SM-EFT requires more free parameters and the analysis loses its statistical significance. In addition, all effects beyond the SM start to contribute in the SM-EFT at next-to-leading order and there is no distinction between the Higgs and the gauge sector. We therefore expect the various effects of new physics to be of comparable size, $\mathcal{O}(v^{2}/\Lambda^{2})$, and we cannot test the SM-Higgs hypothesis as efficient as with the chiral Lagrangian. We also showed that the decoupling SM-EFT describes the low-energy effects of some UV-completions only in parts of the parameter space. In any case, the electroweak chiral Lagrangian describes large effects in the Higgs couplings. If the data of future runs of the LHC reveals only deviations at the percent level or below, the SM-EFT will be more appropriate to use. \\

To summarize, the electroweak chiral Lagrangian is a suitable framework for the current stage of LHC physics. The power counting that we introduced for the electroweak chiral Lagrangian is crucial for a consistent application and interpretation of the EFT in data analysis. Further steps in the development of the electroweak chiral Lagrangian are therefore necessary. One of these steps is a study of the complete one-loop renormalization. This computation will confirm the completeness of the operator basis and provide the running of the Wilson coefficients. The extension of the observables to kinematic distributions requires an extension of the fits to next-to-leading order in the EFT. Even if the size of these effects are, by power counting, expected to be small, it is another important step in the EFT application to LHC data.\\

The upcoming runs of the LHC will collect more data and help to shed light on the dynamics behind electroweak symmetry breaking. If no new states are directly observed, Michelson's pessimistic statement, {\it``the future truths of physical science are to be looked for in the sixth place of decimals''} \cite{Flam:1992jx,university1896annual, wikiquote}, seems to become true. These measurements, however, will reveal indirect signs of new physics and not just more and more digits of fundamental constants. By using effective field theories we will then be able to infer {\it``the future truths of physical science.''}

\cleardoublepage\makeatletter\@openrightfalse\makeatother 

\part{Appendix}
  \appendix
  
  \let\backupskip\chapterheadstartvskip
\let\backupskipp\chapterheadendvskip
\renewcommand*{\chapterheadstartvskip}{\vspace*{-0.5cm}}
\renewcommand*{\chapterheadendvskip}{\vspace{0.5cm}}

\chapter{Generators of $SO(5)$}
\label{ch:appA}
\thispagestyle{fancyplain}
The $SO(5)$ symmetry has ten generators. Six of them are also generators of an $SO(4)$ subgroup. Since $SO(4)$ is isomorphic to $SU(2)\times SU(2)$, we decompose the six generators into the generators of the $SU(2)_{L}$- and $SU(2)_{R}$-subgroup they generate. In the fundamental representation, the generators are \cite{Contino:2010rs}
\begin{equation}
\label{eq:A.1}
\begin{array}{lr}
  t^{1}_{L} = -\frac{i}{2}  \begin{pmatrix} 0 & 0 & 0 & 1 & 0 \\ 0 & 0 & 1 & 0 & 0 \\ 0 & -1 & 0 & 0 & 0 \\ -1 & 0 & 0 & 0 & 0 \\0 & 0 & 0 & 0 & 0 \\ \end{pmatrix}, & 
  t^{1}_{R} = -\frac{i}{2} \begin{pmatrix} 0 & 0 & 0 & -1 & 0 \\ 0 & 0 & 1 & 0 & 0 \\ 0 & -1 & 0 & 0 & 0 \\ 1 & 0 & 0 & 0 & 0 \\ 0 & 0 & 0 & 0 & 0 \\ \end{pmatrix}, \\
  \\
  t^{2}_{L} = -\frac{i}{2}  \begin{pmatrix} 0 & 0 & -1 & 0 & 0 \\ 0 & 0 & 0 & 1 & 0 \\ 1 & 0 & 0 & 0 & 0 \\ 0 & -1 & 0 & 0 & 0 \\ 0 & 0 & 0 & 0 & 0 \\ \end{pmatrix}, &
  t^{2}_{R} = -\frac{i}{2} \begin{pmatrix} 0 & 0 & -1 & 0 & 0 \\ 0 & 0 & 0 & -1 & 0 \\ 1 & 0 & 0 & 0 & 0 \\ 0 & 1 & 0 & 0 & 0 \\ 0 & 0 & 0 & 0 & 0 \\ \end{pmatrix},\\
  \\
  t^{3}_{L} = -\frac{i}{2}  \begin{pmatrix} 0 & 1 & 0 & 0 & 0 \\ -1 & 0 & 0 & 0 & 0 \\ 0 & 0 & 0 & 1 & 0 \\ 0 & 0 & -1 & 0 & 0 \\ 0 & 0 & 0 & 0 & 0 \\ \end{pmatrix}, &
  t^{3}_{R} = -\frac{i}{2} \begin{pmatrix} 0 & 1 & 0 & 0 & 0 \\ -1 & 0 & 0 & 0 & 0 \\ 0 & 0 & 0 & -1 & 0 \\ 0 & 0 & 1 & 0 & 0 \\ 0 & 0 & 0 & 0 & 0 \\ \end{pmatrix}. 
\end{array}
\end{equation}
The remaining four generators span the coset $SO(5)/SO(4)$. They are
\begin{equation}
\label{eq:A.2}
\begin{array}{lr}
  t^{\hat{1}} = -\frac{i }{\sqrt{2}}\begin{pmatrix} 0 & 0 & 0 & 0 & 1 \\ 0 & 0 & 0 & 0 & 0 \\ 0 & 0 & 0 & 0 & 0 \\ 0 & 0 & 0 & 0 & 0 \\ -1 & 0 & 0 & 0 & 0 \\\end{pmatrix}, &
  t^{\hat{2}} = -\frac{i }{\sqrt{2}}\begin{pmatrix} 0 & 0 & 0 & 0 & 0 \\ 0 & 0 & 0 & 0 & 1 \\ 0 & 0 & 0 & 0 & 0 \\ 0 & 0 & 0 & 0 & 0 \\ 0 & -1 & 0 & 0 & 0 \\\end{pmatrix}, \\
  \\
  t^{\hat{3}} = -\frac{i }{\sqrt{2}}\begin{pmatrix} 0 & 0 & 0 & 0 & 0 \\ 0 & 0 & 0 & 0 & 0 \\ 0 & 0 & 0 & 0 & 1 \\ 0 & 0 & 0 & 0 & 0 \\ 0 & 0 & -1 & 0 & 0 \\\end{pmatrix}, &
  t^{\hat{4}} = -\frac{i }{\sqrt{2}}\begin{pmatrix} 0 & 0 & 0 & 0 & 0 \\ 0 & 0 & 0 & 0 & 0 \\ 0 & 0 & 0 & 0 & 0 \\ 0 & 0 & 0 & 0 & 1 \\ 0 & 0 & 0 & -1 & 0 \\\end{pmatrix}. \\
\end{array}
\end{equation}
The generators are normalized to 
\begin{equation}
  \label{eq:A:3}
  \langle t^{i}t^{j}\rangle = \delta_{ij}.
\end{equation}
Further relations involving the generators are discussed in \cite{Contino:2011np}. 
\chapter{Useful Relations for Operator Building}
\label{ch:appB}
\thispagestyle{fancyplain}
In order to reduce the NLO operators of Section~\ref{ch:chiralNLO} to a minimal set, we use the equations of motion, integration by parts, and some $SU(2)$ trace identities. Some of them were discussed in \cite{Nyffeler:1999ap,Grojean:2006nn,Buchalla:2012qq,Buchalla:2013rka}. Here, we list these relations. We consider the electroweak chiral Lagrangian at leading order, Eq.~\eqref{eq:LO}. The equation of motion of $h$ is
\begin{align}
  \begin{aligned}
    \label{eq:B.1}
  \Box h &= \frac{v^2}{4}\ \langle D_\mu U^\dagger D^\mu U\rangle \frac{\delta F_U(h)}{\delta h} - \frac{\delta V(h)}{\delta h}- \frac{1}{\sqrt{2}} \left[ \bar{q}_{L} \sum\limits^\infty_{n=1} n\,Y^{(n)}_u \left(\frac{h}{v}\right)^{n-1} U P_{+}q_{R}\right.\\
  &\left. + \bar{q}_{L} \sum\limits^\infty_{n=1} n\,Y^{(n)}_d \left(\frac{h}{v}\right)^{n-1} U P_{-}q_{R} + \bar{\ell}_{L}  \sum\limits^\infty_{n=1} n\,Y^{(n)}_e \left(\frac{h}{v}\right)^{n-1} U P_{-}\ell_{R} + \text{ h.c.}\right].
  \end{aligned}
\end{align}
The equations of motion of the fermions are
\begin{align}
  \begin{aligned}
    \label{eq:B.2}
    i \slashed{D} q_{L} &=\frac{v}{\sqrt{2}} \left[ \left( Y_u +\sum\limits^\infty_{n=1} Y^{(n)}_u \left(\frac{h}{v}\right)^n \right) U P_{+}q_{R} + \left( Y_d +  \sum\limits^\infty_{n=1} Y^{(n)}_d \left(\frac{h}{v}\right)^n \right) U P_{-}q_{R} \right] ,\\
    i \slashed{D} q_{R} &=\frac{v}{\sqrt{2}} \left[  P_{+}U^{\dagger}\left( Y_u^{\dagger} +\sum\limits^\infty_{n=1} Y^{(n)\dagger}_u \left(\frac{h}{v}\right)^n \right) q_{L} +  P_{-}U^{\dagger}\left( Y_d^{\dagger} +  \sum\limits^\infty_{n=1} Y^{(n)\dagger}_d \left(\frac{h}{v}\right)^n \right) q_{L} \right] ,\\
    i \slashed{D} \ell_{L} &=\frac{v}{\sqrt{2}} \left( Y_e +\sum\limits^\infty_{n=1} Y^{(n)}_e \left(\frac{h}{v}\right)^n \right) U P_{-}\ell_{R}  ,\\
    i \slashed{D} \ell_{R} &=\frac{v}{\sqrt{2}}  P_{-} U^{\dagger}\left( Y_e^{\dagger} +\sum\limits^\infty_{n=1} Y^{(n)\dagger}_e \left(\frac{h}{v}\right)^n \right) \ell_{L} .
  \end{aligned}
\end{align}
The equations of motion of the gauge fields are
\begin{align}
  \begin{aligned}
    \label{eq:B.3}
    \partial_{\mu}B^{\mu\nu}&= \sum\limits_{\Psi\in \{\text{fermions} \}}g'\bar{\Psi}\gamma^{\nu}\mathsf{Y}_{\Psi}\Psi + g' \frac{v^{2}}{2} \langle L^{\nu}\tau_{L}\rangle\left( 1+F_U(h)\right) ,\\
    [D_{\mu}W^{\mu\nu}]_{i} &=  \sum\limits_{\Psi\in \{\text{fermions} \}}g \bar{\Psi}_{L}\gamma^{\nu}T^{i}\Psi_{L}- g \frac{v^{2}}{2} \langle T^{i} L^{\nu}\rangle\left( 1+F_U(h)\right)  ,\\
    [D_{\mu}G^{\mu\nu}]_{A} &=  \sum\limits_{\Psi\in \{\text{quarks} \}}g_{s}\bar{\Psi}\gamma^{\nu}T^{A}\Psi.
  \end{aligned}
\end{align}
We find the equation of motion of the Goldstone bosons by requiring that the variation of the action vanishes, $\delta S = 0$, under the additional constraint $\det{U} = e^{\langle \ln{U}\rangle} = e^{\langle\tfrac{i}{v}\sigma^{a}\varphi^{a}\rangle}=1$. The variations $\delta U$ and $\delta U^{\dagger}$ are not independent,
\begin{equation}
  \label{eq:B.4}
  \delta U^{\dagger}_{ab} = - U^{\dagger}_{ac}\delta U_{cd}U^{\dagger}_{db}.
\end{equation}
The equation of motion of the Goldstone bosons is then \cite{mastersthesis}
\begin{align}
  \begin{aligned}
    \label{eq:B.5}
0&=\frac{v^2}{2}\  D_\mu\left(iL^{\mu} \left( 1+F_U(h)\right)\right)_{ab} \\
&- \frac{v}{\sqrt{2}} \left[ \bar{q}_{L,b} \left( Y_u +\sum\limits^\infty_{n=1} Y^{(n)}_u \left(\frac{h}{v}\right)^n \right) (U P_{+}q_{R})_{a}+ \bar{q}_{L,b} \left( Y_d +  \sum\limits^\infty_{n=1} Y^{(n)}_d \left(\frac{h}{v}\right)^n \right) (U P_{-}q_{R})_{a} \right.\\
 &\left. + \bar{\ell}_{L,b}  \left( Y_e +\sum\limits^\infty_{n=1} Y^{(n)}_e \left(\frac{h}{v}\right)^n \right) (U P_{-}\ell_{R})_{a} - \text{ h.c.}_{ba}\right]\\
&+\frac{v}{\sqrt{2}} \left[ \bar{q}_{L} \left( Y_u +\sum\limits^\infty_{n=1} Y^{(n)}_u \left(\frac{h}{v}\right)^n \right) U P_{+}q_{R} + \bar{q}_{L} \left( Y_d +  \sum\limits^\infty_{n=1} Y^{(n)}_d \left(\frac{h}{v}\right)^n \right) U P_{-}q_{R} \right.\\
  &\left. + \bar{\ell}_{L}  \left( Y_e +\sum\limits^\infty_{n=1} Y^{(n)}_e \left(\frac{h}{v}\right)^n \right) U P_{-}\ell_{R} - \text{ h.c.}\right]\frac{\delta_{ab}}{2}. 
  \end{aligned}
\end{align}
The equations of motion reduce to the equations of motion of the SM, if the polynomials of the Higgs couplings have the form of Eq.~\eqref{eq:5.5}. \\

The traces of two and three generators of $SU(2)$ are
\begin{equation}
  \label{eq:B.6}
  \langle T_{a}T_{b}\rangle =\frac{\delta_{ab}}{2} \qquad \text{and}\qquad \langle T_{a} T_{b}T_{c}\rangle = \frac{i}{4}\varepsilon_{abc}.
\end{equation}
We can express traces of higher numbers of generators through products of the two traces above. For example, consider
\begin{equation}
  \label{eq:B.7}
  \langle T_{a}T_{b}T_{c}T_{d}\rangle = \frac{1}{2} \left(\langle T_{a}T_{b}\rangle \langle T_{c}T_{d}\rangle - \langle T_{a}T_{c}\rangle\langle T_{d}T_{b}\rangle+\langle T_{a}T_{d}\rangle\langle T_{c}T_{b}\rangle\right).
\end{equation}
Further, we use
\begin{equation}
  \label{eq:B.8}
  \langle\tau_{L}AB\rangle \langle\tau_{L}C\rangle = \frac{1}{2} \langle A B C \rangle - \langle \tau_{L}B C\rangle\langle \tau_{L}A\rangle + \langle \tau_{L}AC\rangle\langle\tau_{L}B\rangle,
\end{equation}
with $A,B,C\in\{W_{\mu\nu},\tau_{L},L_{\mu} \}$. We reduce covariant derivatives that act on the building blocks $\tau_{L}$ and $L_{\mu}$ using  
\begin{align}
\begin{aligned}
  \label{eq:B.9}
  D_{\mu}L_{\nu}-D_{\nu}L_{\mu}&= g W_{\mu\nu} - g'B_{\mu\nu}\tau_{L} + i[L_{\mu},L_{\nu}],\\
  D_{\mu}\tau_{L} &= i[L_{\mu},\tau_{L}],\\
  [D_{\mu},D_{\nu}]L_{\rho} &= ig [W_{\mu\nu},L_{\rho}].
\end{aligned}
\end{align}
Operators containing the projectors $P_{12}$ and $P_{21}$ of Eq.~\eqref{eq:5.4} can also be written in terms of $\tau_{L}$ and $L_{\mu}$. We use
\begin{equation}
  \label{eq:B.10}
  2\langle \tau_{L} L_{\mu}L_{\nu}\rangle = \langle U P_{21} U^{\dagger} L_{\mu}\rangle \langle U P_{12} U^{\dagger} L_{\nu}\rangle -  \langle U P_{12} U^{\dagger} L_{\mu}\rangle \langle U P_{21} U^{\dagger} L_{\nu}\rangle
\end{equation}
and
\begin{equation}
  \label{eq:B.11}
  \langle L_{\mu}L_{\nu}\rangle = \langle U P_{21} U^{\dagger} L_{\mu}\rangle \langle U P_{12} U^{\dagger} L_{\nu}\rangle +  \langle U P_{12} U^{\dagger} L_{\mu}\rangle \langle U P_{21} U^{\dagger} L_{\nu}\rangle + 2 \langle \tau_{L}L_{\mu}\rangle \langle \tau_{L}L_{\nu}\rangle.
\end{equation}

\let\chapterheadstartvskip\backupskip
\let\chapterheadendvskip\backupskipp

\cleardoublepage\makeatletter\@openrighttrue\makeatother 
  
  \bibliography{Kapitel/Literatur/literature}{\protect\thispagestyle{fancyplain}}
  \bibliographystyle{unsrturl}
  
  \markboth{}{}

  \chapter*{Acknowledgment --- Danksagung}
\addcontentsline{toc}{chapter}{\protect Acknowledgment / Danksagung}
\thispagestyle{fancyplain}

I would like to thank everyone, who supported me in the past years. First of all, I would like to thank Gerhard Buchalla for being a great supervisor and mentor. Thank you for answering all my questions and giving me advice! Many thanks to Gerhard, Oscar and Alejandro for the great collaboration. \\

Ein ganz gro{\ss}er Dank gilt meiner bezaubernden Frau, Anja, die immer f{\"u}r mich da ist. ``Trust I seek and I find in you; Every day for us something new; Open mind for a different view; And nothing else matters!''  Wo du hingehst, da will ich auch hingehen; wo du bleibst, da bleibe ich auch. Bald sind wir wieder vereint. 

Vielen Dank auch meiner ganzen Familie, besonders meinen Eltern f{\"u}r die Unterst{\"u}tzung, die ich in den vergangenen Jahren erfahren durfte. Danke, dass ich meine Tr{\"a}ume verfolgen durfte. \\

Thank you very much to all my friends, inside and outside of the world of physics, for the joy, fun, support, and inspiration we shared. Special thanks to all the postdocs as well as Ph.D. and Master's students of our chair. Thank you Oscar, Alejandro, Daniel F., Tehseen, Sebastian, Lukas G., Alex P., Alex G., Leila, Nico, Sarah, Jaba, Deb, Alexis, and my office mates of the past years: Rudi, Andr{\'e}, Valentino, Nina, Mischa, Lukas E., Lena, Andrea, and Kepa. I am very happy we shared so many ideas and I got to know all of you. It was a pleasure to have tea time with you.  

I thank Gerhard, Oscar, Alejandro, Anja, Daniel K., Tehseen, Rudi, and Daniel F. for carefully reading my manuscript. 

I am grateful for fruitful and interesting exchanges at the Ph.D. Schools ``Schladming Winter School 2014'' and the ``Herbstschule f{\"u}r Hochenergiephysik'' in Maria Laach, 2015. Thank you for the many discussions and meetings of the LHCHXSWG and the HEFT workshops. Thanks to Zhengkang `Kevin' Zhang for the many discussions we had on Skype and in person. 

I am grateful to Gerhard Buchalla, Alejandro Ibarra, Otmar Biebel, Joachim R{\"a}dler, Thomas Kuhr, and Ilka Brunner for being part of my Ph.D. committee. 

I would like to thank Frank Steffen and the International-Max-Planck-Research-School (IMPRS) for particle physics for the many interesting seminars, block courses and the support during my time as Ph.D. student. I really enjoyed our Young Scientist Workshops at castle Ringberg. I thank our secretaries Gabi Bodenm{\"u}ller and Herta Wiesbeck-Yonis for their support in all administrative questions I had. 

I would also like to thank John Williams, Howard Shore, James Horner, and Hans Zimmer for composing epic movie themes that make working a real pleasure. Special thanks to the Roland-Berger-Foundation for the many seminars and activities during the scholarship and the alumni program that I was able to experience. Thank you to Stephan Paul and Daniel Greenwald for the interesting and useful lecture ``English writing for Physics''. \\

I thank the cluster of excellence ``Origin and Structure of the Universe'' and the DFG under grant BU 1391/2-1 for the funding of my Ph.D. position.
\cleardoublepage
\thispagestyle{empty}
$ $

\end{document}